\documentclass[aps,onecolumn,superscriptaddress,nofootinbib,noshowpacs,longbibliography,preprintnumbers,floatfix]{revtex4-2}
\pdfoutput=1
\usepackage{amsmath}
\usepackage{amssymb}
\usepackage{verbatim}
\usepackage{graphicx}
\usepackage{color}

\definecolor{mymagenta}{RGB}{200, 0, 100}
\definecolor{myblue}{RGB}{45, 48, 146}

\usepackage{ragged2e}
\usepackage{wrapfig}
\usepackage{mathtools}
\usepackage{bbold}
\usepackage{blkarray}
\usepackage[inline]{enumitem}
\usepackage{latexsym}
\usepackage[pdftex,
            colorlinks,
            linkcolor=myblue,
            citecolor=mymagenta,
            urlcolor=myblue]{hyperref}
\usepackage{physics}
\usepackage{qcircuit}
\usepackage{bbm}
\usepackage{float}
\usepackage{longtable}
\usepackage{natbib}
\usepackage{url}

\allowdisplaybreaks

\begin{document}


\title{Lattice Quantum Chromodynamics and Electrodynamics on a Universal Quantum Computer}

\author{Angus Kan}

\affiliation{Institute for Quantum Computing and Department of Physics \& Astronomy, University of Waterloo, Waterloo, Ontario, Canada, N2L 3G1}

\author{Yunseong Nam}

\affiliation{IonQ, Inc., College Park, MD, USA, 20740}
\affiliation{Department of Physics, University of Maryland, College Park, MD, USA, 20742}

\begin{abstract}
It is widely anticipated that a large-scale quantum computer will offer an evermore accurate simulation of nature, opening the floodgates for exciting scientific breakthroughs and technological innovations. Here, we show a complete, instruction-by-instruction rubric to simulate U(1), SU(2), and SU(3) lattice gauge theories on a quantum computer. These theories describe quantum electrodynamics and chromodynamics, the key ingredients that form the fabric of our universe. We further provide a concrete estimate of the quantum computational resources required for an accurate simulation of lattice gauge theories using a second-order product formula. We show that lattice gauge theories in any spatial dimension can be simulated using $\tilde{O}(T^{3/2}N^{3/2}\Lambda/\epsilon^{1/2})$ T gates, where $N$ is the number of lattice sites, $\Lambda$ is the bosonic gauge field truncation, and $T$ is the simulation time.
\end{abstract}
 
\maketitle

\section{Introduction}
\label{sec:Intro}

Development of computers has always directly benefited the advancement of science. Notable examples in the early days of classical computing include the famous Fermi-Pasta-Ulam-Tsingou problem \cite{fermi1955studies}, wherein the ergodic hypothesis was tested by the means of numerical simulations,
and the Metropolis algorithm \cite{metropolis1953equation}, which was employed to perform Monte Carlo simulations of many-body problems.
They went on to become cornerstones of today's technologies, from advancing our capability to better predict weather events and model financial markets to accelerating scientific discovery and design processes in chemistry or materials design. It thus stands to reason that quantum computing, as an emerging computing paradigm, will lead to an era of unexplored, uncharted field of science.

In this paper, we connect two disparate fields, particle physics and quantum computing. In particular, we provide a complete layout of computational instructions at a gate-by-gate level to be run on a quantum computer, to efficiently simulate quantum electrodynamics (QED) and quantum chromodynamics (QCD), in the hopes that it will serve as a stepping stone to future scientific and technological developments. Traditionally, Markov chain Monte Carlo methods are used to simulate QED and QCD on a classical computer. However, despite their tremendous success~\cite{aoki2020flag}, due to the infamous sign problem~\cite{troyer2005computational}, many interesting phenomena such as real-time dynamics are inaccessible classically. Envisioned by Feynman, quantum computers promise to simulate quantum dynamics efficiently \cite{feynman1982simulating}. This motivates the use of quantum computers to simulate particle physics.
Here, we consider U(1), SU(2) and SU(3) lattice gauge theories (LGTs) -- described below, -- based on the fact that (i) the Standard Model (SM) of particle physics is a gauge theory with the symmetry group U(1)$\times$ SU(2)$\times$SU(3) and (ii) LGTs~\cite{wilson1974confinement} are one of the best-known non-perturbative first-principle computational methods for QED and QCD. 

Useful for the foregoing discussion, LGTs may be briefly illustrated as follows. First, we coarse-grain a $d$-dimensional space into a $d$-dimensional lattice. Lattice sites are occupied by fermions or anti-fermions. A link that connects two sites is occupied by the force carriers or the gauge bosons. 
The symmetry group of one's choice determines the types of fermions and bosons, and the physical theory considered. For instance, QED, the theory of the interactions between electric charges and photons, has a U(1) symmetry. In the SM, the U(1)$\times$SU(2) symmetry and SU(3) symmetry correspond to the electroweak theory~\cite{salam1959weak,glashow1959renormalizability,weinberg1967model} and QCD~\cite{ne1961derivation,gell1962symmetries,gell1964schematic}, respectively.
We remark that how to simulate chiral gauge theories such as the electroweak theory on a lattice is a longstanding open problem~\cite{smit2002introduction,preskill2018simulating}. Regardless, U(1) and SU(2) LGTs are simulated separately to model QED~\cite{smit2002introduction} and provide valuable insights to QCD~\cite{feynman1981qualitative}. For the above reasons, we choose to consider quantum simulation of U(1), SU(2), and SU(3) LGTs.

To this end, our main contributions are
\begin{itemize}
\item Explicit, gate-by-gate level construction of quantum circuits that simulate real-time dynamics of U(1), SU(2), and SU(3) LGTs, fully incorporating both fermions and gauge bosons, in an arbitrary spatial dimension $d$.
\item Rigorous upperbounds on the quantum computational resources (quantum gate counts), including quantum-simulation algorithmic errors (second-order Trotter)~\cite{childs2019theory}, circuit-synthesis errors~\cite{bocharov2015efficient}, and quantum arithmetic errors~\cite{bhaskar2016quantum}, relevant for the fault-tolerant regime.
\item Efficient use of quantum computational space and time: Specifically,
\begin{itemize}
\item Exponential reductions in the space requirement (qubit counts), determined by the largest bosonic quantum numbers simulated, for the SU(2) and SU(3) LGTs, by use of a binary encoding of the bosonic quantum numbers (see table~\ref{tab:prior-art}),
\item Superpolynomial savings in the gate counts per time step in quantum simulation by an efficient use of quantum fixed-point arithmetic operations (see table~\ref{tab:yamamoto}).
\end{itemize}
\end{itemize}
We note in passing that there is a wealth of literature related to quantum simulation of LGTs, including analog, variational, or gate-based digital quantum simulations, to name a few. We refer the readers to Supplementary Material (Supp. Mat.) Sec.~\ref{sec:related} for a brief review.
Interested readers are further encouraged to consider~\cite{banuls2020simulating,klco2021standard} for a comprehensive overview.

Our paper is structured as follows. In Sec.~\ref{sec:LGT}, we provide a more comprehensive overview of LGTs, laying out clearly the quantum simulation problem of our interest, i.e., the standard Kogut-Susskind Hamiltonian~\cite{kogut1975hamiltonian}. In Sec.~\ref{sec:Results}, we report our results, detailing our findings on the quantum resource requirements for simulating U(1), SU(2), and SU(3) LGTs on a quantum computer, while also providing a careful comparison to prior art and examples. In Sec.~\ref{sec:Discussion}, we discuss our findings and future work, and in Sec.~\ref{sec:Outlook}, we provide an outlook for quantum simulations of LGTs.

\section{Lattice Gauge Theories}
\label{sec:LGT}

We consider a universe, coarse-grained to form a lattice, capable of hosting fermions and bosons at lattice sites and links, respectively. LGTs then capture the excitation of both the fermions and gauge bosons. Fermionic excitation can readily be encoded in the usual occupation number basis. The basis for bosonic excitation is however carefully chosen based on the considered symmetry group. For instance, in a U(1) LGT, each bosonic link is described by the angular momentum basis of a quantum particle constrained to move along a circle, also known as a planar rotor, with an integer-valued quantum number. The quantum number represents the angular momentum where its sign denotes the two directions of rotation. In a SU(2) LGT, each gauge boson can be interpreted as a rigid rotator, which can be represented by an angular momentum basis with three quantum numbers: total angular momentum, the angular momenta in the space-fixed and body-fixed frames of reference~\cite{kogut1975hamiltonian}. The SU(N) ($\text{N}>2$) generalization of this angular momentum basis is labelled by $\text{N}^2 -1$ (the dimension of SU(N)) quantum numbers~\cite{byrnes2006simulating,zohar2015formulation}. Since the quantum number(s) are unbounded, they need to be encoded into a finite number of qubits on a quantum computer. We truncate them such that each quantum number can only take on $\sim \Lambda$ values.\footnote{In this paper, we considered $\Lambda$ as a parameter in the Hamiltonian. After the completion of this paper, it has been shown in~\cite{tong2021provably} that $\Lambda = \mbox{polylog}(\epsilon^{-1})$, where $\epsilon$ includes errors introduced by the truncation of the gauge fields.}

In the following, we review the standard, Kogut-Susskind Hamiltonian formulation of lattice gauge theories \cite{kogut1975hamiltonian}. The Kogut-Susskind Hamiltonian is defined as 
\begin{equation}
    \hat{H} = \hat{H}_{gauge}+\hat{H}_{matter},
\label{eq:Hamiltonian}    
\end{equation}
where $\hat{H}_{gauge}$ describes the dynamics of the electric and magnetic gauge fields (bosons), and $\hat{H}_{matter}$ describes the dynamics that involve the fermionic matter. The Hamiltonian is defined on a $d$-dimensional spatial lattice, and the temporal direction is continuous. A site on the lattice is denoted by a vector $\vec{n} = \sum_{i=1}^{d} n_i \hat{i}$, where $\hat{i}$ is a unit lattice vector pointing in one of the orthogonal directions. A site is labelled even or odd, depending on if $(-1)^{\vec{n}} = (-1)^{\sum_{i}n_{i}}$ evaluates to 0 or 1, respectively. A link between neighboring sites is denoted by a tuple $(\vec{n},l)$ of the starting site $\vec{n}$ and its direction $\hat{l}$. On the lattice, the fermions and anti-fermions reside on the sites, while the gauge fields, which mediate the interaction between them, occupy the links.

The matter Hamiltonian consists of two terms, mass $\hat{H}_M$ and kinetic $\hat{H}_K$ Hamiltonians defined by~\cite{kogut1975hamiltonian}
\begin{align}
    \hat{H}_{matter} &= \hat{H}_M + \hat{H}_K,\:\text{where} \nonumber \\
    \hat{H}_M &= m\sum_{\vec{n}} \sum_{\alpha} (-1)^{\vec{n}} \hat{\psi}^{\dag}_{\alpha}(\vec{n})\hat{\psi}_{\alpha}(\vec{n}), \nonumber \\
    \hat{H}_K &= \frac{1}{2a} \sum_{\vec{n}} \sum_{l} \sum_{\alpha,\beta}(\hat{\psi}_\alpha (\vec{n})^{\dag}\hat{U}_{\alpha\beta}(\vec{n},l)\hat{\psi}_\beta(\vec{n}+\hat{l}) + \hat{\psi}_\alpha (\vec{n})\hat{U}^{\dag}_{\alpha\beta}(\vec{n},l)\hat{\psi}^{\dag}_\beta(\vec{n}+\hat{l})),
    \label{eq:Hmatter}
\end{align}
where $m$ is the fermionic mass, $a$ is the lattice spacing, $\hat{\psi}_\alpha(\vec{n}), \hat{\psi}^{\dag}_\alpha(\vec{n})$ are vectors of the fermionic annihilation and creation operators, respectively, at site $\vec{n}$ with $\alpha$ labelling different fermionic species, and $\hat{U}_{\alpha\beta}(\vec{n},l),\hat{U}_{\alpha\beta}^\dag(\vec{n},l)$ are matrices known as parallel transporters \cite{smit2002introduction}, of which the elements are gauge field operators that couple fermion-anti-fermion pairs, for instance, a fermion and an anti-fermion of species $\alpha$ and $\beta$ that occupy sites $\vec{n}$ and $\vec{n}+\hat{l}$. The alternating sign $(-1)^{\vec{n}}$ reflects the use of the staggered fermions \cite{kogut1975hamiltonian}. Physically, we can interpret creating (destroying) a particle at an even (odd) site as creating a fermion (anti-fermion). $\hat{H}_M$ computes the mass of all fermionic matter by multiplying the number of fermions and anti-fermions in the lattice by $m$. As such, $\hat{H}_M$ governs the dynamics of free fermions and anti-fermions in the absence of gauge fields. $\hat{H}_K$ describes the dynamics of fermion-anti-fermion pair creation and annihilation, and the corresponding changes in the mediating gauge fields.

As an explicit example, consider the Abelian U(1) LGT. This theory describes quantum electrodynamics (QED) on a lattice, each site can either be occupied by an electron or a positron due to the Pauli exclusion principle. As such, there is only one component each in the vectors $\hat{\psi}_\alpha, \hat{\psi}^{\dag}_\alpha$, and hence, one component each in the parallel transporter matrices. In electrodynamics, Gauss' law implies that point charges emanate electric fluxes. In order to satisfy Gauss' law on a lattice, creating or destroying an electron-positron pair will necessarily generate or remove an electric flux between them. The electric fluxes are raised and lowered by the parallel transporters. 

In SU(N) LGTs, the gauge fields are non-Abelian, and the vectors of fermionic operators have N components and the parallel transporters are ${\rm N}\times {\rm N}$ matrices. As opposed to only one type of electric charge in QED, there are N different species of fermions carrying N types of charges known as colors. On a lattice, the N species are represented by N-component vectors $\hat{\psi}_\alpha, \hat{\psi}^{\dag}_\alpha$. Similar to QED, SU(N) LGTs allow for pair creation and annihilation. However, there are ${\rm N}^2$ different types of fermion-anti-fermion pair creation or annihilation, to account for all possible combinations of colors. As a consequence of the non-Abelian Gauss' law, there are ${\rm N}^2$ different types of electric fluxes that can be generated in or removed from the non-Abelian chromoelectric field. Those changes in the chromoelectric field are induced by the ${\rm N}^2$ elements of the parallel transporters.

In the absence of matter, the Kogut-Susskind Hamiltonian is reduced to a pure gauge Hamiltonian, which consists of an electric $\hat{H}_E$ and magnetic $\hat{H}_B$ Hamiltonian given by~\cite{kogut1975hamiltonian}
\begin{align}
    \hat{H}_{gauge} &= \hat{H}_E + \hat{H}_B,\; {\rm where} \nonumber \\
    \hat{H}_E &= \frac{g^2}{2a^{d-2}} \sum_{\vec{n},l}\sum_{b} [\hat{E}^b(\vec{n},l)]^2, \nonumber \\
    \hat{H}_B &= -\frac{1}{2a^{4-d}g^2}\sum_{\Box} {\rm Tr}[\hat{P}_{\Box} + \hat{P}_{\Box}^{\dag}],
\label{eq:gauge}
\end{align}
where $g$ denotes the bare coupling strength, ${\rm Tr}$ denotes a trace operator, and $\Box$ represents the elementary square cells, called plaquettes, of a lattice. The magnetic Hamiltonian $\hat{H}_{B}$ is formed from products of the parallel transporters around each plaquette, called plaquette operators
\begin{equation}
    {\rm Tr}[\hat{P}_{\Box}] = \sum_{\alpha, \beta, \gamma, \delta}\hat{U}_{\alpha \beta}(\vec{n},i)\hat{U}_{\beta \gamma}(\vec{n}+\hat{i},j)\hat{U}_{\gamma \delta}^{\dag}(\vec{n}+\hat{j},i)\hat{U}_{\delta \alpha}^{\dag}(\vec{n},j),
\end{equation}
traced over the matrix elements. In particular, the plaquette operators create loops of electric fluxes surrounding the plaquettes, and generate magnetic fluxes. The electric flux at each link is measured by a vector of electric field operators $\hat{E}^b$, where the number of components $b$ is the same as the number of generators of the considered gauge group. Moreover, depending on the gauge group, the electric field operators and parallel transporters satisfy a specific set of commutation relations.

Once again, considering U(1) LGT as an instance, the gauge fields are the usual Maxwell electric and magnetic fields.
In the case of SU(N) LGTs, the gauge fields are non-Abelian chromoelectric and chromomagnetic fields, which are considerably more complicated. Let us use SU(3) LGT, which describes QCD, the theory of strong interaction, on a lattice as an example. Here, the non-Abelian electric fluxes are excitations of the underlying gauge field, known as the gluon field, which is eponymously named after the gauge boson for the strong force. Gluons are analogous to photons in QED. However, an important distinction between them is that gluons can self-interact, due to their non-Abelian nature, and may form composite particles, called glueballs, among themselves. This self-interacting nature makes QCD significantly more difficult to analyze than QED. We note that, SU(3) LGT is perhaps the most interesting one in the context of particle physics, due to its relation to QCD. Therefore, we choose to investigate up to and including SU(3) LGT simulation on a quantum computer in detail. A straightforward generalization of our simulation methods may be done for ${\rm N} > 3$.

\section{Results}
\label{sec:Results}

\subsection{Gate complexity}
\label{sec:gate}

We employ the second-order Suzuki-Trotter formula~\cite{suzuki1991general} to implement the evolution operator $e^{i\hat{H}T}$, where $\hat{H}$ is the Hamiltonian in (\ref{eq:Hamiltonian}) and $T$ is the total evolution time. In particular, according to the formula, we approximate
\begin{equation}
e^{i\hat{H}T} = (e^{i\hat{H}T/r})^r \approx \left\{\left[\prod_{s=1}^{N_{s}} \hat{H}_s T/(2r)\right]\left[\prod_{s=N_s}^{1} \hat{H}_s T/(2r)\right]\right\}^r,
\end{equation}
where $\hat{H} = \sum_{s} \hat{H}_s$, $N_s$ is the number of Hamiltonian subterms $\hat{H}_s$ used for Trotter formula, and the number of Trotter steps $r$ is chosen so that $T/r$ is small. The ordering of all individual $\hat{H}_s$ we used follows the ordering of mass, electric, kinetic, and magnetic Hamiltonians, up to subdivisions of each Hamiltonian into multiple subterms if applicable. This choice is used for all the different symmetries, U(1), SU(2), and SU(3) LGTs, we considered.
We chose the second-order formula based on \cite{shaw2020quantum}, wherein it has been shown that the second-order formula achieves a quadratically better gate complexity 
in the truncation parameter $\Lambda$, when compared to other algorithms, such as quantum signal processing \cite{low2017optimal}. 

We next describe the way qubits are used in our simulations. For SU(N) LGTs, we use N qubits per lattice site to encode the occupation by the N different fermions or anti-fermions, each with a distinct color charge. For each lattice link, where we encode the quantum numbers of the gauge-field bosons, we use approximately $({\rm N}^2-1)\log_2(\Lambda)$ qubits, with the logarithmic dependence arising from the binary encoding of the quantum numbers, i.e., a binary number $b$ is used to denote the quantum number $b$ in our quantum computer. This may be contrasted to an unary encoding, where $\sim\Lambda$ many qubits would be used to encode the quantum number, with $b$'th qubit in the state of $|1\rangle$ and the rest in the state of $|0\rangle$ corresponding to the number $b$.

With the particular encoding chosen above, each Trotter term $\exp[i\hat{H}_s T/(2r)]$ can be implemented as follows: 
\begin{itemize}
\item {\it Mass term} -- The mass term is implemented by an application of $R_z$ rotations applied to the lattice-site qubit registers, since it induces phases based on the occupation of the fermions or anti-fermions. \item {\it Electric term} -- In the angular momentum bases described above, the electric term is diagonal and thus, is implemented in a similar fashion to the mass term for the link qubit registers, except we precompute, using quantum integer arithmetic operations, the eigenvalue expression of the electric Hamiltonian operator. This way, for the eigenvalue expression computed, one can use $R_z$ rotations to induce and accumulate appropriate phases implied by the electric term. In some parameter regimes, we provide an alternative method that can more efficiently induce the phases using an adder circuit instead of $R_z$ rotations~\cite{gidney2018halving}. 
\item {\it Kinetic term} -- For the kinetic term, we simultaneously operate over a neighboring fermion--anti-fermion pair and the bosonic link that connects the pair. Further, we observe that the distance between the bosonic quantum numbers that interact is limited by the fermion-boson interactions. As such, the kinetic term can always be written as a multiply-banded matrix, where the bandwidth of the matrix scales as $\sim 2^{{\rm N}^2 -1}$ in SU(N) LGTs. We then decompose the matrix into a sum of a number, constant with respect to the truncation parameter $\Lambda$, of matrices that individually encode interactions within a selected subspace. The decomposition is chosen carefully such that each constituent matrix appears identical to one another in the non-zero element locations, but differs by some constant shifts in the quantum numbers, accessing different parts of the bands. For a decomposed, constituent matrix, we use a CNOT-and-Hadamard network to efficiently diagonalize the evolution implied by it.
This makes the simulation problem into a diagonal phase oracle construction problem. 
We implement each phase oracle by first computing the state-dependent phases with quantum fixed-point arithmetics~\cite{bhaskar2016quantum} and then inducing the appropriate phases with a layer of $R_z$ gates, similar to the electric term.
\item {\it Manetic term} -- As for the magnetic term, essentially the same approach is used as in the kinetic case, except, here, we drop the fermion-anti-fermion parts and consider four links that form a plaquette at the same time. The bandwidth is $\sim 2^{4({\rm N}^2 -1)}$. 
\end{itemize}

Contrasted to SU(N) LGTs, U(1) LGTs can be more easily simulated because the U(1) kinetic and magnetic terms impart phases that depend trivially on the input state. As a result, the phase oracles can be executed with only $R_z$ rotations and without the use of fixed-point arithmetics. This is reflected in the separation between the gate-complexities of U(1), and SU(2) and SU(3) LGT simulations, as shown in Eqs.~(\ref{eq:comp_1}) and (\ref{eq:comp_2}) down below. We in fact propose an alternative approach for the U(1) LGTs, where we use the quantum Fourier transform to diagonalize (the bosonic part of) the U(1) kinetic term and magnetic term. Then, the phase oracles needed to induce the evolution implied by the two terms can be efficiently implemented using quantum signal processing techniques~\cite{low2017optimal}. 

We direct the readers to various Supp. Mat. sections for the exact division of $\hat{H}$ into the sum of $\hat{H}_s$ for any $d$-dimensional U(1), SU(2), and SU(3) LGTs. In particular, we provide the full details of the chosen bases for the U(1), SU(2), and SU(3) bosonic gauge fields, and their relations to the corresponding gauge field operators in Supp. Mat. Secs. \ref{sec:U1_prelim}, \ref{sec:SU2_prelim}, and \ref{sec:SU3_prelim}, respectively. Furthermore, we describe the circuit implementations of the U(1) mass, electric, kinetic, and magnetic subevolutions in Supp. Mat. Secs. \ref{sec:U1_mass}, \ref{sec:U1_electric}, \ref{sec:U1_kin}, and \ref{sec:U1_mag}, respectively. In Supp. Mat. Secs. \ref{sec:SU2_mass}, \ref{sec:SU2_electric}, \ref{sec:SU2_kin}, and \ref{sec:SU2_mag}, we lay out the circuit implementations of the SU(2) mass, electric, kinetic, and magnetic subevolutions, respectively. Finally, the circuit implementations of the SU(3) mass, electric, kinetic, and magnetic subevolutions are discussed in Supp. Mat. Secs. \ref{sec:SU3_mass}, \ref{sec:SU3_electric}, \ref{sec:SU3_kin}, and \ref{sec:SU3_mag}, respectively.

As to the simulation errors, we allocate a total error budget of $\epsilon$ for the entire simulation. In our circuit synthesis, we evenly split this budget into two halves for the U(1) LGT case, spending each on the algorithmic (Trotter) error and the $R_z$ gate synthesis error (incurred for the approximation using single-qubit Clifford and T gates). For SU(2) and SU(3) LGTs, we split the budget three ways, two of which are used on the algorithmic and $R_z$ synthesis errors as in U(1) case. The third is used on fixed-point quantum arithmetic operations. 

Whenever possible, we carefully choose the subdivision of the Trotter term implementations over the lattice topology to maximize parallelism. This includes considering different spatial dimensions and choosing different lattice sites, links to include in one time step.
A careful consideration here comes with two benefits, in addition to the obvious circuit depth or execution time reduction. One is the applicability of efficient quantum circuit constructions. More specifically, we employ the weight-sum trick, reported in~\cite{gidney2018halving,nam2019low}, to synthesize a layer of same-angle $R_z$ gates in parallel. This leads to an exponential reduction in $R_z$-gate count at the cost of a modest increase in T-gate and ancilla-qubit counts. Therefore, we take a full advantage of this trick by arranging as many same-angle $R_z$ gates as possible into layers across the entire lattice. This optimization is performed for the mass and electric terms, as well as the kinetic and magnetic terms in the U(1) case.
The other is the ability to enable streamlined Trotter error analysis. For instance, terms applied in parallel commute with one another, and thus, the commutators to be evaluated for the Trotter error can be more tightly bounded by considering Trotter term collisions at the lattice topology level.

Note the system to be simulated on a quantum computer is a $d$-dimensional lattice with periodic boundary condition, with $L$ lattice points along each dimension. We therefore study the gate complexity in the electric (bosonic) truncation $\Lambda$, the total error budget $\epsilon$, and $L$, assuming the dimension $d$ and the Hamiltonian parameters $g$, $a$, and $m$ [see Eqs.~(\ref{eq:Hmatter}) and (\ref{eq:gauge})] are fixed. 
We use a standard library of controlled-NOT, single-qubit Clifford, and T gates as our basis gate set and consider T gates as our metric of resource requirement as they are the most expensive operations in a fault-tolerant quantum computer and thus are widely used to be a good proxy for the quantum computational resource requirement in the fault-tolerant regime. We obtain simulations that are efficient in the simulation time $T$, number of lattice sites $N=L^d$, $\Lambda$ and $\epsilon$. In particular, the gate complexities for the three considered LGTs are (see Supp. Mat. Secs.~\ref{sec:U1_res}, \ref{sec:SU2_res}, and \ref{sec:SU3_res} for complete, detailed derivations, including the Trotter, fixed-point arithmetic, and $R_z$ gate synthesis errors, for SU(1), SU(2), and SU(3), respectively)

\begin{align}
\text{U(1) complexity: }&O\Bigg(\frac{T^{3/2}d\Lambda L^{d/2}}{\epsilon^{1/2}}\Big[d(\log(\Lambda))^2L^d +\log(\Lambda)\log(dL^d) {\mathcal C}\Big]\Bigg), \nonumber \\
&{\rm where}\,\,{\mathcal C} = \log\left(\frac{T^{3/2}d\Lambda L^{d/2}\log(\Lambda)\log(dL^d)}{\epsilon^{3/2}}\right),
\label{eq:comp_1}
\end{align}
and
\begin{align}
\text{SU(2) or SU(3) complexity: }&O\Bigg(\frac{T^{3/2}d\Lambda L^{d/2}}{\epsilon^{1/2}}\Big[d^2L^d\mathcal{K}^2\log(\mathcal{K}) +\log(\Lambda)\log(dL^d) {\mathcal C}\Big]\Bigg), \nonumber \\
&{\rm where}\,\,{\mathcal{K}} = \log(\frac{T^{3/2}d^3\Lambda L^{3d/2}}{\epsilon^{3/2}})
+\log\log(\frac{T^{3/2}d^3\Lambda L^{3d/2}}{\epsilon^{3/2}}) \nonumber\\
&{\rm and}\,\, {\mathcal C} = \log\left(\frac{T^{3/2}d^3\Lambda L^{3d/2}\mathcal{K}}{\epsilon^{3/2}}\right).
\label{eq:comp_2}
\end{align}

\subsection{Comparison to prior art}

Table~\ref{tab:prior-art} shows the comparison between our work and the state of the art reported in the literature. In particular we compare our work with those reported in Refs.~\cite{byrnes2006simulating} and \cite{shaw2020quantum}, since they both simulate LGTs over unitary groups on a gate-based, universal quantum computer. Our work is of the most general and complete kind to date in that we
\begin{enumerate}
\item explore U(1) and SU($N$) groups with $N=2,3$ being explicitly worked out,
\item consider a lattice in an arbitrary dimension $d$, 
\item include the complete Kogut-Susskind Hamiltonian, i.e., both fermionic (mass, kinetic) and bosonic (electric, magnetic) terms,
\item use an efficient, binary gauge field encoding, and 
\item work out the quantum-gate-by-quantum-gate construction of the entire simulation with a full visibility into the gate complexity.
\end{enumerate}

In~\cite{byrnes2006simulating}, methods to simulate U(1), SU(2) and SU(3) LGTs without fermions on a universal quantum computer were proposed.
The work reported therein also lacks a rigorous analysis of the required gate or qubit counts. Therefore, we analyze the T-gate complexities according to the methods proposed, adapting them to the fault-tolerant setting (see Supp. Mat. Sec.~\ref{sec:yamamoto} for details). 
A comparison between the gate complexities of our work and \cite{byrnes2006simulating} is summarized in Table~\ref{tab:yamamoto}.
We briefly describe the main methods of~\cite{byrnes2006simulating} for completeness.
First, the bosonic quantum numbers are represented in unary encoding, which requires exponentially more logical qubits than our binary encoding. Then, both the electric and magnetic Hamiltonian operators are first expanded element-wise into linear combinations of Pauli operators.\footnote{Each element in a diagonal operator, i.e., $e_{j,j}\ketbra{j}{j}$, and each pair of elements in the off-diagonal magnetic operator, i.e., $e_{j,k}\ketbra{j}{k}+ h.c.$ can be expanded bit-wise into $e_{j,j} \otimes_{i} \ketbra{j_i}$ and $e_{j,k} \otimes_{i} \ketbra{j_i}{k_i} + h.c.$, respectively, where $j_i, k_i$ are the $i$th bits of $j,k$. Using the relations $\ketbra{0} = (\hat{I}+\hat{\sigma^z})/2$, $\ketbra{1} = (\hat{I}-\hat{\sigma^z})/2$, $\ketbra{0}{1} = \hat{\sigma}^-$ and $\ketbra{1}{0} = \hat{\sigma}^+$, $e_{j,j} \otimes_{i} \ketbra{j_i}$ and $e_{j,k} \otimes_{i} \ketbra{j_i}{k_i} + h.c.$ can be expressed as a linear combination of the identity and Pauli-$z$ operators, and that of the identity and Pauli ladder operators, respectively.}. As such, for every simulation time (Trotter) step, using well-known circuit templates for Pauli-evolution operators $e^{it \otimes_k \hat{\sigma}^z_k}$ and $e^{it\otimes_i \hat{\sigma}^\pm_i + h.c.}$, our analysis shows that the electric evolution requires $\tilde{O}(\Lambda)$ T gates for U(1) and SU(2) LGTs, and $\tilde{O}(\Lambda^2)$ T gates for SU(3) LGTs, whereas the magnetic evolution requires $\tilde{O}(\Lambda^4)$, $\tilde{O}(\Lambda^{12})$, and $\tilde{O}(\Lambda^{32})$ T gates for U(1), SU(2), and SU(3) LGTs, respectively.
While the polynomial T-gate complexities are technically efficient, quantum simulations, particularly of SU(2) and SU(3) LGTs, will likely be prohibited in practice by the large degrees of the polynomials.
In comparison, our implementation requires at most $\tilde{O}(\log^2(\Lambda))$ T gates per time step for both electric and magnetic terms. 
This superpolynomial improvement (Table~\ref{tab:yamamoto}) can be attributed to an efficient use of integer and fixed-point arithmetic circuits for the electric and magnetic terms, respectively. Furthermore, for the magnetic term, in contrast to the element-wise approach here, our approach requires only a constant number of queries, with respect to $\Lambda$, to the state-dependent phase oracles (see section~\ref{sec:gate}). 

To compare with~\cite{shaw2020quantum}, we apply our methods to the one-dimensional U(1) LGT, and demonstrate that we achieve a better gate complexity for all terms (see Supp. Mat. Sec.~\ref{sec:U1compare} for details), which can be attributed to the $R_z$ gate optimization discussed in section~\ref{sec:gate}.

\begin{table}[ht]
    \centering
    \begin{tabular}{ c | c | c  | c|}
        { } &  \: \: \cite{byrnes2006simulating} \:\:  & \:\: \cite{shaw2020quantum} \: \:  & \:\ Ours \: \\
        \hline 
        Groups & \begin{tabular}{@{}c@{}} U(1),SU(2),SU(3) \\ Extensible to SU($N$) \end{tabular} & U(1) & \begin{tabular}{@{}c@{}} U(1),SU(2),SU(3) \\ Extensible to SU($N$) \end{tabular} \\
         \hline 
        \begin{tabular}{@{}c@{}} Lattice \\ dimension \end{tabular} & $d$ & $1$ & $d$ \\
        \hline 
        Hamiltonian & \begin{tabular}{@{}c@{}} Electric, Magnetic \\ (fermion free) \end{tabular} & \begin{tabular}{@{}c@{}} Mass, Kinetic, \\ Electric \end{tabular} & \begin{tabular}{@{}c@{}} Mass, Kinetic, \\ Electric, Magnetic \end{tabular} \\
        \hline
        \begin{tabular}{@{}c@{}} Gauge-field \\ encoding \end{tabular} & Unary & Binary & Binary \\
        \hline
        Gate complexity & Unavailable & $\tilde{O}\left(\frac{L^{3/2}T^{3/2}\Lambda}{\epsilon^{1/2}}\right)$ & See Eqs.~(\ref{eq:comp_1}) and (\ref{eq:comp_2}) \\
        \hline
    \end{tabular}
    \caption{\textbf{Comparison to prior art.} This table displays the lattice gauge theories considered in~\cite{byrnes2006simulating,shaw2020quantum} and our work, specified by their unitary symmetry groups, spatial dimension, and the Hamiltonian terms included, as well as the type of gauge-field encoding employed and the resulting gate complexity. Our work is the most general and complete to date, as we include the complete Kogut-Susskind Hamiltonian with both fermionic and bosonic terms in arbitrary spatial dimensions for the Abelian U(1) and non-Abelian SU(N) groups. Compared to~\cite{byrnes2006simulating}, our algorithms require exponentially fewer qubits by using an efficient binary gauge-field encoding, and include much more detailed circuit constructions, down to a quantum-gate-by-quantum-gate level, along with the gate complexity.}
    \label{tab:prior-art}
\end{table}

\begin{table}[ht]
\centering
\begin{tabular}{c|c|c|c|c|c|}
  & \multicolumn{3}{c|}{\cite{byrnes2006simulating}} & \multicolumn{2}{c|}{Ours} \\ \cline{2-6} 
                  & U(1)    & SU(2)    & SU(3)    & U(1)    & SU(2), SU(3)    \\ \hline
Electric          &    $\tilde{O}(\Lambda)$     &      $\tilde{O}(\Lambda)$    &     $\tilde{O}(\Lambda^2)$     &     ${O}(\log^2(\Lambda))$    &          $\tilde{O}(\log^2(\Lambda))$       \\ \hline
Magnetic          &    $\tilde{O}(\Lambda^4)$     &     $\tilde{O}(\Lambda^{12})$     &     $\tilde{O}(\Lambda^{32})$     &     $\tilde{O}(\log^2(\Lambda))$    &    $\tilde{O}(\log^2(\Lambda))$            
\end{tabular}
\caption{\textbf{Superpolynomial reduction in the gate complexity per simulation time step.} We work out the fault-tolerant circuit construction of the methods proposed in~\cite{byrnes2006simulating}, and analyze their gate complexity per simulation time step in Supp. Mat.~\ref{sec:yamamoto}. $\Lambda$ is the bosonic truncation parameter discussed in Sec.~\ref{sec:LGT}. The results summarized in this table show that our circuits, by making use of efficient quantum fixed-point arithmetic circuits, achieve superpolynomial improvements over~\cite{byrnes2006simulating}.}
\label{tab:yamamoto}
\end{table}

\subsection{Examples}

We now apply our complexity results to example realistic physical systems in three dimensions. Note that the choices of simulation parameters considered for the following examples are reasonable, but not rigorous. Our work is applicable for future examples that may further clarify the parameter choices. 

First, our algorithm can be used to compute transport coefficients in gauge theories \cite{cohen2021quantum}. As reported in \cite{cohen2021quantum}, these coefficients are non-perturbative inputs to theoretical models of heavy ions, relevant to the study of quark-gluon plasmas. It was further reported in \cite{cohen2021quantum} that the required lattice parameters are $a=0.1$ and $L=10$, respectively, and the evolution time $T$ is around 1 for the energy scale relevant to quark-gluon plasmas. To be concrete, we assume a total error budget of $\epsilon = 10^{-8}$ and $\Lambda=10$. Furthermore, we consider a range of masses and coupling regimes, i.e., $m,g \in \{0.1,1,10\}$. 
The expected T-gate counts for computing transport coefficients in U(1), SU(2), and SU(3) LGTs are then $7.35 \times 10^{17}$, $2.83\times 10^{34}$, and $3.01\times 10^{49}$, respectively. The qubit-counts for U(1), SU(2), and SU(3) are $7.3 \times 10^4$, $2.2 \times 10^5$, and $5.5 \times 10^5$, respectively.

Further in \cite{cohen2021quantum}, the authors estimate that a simulation of heavy-ion collisions will require $a=0.1$, $L=100$, and $T=10$. Once again, we assume $\epsilon=10^{-8}$, $\Lambda=10$, and $m,g \in \{0.1,1,10\}$. For such simulations of U(1), SU(2), and SU(3) LGTs, the T-gate and qubit counts are at most $7.19\times 10^{23}$, $3.71\times 10^{40}$, and $3.22\times 10^{55}$, and $7.3 \times 10^{7}$, $1.0 \times 10^8$, and $2.6 \times 10^{8}$, respectively. 

Another application of our algorithm is the calculation of the hadronic tensor of a proton \cite{lamm2020parton}. As discussed in \cite{lamm2020parton}, hadronic tensors provide non-perturbative inputs to deep inelastic-scattering cross-sections and the initial conditions for heavy-ion experiments. According to \cite{lamm2020parton}, $O(L^3)$ time steps are needed to obtain the hadronic tensor of a proton. It was further suggested that the lattice parameters $a=0.1$ and $L=20$ suffice to compute the hadronic tensor of a proton with moderate finite-volume effects. We here again assume that $\epsilon=10^{-8}$. By choosing $T=L^3$ and $\Lambda=10$, we explore a range of masses and couplings, i.e., $m,g \in \{0.1,1,10\}$. The T-gate requirement is then at most $5.28\times 10^{56}$, and the number of qubits required is $2.0 \times 10^6$. 

\section{Discussion}
\label{sec:Discussion}

In this paper, we explored quantum simulations of U(1), SU(2), and SU(3) LGTs with mass, kinetic, electric, and magnetic terms. Note LGT simulations, to be performed on a quantum computer, has massive potential to advance fundamental particle physics. A non-limiting list of furtherance to be explored beyond our work shown herein includes investigating exotic matter at extreme conditions such as the core of a neutron star, probing physics related to the matter-anti-matter asymmetry essential to our very existence, as well as guiding experimental particle physics beyond our current theoretical understanding.

To be more specific, quantum computers can help us explore the phase diagram of QCD matter~\cite{bazavov2019hot} at finite chemical potential, paramount to investigating hot, dense matter within the core of a neutron star. We may expect to observe more exotic phenomena, unseen on a classical computer, such as the theoretically predicted color-superconducting phase~\cite{RevModPhys.80.1455}. 

Furthermore, simulating lattice QCD, enabled by a quantum computer, may shed light on to why we have imbalance in the amount of matter and that of anti-matter. Note, physical theories that violate the Charge-Parity (CP) symmetry, such as the electroweak theory, is thought to be essential for the observed matter-anti-matter asymmetry. However, experiments show us that the source of CP-violation, known in the community as the $\theta$-term, in QCD is vanishingly small. This so-called Strong CP problem has been one of the most puzzling problems in particle physics for the past decades~\cite{kim2010axions}. Therefore, simulating lattice QCD with a topological $\theta$-term~\cite{di1981preliminary,kan2021investigating} could reveal yet-to-be considered mechanisms to address the Strong CP problem by enabling access to the $\theta$-dependent phase structure and dynamics non-perturbatively.

We believe quantum simulation of lattice QCD will play a critical role in guiding future experimental high-energy physics. 
Note the recent back-and-forth regarding the $g-2$ experiment at Fermi lab: The experimental results were announced to suggest physics beyond the SM, as they were far away from the best-known lattice QCD predictions~\cite{abi2021measurement}. Immediately afterwards, a new theoretical lattice QCD calculation in line with the experimental results~\cite{borsanyi2021leading} was announced, placing the original claim of physics beyond the SM in doubt. Irrespective of the eventual outcome, it is clear that the progress in physics relies critically on lattice QCD simulation. Therefore, it stands to reason that quantum lattice-QCD simulation, which supersedes its classical counterpart, will play no less of a role.

As an example, consider the fact that quantum lattice-QCD simulation can directly access high-energy collisions between hadrons in real time. It can extract dynamical quantities such as scattering amplitudes and cross-sections. These are all valuable inputs to guide experimental searches for physics within and beyond the SM. They will for instance help us hone in on the experimental setups and parameters to be used in the particle accelerators~\cite{preskill2018simulating}, such as the current and next-generation CERN colliders. 

Shifting the gear, we remark that our work opens up the possibility of quantum computational nuclear physics and engineering.
With efficient lattice QCD simulations enabled on a quantum computer, we may one day hope to explore complex nuclear dynamics on a quantum computer, via ab-initio simulations. Simulating nuclear dynamics could, for instance, help elucidate fusion processes. Knowing today the success and impact of computational chemistry and chemical engineering, it may then not be an overstatement to say that expected technological breakthroughs of the future by the means of quantum computational nuclear engineering could have a huge impact to the broad community and society.

\section{Outlook}
\label{sec:Outlook}

As an outlook to be taken with a healthy dose of skepticism, we conclude this paper with an optimistic speculation on when we may expect to see LGT simulations on a quantum computer and how demanding they would be. Throughout, we rely on publicly available information, different pieces having varying degrees of evidences. 
{\it Hardware}: In the absence of large-scale quantum computer hardware, we predict its development based on the extrapolation of the technological road maps made available to public. {\it Software}: We assume that there will inevitably be a continued advancement in reducing quantum simulation resource requirement, evidenced starkly in the particular example of Femoco in quantum chemistry simulations~\cite{reiher2017elucidating,li2019electronic,berry2019qubitization,von2020quantum,lee2020even}. We assume a similar rate of reduction would occur for our LGT simulations. Combining together, we first estimate in which years we expect to have sufficient number of qubits in a quantum computer to be able to run various LGT simulations. Then, based on the years, the expected gate counts may be estimated by considering the reduction rate. We stress that this speculation is not to distract the readers from the technical results of this paper, but is simply an outlook.

\begin{figure}[ht]
    \centering
    \includegraphics[width=0.47\linewidth]{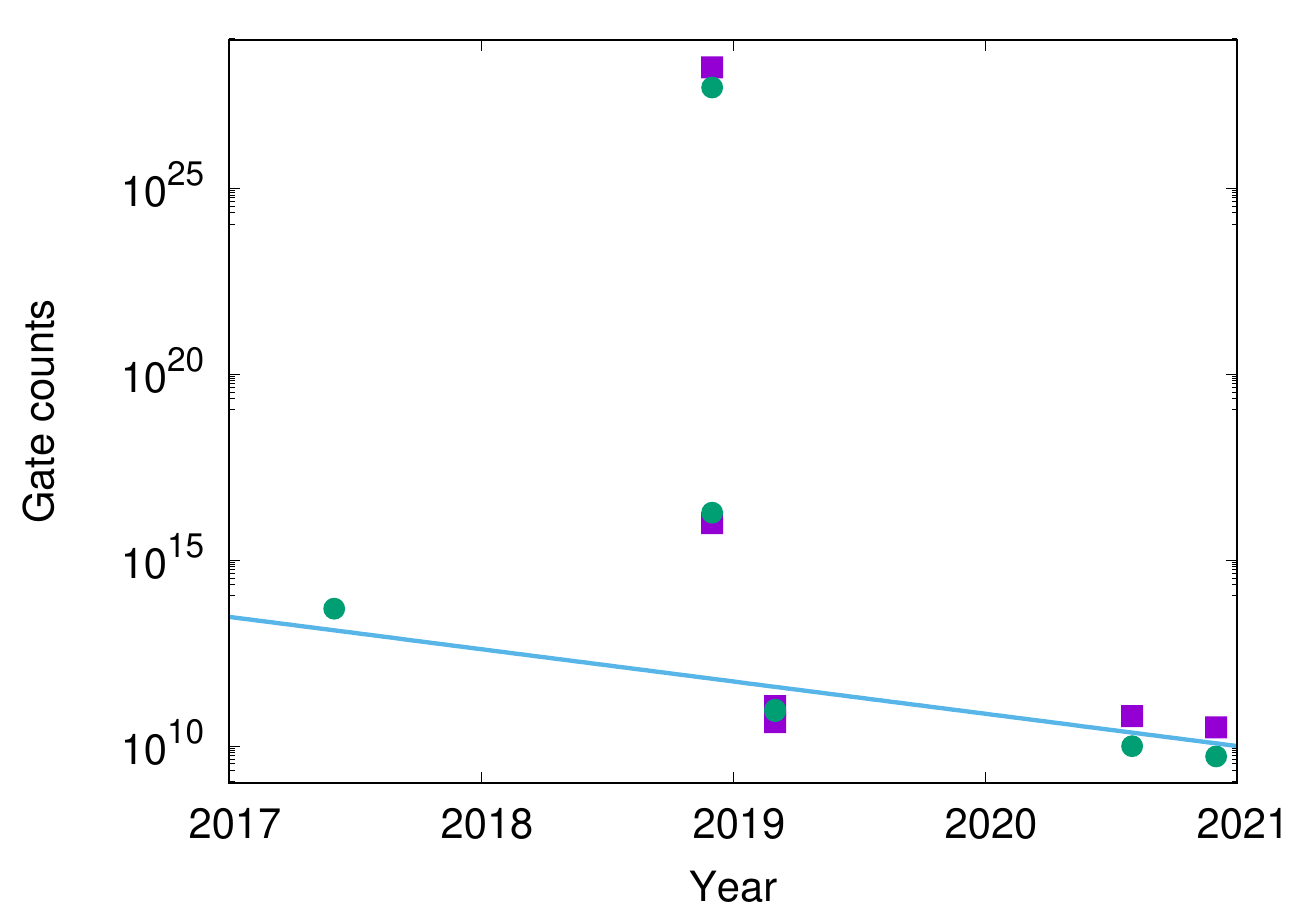}
    \includegraphics[width=0.47\linewidth]{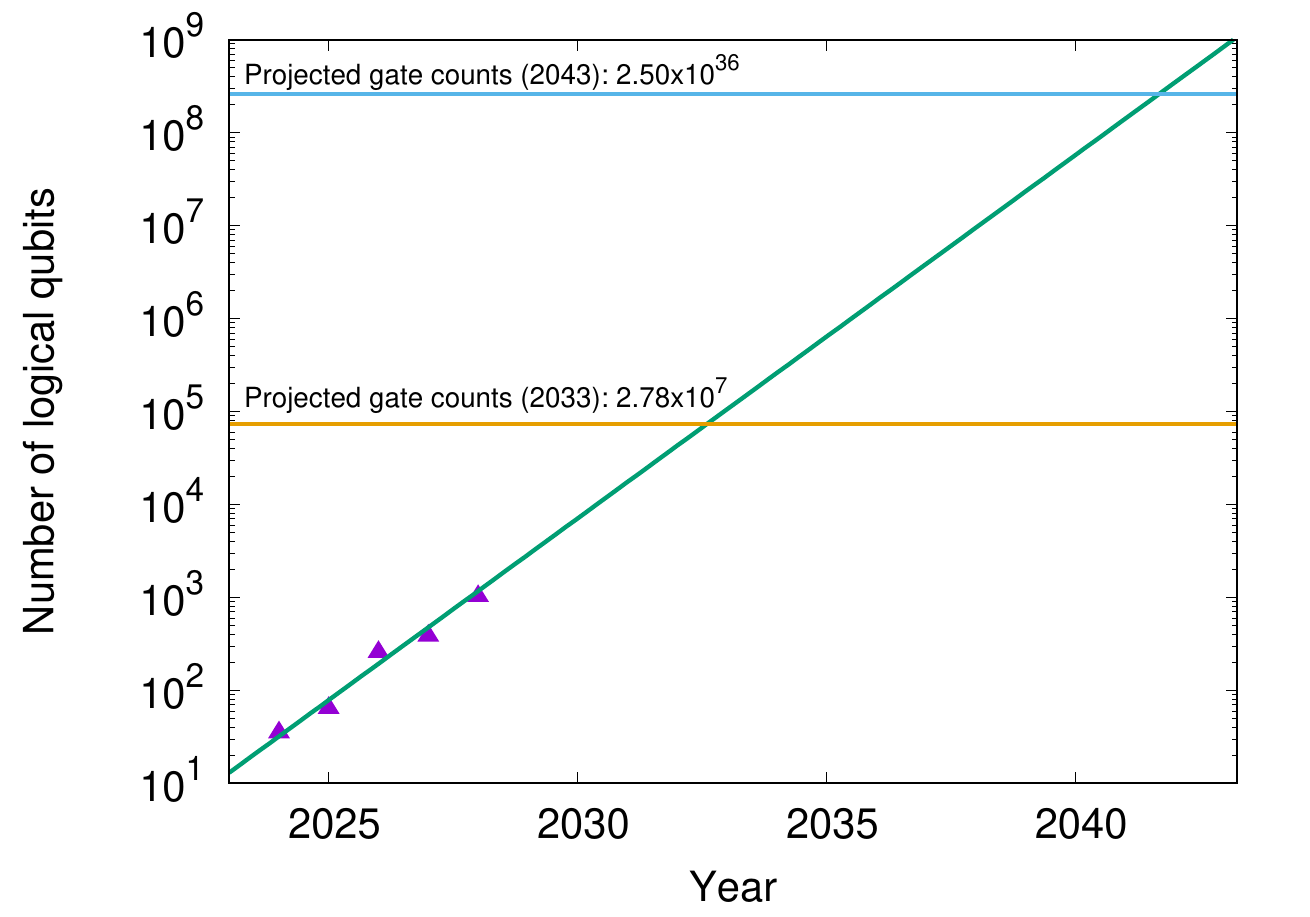}
    \caption{\label{Projection} Projection-based prediction of LGT simulations on a quantum computer. \textbf{Left}: We show the history of resource requirement reduction for Femoco simulation, directly imported from \cite{lee2020even}. The fit function used is $G \approx 3\cdot10^{13}\exp(-2(t-2017))$, obtained by discarding the qDRIFT-based estimates, which appear as the points far from the fit line, which has quadratically worse scaling in the error budget than the rest.
    \textbf{Right}:
    Among scores of industry entities that have published technology road maps, such as IonQ, IBM, Honeywell, and Google~\cite{ionq_2020,ibm_2020,honeywell_2020, google_2020}, we use IonQ's as it states the number of logical qubits, most relevant to our work. The fit function used is $n \approx 32\exp(0.9(t-2024))$, where $n$ is the number of logical qubits and $t$ is the year. For the number of gates, we apply the fit function obtained in the left panel to the current resource requirements for LGT simulations, determined in this paper.}
\end{figure}

Figure~\ref{Projection} shows the relevant projections: U(1) transport problem in year 2033 and SU(3) heavy-ion collision problem in year 2043. The former is expected to require $2.78\times 10^7$ gates. The latter is expected to require far more, exceeding $10^{36}$, an admittedly astronomical number. Hence, to facilitate a more informative foregoing discussion, we consider year 2045 for reasons to become clear. Further, we hereafter use the number of quantum fixed point arithmetic operations (QFOPs) as the metric of interest, motivated by the use of floating point operations in measuring the performance of conventional supercomputers. Our analysis shows 99.998\% of the gate counts stem from QFOPs, justifying its use. We count a QFOP as one round of quantum multiplication followed by quantum addition. The SU(3) heavy-ion collision problem is then expected to require $9.04 \times 10^{25}$ QFOPs. This equates to less than three years of runtime on an exa-scale quantum supercomputer. The runtime is comparable to that required for the state-of-the-art lattice QCD simulations performed on a near exa-scale conventional supercomputer.

While we make no claims on how probable the extrapolated quantum hardware and software advancements are, in view of the history of conventional computing, we remain optimistic that the aforementioned challenges, while ambitious, can eventually be surmounted. Indeed, by comparing Summit, which has a power efficiency of 14.7 gigaFLOPS/watt, and ENIAC, which could only perform 0.00238 multiplication per second per watt, we see a huge leap in efficiency by 6.2 teraFLOPS/watt over only 70 years. The technological advancement brought about by the hardware improvements over the decades for conventional computing cannot be overstated. We anticipate the quantum computing technology advancement will follow a similar trend to one day deliver first-principle simulations of fundamentals of nature of our universe.

\section*{Acknowledgement}
The authors thank Prof. Chris Monroe at Duke University, Prof. Zohreh Davoudi at the University of Maryland, and Prof. Reinhold Bl\"umel at Wesleyan University for their helpful comments.

\bibliographystyle{apsrev4-2}
\bibliography{bibliography.bib}

\appendix

\section{Related Works}
\label{sec:related}

In recent years, there have been significant advances in quantum simulation of various quantum field theories (QFTs), such as scalar field theory, nuclear effective field theory, and LGTs. On the experimental front, a plethora of proof-of-concept simulations of QFTs have been implemented on currently available quantum devices, using approaches including analog quantum simulation \cite{gorg2019realization,schweizer2019floquet,mil2020scalable,yang2020observation}, variational quantum simulation \cite{klco2018quantum,kokail2019self,shehab2019toward,bauer2021simulating,gustafson2021real,kreshchuk2021light,atas20212,nachman2021quantum,Kreshchuk2021simulating}, quantum frequency processing \cite{lu2019simulations}, quantum annealing \cite{rahman20212}, and digital quantum simulation \cite{martinez2016real,klco20202,roggero2020quantum,ciavarella2021trailhead}. On the theoretical front, there exists a host of methods to realize QFTs on quantum hardware \cite{byrnes2006simulating,jordan2012quantum,banerjee2012atomic,tagliacozzo2013simulation,zohar2013cold,zohar2013quantum,jordan2014quantum,jordan2014quantum2,mezzacapo2015non,zohar2015formulation,gonzalez2017quantum,zohar2017digital,moosavian2018faster,zache2018quantum,bender2018digital,klco2019digitization,lamm2019general,alexandru2019gluon,klco2020minimally,klco2020fixed,buser2020state,davoudi2020towards,klco2020systematically,lamm2020parton,mueller2020deeply,kharzeev2020real,banuls2020simulating,kreshchuk2020quantum,ji2020gluon,shaw2020quantum,paulson2020towards,cohen2021quantum,davoudi2021toward,chakraborty2021digital,ferguson2021measurement,stryker2021shearing,klco2021standard}.

In our work, we focus on the algorithms for simulating $d$-dimensional U(1), SU(2), and SU(3) LGTs with fermions on a gate-based fault-tolerant universal quantum computer, and provide rigorous T-gate complexities. To the best of our knowledge, there are two previous works that are directly relevant to our work, \cite{byrnes2006simulating}, where unary encodings of U(1), SU(2), and SU(3) LGTs without fermions on a universal quantum computer were proposed, and \cite{shaw2020quantum}, where an algorithm, along with its gate complexities, for simulating a one-dimensional U(1) LGT with fermions was proposed.

\section{U(1) Lattice Gauge Theory}
\label{sec:U1}
In this section, we detail the methodology to implement the Abelian U(1) lattice gauge theory, of which the continuum describes quantum electrodynamics (QED), on a quantum computer. Due to its Abelian nature, the theory is more straightforward than non-Abelian theories, such as SU(2) or SU(3) that we investigate in detail in the subsequent sections. We therefore take advantage of the simplicity to lay out the ground work useful for SU(2) and SU(3) discussion.

\subsection{Preliminaries}
\label{sec:U1_prelim}

As discussed in the main text, there are four Hamiltonian terms of interest: The electric Hamiltonian $\hat{H}_E$, magnetic Hamiltonian $\hat{H}_B$, mass Hamiltonian $H_M$, and kinetic Hamiltonian $\hat{H}_K$. Inspecting these terms, the first two operate on the links that connect two sites, the mass Hamiltonian operates on the sites themselves, and the kinetic Hamiltonian acts on nearest pairs of sites and the links that connect the pairs. Thus, it is natural to consider two different types of qubit registers, one for the gauge fields ($\hat{H}_E$, $\hat{H}_B$, $\hat{H}_K$) and the other for the fermionic fields ($\hat{H}_M$, $\hat{H}_K$).

To simulate this system, we need to first choose a good basis for each register. For the fermionic register, we consider an occupation basis. The use of an occupation basis of fermionic particles to simulate on a quantum computer is well studied in the literature~\cite{ortiz2001quantum}. For concreteness and simplicity, we use the Jordan-Wigner (JW) transformation~\cite{wigner1928paulische} for the rest of the paper.

As for the gauge-field register, extra care needs to be taken. In particular, we need to ensure the U(1) gauge invariance is satisfied. Gauge invariance is generated by a local constraint, known as Gauss' law, i.e.,
\begin{equation}
    \hat{G}({\vec{n}}) = \sum_{k}(\hat{E}({\vec{n},k}) - \hat{E}({\vec{n}-\hat{k},k})) - \hat{Q}({\vec{n}}), \: \forall \vec{n},
\end{equation}
where $\hat{G}({\vec{n}})$ is the Gauss operator, 
$\hat{E}({\vec{n},k})$ is the electric field of the link that starts from site $\vec{n}$ in direction $k$, and $\hat{Q}({\vec{n}})$ is the charge operator defined according to
\begin{equation}
    \hat{Q}({\vec{n}}) = \hat{\psi}({\vec{n}})^{\dag}\hat{\psi}({\vec{n}}) - \frac{\hat{I}}{2}[1-(-1)^{\vec{n}}],
\end{equation}
where $\hat{\psi}({\vec{n}})^{\dag}$ and $\hat{\psi}({\vec{n}})$ are the fermion creation and annihilation operators and $\hat{I}$ denotes an identity operator. On a $d$-dimensional lattice, $\vec{n}=(n_1,n_2,...,n_d)$ is a vector with $d$ coordinates. Further, we define the parity of each site,
\begin{equation}
    (-1)^{\vec{n}} \equiv (-1)^{\sum_{i=1}^{d} n_i}.
\end{equation}
The Gauss operator generates local gauge transformation and must commute with the Hamiltonian. The physical, gauge-invariant Hilbert space $\mathcal{H}_{G}$ is defined through the eigenstates of the Gauss operator:
\begin{equation}
\label{eq:u1gauss}
    \mathcal{H}_{G} = \{ \ket{\Psi} \in \mathcal{H}_G \: |\:  \hat{G}_{\vec{n}} \ket{\Psi} = 0 \:, \forall \: \vec{n} \}.
\end{equation}
In the current case of U(1) lattice gauge theory,
the eigenstates of the electric field operator qualify
for the link-space basis, as the electric field operator 
and the Gauss operator commute. More specifically,
the electric field operator forms a complete set of
commuting observables on the link space.
Therefore, the eigenbasis of the complete set of
commuting observables, i.e., 
the eigenbasis of the electric field operator,
is a good basis or a good quantum number.

As the last step of preliminaries to a quantum simulation
of U(1) lattice gauge theory, we note that 
\begin{align}
    [\hat{E}({\vec{n},k}),\hat{U}(\vec{n}',k')] &= \delta_{\vec{n}',\vec{n}} \delta_{k,k'}\hat{U}({\vec{n},k}), \nonumber \\
    [\hat{E}({\vec{n},k}),\hat{U}(\vec{n}',k')^\dag] &= -\delta_{\vec{n}',\vec{n}} \delta_{k,k'}\hat{U}({\vec{n},k})^\dag,
\end{align}
where $\hat{U}(\vec{n},k)$ is the parallel transporter operator.
In the electric field basis $\ket{E}$, defined according to
\begin{equation}
\hat{E} = \sum_{E\in \mathbbm{Z}} E \ketbra{E}{E},
\end{equation}
we have
\begin{equation}
\hat{U} = \sum_{E\in \mathbbm{Z}} \ketbra{E+1}{E},\:\: \hat{U}^{\dag} = \sum_{E\in \mathbbm{Z}} \ketbra{E-1}{E}.
\end{equation}
Note that we dropped site and direction indices for notational convenience.

\subsection{Simulation circuit synthesis}
\label{sec:SimCircSynth}

In order to represent the infinite-dimensional gauge-field operator on each link on a finite-size quantum computer, its Hilbert space must be truncated at a cutoff, $\Lambda$. The electric field operator then becomes
\begin{equation}
    \hat{E} = \sum_{E=-\Lambda}^{\Lambda-1} E \ketbra{E}{E}.
\end{equation}
A non-negative integer $0\leq j < 2^{\eta}$ is represented on the binary $\eta$-qubit register as 
\begin{equation}
    \ket{j} = \ket{\sum_{n=0}^{\eta-1} j_{n} 2^n} = \bigotimes_{n=0}^{\eta -1}\ket{j_n}.  
\end{equation}
Using this binary computational basis, the eigenbasis $\ket{E}$ is encoded via $E = j - \Lambda$. Then, the number of qubits on the link register for each link is given by $\eta = \log(2\Lambda)$, where $\Lambda$ is assumed to be a non-negative power of two, and we have and will continue to assume all logarithms are base two, unless otherwise specified.

In \cite{shaw2020quantum}, the authors periodically wrapped the electric fields at $\Lambda$ such that
\begin{equation}
    \hat{U}\ket{\Lambda - 1} = \ket{-\Lambda}, \: \hat{U}^\dag \ket{-\Lambda} = \ket{\Lambda-1}.
\end{equation}
This spoils the on-link commutator at the cutoff to give
\begin{align}
    [\hat{E},\hat{U}] &= \hat{U} - 2\Lambda\ketbra{-\Lambda}{\Lambda-1}, \\
    [\hat{E},\hat{U}^\dag] &= -\hat{U}^\dag + 2\Lambda\ketbra{\Lambda-1}{-\Lambda}.
\end{align}
However, as is also explicitly discussed in \cite{shaw2020quantum}, for a truncation with a large cutoff value, the states $\ket{-\Lambda}$ and $\ket{\Lambda-1}$ are energetically unfavorable, and hence, will hardly be populated at all. Therefore, the spoiled commutator will likely not be a problem.

Equipped with all the necessary tools, we now write the Hamiltonian in the qubit space as
\begin{equation}
\hat{H} = \sum_{\vec{n}}\left[\hat{D}^{(M)}_{\vec{n}} + \hat{D}^{(E)}_{\vec{n}} + \hat{T}^{(K)}_{\vec{n}} + \hat{L}^{(B)}_{\vec{n}} \right],
\end{equation}
where
\begin{align}
\hat{D}_{\vec{n}}^{(M)} &= -\frac{m}{2}(-1)^{\vec{n}} \hat{Z}({\vec{n}}), \label{eq:mass_U1} \\
\hat{D}_{\vec{n}}^{(E)} &= \frac{g^2}{2a^{d-2}}\sum_{l=1}^{d}\hat{E}^{2}({\vec{n}},l),
\label{eq:U1Diagonal}
\end{align}
are diagonal operators, where $(-1)^{\vec{n}}$ is either $+1$ or $-1$ depending on whether $\vec{n}$ is a fermion or anti-fermion site, respectively, reflective of the use of staggered-fermions \cite{kogut1975hamiltonian}, and
\begin{align}
    \hat{T}^{(K)}_{\vec{n}} &= \sum_{l=1}^d \frac{1}{8a} [(\hat{U}({\vec{n}},l)+\hat{U}^{\dag}({\vec{n}},l))(\hat{X}({\vec{n}}) \hat{X}({\vec{n}+\hat{l}}) + \hat{Y}({\vec{n}}) \hat{Y}({\vec{n}+\hat{l}}))\hat{\zeta}_{\vec{n},l}
    \nonumber \\
    &\quad + i(\hat{U}(\vec{n},l)-\hat{U}^{\dag}({\vec{n}},l))
    (\hat{X}({\vec{n}}) \hat{Y}({\vec{n}+\hat{l}}) - \hat{Y}({\vec{n}}) \hat{X}({\vec{n}+\hat{l}}))\hat{\zeta}_{\vec{n},l}]
    \label{eq:U1kinetic}
\end{align}
is an off-diagonal operator. We use $\hat{X}$, $\hat{Y}$, and $\hat{Z}$ as the Pauli $x$, $y$, and $z$ matrices, respectively. We abuse the notation $\hat{l}$ to denote a unit vector in direction $l$. The operators $\hat{\zeta}_{\vec{n},l}$ are tensor products of $\hat{Z}$, which arise from the JW transformation. We consider a $d$-dimensional $L^d$-site lattice, where there are $L$ sites in each direction. The length of each $\hat{\zeta}_{\vec{n},l}$ is $O(L^{d-1})$. For brevity, we suppress the $\hat{\zeta}_{\vec{n},l}$ operators in the remaining part of the section. The second off-diagonal operator due to the magnetic contribution is given by
\begin{equation}
\hat{L}^{(B)}_{\vec{n}} = \frac{-1}{2a^{4-d}g^2}\sum_{j\neq i;j,i=1}^{d}[\hat{U}(\vec{n},i)\hat{U}(\vec{n}+\hat{i},j)\hat{U}^{\dag}(\vec{n}+\hat{j},i)\hat{U}^{\dag}(\vec{n},j)+ h.c.],
\label{eq:U1_mag}
\end{equation}
where $h.c.$ denotes Hermitian conjugate. We use Suzuki-Trotter formula \cite{suzuki1991general} as our simulation method. The Trotter terms to be implemented are of the form
$
e^{i\hat{D}_{\vec{n}}^{(M)}t},
e^{i\hat{D}_{\vec{n}}^{(E)}t},
e^{i\hat{T}_{\vec{n}}^{(K)}t},
e^{i\hat{L}_{\vec{n}}^{(B)}t},
$
where $t$ is a sufficiently small number to ensure the Trotter error incurred is within a pre-specified tolerance.
In the remaining part of this subsection, we discuss synthesizing circuits for each of the four Trotter terms.

\subsubsection{Mass term \texorpdfstring{$e^{i\hat{D}_{\vec{n}}^{(M)}t}$}{}}
\label{sec:U1_mass}

The implementation of this term is straightforward. A single-qubit $R_{z}(\theta) = \exp(-i\theta\hat{Z}/2)$ gate, where $\theta = -m(-1)^{\vec{n}}t$, applied to the qubit that corresponds to site $\vec{n}$ in the site register suffices. Note that the angles of rotation are the same for all even and odd sites, respectively, up to a sign. The sign difference can be rectified by conjugating the $z$-rotations with NOT gates. Then, a circuit with one layer of $R_z$ gates with the same angle of rotation results. This circuit can be implemented efficiently using the weight-sum trick in \cite{gidney2018halving,nam2019low}. Briefly, consider applying the same angle $R_z$ gates on $p$ qubits simultaneously. This imparts a phase to an input state with the phase angle being proportional to the Hamming weight of the input. This can thus alternatively be implemented by first computing ${\rm Weight}(p)$ into an ancilla register, while incurring $p-{\rm Weight}(p)$ ancilla qubits and at most $4(p - {\rm Weight}(p))$ T gates, where Weight denotes the number of ones in the binary expansion of the integer number $p$. Finally, we apply $\lfloor \log(p) +1\rfloor$ $R_z$ rotations to the ancilla register to impart the correct phase, and then uncompute the weight on the ancilla register. For a $d$-dimensional lattice with $L^d$ lattice sites, $p = L^d$. 

\subsubsection{Electric term \texorpdfstring{$e^{i\hat{D}_{\vec{n}}^{(E)}t}$}{}}

\label{sec:U1_electric}
Here, we present a method to implement the electric term. The method modifies that presented in \cite{shaw2020quantum}, and provides an improvement in gate counts. We import the steps detailed in \cite{shaw2020quantum} for the convenience of the readers. $\hat{D}_{\vec{n}}^{(E)}$ is a sum of $d$ commuting terms, and hence, its evolution can be implemented exactly as a product of $d$ sub-evolutions,
\begin{equation}
    e^{it\hat{D}_{\vec{n}}^{(E)}} = \prod_{l=1}^{d} e^{i \frac{g^2t}{2a^{d-2}}\hat{E}^2(\vec{n},l)}.
\end{equation}
We will discuss the implementation of only one sub-evolution without loss of generality. We herein drop the link location index for notational convenience. The electric field operator and a qubit-encoded gauge field state obeys the eigenvalue relation
\begin{equation}
    \hat{E} \ket{j} = (j-2^{\eta-1}) \ket{j}.
\end{equation}
As such, the evolution of the electric part $e^{it \frac{g^2}{2a^{d-2}}\hat{E}^2}$ is given by
\begin{equation}
    \ket{j} \mapsto e^{it \frac{g^2}{2a^{d-2}}(j-2^{\eta-1})^2} \ket{j}.
\end{equation}
To implement the term for each link, we first compute $(j-2^{\eta-1})^2$ into an ancilla register, and then, impart the phase by applying an $R_z$ gate on every qubit in the ancilla register. We perform the arithmetic operations by first computing $j-2^{\eta-1}$, using an out-of-place adder, which incurs $4(\eta-2)$ T gates and $\eta$ reusable ancilla qubits \cite{gidney2018halving}, and then squaring the $(\eta+1)-$bit ancilla state, which costs $4\eta(12\eta -3\lfloor \log(\eta+1)\rfloor - 2)$ T gates with the multiplier proposed in \cite{shaw2020quantum}. We induce approximate phases (described below) and then finally uncompute the ancilla register. Therefore, the entire arithmetic operations cost $8(\eta-2)+8\eta(12\eta -3\lfloor \log(\eta+1)\rfloor - 2)$ T gates. Here, we choose to perform the arithmetic operations in series to reduce the ancilla-qubit count. Since there are $dL^d$ links on an $L^d$-site $d$-dimensional lattice, the arithmetic operations on all links cost at most $8d L^d[(\eta-2)+\eta(12\eta -3\lfloor \log(\eta+1)\rfloor - 2)]$ T gates, $3(\eta + 1)dL^d$ ancilla qubits to store $\ket{j-2^{\eta +1}}$ and $\ket{(j-2^{\eta +1})^2}$, and $3(\eta + 1) - \lfloor \log(\eta + 1)\rfloor - 1$ reusable workspace ancilla qubits \cite{shaw2020quantum}. If we choose to optimize the T-depth, we can parallelize the squaring operations, at the cost of increasing the workspace ancilla-qubit count.

We now discuss the phase induction. The correct phase can be induced by applying $R_z(2^k \theta)$, where $\theta = \frac{g^2t}{2a^{d-2}}$, on the $k$th qubit of the $2(\eta + 1)-$bit ancilla state, $\ket{(j-2^{\eta +1})^2}$. Hence, there are $2(\eta+1)$ sets of $dL^d$ same-angle $R_z$ rotations to implement, where each set can be effected using the weight-sum trick. Once again, we first compute Weight$(dL^d)$ into the ancilla register, incurring $4(dL^d - \text{Weight}(dL^d))$ T gates and $dL^d - \text{Weight}(dL^d)$ ancilla qubits, and then, applying $\lfloor \log(dL^d) + 1 \rfloor$ $R_z$ gates to the ancilla register to induce the right phase. 

There is an alternative method for simulations with a fixed Trotter step $t$, $d$ and $g^2$, where $a$ can be chosen such that $\frac{g^2 t}{2a^{d-2}} = \frac{\pi}{2^{M}}$ with $M > \eta$. The electric evolution is then given by
\begin{equation}
    \ket{j} \mapsto e^{i \frac{\pi}{2^{M}}(j-2^{\eta-1})^2} \ket{j}.
\end{equation}
Once again, we first compute $(j-2^{\eta-1})^2$ into the ancilla register. Then, we impart the phase by a phase gradient operation, which consists of an $M$-bit addition on a specially prepared phase gradient state \cite{kitaev2002classical}
\begin{equation}
    \ket{\psi_M} = \frac{1}{\sqrt{2^{M}}}\sum_{b=0}^{2^{M}-1}e^{-2\pi ib/2^{M}}\ket{b},
    \label{eq:phgradstate}
\end{equation}
incurring $4M+O(1)$ T gates due to the $M-$bit adder \cite{gidney2018halving}. Here, we perform the arithmetic operations and phase gradient operation on one link at a time. Since $M=\log(\frac{2\pi a^{d-2}}{g^2 t})$, the number of T gates required by the adders operations on all the links is $4dL^d\log(\frac{2\pi a^{d-2}}{g^2 t})+O(dL^d)$. In order to synthesize the phase gradient state, which can be reused for all phase gradient operations, $M-1$ $Z^\alpha$ phase-shift rotation gates, defined by
\begin{equation}
    Z^\alpha = \begin{pmatrix}
    1 & 0 \\
    0 & e^{i\pi \alpha} 
    \end{pmatrix},
\end{equation}
are needed \cite{nam2019low}. Each $Z^\alpha$ can be synthesized by using RUS circuits \cite{bocharov2015efficient}. 

\subsubsection{Kinetic term \texorpdfstring{$e^{i\hat{T}_{\vec{n}}^{(K)}t}$}{}}
\label{sec:U1_kin}

Here we present two different methods to implement the kinetic term.
The first method is a small modification of the method in \cite{shaw2020quantum}, 
so we import the steps detailed in \cite{shaw2020quantum} for the convenience of the readers. 
The second method is based on the diagonalization of $\hat{U}$ operators. Herein, we drop the exact site and link position dependence and instead use $r$ and $r+1$ to denote two sites without loss of generality.

{\bf Method 1: Block-diagonal decomposition ---}
In order to decompose the off-diagonal term into elementary gates, we write
\begin{align}
    \hat{U} + \hat{U}^{\dag} = \hat{A} + \hat{\tilde{A}},
\end{align}
where $\hat{A} = \hat{I} \otimes \hat{I} ...\otimes \hat{X}$ and $\hat{\tilde{A}} = \hat{U}^{\dag} \hat{A} \hat{U}$, and similarly,
\begin{align}
    i(\hat{U} - \hat{U}^{\dag}) = \hat{B} + \hat{\tilde{B}},
\end{align}
where $\hat{B} = \hat{I}\otimes\hat{I} ... \otimes \hat{Y}$ and $\hat{\tilde{B}} = \hat{U}^{\dag} \hat{B} \hat{U}$.

\begin{figure}[ht]
\[
\Qcircuit @C=1em @R=.7em {
     &\qw & \targ & \ctrl{1}&\targ &\qw &\qw \\
     &\targ & \qw & \ctrl{1}&\qw &\targ &\qw\\
     &\ctrl{-1} & \ctrl{-2} & \gate{R_x} & \ctrl{-2}&\ctrl{-1} &\qw
}
\]
\caption{Circuit for an individual kinetic term. Top: the zeroth qubit of the bosonic gauge field between fermionic sites $r$ and $r+1$. Middle/Bottom: fermionic sites $r$ and $r+1$. Conjugation of this circuit by controlled-$Z$ gates with the target on the bottom qubit addresses the JW string, and with the control on a qubit in the JW string.}
\label{fig:singlekinetic_no_z}
\end{figure}
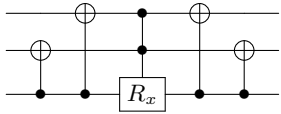

Furthermore, we define
\begin{align}
    \hat{P}_{r} &= \hat{X}_r \hat{X}_{r+1} + \hat{Y}_r \hat{Y}_{r+1}, \nonumber \\
    \hat{\tilde{P}}_{r} &= \hat{X}_r \hat{Y}_{r+1} - \hat{Y}_r \hat{X}_{r+1}.
    \label{eq:ising}
\end{align}
To simulate the off-diagonal term, we approximate
\begin{equation}
    e^{-i\frac{t}{8a}[(\hat{A} + \hat{\tilde{A}})\otimes \hat{P}_r + (\hat{B}+\hat{\tilde{B}})\otimes \hat{\tilde{P}}_r]}
    \approx e^{-i t (\hat{T}_r^{(2)}+\hat{T}_r^{(3)})/2}e^{-i t (\hat{T}_r^{(1)}+\hat{T}_r^{(4)})/2},
    \label{eq:KineticU1}
\end{equation}
where 
\begin{align}
    \hat{T}_r^{(1)} &= (\hat{A}\otimes \hat{P}_r)/4a, \label{eq:T1_U1}\\
    \hat{T}_r^{(2)} &= (\hat{\tilde{A}}\otimes \hat{P}_r)/4a \nonumber \\
            &= \hat{U}^{\dag}T_r^{(1)}\hat{U}, \label{eq:T2_U1} \\
    \hat{T}_r^{(3)} &= (\hat{\tilde{B}}\otimes \hat{\tilde{P}}_r)/4a \nonumber \\
            &= \hat{U}^{\dag} \hat{T}_r^{(4)}  \hat{U}, \label{eq:T3_U1} \\
    \hat{T}_r^{(4)} &= (\hat{B}\otimes \hat{\tilde{P}}_r)/4a \label{eq:T4_U1}.
\end{align}
In contrast to \cite{shaw2020quantum}, wherein the Trotterization was performed for individual $\hat{T}$ terms in the order of $\hat{T}^{(1)}_r$, $\hat{T}^{(2)}_r$, $\hat{T}^{(3)}_r$, and $\hat{T}^{(4)}_r$, here, we Trotterize them into two terms (see (\ref{eq:KineticU1})). We do this since there is an efficient circuit known to implement $e^{-i t (\hat{T}^{(1)}_r+\hat{T}^{(4)}_r)/2}$ and $e^{-i t (\hat{T}_r^{(2)}+\hat{T}_r^{(3)})/2}$~\cite{wang2020resource}. Briefly, the circuit is a doubly-controlled $R_x$ gate whose angle is four times the angle of rotation in the Trotter term written in the Pauli basis, conjugated by a simple CNOT network -- in this particular case the network is a CNOT gate with control on fermion site $r+1$ and target on fermion site $r$, followed by another CNOT gate with the same control but the target being the zeroth bit of the Bosonic link in between. See Fig.~\ref{fig:singlekinetic_no_z}. The doubly-controlled $R_x$ gate can be implemented using two uncontrolled $R_z$ gates and two relative-phase Toffoli gates, which cost $4$ T gates each \cite{maslov2016advantages}. Note that the angles of rotation here, one minus and one plus per $r$ are the same, for all choices of $r$. Conjugated by a pair of NOT gates, the negative angle rotations become positive. As such, the emergent two subcircuits, each being a layer of individual $R_z$ gates associated with each $r$, have the same angle of rotation. As in the mass term, we can use the weight-sum trick to implement the kinetic term.

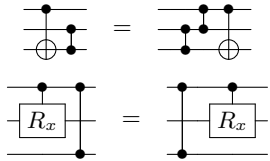
\begin{figure}[ht]
    \centering
        \[
        \Qcircuit @C=.5em @R=0em @!R {
        & \ctrl{2}&\qw      & \qw & & & \qw &\qw&\ctrl{1} &\ctrl{2} &\qw\\
        & \qw     &\ctrl{1} & \qw &\push{\rule{.3em}{0em}=\rule{.3em}{0em}} & &\qw&\ctrl{1}&\ctrl{-1}&\qw&\qw \\
        & \targ   &\ctrl{-1}& \qw & & & \qw &\ctrl{-1} &\qw&\targ&\qw
        }
        \]
        \[
        \Qcircuit @C=.5em @R=0em @!R {
        & \ctrl{1}&\ctrl{2}      & \qw & & & \ctrl{2} &\qw&\ctrl{1} &\qw\\
        & \gate{R_x}     &\qw & \qw &\push{\rule{.3em}{0em}=\rule{.3em}{0em}} & &\qw&\qw&\gate{R_x}&\qw \\
        & \qw   &\ctrl{-2}& \qw & & & \ctrl{-2} &\qw &\qw&\qw
        }
        \]
    \caption{Commutation and cancellation rules used to parallelize the $U(1)$ kinetic term implementation. Note two closed circles connected by a line indicates a controlled-$Z$ gate. A collision of two controlled-$Z$ gates on the same two-qubit lines cancel each other (not shown). Two gates acting on two disjoint set of qubits commute (not shown).}
    \label{fig:rules}
\end{figure}

However, unlike the mass term, we cannot implement the kinetic terms of all the sites in parallel due to two reasons. First, the kinetic terms for nearest-neighbors do not commute. As such, we have to evolve the odd and even sites separately. Second, the evolution operator includes the multi-site Pauli-$z$ operators $\hat{\zeta}$, shown in (\ref{eq:U1kinetic}), due to the JW transformation. Unless two nearest-neighbor sites are connected on the JW path, the multi-site Pauli-$z$ operators need to be taken into account. Note on a $d$-dimensional $L^d$-site lattice, there are $dL^d$ individual kinetic terms to implement. In the following, we construct a circuit that implements these individual terms in parallel whenever possible.

We start with the JW transformation. Specifically, we follow a zigzag pattern. On a one-dimensional lattice, this path is simply a line. On a two-dimensional $L^2$-site lattice, one can draw $L$ lines of length $L$ in the $x$-direction, for instance, and connect the neighboring lines to form a zigzagging path. This zigzagging JW path can be generalized to $d$-dimensional lattices. 

We next consider the terms in the bulk and on the edges of the lattice separately. The edges are $(d-1)$-dimensional hyperplanes, on which the terms are connected with periodic boundary conditions, but not with open boundary conditions. There are $dL^{d-1}$ terms on the edges. The terms in the bulk are connected in both periodic and open boundary conditions. There are $d(L^d-L^{d-1})$ terms in the bulk. 

We implement the evolution of the bulk- and edge-terms one orthogonal direction at a time. 
For each direction, we use a template circuit in Fig.~\ref{fig:singlekinetic_no_z}, modified to accommodate for $\hat{\zeta}$. This is straightforwardly done by a conjugation of the circuit by controlled-Z gates~\cite{wang2020resource}. 

We parallelize the circuit that implements the kinetic term for each direction as follows. Denote the subcircuit that is to the left of the doubly controlled $R_x$, including the aforementioned controlled-Z gates, as a circuit prefix $P$. Denote the subcircuit that appears to the right of the doubly controlled $R_x$, including the aforementioned controlled-Z gates, as a circuit suffix $S$. We can gather all of the $P$s, in the order of their appearance, to the left end of the circuit. Similarly, we can collect all of the $S$s to the right end of the circuit. Note that this process requires applications of gate commutation and cancellation rules. We report these in Fig.~\ref{fig:rules}. 

The resulting, parallelized circuit has a layer of doubly controlled $R_x$ gates in the middle. We discussed earlier how to implement a doubly-controlled $R_x$ gate using two $R_z$ gates in parallel. With all angles of rotation being the same, we can employ the weight-sum trick to reduce the gate complexity. 

We now gather everything. 
Recall we have four different levels to consider: 
(i) $\hat{T}^{(1/4)}$ vs $\hat{T}^{(2/3)}$,
(ii) even vs odd sites,
(iii) bulk vs edge terms, and
(iv) each orthogonal directions $i = 1,..,d$
in the $d$-dimensional lattice.
Levels (i) and (ii) imply that we apply four
stages of circuits per given levels (iii) and (iv).
Consider a single stage. 
For the bulk terms, for each direction $i$,
the number of controlled-$Z$ gates
required in the parallelized circuit is $O(L^{i-1})$, assuming
our zigzag pattern is formed in the ascending order of $i$.
The number of doubly-controlled $R_x$ gates implemented 
in parallel is $N_B = (L^d-L^{d-1})/2$.
Our implementation requires 
$8N_B$ T gates to detach the
double controls from $R_x$ gates,
$N_B$ ancilla qubits to parallelize
the $R_z$ implementations, and
additional $2N_B - {\rm Weight}(2N_B)$ ancilla qubits
and $4(2N_B- {\rm Weight}(2N_B))$ T gates 
to compute the weight into the ancilla register, after which $\lfloor \log(2N_B) + 1 \rfloor$ $R_z$ gates are applied to the ancilla register.
For the edge terms, for each direction $i$,
the number of controlled-$Z$ gates
required in the parallelized circuit is $O(L^{i})$.
The number of doubly-controlled $R_x$ gates implemented 
in parallel is $N_E = L^{d-1}/2$.
The resource required is the same as above,
where all occurrences of $N_B$ is replaced by $N_E$.

To be complete, for level (i), note $e^{-i t (\hat{T}^{(2)}_r+\hat{T}^{(3)}_r)/2}$ has an extra incrementer and decrementer. This incurs an additional cost of $8(\eta-2)$ T gates and $\eta$ reusable ancilla qubits \cite{shaw2020quantum}, when using the compute-uncompute trick for logical ANDs in \cite{gidney2018halving} and the Toffoli construction in \cite{jones2013low}. 

{\bf Method 2: Diagonalization of $\hat{U}$ ---}
Here we consider an alternative method to implement the kinetic term. We first consider an $\eta$-qubit quantum Fourier transform (QFT) ${\mathcal F}$ defined according to
\begin{equation}
{\mathcal F} = \frac{1}{\sqrt{2^\eta}} \sum_{j,k=0}^{2^\eta-1} e^{\frac{2\pi i jk}{2^\eta}} \ket{k}\bra{j}.
\end{equation}
Conjugating $\hat{U}^\dag$ with the QFT, we obtain
\begin{align}
    {\mathcal F}(\hat{U}^{\dag}){\mathcal F}^{\dag} &= \frac{1}{{2^\eta}} \sum_{l,j,k,j',k'=0}^{2^\eta-1} e^{\frac{2\pi i (jk-j'k')}{2^\eta}} \ket{k}\braket{j}{l-1} \braket{l}{j'} \bra{k'} \nonumber \\ 
    &= \frac{1}{2^\eta} \sum_{l,j,k,j',k'=0}^{2^\eta-1} e^{\frac{2\pi i (jk-j'k')}{2^\eta}} \ketbra{k}{k'}\delta_{j,l-1}\delta_{l,j'} \nonumber \\
    &= \frac{1}{2^\eta}\sum_{l,k,k'=0}^{2^\eta-1} e^{\frac{2\pi i (l-1)k}{2^\eta}} e^{\frac{-2\pi i lk'}{2^\eta}} \ketbra{k}{k'} \nonumber \\
    &= \frac{1}{2^\eta}\sum_{l,k,k'=0}^{2^\eta-1} e^{\frac{2\pi i (k-k')l}{2^\eta}}e^{\frac{-2\pi i k}{2^\eta}}\ketbra{k}{k'} \nonumber \\
    &= \sum_{k,k'=0}^{2^\eta-1}\delta_{k,k'}e^{\frac{-2\pi i k}{2^\eta}}\ketbra{k}{k'} \nonumber \\
    &= \sum_{k=0}^{2^\eta-1} e^{\frac{-2\pi i k}{2^\eta}}\ketbra{k}{k}.
    \label{eq:udag_qft}
\end{align}
Taking the Hermitian conjugate,
\begin{equation}
    {\mathcal F}(\hat{U}){\mathcal F}^{\dag} = \sum_{k=0}^{2^\eta-1} e^{\frac{2\pi i k}{2^\eta}}\ketbra{k}{k}.
    \label{eq:u_qft}
\end{equation}
By linearity, we obtain
\begin{align}
    {\mathcal F}(\hat{U}+\hat{U}^{\dag}){\mathcal F}^{\dag} &= \sum_{k=0}^{2^\eta-1} (e^{\frac{2\pi i k}{2^\eta}}+ e^{\frac{-2\pi i k}{2^\eta}})\ketbra{k}{k} \nonumber \\
    &= \sum_{k=0}^{2^\eta-1} 2\cos(\frac{2\pi k}{2^\eta}) \ketbra{k}{k}\equiv \hat{D}_c, \\
    {\mathcal F}i(\hat{U}-\hat{U}^{\dag}){\mathcal F}^{\dag} &= \sum_{k=0}^{2^\eta-1} i(e^{\frac{2\pi i k}{2^\eta}}- e^{\frac{-2\pi i k}{2^\eta}})\ketbra{k}{k} \nonumber \\
    &= \sum_{k=0}^{2^\eta-1} -2\sin(\frac{2\pi k}{2^\eta}) \ketbra{k}{k} \equiv \hat{D}_s,
\end{align}
where $\hat{D}_c$ and $\hat{D}_s$ are diagonal.
The fermionic operators $\hat{P}_r, \hat{\tilde{P}}_r$ can also be diagonalized. Define the basis as,
\begin{equation}
    \ket{\beta_{ab}} = \frac{1}{\sqrt{2}}(\ket{0b} + (-1)^a\ket{1\bar{b}}),
    \label{eq:bell}
\end{equation}
where $\bar{b}$ is the binary negation of $b$.
Let $\hat{U}_{Bell}$ be
\begin{equation}
    \sum_{a,b=0}^{1}\ketbra{\beta_{ab}}{ab}.
\end{equation}
Then,
\begin{align}
    \hat{P}_r &= \hat{U}_{Bell} \sum_{a,b=0}^{1} 2b(-1)^a \ketbra{ab}{ab}\hat{U}_{Bell}^\dag \nonumber \\
    &\equiv \hat{U}_{Bell} \hat{D}^f \hat{U}_{Bell}^\dag
\end{align}
and 
\begin{align}
    \hat{\tilde{P}}_r &= (\hat{S}^\dag \otimes \hat{I})\hat{P}_r(\hat{S} \otimes \hat{I}) \nonumber \\
    &\equiv (\hat{S}^\dag \otimes \hat{I}) \hat{U}_{Bell} \hat{D}^f \hat{U}_{Bell}^\dag (\hat{S} \otimes \hat{I}).
\end{align}
Note $\hat{D}^f$ is diagonal. As such, the kinetic term can be written as a sum of two diagonalizable operators
\begin{align}
    &\quad \frac{1}{8a} [(\hat{U}+\hat{U}^{\dag})\otimes\hat{P}_r + i(\hat{U}-\hat{U}^{\dag})\otimes\hat{\tilde{P}}_r] \nonumber \\
    &= \frac{1}{8a}[{\mathcal F}^\dag \otimes \hat{U}_{Bell} (\hat{D}_c\otimes \hat{D}^f) {\mathcal F} \otimes \hat{U}_{Bell}^\dag + {\mathcal F}^\dag \otimes (\hat{S}_r^{f\dag}\hat{U}_{Bell}) (\hat{D}_s \otimes \hat{D}^f) {\mathcal F} \otimes   (\hat{U}_{Bell}^\dag \hat{S}_r^f)].
\end{align}
To first order, the Trotterization of the kinetic term is then
\begin{align}
    &\quad e^{i\frac{t}{8a} [(\hat{U}+\hat{U}^{\dag})\otimes\hat{P}_r + i(\hat{U}-\hat{U}^{\dag})\otimes\hat{\tilde{P}}_r]} \nonumber \\
    &\approx e^{i\frac{t}{8a} (\hat{U}+\hat{U}^{\dag})\otimes\hat{P}_r} e^{i\frac{t}{8a}i(\hat{U}-\hat{U}^{\dag})\otimes\hat{\tilde{P}}_r}\nonumber \\
    &= [{\mathcal F}^\dag \otimes \hat{U}_{Bell}e^{i\frac{t}{8a}(\hat{D}_c\otimes \hat{D}^f)}{\mathcal F} \otimes \hat{U}_{Bell}^\dag][{\mathcal F}^\dag \otimes (\hat{S}_r^{f\dag}\hat{U}_{Bell}) e^{i\frac{t}{8a}(\hat{D}_s \otimes \hat{D}^f)} {\mathcal F} \otimes   (\hat{U}_{Bell}^\dag \hat{S}_r^f)] \nonumber \\
    &= {\mathcal F}^\dag \otimes \hat{U}_{Bell}e^{i\frac{t}{8a}(\hat{D}_c\otimes \hat{D}^f)}\hat{I} \otimes (\hat{U}_{Bell}^\dag \hat{S}_r^{f\dag}\hat{U}_{Bell}) e^{i\frac{t}{8a}(\hat{D}_s \otimes \hat{D}^f)} {\mathcal F} \otimes   (\hat{U}_{Bell}^\dag \hat{S}_r^f),
\end{align}
where the first equality is due to $e^{it \hat{U}\hat{D}\hat{U}^\dag} = \hat{U}e^{it \hat{D}}\hat{U}^\dag$, with $\hat{U}$ and $\hat{D}$ being a unitary and a diagonal operator, respectively. Since the circuit implementation of the QFT and its inverse is known \cite{nielsen2002quantum}, the implementation of the kinetic term hinges upon the syntheses of $e^{i\frac{t}{8a}(\hat{D}_c\otimes \hat{D}^f)}$ and $e^{i\frac{t}{8a}(\hat{D}_s \otimes \hat{D}^f)}$. Expanding $\hat{D}_c\otimes \hat{D}^f$, we obtain the relation
\begin{equation}
    e^{i\frac{t}{8a}(\hat{D}_c\otimes \hat{D}^f)} = \hat{I}\otimes \ketbra{00}{00} + e^{i\frac{t}{4a}\hat{D}_c} \otimes \ketbra{01}{01}
    + \hat{I}\otimes \ketbra{10}{10} + e^{-i\frac{t}{4a}\hat{D}_c} \otimes \ketbra{11}{11},
\label{eq:c_Uc}
\end{equation}
which can be implemented via applications of $\hat{U}_{c}\equiv e^{i\frac{t}{4a}\hat{D}_c}$ controlled upon the fermionic state. Similarly, $e^{i\frac{t}{4a}(\hat{D}_s \otimes \hat{D}^f)}$ can be implemented by controlled applications of $\hat{U}_{s}\equiv e^{i\frac{t}{4a}\hat{D}_s}$. 

We next show that both $\hat{U}_{c}$ and $\hat{U}_{s}$ can be implemented efficiently via Quantum Signal Processing (QSP) \cite{low2017optimal}. First, we rewrite them in the form of $\sum_{\lambda} e^{-i\tau \sin(\theta_{\lambda})}\ketbra{u_{\lambda}}{u_{\lambda}}$:
\begin{align}
    \hat{U}_{s} &= \sum_{k}e^{-i\frac{t}{2a}\sin(\frac{2\pi k}{2^\eta})}\ketbra{k}{k}, \\
    \hat{U}_{c} &= \sum_{k}e^{-i\frac{t}{2a}\sin(\frac{2\pi k}{2^\eta} + \frac{\pi}{2})}\ketbra{k}{k},
\end{align}
where $\tau = \frac{t}{2a}$. QSP implements an operator in such form with $N$ queries of an oracle $\hat{U}_{\phi}$, which is defined as
\begin{align}
    \hat{U}_{\phi} &= (e^{-i\phi \hat{Z}/2} \otimes \hat{I})\hat{U}_0 (e^{i\phi \hat{Z}/2} \otimes \hat{I}), \label{eq:QSP_phase}\\
    \hat{U}_0 &= \ketbra{+}{+} \otimes \hat{I} + \ketbra{-}{-}\otimes \hat{W}, \label{eq:QSP_oracle}\\
    \hat{W} &= \sum_{\lambda} e^{i\theta_\lambda}\ketbra{u_{\lambda}}{u_\lambda}.
\end{align}
In this case, for both $\hat{U}_c$ and $\hat{U}_s$, the $\hat{W}$ oracle is given by
\begin{equation}
    \hat{W} = \sum_{k=0}^{2^\eta -1} e^{i\frac{2\pi k}{2^\eta}} \ketbra{k}{k},
\end{equation}
which can be implemented with an $\eta-$qubit phase gradient at the cost of $4\eta + O(1)$ T gates. The $\hat{W}$ oracle needs to be controlled by three qubits, one due to the QSP oracle in (\ref{eq:QSP_oracle}) and two from the fermionic register. See (\ref{eq:c_Uc}). This can be accomplished by applying a relative phase Toffoli on an ancilla, controlled by the three qubits, then applying a $\hat{W}$ controlled by the ancilla, and then uncomputing the Toffoli gate. The controlled-$\hat{W}$ operation can be effected by a controlled phase gradient, which can be synthesized with $8\eta+O(1)$ T gates \cite{nam2020approximate}, and each triply-controlled relative-phase Toffoli costs $8$ T gates \cite{maslov2016advantages}, which adds a constant overhead to the triply-controlled-$\hat{W}$. 

Now we consider the simulation of an individual kinetic term, which is a diagonal norm-one Hamiltonian. In this case, the QSP query complexity depends on only the simulation length $\tau$, and error $\epsilon$. Since the simulation length is fixed to be $\tau=\frac{t}{2a}=\frac{\pi}{g^2 2^M}$, which is much smaller than one, as required by the phase gradient operation for the electric term, we expect that a small number of query is enough to implement $\hat{\tilde{U}}_{g}$ with $g\in\{c,s\}$. In particular, since our error is bounded by $\frac{4\tau^q}{2^q q!}$, where $t\leq q-1$ and the number of queries is $2(q-1)$ \cite{low2017optimal}, with small $\tau$, $q=2$ is likely sufficient.

We emphasize in passing that, when the cutoff is severe, the effects of the unwanted periodic wrapping terms in our implementation of $\hat{U}_r$ and $\hat{U}_r^\dag$, which connect $\ket{-\Lambda}=\ket{00...0}, \ket{\Lambda-1}=\ket{11...1}$, are no longer negligible. For instance, an application of the kinetic term, which can be expressed as $e^{\frac{it}{2a}(\hat{U}_r\hat{\sigma}_r^-\hat{\sigma}_{r+1}^+ + h.c.)}$, on a link in the state $\ket{\Lambda-1}=\ket{11...1}$ will introduce a superposition between $\ket{-\Lambda}$ and $\ket{\Lambda-2}$ because $\hat{U}_r \ket{\Lambda-1} = \ket{-\Lambda}$ and $\hat{U}_r^\dag\ket{\Lambda-1} = \ket{\Lambda-2}$. However, in the original U(1) LGT, $\ket{-\Lambda}$ will not arise because $\ket{\Lambda-1}$ is in the null space of $\hat{U}_r$. This effect can be reversed by the evolution $e^{\frac{-it}{2a}(\otimes_{i=0}^{\eta-1}\hat{\sigma}_i^- \hat{\sigma}_r^-\hat{\sigma}_{r+1}^+ + h.c.)}$. This can be implemented using $2(\eta + 1)$ CNOTs, two Hadamard gates, and two $R_z$ gates.

\subsubsection{Magnetic term \texorpdfstring{$e^{i\hat{L}_{\vec{n}}^{(B)}t}$}{}}
\label{sec:U1_mag}

Here, we extend the two methods used to implement the kinetic term to implement the magnetic term. Once again, we drop the location indices for brevity.

{\bf Method 1: Block-diagonal decomposition ---}
First, we decompose the ladder operators into the following off-diagonal operators:
\begin{align}
    \hat{U} = \hat{R} + \hat{U}^{\dag}\hat{R}\hat{U}, \label{eq:U1_U_decomp}\\
    \hat{U}^\dag = \hat{R}^\dag + \hat{U}^{\dag}\hat{R}^\dag\hat{U},
\end{align}
where $\hat{R} = \hat{I}\otimes \hat{I} \otimes ...\otimes \hat{\sigma}^+$ and $\hat{U}^{\dag}\hat{R}\hat{U}$ are raising operators for $\ket{E}$ when $E$ is odd and even, respectively, and similarly, $\hat{R}^\dag = \hat{I}\otimes \hat{I} ...\otimes \hat{\sigma}^-$ and $\hat{U}^{\dag}\hat{R}^\dag\hat{U}$ are lowering operators for $\ket{E}$ when $E$ is even and odd, respectively.

Defining 
\begin{equation}
    \hat{R}_{\Box} = \hat{R}\hat{R}\hat{R}^\dag\hat{R}^\dag+\hat{R}^\dag\hat{R}^\dag\hat{R}\hat{R},
\end{equation}
each plaquette operator in $\hat{L}^{(B)}$ can then be expressed as
\begin{align}
    &\quad \hat{U}^{(1)} \hat{U}^{(2)} \hat{U}^{(3)\dag} \hat{U}^{(4)\dag} + \hat{U}^{{(1)}\dag} \hat{U}^{{(2)}\dag} \hat{U}^{(3)} \hat{U}^{(4)} \nonumber \\
    &= (\hat{R} + \hat{U}^{{(1)}\dag}\hat{R}\hat{U}^{(1)})(\hat{R} + \hat{U}^{(2)\dag}\hat{R}\hat{U}^{(2)})(\hat{R}^\dag + \hat{U}^{(3)\dag}\hat{R}^\dag\hat{U}^{(3)}) (\hat{R}^\dag + \hat{U}^{(4)\dag}\hat{R}^\dag\hat{U}^{(4)}) + h.c. \nonumber \\
    &= \sum_{i,j,k,l=0}^1 \hat{U}^{(1)^\dag i}\hat{U}^{(2)\dag j} \hat{U}^{(3)\dag k} \hat{U}^{(4)^\dag l} \hat{R}_{\Box}\hat{U}^{(1) i} \hat{U}^{(2) j} \hat{U}^{(3) k}\hat{U}^{(4) l},
\end{align}
where the superscripts in the parentheses are used to denote the links around a plaquette, and the sum is over the powers to which the gauge field ladder operators are raised. In order to maximize the cancellation of $\hat{U}$ and $\hat{U}^\dag$ in the plaquette operator's first-order Trotter evolution, we order the sum using the Gray code. The Gray code is a binary encoding where two successive values differ in only one bit. In particular, we first label each term by a vector of the ladder operators' powers $(i,j,k,l)$, and then, arrange the labels in the Gray code ordering of the integers $0-15$. The first-order Trotterization of the evolution is thus given by
\begin{align}
    &\quad e^{-i\frac{t}{2a^{4-d}g^2}(\hat{U}^{(1)} \hat{U}^{(2)} \hat{U}^{(3)\dag} \hat{U}^{(4)\dag} + h.c.)} \nonumber \\
    &\approx \prod_{(i,j,k,l)=GC(0)}^{GC(15)}\hat{U}^{(1)^\dag i}\hat{U}^{(2)\dag j} \hat{U}^{(3)\dag k} \hat{U}^{(4)^\dag l} e^{-i\frac{t}{2a^{4-d}g^2} \hat{R}_{\Box}} \hat{U}^{(1) i} \hat{U}^{(2) j} \hat{U}^{(3) k}\hat{U}^{(4) l},
    \label{eq:U1_mag_trot}
\end{align}
where $GC(n)$ is the Gray code encoding of an integer $n$. There are $16$ sub-evolutions, and between each consecutive sub-evolutions, there will be one $\hat{U}$ or $\hat{U}^\dag$ that needs to be implemented in this ordering. Including the one $\hat{U}^\dag$ operator at the end, there will be $16$ $\hat{U}$ and $\hat{U}^\dag$ operators in the evolution. $\hat{U}$ and $\hat{U}^\dag$ can be implemented as an $\eta-$qubit binary incrementer and decrementer, respectively, each of which costs $4(\eta-2)$ T gates and $\eta$ reusable ancilla qubits \cite{shaw2020quantum}. Furthermore, each $e^{-i\frac{t}{2a^{4-d}g^2} \hat{R}_{\Box}}$ operator costs two relative-phase triply-controlled Toffoli gates, which take $16$ T gates in total to construct, and two $R_z$ gates \cite{wang2020resource}.

Briefly, any operator $U$ that is of the form $e^{-i(\theta \otimes_k \hat{\sigma}_k+h.c.)/2}$, where $\hat{\sigma}_k \in \{\hat{\sigma}^+,\hat{\sigma}^-\}$ and $k=1,2,..,k_{\max}$ can be diagonalized by $C$ as $CDC^{-1}$, where $C$ is composed of a CNOT network with $k-1$ CNOTs and a Hadamard gate and $D$ is a diagonal operator. Without loss of generality, consider a case where there are $p$ $\hat{\sigma}^+$'s and $k_{\max}-p$ $\hat{\sigma}^-$'s. Further pick an arbitrary qubit index, say, $k_{\max}$ and $\hat{\sigma}_{k_{\max}} = \hat{\sigma}^+$. We apply NOTs to $k_{\max}-p$ $\hat{\sigma}^-$, all controlled on the same qubit $k_{\max}$. We now pick one of the $k_{\max}-p$ qubits with $\hat{\sigma}^-$, say $k'$. We apply NOTs to the $\hat{\sigma}^+$ qubits except for the $k_{\max}$ qubit, of which there are $p-1$, all controlled on $k'$. Applying a Hadamard gate on the $k_{\max}$'th qubit diagonalizes $U$, i.e., $D$ would now be an $k_{\max}-1$-controlled $R_z$ gate with target on $k_{\max}$. A standard method to detach the controls results in two uncontrolled $R_z$ gates.

Note the $R_z$ gates have the same rotation angles, up to a sign, for each and every plaquette. This means, once again, just as in mass and kinetic terms, we can use the weight-sum trick to reduce the number of $R_z$ gates to be implemented. Since the magnetic terms for nearest-neighbors act on overlapping links, their evolutions are difficult to implement in parallel. Similar to the case of the kinetic term, we implement the magnetic evolutions for the even and odd sites along each two-dimensional plane, separately, in parallel.

Without loss of generality, we assume the number of odd and even sites are $\frac{L^d}{2}$ each. Hence, for each two-dimensional plane, there are $\frac{L^d}{2}$ plaquette evolutions to apply in parallel for even and odd sites, respectively. Consider just the even sites on one plane, the relative-phase triply-controlled Toffoli gates contribute $\frac{L^d}{2} \cdot 16 \cdot 16$ T gates. Further, there are $L^d$ equal-angle $R_z$ gates to implement in parallel for $16$ times. Once again, we employ the weight-sum trick to effect the equal-angle $R_z$ gates. The first step of computing Weight$(L^d)$ costs $4(L^d-\text{Weight}(L^d))$ T gates and $L^d -\text{Weight}(L^d)$ ancilla qubits. Then, we apply $\lfloor \log(L^d) +1 \rfloor$ $R_z$ rotations for $16$ times. Lastly, the incrementers and decrementers cost $\frac{L^d}{2}\cdot 16\cdot 4(\eta-2)$ T gates and $\eta$ ancilla qubits \cite{shaw2020quantum}. 

{\bf Method 2: Diagonalization of $\hat{U}$ ---}
We can diagonalize the plaquette operator $\hat{L}^{(B)}$ by taking tensor products of Eqs. (\ref{eq:udag_qft}) and (\ref{eq:u_qft}), i.e.,
\begin{align}
    &\quad {\mathcal F}^{\otimes 4}(\hat{U}^{(1)} \hat{U}^{(2)} \hat{U}^{(3)\dag} \hat{U}^{(4)\dag}){\mathcal F}^{\dag \otimes 4} \nonumber \\ 
    &={\mathcal F}(\hat{U}^{(1)}){\mathcal F}^{\dag}\otimes {\mathcal F}(\hat{U}^{(2)}){\mathcal F}^{\dag} \otimes {\mathcal F}(\hat{U}^{(3)\dag}){\mathcal F}^{\dag} \otimes {\mathcal F}(\hat{U}^{(4)\dag}){\mathcal F}^{\dag} \nonumber \\
    &= \sum_{k_1, k_2, k_3, k_4=0}^{2^\eta-1} e^{\frac{-2\pi i k_1}{2^\eta}}\ketbra{k_1}{k_1} \otimes e^{\frac{-2\pi i k_2}{2^\eta}}\ketbra{k_2}{k_2} \otimes e^{\frac{2\pi i k_3}{2^\eta}}\ketbra{k_3}{k_3}\otimes e^{\frac{2\pi i k_4}{2^\eta}}\ketbra{k_4}{k_4} \nonumber \\
    &= \sum_{k_1, k_2, k_3, k_4=0}^{2^\eta-1} e^{\frac{-2\pi i (k_1+k_2 - k_3 - k_4)}{2^\eta}} \ketbra{k_1,k_2,k_3,k_4}{k_1,k_2,k_3,k_4}.
\end{align}
Again, by linearity,
\begin{align}
    &{\mathcal F}^{\otimes 4}(\hat{U}^{(1)} \hat{U}^{(2)} \hat{U}^{(3)\dag} \hat{U}^{(4)\dag} + \hat{U}^{(1)\dag} \hat{U}^{(2)\dag} \hat{U}^{(3)} \hat{U}^{(4)}){\mathcal F}^{\dag \otimes 4} \nonumber \\
    &= \sum_{k_1, k_2, k_3, k_4=0}^{2^\eta-1} 2\cos(\frac{2\pi (k_1+k_2 - k_3 - k_4)}{2^\eta}) \ketbra{k_1,k_2,k_3,k_4}{k_1,k_2,k_3,k_4} \equiv  D_{\Box} .
\end{align}

Now we show that the evolution operator implied by the magnetic term, all of which are of the form, 
\begin{equation}
    e^{-i\frac{t}{2a^{4-d}g^2}(\hat{U}^{(1)} \hat{U}^{(2)} \hat{U}^{(3)\dag} \hat{U}^{(4)\dag} + h.c.)}, \nonumber
\end{equation}
where $h.c.$ denotes the Hermitian conjugate, can also be diagonalized by a tensor product of four QFTs. Taylor expanding the evolution operator, we get
\begin{align}
    &\quad e^{-i\frac{t}{2a^{4-d}g^2}(\hat{U}^{(1)} \hat{U}^{(2)} \hat{U}^{(3)\dag} \hat{U}^{(4)\dag} + h.c.)} \nonumber \\
    &= 1 +(-i\frac{t}{2a^{4-d}g^2}){\mathcal F}^{\dag\otimes 4} \hat{D}_{\Box} {\mathcal F}^{ \otimes 4}  +\frac{1}{2} (-i\frac{t}{2a^{4-d}g^2})^2{\mathcal F}^{\dag\otimes 4} \hat{D}_{\Box} {\mathcal F}^{\otimes 4} {\mathcal F}^{\dag \otimes 4} \hat{D}_{\Box} {\mathcal F}^{ \otimes 4} + ... \nonumber \\
    &= 1+ (-i\frac{t}{2a^{4-d}g^2}){\mathcal F}^{\dag\otimes 4} \hat{D}_{\Box} {\mathcal F}^{ \otimes 4} +\frac{1}{2} (-i\frac{t}{2a^{4-d}g^2})^2 {\mathcal F}^{\dag\otimes 4} \hat{D}_{\Box}^2 {\mathcal F}^{\otimes 4} +\frac{1}{3!}(-i\frac{t}{2a^{4-d}g^2})^3 {\mathcal F}^{\dag\otimes 4} \hat{D}_{\Box}^3 {\mathcal F}^{ \otimes 4} + ... \nonumber \\
    &={\mathcal F}^{\dag \otimes 4} e^{-i\frac{t}{2a^{4-d}g^2}\hat{D}_{\Box}}{\mathcal F}^{ \otimes 4} \equiv {\mathcal F}^{\dag \otimes 4} \hat{U}_{\Box} {\mathcal F}^{ \otimes 4}.
\label{eq:u1_plaq}
\end{align}
Since the circuit implementation of the QFT and its inverse is known \cite{nielsen2002quantum}, all that remains is to find a circuit that implements $\hat{U}_{\Box}$.

To implement $\hat{U}_{\Box}$, 
we consider its action on the input state $\ket{k_1,k_2,k_3,k_4}$, i.e.,
\begin{equation}
\hat{U}_{\Box}\ket{k_1,k_2,k_3,k_4} = e^{-i\frac{t}{2a^{4-d}g^2}\hat{D}_{\Box}}\ket{k_1,k_2,k_3,k_4}
=e^{-i\frac{t}{a^{4-d}g^2}\cos(\frac{2\pi (k_1+k_2 - k_3 - k_4)}{2^\eta})} \ket{k_1,k_2,k_3,k_4}.
\label{eq:diagonalgate}
\end{equation}
Similar to the kinetic term implementation, we use QSP to implement the evolution of the plaquette term. As explained in the kinetic term section, it boils down to synthesizing a controlled $\hat{W}$ operator, which is given by
\begin{equation}
    \hat{W} = \sum_{k_1,k_2,k_3,k_4}e^{i \frac{2\pi (k_1+k_2 - k_3 - k_4)}{2^\eta}} \ketbra{k_1,k_2,k_3,k_4}{k_1,k_2,k_3,k_4}.
\end{equation}
The controlled $\hat{W}$ operator can be effected by first computing $k_1+k_2 - k_3 - k_4$ into an ancilla register, then performing a controlled phase gradient operation on the ancilla, and finally uncomputing the ancilla. We can compute $k_1+k_2 - k_3 - k_4$ using two out-of-place $\eta-$bit adders and one out-of-place $(\eta+1)$-bit adder proposed in \cite{draper2006logarithmic}. The two $\eta-$bit adders compute $\ket{k_1}\ket{k_2}\ket{k_3}\ket{k_4}\mapsto \ket{k_1}\ket{k_2}\ket{k_3}\ket{k_4}\ket{k_1+k_2}\ket{k_3+k_4}$, using $2\cdot (5\eta-3\lfloor \log(\eta) \rfloor)-4)$ Toffoli gates and at most $\eta- \lfloor \log(\eta) \rfloor - 1$ ancillas. The $(\eta+1)$-bit adder computes $\ket{k_1}\ket{k_2}\ket{k_3}\ket{k_4}\ket{k_1+k_2}\ket{k_3+k_4}$ $\mapsto$ $\ket{k_1}\ket{k_2}$ $\ket{k_3}\ket{k_4}$ $\ket{k_1+k_2}\ket{k_3+k_4}\ket{k_1+k_2-k_3-k_4}$, where $\ket{k_1+k_2}$ and $\ket{k_3+k_4}$ are $(\eta+1)-$qubit registers, and $\ket{k_1+k_2-k_3-k_4}$ is an $(\eta+2)-$qubit registers. This operation requires $5(\eta+1)-3\lfloor \log(\eta+1) \rfloor -4$ Toffoli gates and $(\eta+1)-\lfloor \log(\eta+1) \rfloor-1$ ancillas. As such, the computation and uncomputation of $k_1+k_2 - k_3 - k_4$ each costs $4\cdot (15\eta - 3\lfloor \log(\eta+1) \rfloor - 6\lfloor \log(\eta) \rfloor - 7)$ T gates, using the Toffoli construction in \cite{jones2013low}, where each Toffoli gate costs four T gates and an ancilla qubit. As such, $\eta-\lfloor \log(\eta+1) \rfloor$ workspace ancilla qubits are used, and $3\eta+4$ qubits are needed to store the outputs $\ket{k_1+k_2-k_3-k_4}\ket{k_3+k_4}$.

We consider the simulation of an individual magnetic term, which is a diagonal norm-one Hamiltonian. As in the case of kinetic term, the QSP query complexity is determined by the simulation length $\tau = \frac{t}{a^{4-d}g^2}$, shown in (\ref{eq:diagonalgate}). In order to keep the trotter error small, $t$, and hence $\frac{t}{a^{4-d}g^2}$, must be much smaller than one. Once again, we expect that two queries are enough to implement $\hat{\tilde{U}}_{\Box}$. Therefore, the evolution of a magnetic term requires three serial $R_z$ gates due to the concatenated phase oracle, which is an improvement over method $1$, which needs $16$ serial pairs of parallel $R_z$ gates. As for the entire magnetic Hamiltonian evolution, we adopt the same parallelization strategy as in method $1$, and achieve factors of $32/3$ and $16/3$ improvement in the total number and layers of $R_z$ gates, respectively.

As in the kinetic term, when the cutoff is severe, the effects of the unwanted periodic wrapping terms of the operators $\hat{U}^{(\dag)}$, which connect $\ket{-\Lambda}=\ket{00...0}, \ket{\Lambda-1}=\ket{11...1}$, are no longer negligible. Suppose we want to undo the periodic wrapping effect on the $i$th link of the plaquette. Then, we implement the evolution $e^{\frac{it}{2a^{4-d}g^2}(\otimes_{q=0}^{\eta-1}\hat{\sigma}_q^- \hat{U}^{(j)}\hat{U}^{(k)\dag }\hat{U}^{(l)\dag} + h.c.)}$, where the Pauli ladder operators act on the $i$th link. This evolution can be implemented using both the block-diagonal decomposition method and the diagonalization method. Using the block-diagonal method, the evolution can be implemented by
\begin{align}
    &\quad e^{\frac{it}{2a^{4-d}g^2}(\otimes_{q=0}^{\eta-1}\hat{\sigma}_q^- \hat{U}^{(j)}\hat{U}^{(k)^\dag} \hat{U}^{(l)^\dag} + h.c.)} \nonumber \\
    &\approx \prod_{(\alpha,\beta,\gamma)=GC(0)}^{GC(7)} \hat{U}^{(j)\dag\alpha}\hat{U}^{(k)\dag\beta} \hat{U}^{(l)\dag\gamma}   e^{\frac{it}{2a^{4-d}g^2}(\otimes_{q=0}^{\eta-1}\hat{\sigma}_q^-\hat{R}\hat{R}^\dag \hat{R}^\dag + h.c.)} \hat{U}^{(j)\alpha}\hat{U}^{(k)\beta} \hat{U}^{(l)\gamma},
\end{align}
where $\hat{R}$ is defined in (\ref{eq:U1_U_decomp}), and $e^{\frac{it}{2a^{4-d}g^2}(\otimes_{q=0}^{\eta-1}\hat{\sigma}_q^-\hat{R}\hat{R}^\dag \hat{R}^\dag + h.c.)}$ can be effected using $2(\eta + 2)$ CNOTs, two Hamadard gates and two $R_z$ gates. Using the diagonalization method, we first diagonalize the evolution by applying CNOTs and Hadamards on the $i$th link to diagonalize the Pauli ladder operators. Next, we apply QFT on links $j,k,l$ to diagonalize the remaining part of the evolution, and obtain
\begin{align}
    &\quad \sum_{m_i,m_j,m_k,m_l}e^{\frac{it}{2a^{4-d}g^2}[(-1)^{b_0}\prod_{q=1}^{\eta-1}b_q \cdot 2\cos(\frac{2\pi (m_j-m_k-m_l)}{2^{\eta}}) ]}\ketbra{m_i,m_j,m_k,m_l}{m_i,m_j,m_k,m_l} \nonumber \\
    &= \ketbra{11...1}{11...1}\otimes \sum_{m_j,m_k,m_l}e^{\frac{-it}{a^{4-d}g^2}\cos(\frac{2\pi (m_j-m_k-m_l)}{2^{\eta}})}\ketbra{m_j,m_k,m_l}{m_j,m_k,m_l}\nonumber \\
    &\quad + \ketbra{01...1}{01...1}\otimes \sum_{m_j,m_k,m_l}e^{\frac{it}{a^{4-d}g^2}\cos(\frac{2\pi (m_j-m_k-m_l)}{2^{\eta}})}\ketbra{m_j,m_k,m_l}{m_j,m_k,m_l},
\end{align}
where $b_q$ is the bit-value of the $q$th qubit of the $i$th link register $\ket{m_i}$. Once again, the evolution can be implemented using two controlled QSP, as in the diagonalization method for the kinetic term. Now suppose we are to undo the periodic wrapping effect on two links, e.g. $j$th and $k$th links. Then, we implement the evolution $e^{\frac{it}{2a^{4-d}g^2} (\hat{U}^{(i)}\otimes_{q=0}^{\eta-1}\hat{\sigma}_q^- \otimes_{r=0}^{\eta-1}\hat{\sigma}_r^+ \hat{U}^{(l)\dag} + h.c.)}$, using either the block-diagonal or diagonalization method. This technique can be straightforwardly generalized to undo the periodic wrapping effect on all combinations of links on each plaquette.

\subsection{Resource requirement estimates}
\label{sec:U1_res}

In this section, we analyze the algorithmic and synthesis errors for our simulations. In Sec.~\ref{subsubsec:Trotter_U1} we compute the algorithmic error for the Suzuki-Trotter formula for our U(1) Hamiltonian. Therein we show our result first, then show a full derivation of it for completeness. In Sec.~\ref{subsubsec:Synth_U1} we compute the $R_z$ synthesis error. In Sec.~\ref{subsubsec:Analysis_U1} we combine the two errors discussed in Secs.~\ref{subsubsec:Trotter_U1} and~\ref{subsubsec:Synth_U1} to report the gate complexity and ancilla requirements.

\subsubsection{Trotter errors}
\label{subsubsec:Trotter_U1}

Simulating quantum dynamics can be boiled down to compiling the evolution generated by a Hamiltonian into a sequence of implementable quantum gates. Commonly used efficient quantum simulation algorithms include Trotter-Suzuki product formulas (PF) \cite{suzuki1991general, lloyd1996universal}, linear combinations of unitaries (LCU) \cite{childs2012hamiltonian, berry2015simulating}, and quantum signal processing (QSP) \cite{low2017optimal}. Here, we choose to employ the second-order PF over LCU and QSP. The reasons for this choice is given in \cite{shaw2020quantum}, and we import them here for the convenience of the readers. The main reason is that the second-order PF achieves a quadratically better scaling, up to polylogarithmic factors, in the electric truncation $\Lambda$, when compared to the other algorithms.

The first step for any PF algorithms is to divide up a total simulation time $T$ into $r$ segments. Then, the evolution generated by a Hamiltonian $\hat{H}$ can be expressed as $e^{-i\hat{H}T} = (e^{-i\hat{H}T/r})^r \equiv (e^{-i\hat{H}t})^r$. Further, the Hamiltonian $\hat{H}$ is decomposed into a sum of simpler Hamiltonians, $\hat{H}=\sum_{j=1}^{l}\hat{H}_j$, where each $e^{-i\hat{H}_j t}$ can be simulated with an efficient quantum circuit. The first-order PF is given by
\begin{equation}
    \hat{U}_1(t) = \prod_{j=1}^{l} e^{-i\hat{H}_j t}.
\end{equation}
The error is bounded by \cite{childs2019theory}
\begin{equation}
    || e^{-i\hat{H}T} - \hat{U}_1^r(t) || \leq \frac{1}{2}\sum_{j}|| [\sum_{i>j}\hat{H}_i, \hat{H}_j] ||\frac{T^2}{r}.
\end{equation}
Therefore, for a given spectral-norm error $\epsilon$, the value of $r$ grows as $O(T^2/\epsilon)$. This scaling can be improved by using a higher-order PF. For instance, consider the second-order PF, defined as
\begin{equation}
    \hat{U}_2(t) = \prod_{j=1}^{l} e^{-i\hat{H}_j t/2}\prod_{j=l}^{1} e^{-i\hat{H}_j t/2}.
    \label{eq:2nd_PF}
\end{equation}
In this case, the error bound is \cite{childs2019theory}
\begin{equation}
    || e^{-i\hat{H}t} - \hat{U}_2(t) || \leq \frac{1}{12}\sum_{i}|| [[\hat{H}_i, \sum_{j>i}\hat{H}_j],\hat{H}_i] ||t^3
    + \frac{1}{24}\sum_{i}|| [[\hat{H}_i, \sum_{j>i}\hat{H}_j],\sum_{k>i}\hat{H}_k] ||t^3,
    \label{eq:U1_trotter_err}
\end{equation}
which leads to a Trotter-step $r$ scaling of $O(T^{3/2}/\epsilon^{1/2})$. Further improvements can be achieved with high-order PF, which can be constructed recursively \cite{suzuki1991general}. However, it is reported in \cite{shaw2020quantum} the higher-order PFs are rarely preferred for quantum simulations due to the fact that the second-order formula can actually cost fewer computational resources and outperform asymptotically more efficient methods such as LCU and QSP \cite{childs2018toward}. Furthermore, it is reported in \cite{shaw2020quantum}, compared to the error bounds for the first- and second-order PF, those for higher-order PF are unlikely to be tight and are more difficult to compute \cite{childs2019theory}. 

Therefore, we as well choose to use the second-order PF as our simulation algorithm, just as in \cite{shaw2020quantum}, and evaluate the commutator bound for the error given in (\ref{eq:U1_trotter_err}). The result is
\begin{align}
    || e^{-i\hat{H}T}-\hat{U}_2^r(t)||
    \leq r\left(\frac{T}{r}\right)^3 \rho
    \equiv \epsilon_{Trotter},
\end{align}
where
\begin{align}
    \rho &= \frac{1}{12}[\frac{4 dL^d m^2}{a} +dL^d\frac{g^4}{4a^{2d-3}}(4\Lambda^2-1)+\frac{2L^d d(d-1)g^2(2\Lambda-1)^2}{a^d}+\frac{4d(d-1)L^d}{a^{6-d}g^2} \nonumber \\
    &\quad +\frac{(8d^2-3d)L^d + (16d^2-8d) L^{d-1}}{2a^3}] +\frac{1}{24}[\frac{mg^2}{2a^{d-1}} (2\Lambda - 1)dL^d+\frac{mL^d(16d^2-8d)}{a^2} \nonumber \\
    &\quad + \frac{(4d^2-2d)L^d g^2 (2\Lambda +1)}{a^d}+\frac{mL^d 8d(d-1)}{g^2a^{5-d}}+\frac{2d(d-1)L^d}{a^3}(2\Lambda + 1)+\frac{L^d d(d-1) (16\Lambda - 8)}{a^3} \nonumber \\
    &\quad +\frac{L^d d(d-1)(8d - 11)(4\Lambda -2)}{g^2a^{6-d}}+\frac{L^d}{a^3}(\frac{32}{3} d^3 - 4 d^2 + \frac{11}{6} d) + \frac{L^{d-1}}{a^3}(\frac{160}{3}d^3 - 20 d^2 - \frac{16}{3}d)\nonumber\\ 
    &\quad +\frac{L^{d-2}}{a^3}(2d^2-2d)+\frac{2L^{d-3}}{3a^3}(d^3-3d^2+2d) 
    +\frac{L^d}{g^2 a^{6-d}}(48d^3-102d^2+54d) \nonumber \\
    &\quad + \frac{L^{d-1}}{g^2 a^{6-d}}(96d^3-232d^2+136d) +\frac{L^d}{g^2a^{6-d}}(16 d^3 - 10 d^2 -6 d)  + \frac{L^{d-1}}{g^2a^{6-d}}(32 d^3 - 56 d^2+24d) \nonumber \\
    &\quad +\frac{L^d(224 d^3-544 d^2 + 320 d)}{a^{9-2d} g^4}]+L^d\frac{d(d-1)}{2}\frac{8}{a^{12-3d} g^6}.
\end{align}

For completeness, we show below a full derivation of the results shown above. Readers interested in how the results compare with the size of the synthesis error and how, together, they affect our simulation gate complexity should proceed to Secs.~\ref{subsubsec:Synth_U1} and~\ref{subsubsec:Analysis_U1}.

We start our derivation by first ordering the terms in the Hamiltonian $\hat{H}$. Consider an ordered list $\{ \hat{H}_x \}_{x=1}^{x_{\max}}$ of $x_{\max}$-many individual Hamiltonian terms $H_x$, i.e.,
\begin{multline}
    \{\hat{H}_x\}_{x=1}^{d^2+7d+2}=\{ \sum_{\vec{n}}\hat{D}^{(M)}_{\vec{n}},\sum_{\vec{n}}\hat{D}^{(E)}_{\vec{n}},\sum_{\{ \vec{n}_{e,1}\}}\frac{\hat{T}^{(1)}_{\vec{n}_{e,1}}+\hat{T}^{(4)}_{\vec{n}_{e,1}}}{2}
    ,\sum_{\{ \vec{n}_{e,1}\}}\frac{\hat{T}^{(2)}_{\vec{n}_{e,1}}+\hat{T}^{(3)}_{\vec{n}_{e,1}}}{2},\sum_{\{ \vec{n}_{o,1}\}}\frac{\hat{T}^{(1)}_{\vec{n}_{o,1}}+\hat{T}^{(4)}_{\vec{n}_{o,1}}}{2}\\
    ,\sum_{\{ \vec{n}_{o,1}\}}\frac{\hat{T}^{(2)}_{\vec{n}_{o,1}}+\hat{T}^{(3)}_{\vec{n}_{o,1}}}{2},...,\sum_{\{ \vec{n}_{e,2d}\}}\frac{\hat{T}^{(1)}_{\vec{n}_{e,2d}}+\hat{T}^{(4)}_{\vec{n}_{e,2d}}}{2} 
     ,\sum_{\{ \vec{n}_{e,2d}\}}\frac{\hat{T}^{(2)}_{\vec{n}_{e,2d}}+\hat{T}^{(3)}_{\vec{n}_{e,2d}}}{2},\sum_{\{ \vec{n}_{o,2d}\}}\frac{\hat{T}^{(1)}_{\vec{n}_{o,2d}}+\hat{T}^{(4)}_{\vec{n}_{o,2d}}}{2}\\
    ,\sum_{\{ \vec{n}_{o,2d}\}}\frac{\hat{T}^{(2)}_{\vec{n}_{o,2d}}+\hat{T}^{(3)}_{\vec{n}_{o,2d}}}{2},\sum_{\{ \vec{n}_{e,1}\}}\hat{L}^{(B)}_{\vec{n}_{e,1}},\sum_{\{ \vec{n}_{o,1}\}}\hat{L}^{(B)}_{\vec{n}_{o,1}} 
    ,...,\sum_{\{ \vec{n}_{e,\frac{d(d-1)}{2}}\}}\hat{L}^{(B)}_{\vec{n}_{e,\frac{d(d-1)}{2}}},\sum_{\{ \vec{n}_{o,\frac{d(d-1)}{2}}\}}\hat{L}^{(B)}_{\vec{n}_{o,\frac{d(d-1)}{2}}}\},
    \label{eq:order_U1}
\end{multline}
where the grouping of terms as they appear in the list is motivated by the commutation so that each element in the list does not commute with at least one of the elements in the ordered set. For convenience, we group all the mass terms and electric terms (the first two in the list) into one term each, since the grouping incurs no Trotter error as the individual mass / electric terms commute with one another. The set of subindices on each kinetic term $\{\vec{n}_{i,p} \}$ denotes the set of even or odd sites, i.e., $i \in \{e, o\}$, and the different directions, i.e., $p = 1,2, .., 2d$. Note $p=1,...,d$ are the directions for the edge terms, and $p=d+1, ..., 2d$ are the directions for the bulk terms. In relation to (\ref{eq:T1_U1}-\ref{eq:T4_U1}), each kinetic term $\hat{T}^{(a)}_{\vec{n}_{i,p}}$, where $a\in \{1,2,3,4\}$, acts on a link in direction $p$, which originate from either an odd or even site, depending on $i$, in either the bulk or edge, depending on the value of $p$. Furthermore, the sum of all kinetic terms is equivalent to $\sum_{\vec{n}}\hat{T}^{(K)}_{\vec{n}}$, defined in (\ref{eq:U1kinetic}). The set of subindices on each magnetic term $\{\vec{n}_{k,\Box} \}$, with $k \in \{e,o \}$, stands for the set of even or odd sites, respectively, on a two-dimensional plane denoted by $\Box$. In relation to (\ref{eq:U1_mag}), each $\Box$ corresponds to a specific pair of directions $(i,j)$. For a $d$-dimensional lattice, there are $\frac{d(d-1)}{2}$ two-dimensional planes and plaquette operators in the magnetic term at each site. Moreover, note that the sum of all magnetic terms yields $\sum_{\vec{n}} \hat{L}_{\vec{n}}^{(B)}$, as defined in (\ref{eq:U1_mag}).

We now proceed to evaluate the Trotter error incurred by the second-order PF. First, we focus on the first sum in (\ref{eq:U1_trotter_err}), i.e.,
\begin{align}
\label{eq:U1_trotter_err_1}
\sum_{i} \lvert\lvert [[\hat{H}_i, \sum_{j>i}\hat{H}_j],\hat{H}_i] \rvert\rvert \leq \sum_{k=1}^{7} \lvert\lvert C_{1,k} \rvert\rvert,
\end{align}
where
\begin{align}
C_{1,1} =& [[\sum_{\vec{n}}\hat{D}_{\vec{n}}^{(M)},\sum_{\vec{n}'}\hat{D}_{\vec{n}'}^{(E)}],\sum_{\vec{n}}\hat{D}_{\vec{n}}^{(M)}], \nonumber \\
C_{1,2} =& 
         [[\sum_{\vec{n}}\hat{D}_{\vec{n}}^{(M)}, \sum_{\vec{n}'}\hat{T}^{(K)}_{\vec{n}'} ],\sum_{\vec{n}}\hat{D}_{\vec{n}}^{(M)}], \nonumber \\
C_{1,3} =&  [[\sum_{\vec{n}}\hat{D}_{\vec{n}}^{(M)},\sum_{\vec{n}'}\hat{L}_{\vec{n}'}^{(B)}],\sum_{\vec{n}}\hat{D}_{\vec{n}}^{(M)}] , \nonumber \\
C_{1,4} =& [[\sum_{\vec{n}}\hat{D}_{\vec{n}}^{(E)}, \sum_{ \vec{n}'}\hat{T}^{(K)}_{\vec{n}'}] ,\sum_{\vec{n}}\hat{D}_{\vec{n}}^{(E)}] , \nonumber \\
C_{1,5} =&  [[\sum_{\vec{n}}\hat{D}_{\vec{n}}^{(E)},\sum_{\vec{n}'}\hat{L}_{\vec{n}'}^{(B)}],\sum_{\vec{n}}\hat{D}_{\vec{n}}^{(E)}] , \nonumber \\
C_{1,6} =& \sum_{i=e}^{o}\sum_{p=1}^{2d}\sum_{(a,b)          =(1,4)}^{(2,3)}
         [[\sum_{\{ \vec{n}_{i,p}\}}\frac{\hat{T}^{(a)}_{\vec{n}_{i,p}}+\hat{T}^{(b)}_{\vec{n}_{i,p}}}{2},\sum_{\vec{n}'}\hat{L}_{\vec{n}'}^{(B)} ],
         \sum_{\{ \vec{n}_{i,p}\}}\frac{\hat{T}^{(a)}_{\vec{n}_{i,p}}+\hat{T}^{(b)}_{\vec{n}_{i,p}}}{2}], \nonumber \\
        C_{1,7} =& \sum_{i=e}^{o}\sum_{p=1}^{2d} \sum_{(a,b )=(1,4)}^{(2,3)}[[\sum_{\{ \vec{n}_{i,p}\}}\frac{\hat{T}^{(a)}_{\vec{n}_{i,p}}+\hat{T}^{(b)}_{\vec{n}_{i,p}}}{2},\sum_{\substack{j,p',(c,d),\\ \{\vec{n}_{j,p'}\}}}\frac{\hat{T}^{(c)}_{\vec{n}_{j,p'}}+\hat{T}^{(d)}_{\vec{n}_{j,p'}}}{2} ], \sum_{\{ \vec{n}_{i,p}\}}\frac{\hat{T}^{(a)}_{\vec{n}_{i,p}}+\hat{T}^{(b)}_{\vec{n}_{i,p}}}{2}].
\end{align}
In $C_{1,7}$, $(\hat{T}^{(c)}_{\vec{n}_{j,p'}}+\hat{T}^{(d)}_{\vec{n}_{j,p'}})/{2}$ is listed in (\ref{eq:order_U1}) after and hence, implemented after $(\hat{T}^{(a)}_{\vec{n}_{i,p}}+\hat{T}^{(b)}_{\vec{n}_{i,p}})/{2}$.

It remains to evaluate in the following each term $C_{1,n}$.  
Note the following expressions will be useful
in the foregoing evaluations of the terms:
\begin{equation}
\label{eq:comm_bound}
    ||[[A,B],C]||\leq4||A|| \cdot ||B||\cdot ||C||,
\end{equation}
\begin{equation}
\label{eq:useful1}
||\hat{D}_{\vec{n}}^{(M)}+\hat{D}_{\vec{n}+\hat{l}}^{(M)}|| \leq m,
\end{equation}
which follows from (\ref{eq:mass_U1}),
\begin{equation}
\label{eq:useful2}
||\hat{T}^{(b)}_{\vec{n}_{i,p}}|| \leq \frac{1}{2a},
\end{equation}
which is due to (\ref{eq:T1_U1}-\ref{eq:T4_U1}),
\begin{equation}
\label{eq:useful2_2}
||\sum_{b=1}^{4}\frac{\hat{T}^{(b)}_{\vec{n},l}}{2}|| =|| \hat{K}(\vec{n},l) ||\leq \frac{1}{a},
\end{equation}
where we define
\begin{align}
    \hat{K}(\vec{n},l)&=\frac{1}{8a} [(\hat{U}({\vec{n}},l)+\hat{U}^{\dag}({\vec{n}},l))(\hat{X}({\vec{n}}) \hat{X}({\vec{n}+\hat{l}}) + \hat{Y}({\vec{n}}) \hat{Y}({\vec{n}+\hat{l}}))\hat{\zeta}_{\vec{n},l} \nonumber \\
    &\quad+ i(\hat{U}(\vec{n},l)-\hat{U}^{\dag}({\vec{n}},l))
    (\hat{X}({\vec{n}}) \hat{Y}({\vec{n}+\hat{l}}) - \hat{Y}({\vec{n}}) \hat{X}({\vec{n}+\hat{l}}))\hat{\zeta}_{\vec{n},l}],
\end{align}
and use
\begin{equation}
\label{eq:useful3}
||\hat{U}|| = ||\hat{U}^\dagger || \leq 1,
\end{equation}
\begin{equation}
\label{eq:useful4}
||\hat{U}+\hat{U}^\dagger || \leq 2.
\end{equation}
Lastly,
\begin{equation}
\lvert\lvert \hat{L}^{(B)}_{\vec{n}_k,\Box} \rvert\rvert = || \frac{1}{2g^2a^{4-d}}(\hat{U}\hat{U}\hat{U}^\dag \hat{U}^\dag + h.c.)||=\frac{1}{g^2a^{4-d}},
\end{equation}
where $\vec{n}_k$ may be an arbitrary position vector.
Whenever these useful expressions are used, we use them
without explicit references for brevity.

$C_{1,1}$ is straightforward to evaluate, since the mass and electric terms commute, i.e.,
\begin{equation}
    [[\sum_{\vec{n}}\hat{D}_{\vec{n}}^{(M)},\sum_{\vec{n}'}\hat{D}_{\vec{n}'}^{(E)}],\sum_{\vec{n}'}\hat{D}_{\vec{n}'}^{(M)}] = 0.
\end{equation}

$C_{1,2}$ is bounded by
\begin{align}
    &\quad|| [[\sum_{\vec{n}}\hat{D}_{\vec{n}}^{(M)}, \sum_{ \vec{n}}\hat{T}^{(K)}_{\vec{n}}] ,\sum_{\vec{n}}\hat{D}_{\vec{n}}^{(M)}] || \nonumber \\
    &=|| [\sum_{ \vec{n}}\sum_{l=1}^{d}[\hat{D}_{\vec{n}}^{(M)}+\hat{D}_{\vec{n}+\hat{l}}^{(M)}, \hat{K}(\vec{n},l)] ,\hat{D}_{\vec{n}}^{(M)}+\hat{D}_{\vec{n}+\hat{l}}^{(M)}] || \nonumber \\
    &\leq\sum_{ \vec{n}}\sum_{l=1}^{d}|| [[\hat{D}_{\vec{n}}^{(M)}+\hat{D}_{\vec{n}+\hat{l}}^{(M)}, \hat{K}(\vec{n},l)] ,\hat{D}_{\vec{n}}^{(M)}+\hat{D}_{\vec{n}+\hat{l}}^{(M)}] || \nonumber \\
    &\leq  \frac{4 dL^d m^2}{a}.
    \label{eq:second_term_iji_U1}
\end{align}
The first equality in (\ref{eq:second_term_iji_U1}) is due to the fact that each kinetic term at site $\vec{n}$ couples two sites, $\vec{n}$ and $\vec{n}+\hat{l}$ with $l$ denoting the direction considered. The inequality that immediately follows from it is due to the triangle inequality. This is because, although the kinetic operators have $\hat{\zeta}$ JW strings, mass terms are diagonal, and thus they commute with the JW strings. The second inequality is due to (\ref{eq:comm_bound}).
In the last inequality, the bound for the mass and kinetic terms are due to (\ref{eq:mass_U1}) and (\ref{eq:T1_U1}-\ref{eq:T4_U1}), respectively. 

$C_{1,3}$ is straightforward to evaluate since mass and magnetic terms commute and hence,
\begin{equation}
    [[\sum_{\vec{n}}\hat{D}_{\vec{n}}^{(M)},\sum_{\vec{n}}\hat{L}_{\vec{n}}^{(B)}],\sum_{\vec{n}}\hat{D}_{\vec{n}}^{(M)}] = 0.
\end{equation}

Before we evaluate the commutator between the electric and kinetic terms ($C_{1,4}$), we provide a couple useful properties about the kinetic operators $\hat{K}(\vec{n},l)$. Acting on the fermionic space, a kinetic operator takes a computational basis state to another basis state. Acting on the gauge field on a link, it takes $\ket{E}\mapsto \ket{E\pm 1}$, where $E\in [-\Lambda,\Lambda-1]$, up to a multiplicative constant. Therefore, if we consider an electric and a kinetic operator acting on the same link, we obtain
\begin{align}
    &\quad||[\frac{g^2}{2a^{d-2}}\hat{E}^2(\vec{n},l),\hat{K}(\vec{n},l)]||\nonumber\\
    &=\frac{g^2}{2a^{d-2}}||\hat{E}^2(\vec{n},l)\hat{K}(\vec{n},l) - \hat{K}(\vec{n},l)\hat{E}^2(\vec{n},l)||\nonumber\\
    &\mapsto \frac{g^2}{2a^{d-2}}|| (E\pm 1)^2 \hat{K}(\vec{n},l) - \hat{K}(\vec{n},l) l^2||, \nonumber\\
    &=\frac{g^2}{2a^{d-2}}||\hat{K}(\vec{n},l)(\pm2E+1)|| \nonumber\\
    &\leq \frac{g^2}{2a^{d-2}}(2\Lambda+1)||\hat{K}(\vec{n},l)|| \nonumber \\
    &\leq \frac{g^2}{2a^{d-2}}(2\Lambda+1) \frac{1}{a} = \frac{g^2}{2a^{d-1}}(2\Lambda+1),
\label{eq:U1_EK_comm}    
\end{align}
where we used $\hat{E}^2(\vec{n},l)$ to denote an electric term for the link $(\vec{n},l)$ up to a multiplicative constant,
and
\begin{align}
    &\quad || [[\frac{g^2}{2a^{d-2}}\hat{E}^2,\hat{K}(\vec{n},l)], \frac{g^2}{2a^{d-2}}\hat{E}^2 ] || \nonumber \\
    &\mapsto ||\frac{g^2}{2a^{d-2}}(E^2-\hat{E}^2) [\frac{g^2}{2a^{d-2}}\hat{E}^2,\hat{K}(\vec{n},l)]|| \nonumber \\
    &\mapsto ||\frac{g^2}{2a^{d-2}}(E^2-(E\pm 1)^2) [\frac{g^2}{2a^{d-2}}\hat{E}^2,\hat{K}(\vec{n},l)]|| \nonumber \\
    &\mapsto ||\frac{g^2}{2a^{d-2}}(\mp2E-1) [\frac{g^2}{2a^{d-2}}\hat{E}^2,\hat{K}(\vec{n},l)]|| \nonumber \\
    &\leq \frac{g^2}{2a^{d-2}}(2\Lambda-1) ||[\frac{g^2}{2a^{d-2}}\hat{E}^2,\hat{K}(\vec{n},l)]|| \nonumber \\
    &\leq \frac{g^2}{2a^{d-2}}(2\Lambda-1)\frac{g^2}{2a^{d-1}}(2\Lambda+1) \nonumber \\
    & = \frac{g^4}{4a^{2d-3}}(4\Lambda^2-1).
\end{align}
Using these equations, we evaluate the bound of $C_{1,4}$ and obtain
\begin{align}
    &\quad || [[\sum_{\vec{n}}\hat{D}_{\vec{n}}^{(E)}, \sum_{ \vec{n} } \hat{T}_{\vec{n}'}^{(K)}] ,\sum_{\vec{n}}\hat{D}_{\vec{n}}^{(E)}] || \nonumber \\
    &= || [[\sum_{\vec{n}}\hat{D}_{\vec{n}}^{(E)}, \sum_{ \vec{n} }\sum_{l=1}^{d}\hat{K}(\vec{n},l)],\sum_{\vec{n}}\hat{D}_{\vec{n}}^{(E)}] || \nonumber \\
    &\leq\sum_{ \vec{n} }\sum_{l=1}^{d}|| [[\frac{g^2}{2a^{d-2}}\hat{E}^2(\vec{n},l), \hat{K}(\vec{n},l)],\frac{g^2}{2a^{d-2}}\hat{E}^2(\vec{n},l)] || \nonumber \\
    &\leq dL^d\frac{g^4}{4a^{2d-3}}(4\Lambda^2-1).
\end{align}

Next, we evaluate the commutators between the electric and plaquette operators ($C_{1,5}$), which are trivial unless the operators act on a common link. For the sake of brevity, we let the plaquette operator act on links $1,2,3,4$, and let the electric field operator act on link $1$. The commutators are then
\begin{align}
    &\quad ||[\frac{g^2}{2a^{d-2}} \hat{E}^2_{1},\frac{-1}{2g^2 a^{4-d}}\hat{U}_{1}\hat{U}_{2}\hat{U}_{3}^\dag\hat{U}_{4}^\dag]|| \nonumber \\
    &= ||[\frac{g^2}{2a^{d-2}} \hat{E}^2_{1},\frac{-1}{2g^2 a^{4-d}}\hat{U}_{1}]||\cdot||\hat{U}_{2}\hat{U}_{3}^\dag\hat{U}_{4}^\dag|| \nonumber \\
    &\leq ||[\frac{g^2}{2a^{d-2}} \hat{E}^2_{1},\frac{-1}{2g^2 a^{4-d}}\hat{U}_{1}]|| \nonumber \\
    &=|| \frac{-1}{4a^2}(\hat{E}^2\hat{U}-\hat{U}\hat{E}^2) || \nonumber \\
    &=|| \frac{-1}{4a^2}(\sum_{E}E^2\ketbra{E}{E}\hat{U}-\hat{U}\sum_{E}E^2\ketbra{E}{E}) || \nonumber \\
    &=|| \frac{-1}{4a^2} (\sum_{E}E^2\ketbra{E}{E-1}-\sum_{E}E^2\ketbra{E+1}{E}) || \nonumber \\
    &=|| \frac{-1}{4a^2} \sum_{E}\left[(E+1)^2-E^2\right]\ketbra{E+1}{E} || \nonumber \\
    &=||\frac{-1}{4a^2} \sum_{E}(2E+1)\ketbra{E+1}{E} ||  \nonumber \\
    &\leq || \frac{-1}{4a^2}(-2\Lambda + 1) || = \frac{2\Lambda-1}{4a^2}
\end{align}
and
\begin{align}
    &\quad ||[\frac{g^2}{2a^{d-2}} \hat{E}^2_{1},\frac{-1}{2g^2 a^{4-d}}\hat{U}_{1}^\dag\hat{U}_{2}^\dag\hat{U}_{3}\hat{U}_{4}]|| \nonumber \\
    &\leq ||[\frac{g^2}{2a^{d-2}} \hat{E}^2_{1},\frac{-1}{2g^2 a^{4-d}}\hat{U}_{1}^\dag]|| \nonumber \\
    &=|| \frac{-1}{4a^2}(\hat{E}^2\hat{U}^\dag-\hat{U}^\dag\hat{E}^2) || \nonumber \\
    &=|| \frac{-1}{4a^2} (\sum_{E}E^2\ketbra{E}{E+1}-\sum_{E}(E+1)^2\ketbra{E}{E+1}) || \nonumber \\
    &=||\frac{1}{4a^2} \sum_{E}(2E+1)\ketbra{E}{E+1} ||  \nonumber \\
    &\leq || \frac{1}{4a^2}(2(\Lambda-1) + 1) || = \frac{2\Lambda-1}{4a^2}.
\end{align}
Combining the two commutators, we obtain
\begin{align}
    &\quad||[\frac{g^2}{2a^{d-2}}\sum_{i=1}^{4}\hat{E}^2_{i},\frac{-1}{2g^2 a^{4-d}}(\hat{U}_{1}\hat{U}_{2}\hat{U}_{3}^\dag\hat{U}_{4}^\dag+h.c.)]|| \nonumber \\
    &\leq 4||[\frac{g^2}{2a^{d-2}} \hat{E}^2_{1},\frac{-1}{2g^2 a^{4-d}}(\hat{U}_{1}\hat{U}_{2}\hat{U}_{3}^\dag\hat{U}_{4}^\dag+h.c.)]|| \nonumber \\
    &= 4||\frac{1}{4a^2} [\sum_{E}(2E+1)\ketbra{E}{E+1} -(2E+1)\ketbra{E+1}{E}] || \nonumber \\
    &\leq \frac{4\Lambda-2}{a^2}.
\label{eq:U1_EB_comm}
\end{align}
We are now equipped to evaluate the bound of $C_{1,5}$. The bound is
\begin{align}
    &\quad ||[[\sum_{\vec{n}}\hat{D}_{\vec{n}}^{(E)},\sum_{\vec{n}}\hat{L}_{\vec{n}}^{(B)}],\sum_{\vec{n}}\hat{D}_{\vec{n}}^{(E)}]|| \nonumber \\
    &\leq \frac{L^d d(d-1)}{2}  || [[\frac{g^2}{2a^{d-2}}\sum_{i=1}^{4}\hat{E}^2_{i},\frac{-1}{2g^2 a^{4-d}}(\hat{U}_{1}\hat{U}_{2}\hat{U}_{3}^\dag\hat{U}_{4}^\dag+h.c.)] ,\frac{g^2}{2a^{d-2}}\sum_{i=1}^{4}\hat{E}^2_{i}] ||\nonumber\\
    &= \frac{L^d d(d-1)}{2}  ||\frac{g^2}{2a^{d-2}}\sum_{i=1}^{4}(E_{i}^2 - \hat{E}^2_{i}) [\frac{g^2}{2a^{d-2}}\sum_{i=1}^{4}\hat{E}^2_{i} ,\frac{-1}{2g^2 a^{4-d}}(\hat{U}_{1}\hat{U}_{2}\hat{U}_{3}^\dag\hat{U}_{4}^\dag+h.c.)]|| \nonumber \\
    &\mapsto \frac{L^d d(d-1)}{2}  ||\frac{g^2}{2a^{d-2}}\sum_{i=1}^{4}(E_{i}^2 - (E_i \pm 1)^2) [\frac{g^2}{2a^{d-2}}\sum_{i=1}^{4}\hat{E}^2_{i},\frac{-1}{2g^2 a^{4-d}}(\hat{U}_{1}\hat{U}_{2}\hat{U}_{3}^\dag\hat{U}_{4}^\dag+h.c.)]|| \nonumber \\
    &=\frac{L^d d(d-1)}{2}  ||\frac{g^2}{2a^{d-2}}\sum_{i=1}^{4}(\mp 2 E_i-1) [\frac{g^2}{2a^{d-2}}\sum_{i=1}^{4}\hat{E}^2_{i} ,\frac{-1}{2g^2 a^{4-d}}(\hat{U}_{1}\hat{U}_{2}\hat{U}_{3}^\dag\hat{U}_{4}^\dag+h.c.)]|| \nonumber \\
    &\leq \frac{L^d d(d-1)}{2}  \frac{g^2}{2a^{d-2}}4(2\Lambda -1) ||[\frac{g^2}{2a^{d-2}}\sum_{i=1}^{4}\hat{E}^2_{i},\frac{-1}{2g^2 a^{4-d}}(\hat{U}_{1}\hat{U}_{2}\hat{U}_{3}^\dag\hat{U}_{4}^\dag+h.c.)]|| \nonumber \\
    &\leq \frac{L^d d(d-1)}{2}  \frac{g^2}{2a^{d-2}}4(2\Lambda -1) \frac{4\Lambda-2}{a^2} \nonumber \\
    &= \frac{2L^d d(d-1)g^2(2\Lambda-1)^2}{a^d},
\end{align}
where in the second line, we have used the fact that there are $L^d \frac{d(d-1)}{2}$ plaquette terms in the sum $\sum_{\vec{n}}\hat{L}_{\vec{n}}^{(B)}$.

Considering $C_{1,6}$, we first fix the evenness and oddness of the sites. We consider the bulk and edge terms separately, although the required analysis is similar. We use $p$ to denote the direction of the links acted on by the kinetic term. The commutator is trivially zero if the magnetic part and kinetic part act on different links. Thus we consider the case where each plaquette operator in the magnetic part acts on links that the kinetic operators also act on. For direction $p$, we then have
\begin{align}
    &\quad ||[[\sum_{\{ \vec{n}_{i,p}\}}\frac{\hat{T}^{(a)}_{\vec{n}_{i,p}}+\hat{T}^{(b)}_{\vec{n}_{i,p}}}{2},\sum_{\vec{n}_{j,\Box}}\hat{L}_{\vec{n}_{j,\Box}}^{(B)} ],\sum_{\{ \vec{n}_{i,p}\}}\frac{\hat{T}^{(a)}_{\vec{n}_{i,p}}+\hat{T}^{(b)}_{\vec{n}_{i,p}}}{2}]|| \nonumber \\
    &= ||[\sum_{\{ \vec{n}_{i,p}\}}[\frac{\hat{T}^{(a)}_{\vec{n}_{i,p}}+\hat{T}^{(b)}_{\vec{n}_{i,p}}}{2},\sum_{\vec{n}_{j,\Box}}\hat{L}_{\vec{n}_{j,\Box}}^{(B)} ],\sum_{\{ \vec{n}_{i,p}\}}\frac{\hat{T}^{(a)}_{\vec{n}_{i,p}}+\hat{T}^{(b)}_{\vec{n}_{i,p}}}{2}]|| \nonumber \\
    &= ||[N_k[\frac{\hat{T}_{\vec{n}_{i,p}}^{(a)}+\hat{T}_{\vec{n}_{i,p}}^{(b)}}{2},\hat{L}_{\vec{n}_{i,\Box}}^{(B)} ],\frac{\hat{T}_{\vec{n}_{i,p}}^{(a)}+\hat{T}_{\vec{n}_{i,p}}^{(b)}}{2}]|| \nonumber \\
    &\leq 4N_k || \frac{\hat{T}_{\vec{n}_{i,p}}^{(a)}+\hat{T}_{\vec{n}_{i,p}}^{(b)}}{2} ||^2 || \frac{-1}{2a^{4-d}g^2}(\hat{U} \hat{U} \hat{U}^\dag \hat{U}^\dag + h.c.) || \nonumber \\
    &\leq 4N_k \frac{1}{4a^2} \frac{1}{a^{4-d}g^2} = \frac{N_k}{a^{6-d} g^2},
    \label{eq:sixth_term_iji_U1}
\end{align}
where $N_k \in \{N_B,N_E\}$ is the number of odd or even sites in the bulk or on the edges in each direction. Recall that $N_B=(L^d-L^{d-1})/2$, and $N_E=L^{d-1}/2$. In the third line of (\ref{eq:sixth_term_iji_U1}), we have used the fact that for each kinetic term, there is a plaquette operator, of the type $\vec{n}_{i,\Box}$, acting on the same link. Expanding this to include all sites, all directions, and all plaquette and kinetic operators, we obtain
\begin{align}
    &\quad C_{1,6} \nonumber \\
    &\leq \sum_{i,j=e}^{o}\sum_{p=1}^{2d}\sum_{\Box=1}^{d(d-1)/2}\sum_{(a,b)=(1,4)}^{(2,3)}||[[\sum_{\{ \vec{n}_{i,p}\}}\frac{\hat{T}^{(a)}_{\vec{n}_{i,p}}+\hat{T}^{(b)}_{\vec{n}_{i,p}}}{2},\sum_{\vec{n}_{j,\Box}}\hat{L}_{\vec{n}_{j,\Box}}^{(B)} ],\sum_{\{ \vec{n}_{i,p}\}}\frac{\hat{T}^{(a)}_{\vec{n}_{i,p}}+\hat{T}^{(b)}_{\vec{n}_{i,p}}}{2}]|| \nonumber \\
    &\leq 8d(d-1)(N_E+N_B)\frac{1}{a^{6-d} g^2} = \frac{4d(d-1)L^d}{a^{6-d}g^2},
\end{align}
where in the last inequality, the factor $d$ comes from the fact that there are $d$ directions each for the bulk and edges of the lattice. Further, for each direction $p$, there are $d-1$ two-dimensional planes $\Box$ that contain links in that direction, hence the factor $d-1$.

Now, we compute $C_{1,7}$, which consists of the commutators between kinetic terms. There are three types of commutators; those between (i) two bulk terms, (ii) two edge terms, and (iii) a bulk and an edge term. We analyze case (i) first. We reiterate that there are $4d$ bulk terms of the form $\mathcal{T}^{(a,b)}_{\vec{n}_{i,p}}=\sum_{\{ \vec{n}_{i,p}\}}\frac{\hat{T}^{(a)}_{\vec{n}_{i,p}}+\hat{T}^{(b)}_{\vec{n}_{i,p}}}{2}$ in total (we remind the readers that there are even / odd sites, then $\hat{T}^{(1)}+\hat{T}^{(4)}$ and $\hat{T}^{(2)}+\hat{T}^{(3)}$), where $\{ \vec{n}_{i,p}\}$ runs over all sites for a given parity $i$, with fixed direction $p$. Notice that, in a non-vanishing commutator, two bulk terms must either act on the same set of links and sites, or they overlap on one of the fermionic sites. This is so since, in all other scenarios, either the two kinetic terms simply act on disjoint Hilbert spaces (they act on two disjoint sets of qubit registers) or a kinetic term's JW string $\hat{\zeta}$ always commutes with the other kinetic term in our zig-zag JW path. To illustrate, consider a number system with basis $d$, where a number here encodes a fermionic site. A kinetic term is defined over picking a pair of two numbers that are different by one digit, with the difference of that digit being one. Consider two pairs of these numbers. The two pairs have two different ranges of numbers, over which the JW string acts. The former scenario of disjointedness arises when the two ranges do not overlap. The latter scenario arises when one range is inside the other range. Since operators $\hat{P}_r$ and $\hat{\tilde{P}}_r$ in (\ref{eq:ising}) that act on fermion registers that collide with the JW string commute with the string, the latter scenario does not contribute to the commutator bound.

Since we always implement $\sum_{\{ \vec{n}_{i,p}\}}\frac{\hat{T}^{(1)}_{\vec{n}_{i,p}}+\hat{T}^{(4)}_{\vec{n}_{i,p}}}{2}$ before $\sum_{\{ \vec{n}_{i,p}\}}\frac{\hat{T}^{(2)}_{\vec{n}_{i,p}}+\hat{T}^{(3)}_{\vec{n}_{i,p}}}{2}$, as shown in (\ref{eq:order_U1}), for each $\vec{n}_{i,p}$, we know that $\hat{H}_i = \sum_{\{ \vec{n}_{i,p}\}}\frac{\hat{T}^{(1)}_{\vec{n}_{i,p}}+\hat{T}^{(4)}_{\vec{n}_{i,p}}}{2}$ and $\hat{H}_j = \sum_{\{ \vec{n}_{i,p}\}}\frac{\hat{T}^{(2)}_{\vec{n}_{i,p}}+\hat{T}^{(3)}_{\vec{n}_{i,p}}}{2}$. Hence, the commutator for the case where two bulk terms act on the same set of links and sites is
\begin{align}
    &\quad ||[[\sum_{\{ \vec{n}_{i,p}\}}\frac{\hat{T}^{(1)}_{\vec{n}_{i,p}}+\hat{T}^{(4)}_{\vec{n}_{i,p}}}{2},\sum_{\{ \vec{n}_{i,p}\}}\frac{\hat{T}^{(2)}_{\vec{n}_{i,p}}+\hat{T}^{(3)}_{\vec{n}_{i,p}}}{2} ] ,\sum_{\{ \vec{n}_{i,p}\}}\frac{\hat{T}^{(1)}_{\vec{n}_{i,p}}+\hat{T}^{(4)}_{\vec{n}_{i,p}}}{2}]|| \nonumber \\
    &\leq ||\sum_{\{ \vec{n}_{i,p}\}}[[\frac{\hat{T}^{(1)}_{\vec{n}_{i,p}}+\hat{T}^{(4)}_{\vec{n}_{i,p}}}{2},\frac{\hat{T}^{(2)}_{\vec{n}_{i,p}}+\hat{T}^{(3)}_{\vec{n}_{i,p}}}{2} ] ,\frac{\hat{T}^{(1)}_{\vec{n}_{i,p}}+\hat{T}^{(4)}_{\vec{n}_{i,p}}}{2}]|| \leq 4 N_B || \frac{\hat{T}^{(a)}_{\vec{n}_{i,p}}+\hat{T}^{(b)}_{\vec{n}_{i,p}}}{2} ||^3 \nonumber \\
    &\leq 4N_B (\frac{1}{2a})^3 = \frac{N_B}{2a^3}.
    \label{eq:U1_bulk_comm2}
\end{align}
The first inequality is due to the fact that the kinetic terms acting on different sites and links commute. We used (\ref{eq:comm_bound}) for the second inequality. In the case where the two bulk terms overlap on one of the fermionic sites, the commutator is given by
\begin{align}
\label{eq:U1_bulk_comm}
    &\quad ||[[\sum_{\{ \vec{n}_{j,p'}\}}\frac{\hat{T}^{(a)}_{\vec{n}_{j,p'}}+\hat{T}^{(b)}_{\vec{n}_{j,p'}}}{2},\sum_{\{ \vec{n}_{i,p}\}}\sum_{c=1}^{4}\frac{\hat{T}^{(c)}_{\vec{n}_{i,p}}}{2} ],\sum_{\{ \vec{n}_{j,p'}\}}\frac{\hat{T}^{(a)}_{\vec{n}_{j,p'}}+\hat{T}^{(b)}_{\vec{n}_{j,p'}}}{2}]|| \nonumber \\
    &\leq \sum_{\{ \vec{n}_{i,p}\}} 4|| 2 \frac{\hat{T}^{(a)}_{\vec{n}_{i,p}}+\hat{T}^{(b)}_{\vec{n}_{i,p}}}{2} ||^2 || \hat{K}(\vec{n}_{i},p) || \nonumber \\
    &\leq \frac{4N_B}{a^3},
\end{align}
where we once again used (\ref{eq:comm_bound}) and the fact that each term at site $\vec{n}_{i,p}$ has two neighboring kinetic terms acting on the sites $\vec{n}_{i,p}$ or $\vec{n}_{i,p}+\hat{p}$. We now compute the number of occurrences of the commutators considered in (\ref{eq:U1_bulk_comm2}) and (\ref{eq:U1_bulk_comm}). Since there are $d$ directions, and two parities, there are $2d$ terms of the form (\ref{eq:U1_bulk_comm2}). For (\ref{eq:U1_bulk_comm}), since each term $\mathcal{T}^{(a,b)}_{\vec{n}_{i,p}}$ commutes with itself, but not with the terms that are implemented afterwards, there are
\begin{equation}
\label{eq:U1_tt_pair}
2[(2d-1)+(2d-2)+...+1+0]=(2d-1)2d=4d^2-2d,
\end{equation}
where a factor of $2$ is due to the fact that there are two $(a,b)$-combinations, non-vanishing commutators in total. We have also used the fact that there are $2d$ combinations of parity and direction, i.e., $i$ and $p$, in the bulk. As such, the sum of the commutators between all the bulk terms is bounded by
\begin{equation}
    (4d^2-2d)\frac{4N_B}{a^3} + 2d\frac{N_B}{2a^3} = (16d^2-6d)\frac{N_B}{a^3}.
\end{equation}
Using similar arguments, we obtain the bound for the sum of the commutators between all the edge terms, $(16d^2-6d)\frac{N_E}{a^3}$. 

Lastly, we evaluate the commutators between a term $\hat{K}(\vec{n}_{j},p')$ from the bulk and a term $\mathcal{T}^{(a,b)}_{\vec{n}_{i,p}}$ from the edge. Each $\mathcal{T}^{(a,b)}_{\vec{n}_{i,p}}$ on an edge link has two $\hat{K}(\vec{n}_{j},p')$ terms from the bulk acting on the same sites, and hence, the commutator between a bulk and edge term is bounded by $\frac{4N_E}{a^3}$, where we have replaced $N_B$ with the smaller $N_E$ in (\ref{eq:U1_bulk_comm}). Since there are $2d$ $\hat{K}(\vec{n}_{j},p')$ terms from the bulk, and $4d$ $\mathcal{T}^{(a,b)}_{\vec{n}_{i,p}}$ terms from the edge, there are in total $8d^2$ commutators between bulk and edge terms. The bulk and edge terms that act on links along the same direction and sites with the same parity commute, and there are $4d$ pairs of such terms. Hence, the sum of such commutators is upper-bounded by $(8d^2-4d)\frac{4N_E}{a^3}$. In total, $C_{1,7}$ is bounded from above by
\begin{equation}
    (16d^2-6d)\frac{N_E+N_B}{a^3} + (32d^2-16d)\frac{N_E}{a^3} = \frac{(8d^2-3d)L^d + (16d^2-8d) L^{d-1}}{a^3}.
\end{equation}

Next, we analyze the second sum in (\ref{eq:U1_trotter_err}), which is given by
\begin{align}
\label{eq:U1_trotter_err_2}
\sum_{i}|| [[\hat{H}_i, \sum_{j>i}\hat{H}_j],\sum_{k>i}\hat{H}_k] || \leq
\sum_{n=1}^{11} || C_{2,n} ||,
\end{align}
where
\begin{align}
C_{2,1} =&  [[\sum_{\vec{n}}\hat{D}_{\vec{n}}^{(M)}, \sum_{\vec{n}'}\hat{T}^{(K)}_{\vec{n}'}] ,\sum_{\vec{n}''}\hat{D}_{\vec{n}''}^{(E)}]  \nonumber \\
C_{2,2} =&  [[\sum_{\vec{n}}\hat{D}_{\vec{n}}^{(M)}, \sum_{\vec{n}'}\hat{T}^{(K)}_{\vec{n}'},\sum_{\vec{n}''}\hat{T}^{(K)}_{\vec{n''}}]  \nonumber \\
C_{2,3} =&  [[\sum_{\vec{n}}\hat{D}_{\vec{n}}^{(M)}, \sum_{\vec{n}'}\hat{T}^{(K)}_{\vec{n}'}, \sum_{\vec{n}''}\hat{L}^{(B)}_{\vec{n}''}] \nonumber \\
C_{2,4} =&  [[\sum_{\vec{n}}\hat{D}_{\vec{n}}^{(E)},\sum_{\vec{n}'}\hat{T}^{(K)}_{\vec{n}'}],\sum_{\vec{n}''}\hat{T}^{(K)}_{\vec{n''}}]  \nonumber \\
C_{2,5} =&  [[\sum_{\vec{n}}\hat{D}_{\vec{n}}^{(E)}, \sum_{\vec{n}'}\hat{T}^{(K)}_{\vec{n}'}],\sum_{\vec{n}'}\hat{L}_{\vec{n}''}^{(B)}]  \nonumber \\
C_{2,6} =&  [[\sum_{\vec{n}}\hat{D}_{\vec{n}}^{(E)},\sum_{\vec{n}'}\hat{L}_{\vec{n}'}^{(B)}] , \sum_{\vec{n}''}\hat{T}^{(K)}_{\vec{n''}}]  \nonumber \\
C_{2,7} =&  [[\sum_{\vec{n}}\hat{D}_{\vec{n}}^{(E)},\sum_{\vec{n}'}\hat{L}_{\vec{n}'}^{(B)}] , \sum_{\vec{n}''}\hat{L}_{\vec{n}''}^{(B)}]  \nonumber \\
C_{2,8} =& \sum_{i=e}^{o}\sum_{p=1}^{2d}\sum_{(a,b)=(1,4)}^{(2,3)} [[\sum_{\{ \vec{n}_{i,p}\}}\frac{\hat{T}^{(a)}_{\vec{n}_{i,p}}+\hat{T}^{(b)}_{\vec{n}_{i,p}}}{2},\sum_{\substack{j,p',(c,d),\\ \{\vec{n}_{j,p'}\}}}\frac{\hat{T}^{(c)}_{\vec{n}_{j,p'}}+\hat{T}^{(d)}_{\vec{n}_{j,p'}}}{2}],\sum_{\substack{k,p'',(e,f),\\ \{\vec{n}_{k,p''}\}}}\frac{\hat{T}^{(e)}_{\vec{n}_{k,p''}}+\hat{T}^{(f)}_{\vec{n}_{k,p''}}}{2}]  \nonumber \\
C_{2,9} =& \sum_{i=e}^{o}\sum_{p=1}^{2d}\sum_{(a,b)=(1,4)}^{(2,3)} [[\sum_{\{ \vec{n}_{i,p}\}} \frac{\hat{T}^{(a)}_{\vec{n}_{i,p}}+\hat{T}^{(b)}_{\vec{n}_{i,p}}}{2}, \sum_{\substack{j,p',(c,d),\\ \{\vec{n}_{j,p'}\}}}\frac{\hat{T}^{(c)}_{\vec{n}_{j,p'}}+\hat{T}^{(d)}_{\vec{n}_{j,p'}}}{2}],\sum_{\vec{n}''}\hat{L}_{\vec{n}''}^{(B)}]  \nonumber \\
C_{2,10} =& \sum_{i=e}^{o}\sum_{p=1}^{2d}\sum_{(a,b)=(1,4)}^{(2,3)} [[\sum_{\{ \vec{n}_{i,p}\}} \frac{\hat{T}^{(a)}_{\vec{n}_{i,p}}+\hat{T}^{(b)}_{\vec{n}_{i,p}}}{2}, \sum_{\vec{n}'}\hat{L}_{\vec{n}'}^{(B)}],\sum_{\substack{j,p',(c,d),\\ \{\vec{n}_{j,p'}\}}}\frac{\hat{T}^{(c)}_{\vec{n}_{j,p'}}+\hat{T}^{(d)}_{\vec{n}_{j,p'}}}{2}] \nonumber \\
C_{2,11} =& \sum_{i=e}^{o}\sum_{p=1}^{2d}\sum_{(a,b)=(1,4)}^{(2,3)} [[\sum_{\{ \vec{n}_{i,p}\}} \frac{\hat{T}^{(a)}_{\vec{n}_{i,p}}+\hat{T}^{(b)}_{\vec{n}_{i,p}}}{2},\sum_{\vec{n}'}\hat{L}_{\vec{n}'}^{(B)}],\sum_{\vec{n}''}\hat{L}_{\vec{n}''}^{(B)}]],
\end{align}
where we have implicitly assumed in $C_{2,8}$, $C_{2,9}$ and $C_{2,10}$ that $(\hat{T}^{(c)}_{\vec{n}_{j,p'}}+\hat{T}^{(d)}_{\vec{n}_{j,p'}})/{2}$ and $(\hat{T}^{(e)}_{\vec{n}_{k,p''}}+\hat{T}^{(f)}_{\vec{n}_{k,p''}})/{2}$ are listed in (\ref{eq:order_U1}) after and hence, implemented after $(\hat{T}^{(a)}_{\vec{n}_{i,p}}+\hat{T}^{(b)}_{\vec{n}_{i,p}})/{2}$. We have removed any trivially vanishing terms, which involve commutators between mass and electric terms, mass and magnetic terms, and two magnetic terms, since in each respective case the operators commute with each other. In the following, we evaluate the bounds for each $C_{2,n}$.

For $C_{2,1}$, we obtain the bound
\begin{align}
    &\quad || [[\sum_{\vec{n}}\hat{D}_{\vec{n}}^{(M)}, \sum_{\vec{n}'}\hat{T}^{(K)}_{\vec{n}'}],\sum_{\vec{n}''}\hat{D}_{\vec{n}''}^{(E)}]|| \nonumber \\
    &\leq || [\sum_{\vec{n}}\sum_{l=1}^{d}[ \hat{D}_{\vec{n}}^{(M)}+\hat{D}_{\vec{n} + \hat{l}}^{(M)}, \hat{K}(\vec{n},l)], \frac{g^2}{2a^{d-2}} \hat{E}^2(\vec{n}, l) ] || \nonumber \\
    &\mapsto ||\frac{g^2}{2a^{d-2}} (E^2 -\hat{E}^2) \sum_{\vec{n}}\sum_{l=1}^{d}[ \hat{D}_{\vec{n}}^{(M)}+\hat{D}_{\vec{n} + \hat{l}}^{(M)}, \hat{K}(\vec{n},l)] || \nonumber \\
    &\mapsto ||\frac{g^2}{2a^{d-2}} (E^2 -(E\pm 1)^2) \sum_{\vec{n}}\sum_{l=1}^{d}[ \hat{D}_{\vec{n}}^{(M)}+\hat{D}_{\vec{n} + \hat{l}}^{(M)}, \hat{K}(\vec{n},l)] || \nonumber \\
    &= ||\frac{g^2}{2a^{d-2}} (\mp 2E - 1) \sum_{\vec{n}}\sum_{l=1}^{d}[ \hat{D}_{\vec{n}}^{(M)}+\hat{D}_{\vec{n} + \hat{l}}^{(M)}, \hat{K}(\vec{n},l)] || \nonumber \\
    &\leq \frac{g^2}{2a^{d-2}}(2\Lambda - 1) || \sum_{\vec{n}}\sum_{l=1}^{d}[ \hat{D}_{\vec{n}}^{(M)}+\hat{D}_{\vec{n} + \hat{l}}^{(M)}, \hat{K}(\vec{n},l)]|| \nonumber \\
    &\leq \frac{g^2}{2a^{d-2}}(2\Lambda - 1) dL^d ||\hat{D}_{\vec{n}}^{(M)}+\hat{D}_{\vec{n} + \hat{l}}^{(M)}|| \cdot || \hat{K}(\vec{n},l) || \nonumber \\
    &= \frac{mg^2}{2a^{d-1}} (2\Lambda - 1)dL^d,
\end{align}
where in the first inequality, we have used the fact that each kinetic term $\hat{K}(\vec{n},l)$ acts on the sites $\vec{n}$ and $\vec{n}+\hat{l}$, and the link $(\vec{n},l)$. The term does not commute with the mass and electric terms acting on the same space, but commutes with the rest.

For $C_{2,2}$, we obtain the bound
\begin{align}
    &\quad ||[[\sum_{\vec{n}}\hat{D}_{\vec{n}}^{(M)}, \sum_{\vec{n}'}\hat{T}^{(K)}_{\vec{n}'}],\sum_{\vec{n}''}\hat{T}^{(K)}_{\vec{n}''}]||  \nonumber \\
    &\leq ||\sum_{\vec{n}}\sum_{l=1}^{d} [[\hat{D}_{\vec{n}}^{(M)}+\hat{D}_{\vec{n}+\hat{l}}^{(M)}, \hat{K}(\vec{n},l)],\sum_{\vec{n}''}\hat{T}^{(K)}_{\vec{n}''}]|| \nonumber \\
    &\leq 4 dL^d||\hat{D}_{\vec{n}}^{(M)}+\hat{D}_{\vec{n}+\hat{l}}^{(M)}|| \cdot || \hat{K}(\vec{n},l) ||\cdot ||(4d-2)\hat{K}(\vec{n},l)|| \nonumber \\
    &=\frac{mL^d(16d^2-8d)}{a^2},
\end{align}
where the factor of $(4d-2)$ in the third norm term of the second inequality is due to the fact that there are $(4d-2)$ kinetic terms of the type $\hat{K}(\vec{n},l)$ acting on the same fermionic sites as $\hat{K}(\vec{n},l)$.

Using a similar method, we evaluate the bound of the $C_{2,4}$ term (we consider the $C_{2,3}$ term right afterwards). The only difference from the $C_{2,2}$ term is that the mass term is replaced by the electric term. As such, we obtain the upper bound of $C_{2,4}$,
\begin{align}
    &\quad||[[\sum_{\vec{n}}\hat{D}_{\vec{n}}^{(E)}, \sum_{\vec{n}'}\hat{T}^{(K)}_{\vec{n}'}],\sum_{\vec{n}''}\hat{T}^{(K)}_{\vec{n}''}]||  \nonumber \\
    &\leq ||\sum_{\vec{n}}\sum_{l=1}^{d}[[\frac{g^2}{2a^{d-2}}\hat{E}^2(\vec{n},l), \hat{K}(\vec{n},l)],\sum_{\vec{n}''}\hat{T}^{(K)}_{\vec{n}''}]|| \nonumber \\
    &\leq 2 dL^d || [\frac{g^2}{2a^{d-2}}\hat{E}^2(\vec{n},l), \hat{K}(\vec{n},l)] || \cdot|| \sum_{\vec{n}''}\hat{T}^{(K)}_{\vec{n}''} || \nonumber \\
    &\leq 2 dL^d || \frac{g^2}{2a^{d-1}}(2\Lambda + 1) ||\cdot || (4d-2)\hat{K}(\vec{n},l)|| \nonumber \\
    &\leq 2 dL^d \frac{g^2}{2a^{d-1}}(2\Lambda + 1) \frac{4d-2}{a} = \frac{(4d^2-2d)L^d g^2 (2\Lambda +1)}{a^d},
\end{align}
where in the first inequality, we have used the fact that the electric terms acting on different links as the kinetic terms commute, and in the third inequality, we used the bound in (\ref{eq:U1_EK_comm}).

Moving onto the $C_{2,3}$ term in (\ref{eq:U1_trotter_err_2}), we evaluate its bound as follows:
\begin{align}
   &\quad|| [[\sum_{\vec{n}}\hat{D}_{\vec{n}}^{(M)}, \sum_{\vec{n}'}\hat{T}^{(K)}_{\vec{n}'}],\sum_{\vec{n}''}\hat{L}_{\vec{n}''}^{(B)}] || \nonumber \\
   &\leq || [\sum_{\vec{n}}\sum_{l=1}^{d}[\hat{D}_{\vec{n}}^{(M)}+\hat{D}_{\vec{n} + \hat{l}}^{(M)}, \hat{K}(\vec{n},l)],\sum_{\vec{n}''}\hat{L}_{\vec{n}''}^{(B)}] || \nonumber \\
   &\leq \sum_{\vec{n}}\sum_{l=1}^{d} 4 ||  \hat{D}_{\vec{n}}^{(M)}+\hat{D}_{\vec{n} + \hat{l}}^{(M)}|| \cdot || \hat{K}(\vec{n},l)|| \cdot ||2(d-1) \hat{L}_{\vec{n}_{i,\Box}}^{(B)} || \nonumber \\
   &\leq \frac{mL^d 8d(d-1)}{g^2a^{5-d}},
\end{align}
where in the second inequality, we have used the fact that there is one kinetic and $2(d-1)$ magnetic operators acting on the same link.

We bound the $C_{2,5}$ term using a similar method, but with the mass term replaced by the electric term. We evaluate its bound as follows:
\begin{align}
   &\quad|| [[\sum_{\vec{n}}\hat{D}_{\vec{n}}^{(E)}, \sum_{\vec{n}'}\hat{T}^{(K)}_{\vec{n}'}],\sum_{\vec{n}''}\hat{L}_{\vec{n}''}^{(B)}] || \nonumber \\
   &\leq || \sum_{\vec{n}}\sum_{l=1}^{d} [[\frac{g^2}{2a^{d-2}}\hat{E}^2(\vec{n}, \hat{l}), \hat{K}(\vec{n},l)],\sum_{\vec{n}''}\hat{L}_{\vec{n}''}^{(B)}] || \nonumber \\
   &\leq 2dL^d || [\frac{g^2}{2a^{d-2}}\hat{E}^2(\vec{n}, \hat{l}), \hat{K}(\vec{n},l)] || \cdot || 2(d-1) \hat{L}_{\vec{n}_{i,\Box}}^{(B)} || \nonumber \\
   &\leq 2dL^d \frac{g^2}{2a^{d-1}}(2\Lambda + 1) \frac{2(d-1)}{g^2 a^{4-d}} = \frac{2d(d-1)L^d}{a^3}(2\Lambda + 1),
\end{align}
where we have used (\ref{eq:U1_EK_comm}).

Each commutator in the $C_{2,6}$ term contains an electric, magnetic, and kinetic term. Evaluating the bound, we have
\begin{align}
    &\quad|| [[\sum_{\vec{n}}\hat{D}_{\vec{n}}^{(E)},\sum_{\vec{n}'}\hat{L}_{\vec{n}'}^{(B)}] , \sum_{\vec{n}''}\hat{T}^{(K)}_{\vec{n''}}] || \nonumber \\
    &= || \sum_{\vec{n}}\sum_{i,j\neq i} [ \frac{g^2}{2a^{d-2}}(\hat{E}^2(\vec{n},i) + \hat{E}^2(\vec{n}+\hat{i},j)+\hat{E}^2(\vec{n}+\hat{j},i)+\hat{E}^2(\vec{n},j)),\nonumber \\ &\quad \frac{-1}{2a^{4-d}g^2}(\hat{U}(\vec{n},i)\hat{U}(\vec{n}+\hat{i},j)\hat{U}^\dag(\vec{n}+\hat{j},i) \hat{U}^\dag(\vec{n},j) + h.c.)] \nonumber \\
    &\quad ,\hat{K}(\vec{n},i) + \hat{K}(\vec{n}+\hat{i},j)+\hat{K}(\vec{n}+\hat{j},i)+\hat{K}(\vec{n},j)] || \nonumber \\
    &\leq L^d\frac{d(d-1)}{2} 2 ||[ \frac{g^2}{2a^{d-2}}(\hat{E}^2(\vec{n},i) + \hat{E}^2(\vec{n}+\hat{i},j)+\hat{E}^2(\vec{n}+\hat{j},i)+\hat{E}^2(\vec{n},j)),\nonumber \\ &\quad \frac{-1}{2a^{4-d}g^2}(\hat{U}(\vec{n},i)\hat{U}(\vec{n}+\hat{i},j)\hat{U}^\dag(\vec{n}+\hat{j},i) \hat{U}^\dag(\vec{n},j) + h.c.)] || \cdot \nonumber \\
    &\quad ||\hat{K}(\vec{n},i) + \hat{K}(\vec{n}+\hat{i},j)+\hat{K}(\vec{n}+\hat{j},i)+\hat{K}(\vec{n},j) || \nonumber \\
    &\leq L^d\frac{d(d-1)}{2} 2 || \frac{4\Lambda - 2}{a^2}|| \cdot || \frac{4}{a} || \nonumber \\
    &= \frac{L^d d(d-1) (16\Lambda - 8)}{a^3},
\end{align}
where we used the fact that terms not acting on the same links always commute, and in the second inequality, we used the bound in (\ref{eq:U1_EB_comm}).

We now evaluate the bound of the $C_{2,7}$ term, which contains two magnetic terms, using a similar method,
\begin{align}
    &\quad || [[\sum_{\vec{n}}\hat{D}_{\vec{n}}^{(E)},\sum_{\vec{n}'}\hat{L}_{\vec{n}'}^{(B)}],\sum_{\vec{n}''}\hat{L}_{\vec{n}''}^{(B)}] ||\nonumber \\
    &= || \sum_{\vec{n}}\sum_{i,j\neq i} [ \frac{g^2}{2a^{d-2}}(\hat{E}^2(\vec{n},i) + \hat{E}^2(\vec{n}+\hat{i},j)+\hat{E}^2(\vec{n}+\hat{j},i)+\hat{E}^2(\vec{n},j)),\nonumber \\ &\quad \frac{-1}{2a^{4-d}g^2}(\hat{U}(\vec{n},i)\hat{U}(\vec{n}+\hat{i},j)\hat{U}^\dag(\vec{n}+\hat{j},i) \hat{U}^\dag(\vec{n},j) + h.c.)] ,\sum_{\vec{n}''}\hat{L}_{\vec{n}''}^{(B)}]|| \nonumber \\
    &\leq 2\frac{d(d-1)}{2}L^d|| \frac{4\Lambda -2}{a^2} || \cdot || \frac{8d-11}{a^{4-d}g^2}|| \nonumber \\
    &\leq \frac{L^d d(d-1)(8d - 11)(4\Lambda -2)}{g^2a^{6-d}},
    \label{eq:U1_EBB}
\end{align}
where in the first inequality, we used the bound in (\ref{eq:U1_EB_comm}) for the first norm term, and the fact that there are $8d-11$ magnetic operators acting on the same plaquette as the magnetic operator in the inner commutator for the second norm term. The factor $8d-8$ is due to the fact that each of the four links on a plaquette is acted on by $2(d-1)$ magnetic terms, but $3$ out of the $8d-8$ have been overcounted.

$C_{2,8}$ is a sum of commutators between three kinetic terms, which we label as $\hat{H}_i, \hat{H}_j$, and $\hat{H}_k$. We remind the readers that the edge terms are evolved before the bulk terms, as indicated by (\ref{eq:order_U1}). As such, we divide up the tuples $(\hat{H}_i,\hat{H}_j,\hat{H}_k)$ into five types: (i) all three terms are bulk terms, (ii) all three terms are edge terms, (iii) $\hat{H}_i$ and $\hat{H}_j$ are edge terms, and $\hat{H}_k$ is a bulk term, (iv) $\hat{H}_i$ and $\hat{H}_k$ are edge terms, and $\hat{H}_j$ is a bulk term, and (v) $\hat{H}_i$ is an edge term, and $\hat{H}_j$ and $\hat{H}_k$ are bulk terms. 

We consider the type-(i) terms, and further divide it into six separate cases. In the first case, we consider a scenario where $\hat{H}_i$ and $\hat{H}_j$ act on the same links and $\hat{H}_k = \hat{H}_j$. There are a total of $2d$ of such instances with even/odd parities and $d$ directions. Since we always implement $\sum_{\{ \vec{n}_{i,p}\}}\frac{\hat{T}^{(1)}_{\vec{n}_{i,p}}+\hat{T}^{(4)}_{\vec{n}_{i,p}}}{2}$ before $\sum_{\{ \vec{n}_{i,p}\}}\frac{\hat{T}^{(2)}_{\vec{n}_{i,p}}+\hat{T}^{(3)}_{\vec{n}_{i,p}}}{2}$, as shown in (\ref{eq:order_U1}), for each $\vec{n}_{i,p}$, we know that $\hat{H}_i = \sum_{\{ \vec{n}_{i,p}\}}\frac{\hat{T}^{(1)}_{\vec{n}_{i,p}}+\hat{T}^{(4)}_{\vec{n}_{i,p}}}{2}$ and $\hat{H}_j = \sum_{\{ \vec{n}_{i,p}\}}\frac{\hat{T}^{(2)}_{\vec{n}_{i,p}}+\hat{T}^{(3)}_{\vec{n}_{i,p}}}{2}$. Hence, the bound for each of the instance is given by
\begin{align}
    &\quad|| [[\sum_{\{ \vec{n}_{i,p}\}}\frac{\hat{T}^{(1)}_{\vec{n}_{i,p}}+\hat{T}^{(4)}_{\vec{n}_{i,p}}}{2}, \sum_{\{ \vec{n}_{i,p}\}}\frac{\hat{T}^{(2)}_{\vec{n}_{i,p}}+\hat{T}^{(3)}_{\vec{n}_{i,p}}}{2}],\sum_{\{ \vec{n}_{i,p}\}}\frac{\hat{T}^{(2)}_{\vec{n}_{i,p}}+\hat{T}^{(3)}_{\vec{n}_{i,p}}}{2}] || \nonumber \\
    &= || \sum_{\{ \vec{n}_{i,p}\}} [[\frac{\hat{T}^{(1)}_{\vec{n}_{i,p}}+\hat{T}^{(4)}_{\vec{n}_{i,p}}}{2}, \frac{\hat{T}^{(2)}_{\vec{n}_{i,p}}+\hat{T}^{(3)}_{\vec{n}_{i,p}}}{2}],\frac{\hat{T}^{(2)}_{\vec{n}_{i,p}}+\hat{T}^{(3)}_{\vec{n}_{i,p}}}{2}] || \nonumber \\
    &\leq 4 N_B || \frac{\hat{T}^{(1)}_{\vec{n}_{i,p}}+\hat{T}^{(4)}_{\vec{n}_{i,p}}}{2} ||^3 = \frac{N_B}{2a^3},
    \label{eq:U1_TTT}
\end{align}
where the first equality is due to the fact that kinetic operators acting on different links and sites commute, and we have used (\ref{eq:comm_bound}) for the inequality. The second case we consider is when $\hat{H}_i$ and $\hat{H}_j$ act on the same links and $\hat{H}_k \neq \hat{H}_j$. There are in total $2d^2-d$ of such instances. This is so because a link can have $2d$ combinations of parity and direction, and if $\hat{H}_i$ and $\hat{H}_j$ act on the link labelled by the $n$th combination, then $\hat{H}_k$ can act on links with $2d-n$ different combinations, since $k>i$ and $\hat{H}_k \neq \hat{H}_j$. Thus, we obtain $(2d-1)+(2d-2)+...+1=2d^2-d$ for the total number of instances. We consider one of such instance, where we fix $\hat{H}_k = \sum_{\{ \vec{n}_{k,p'}\}}\sum_{a=1}^{4}\frac{\hat{T}^{(a)}_{\vec{n}_{k,p'}}}{2}$, and obtain its bound
\begin{align}
\label{eq:U1_ttt1_comm}
    &\quad|| [[\sum_{\{ \vec{n}_{i,p}\}}\frac{\hat{T}^{(1)}_{\vec{n}_{i,p}}+\hat{T}^{(4)}_{\vec{n}_{i,p}}}{2}, \sum_{\{ \vec{n}_{i,p}\}}\frac{\hat{T}^{(2)}_{\vec{n}_{i,p}}+\hat{T}^{(3)}_{\vec{n}_{i,p}}}{2}],\sum_{\{ \vec{n}_{k,p'}\}}\sum_{a=1}^{4}\frac{\hat{T}^{(a)}_{\vec{n}_{k,p'}}}{2}] || \nonumber \\
    &= || [\sum_{\{ \vec{n}_{i,p}\}}[\frac{\hat{T}^{(1)}_{\vec{n}_{i,p}}+\hat{T}^{(4)}_{\vec{n}_{i,p}}}{2}, \frac{\hat{T}^{(2)}_{\vec{n}_{i,p}}+\hat{T}^{(3)}_{\vec{n}_{i,p}}}{2}],\sum_{\{ \vec{n}_{k,p'}\}}\sum_{a=1}^{4}\frac{\hat{T}^{(a)}_{\vec{n}_{k,p'}}}{2}] || \nonumber \\
    &\leq 4N_B ||\frac{\hat{T}^{(1)}_{\vec{n}_{i,p}}+\hat{T}^{(4)}_{\vec{n}_{i,p}}}{2} ||^2 ||2 \sum_{a=1}^{4}\frac{\hat{T}^{(a)}_{\vec{n}_{k,p'}}}{2} || \nonumber \\
    &= \frac{2N_B}{a^3},
\end{align}
where in the first inequality, the extra factor two in the second norm expression is due to the fact that there are two choices of sites in  $\{ \vec{n}_{k,p'}\}$ for $\hat{H}_k$ that result in non-vanishing commutator with $\hat{H}_i$ and $\hat{H}_j$ due to their collisions on the two fermionic sites that sit at the two ends of a link that $\hat{H}_i$ and $\hat{H}_j$ act on. Similarly, in the third case, $\hat{H}_i$ and $\hat{H}_k$ act on the same links, while $\hat{H}_j\neq \hat{H}_k$ act on links of a different parity in the same direction. Since $(1,4)$ is implemented after $(2,3)$, and even terms are implemented before odd terms, $\hat{H}_i$, $\hat{H}_j$ and $\hat{H}_k$ are of the forms $\sum_{\{ \vec{n}_{e,p}\}}\frac{\hat{T}^{(1)}_{\vec{n}_{e,p}}+\hat{T}^{(4)}_{\vec{n}_{e,p}}}{2}$, $\sum_{\{ \vec{n}_{o,p}\}}\sum_{a=1}^{4}\frac{\hat{T}^{(a)}_{\vec{n}_{o,p}}}{2}$ and $\sum_{\{ \vec{n}_{e,p}\}}\frac{\hat{T}^{(2)}_{\vec{n}_{e,p}}+\hat{T}^{(3)}_{\vec{n}_{e,p}}}{2}$, respectively. There are $d$ such instances. We consider one of such instance, and obtain its bound
\begin{align}
\label{eq:U1_ttt1_comm2}
    &\quad|| [[\sum_{\{ \vec{n}_{e,p}\}}\frac{\hat{T}^{(1)}_{\vec{n}_{e,p}}+\hat{T}^{(4)}_{\vec{n}_{e,p}}}{2},\sum_{\{ \vec{n}_{o,p}\}}\sum_{a=1}^{4}\frac{\hat{T}^{(a)}_{\vec{n}_{o,p}}}{2}],\sum_{\{ \vec{n}_{e,p}\}}\frac{\hat{T}^{(2)}_{\vec{n}_{e,p}}+\hat{T}^{(3)}_{\vec{n}_{e,p}}}{2}
    ] || \nonumber \\
    &= || [\sum_{\{ \vec{n}_{e,p}\}}[\frac{\hat{T}^{(1)}_{\vec{n}_{e,p}}+\hat{T}^{(4)}_{\vec{n}_{e,p}}}{2},\sum_{\{ \vec{n}_{o,p}\}}\sum_{a=1}^{4}\frac{\hat{T}^{(a)}_{\vec{n}_{o,p}}}{2}],
    \frac{\hat{T}^{(2)}_{\vec{n}_{e,p}}+\hat{T}^{(3)}_{\vec{n}_{e,p}}}{2}
    ] || \nonumber \\
    &\leq 4N_B ||\frac{\hat{T}^{(1)}_{\vec{n}_{e,p}}+\hat{T}^{(4)}_{\vec{n}_{e,p}}}{2} || \cdot ||2 \sum_{a=1}^{4}\frac{\hat{T}^{(a)}_{\vec{n}_{o,p}}}{2} || \cdot ||3 \frac{\hat{T}^{(2)}_{\vec{n}_{e,p}}+\hat{T}^{(3)}_{\vec{n}_{e,p}}}{2} || \nonumber \\
    &= \frac{6N_B}{a^3},
\end{align}
where in the first inequality, the factor two in the second norm term is due to the fact that there are two choices of sites in $\{ \vec{n}_{o,p}\}$ for $\hat{H}_j$ that result in collisions on two sites with $\hat{H}_i$. Further, there are three $\hat{H}_k$ terms that collide with the inner commutator, which acts on three links and four sites, i.e. $\hat{H}_k$ and $\hat{H}_i$ act on the same link, and $\hat{H}_k$ collides with $\hat{H}_j$ on two sites. We consider the fourth case where now $\hat{H}_j$ acts on links in a different direction. There are $2d^2-2d$ such instances. This is so because a link can have $d$ different directions, and if $\hat{H}_i$ and $\hat{H}_k$ act on the link with the $n$th direction, then $\hat{H}_j$ can act on links in $d-n$ directions. Thus, we obtain $4[(d-1)+(d-2)...+1]=2d^2-2d$ for the total number of instances, where the factor of $4$ is because there are two parities for $\hat{H}_i$ and $\hat{H}_k$, and $\hat{H}_j$. We consider one of such instance, where we fix $\hat{H}_j = \sum_{\{ \vec{n}_{k,p'}\}}\sum_{a=1}^{4}\frac{\hat{T}^{(a)}_{\vec{n}_{k,p'}}}{2}$, and obtain its bound
\begin{align}
\label{eq:U1_ttt1_comm3}
    &\quad|| [[\sum_{\{ \vec{n}_{i,p}\}}\frac{\hat{T}^{(1)}_{\vec{n}_{i,p}}+\hat{T}^{(4)}_{\vec{n}_{i,p}}}{2},\sum_{\{ \vec{n}_{k,p'}\}}\sum_{a=1}^{4}\frac{\hat{T}^{(a)}_{\vec{n}_{k,p'}}}{2}],\sum_{\{ \vec{n}_{i,p}\}}\frac{\hat{T}^{(2)}_{\vec{n}_{i,p}}+\hat{T}^{(3)}_{\vec{n}_{i,p}}}{2}
    ] || \nonumber \\
    &= || [\sum_{\{ \vec{n}_{i,p}\}}[\frac{\hat{T}^{(1)}_{\vec{n}_{i,p}}+\hat{T}^{(4)}_{\vec{n}_{i,p}}}{2},\sum_{\{ \vec{n}_{k,p'}\}}\sum_{a=1}^{4}\frac{\hat{T}^{(a)}_{\vec{n}_{k,p'}}}{2}],
    \frac{\hat{T}^{(2)}_{\vec{n}_{i,p}}+\hat{T}^{(3)}_{\vec{n}_{i,p}}}{2}
    ] || \nonumber \\
    &\leq 4N_B ||\frac{\hat{T}^{(1)}_{\vec{n}_{i,p}}+\hat{T}^{(4)}_{\vec{n}_{i,p}}}{2} || \cdot ||2 \sum_{a=1}^{4}\frac{\hat{T}^{(a)}_{\vec{n}_{k,p'}}}{2} || \cdot || \frac{\hat{T}^{(2)}_{\vec{n}_{i,p}}+\hat{T}^{(3)}_{\vec{n}_{i,p}}}{2} || \nonumber \\
    &= \frac{2N_B}{a^3},
\end{align}
where the factor two in the second norm expression of the third line is due to the fact that there are two choices of sites in $\{ \vec{n}_{k,p'}\}$ for $\hat{H}_j$ that collide on two sites with $\hat{H}_i$ and $\hat{H}_k$. The fifth case we consider is when $\hat{H}_i$ and $\hat{H}_j$ act on different links, and $\hat{H}_j = \hat{H}_k = \sum_{\{ \vec{n}_{j,p'}\}}\sum_{a=1}^{4}\frac{\hat{T}^{(a)}_{\vec{n}_{j,p'}}}{2}$ act on the same links. In total, there are $4d^2-2d$ of such instances, which is two times that of the second case because $\hat{H}_i$ can be labelled by both $(1,4)$ and $(2,3)$. Each instance is bounded by
\begin{align}
    &\quad|| [[\sum_{\{ \vec{n}_{i,p}\}}\frac{\hat{T}^{(a)}_{\vec{n}_{i,p}}+\hat{T}^{(b)}_{\vec{n}_{i,p}}}{2}, \sum_{\{ \vec{n}_{j,p'}\}}\sum_{a=1}^{4}\frac{\hat{T}^{(a)}_{\vec{n}_{j,p'}}}{2}],\sum_{\{ \vec{n}_{j,p'}\}}\sum_{a=1}^{4}\frac{\hat{T}^{(a)}_{\vec{n}_{j,p'}}}{2}] || \nonumber \\
    &= || [\sum_{\{ \vec{n}_{j,p'}\}}[\sum_{\{ \vec{n}_{i,p}\}}\frac{\hat{T}^{(a)}_{\vec{n}_{i,p}}+\hat{T}^{(b)}_{\vec{n}_{i,p}}}{2}, \sum_{a=1}^{4}\frac{\hat{T}^{(a)}_{\vec{n}_{j,p'}}}{2}],\sum_{a=1}^{4}\frac{\hat{T}^{(a)}_{\vec{n}_{j,p'}}}{2}] || \nonumber \\
    &\leq 4N_B ||2\frac{\hat{T}^{(a)}_{\vec{n}_{i,p}}+\hat{T}^{(b)}_{\vec{n}_{i,p}}}{2} || \cdot || \sum_{a=1}^{4}\frac{\hat{T}^{(a)}_{\vec{n}_{j,p'}}}{2}||^2 \nonumber \\
    &= \frac{4N_B}{a^3},
\end{align}
where, once again, in the first inequality, we have used the fact that there are two kinetic operators in $\hat{H}_i$ acting on the same sites as each term in $\hat{H}_j$ and $\hat{H}_k$. In the last case, $\hat{H}_i$, $\hat{H}_j$ and $\hat{H}_k$ all act on different links. There are $\frac{8}{3}(2d^3-3d^2+d)$ such instances. This is so because if $\hat{H}_i$ acts on links with the $n$th parity-direction label, then $\hat{H}_j$ and $\hat{H}_k$ can act on links with $2d-n$ and $2d-n-1$ different labels, respectively. As such, we obtain $2[(2d-1)(2d-2)+(2d-2)(2d-3)+...+2\cdot 1] = \frac{8}{3}(2d^3-3d^2+d)$ for the total number of instances, where the factor of two is because $\hat{H}_i$ can be labelled by $(1,4)$ and $(2,3)$. Each instance is bounded by
\begin{align}
\label{eq:U1_ttt2_comm}
    &\quad|| [[\sum_{\{ \vec{n}_{i,p}\}}\frac{\hat{T}^{(a)}_{\vec{n}_{i,p}}+\hat{T}^{(b)}_{\vec{n}_{i,p}}}{2}, \sum_{\{ \vec{n}_{j,p'}\}}\sum_{c=1}^{4}\frac{\hat{T}^{(c)}_{\vec{n}_{j,p'}}}{2}],\sum_{\{ \vec{n}_{k,p''}\}}\sum_{e=1}^{4}\frac{\hat{T}^{(e)}_{\vec{n}_{k,p''}}}{2}] || \nonumber \\
    &= || [\sum_{\{ \vec{n}_{i,p}\}}[\frac{\hat{T}^{(a)}_{\vec{n}_{i,p}}+\hat{T}^{(b)}_{\vec{n}_{i,p}}}{2},\sum_{\{ \vec{n}_{j,p'}\}}\sum_{c=1}^{4} \frac{\hat{T}^{(c)}_{\vec{n}_{j,p'}}}{2}],\sum_{\{ \vec{n}_{k,p''}\}}\sum_{e=1}^{4}\frac{\hat{T}^{(e)}_{\vec{n}_{k,p''}}}{2}] || \nonumber \\
    &\leq 4N_B ||\frac{\hat{T}^{(a)}_{\vec{n}_{i,p}}+\hat{T}^{(b)}_{\vec{n}_{i,p}}}{2} || \cdot ||2 \frac{\hat{T}^{(c)}_{\vec{n}_{j,p'}}}{2} ||\cdot ||4 \frac{\hat{T}^{(e)}_{\vec{n}_{k,p''}}}{2} || \nonumber \\
    &= \frac{4N_B}{a^3},
\end{align}
where the factor of two in the second norm term of the first inequality is due to the fact that there are two kinetic operators from $\hat{H}_j$ acting on the same sites as each kinetic operator in $\hat{H}_i$. Note the inner commutator acts on four sites, connecting three links in total. Then, there are four operators in $\hat{H}_k$, acting on the four sites. This explains the factor of four in the third norm term of the first inequality. All together, the type-(i) bound is given by
\begin{align}
    \frac{N_B}{a^3} (\frac{64}{3} d^3 - 8 d^2 + \frac{11}{3} d).
\end{align}

For type-(ii) terms, we divide them into five distinct cases. Case-(i) terms are those where $\hat{H}_i$ and $\hat{H}_j$ act on the same links and $\hat{H}_j=\hat{H}_{k}$. The bound for each of the $2d$ instances in case (i) is $\frac{N_E}{2a^3}$, which we have obtained by replacing the $N_B$ with $N_E$ in (\ref{eq:U1_TTT}). Case (ii) contains terms, where $\hat{H}_i$ and $\hat{H}_j$ act on the same links, and $\hat{H}_k$ acts on links in different directions. In order to obtain a nontrivial commutator, $\hat{H}_k$ must collide with $\hat{H}_i$ and $\hat{H}_j$. Since there is only one link in each direction per site in the edge, $\hat{H}_k$ cannot share a direction with $\hat{H}_i$ and $\hat{H}_j$. Thus, $\hat{H}_k$ shares the same parity with $\hat{H}_i$ and $\hat{H}_j$, because otherwise, $\hat{H}_k$ will act on the bulk connected to the sites acted on by $\hat{H}_i$ and $\hat{H}_j$. As such, $\hat{H}_k$ acts on links in different directions but of the same parity as those acted on by $\hat{H}_i$ and $\hat{H}_j$. There are $d^2-d$ of such instances. This is so because if $\hat{H}_i$ and $\hat{H}_j$ act on links labelled by the $n$th direction, then $\hat{H}_k$ can act on links in $d-n$ different directions. Thus, we obtain $2[(d-1)+(d-2)+...+1] = d^2-d$ for the total number of instances, where the factor of $2$ is because there are two parity degrees of freedom. Moreover, since each edge is a $(d-1)-$dimensional surface, $\hat{H}_i$, $\hat{H}_j$, and $\hat{H}_k$ will collide on a $(d-2)$-dimensional surface, containing $\frac{L^{d-2}}{2}$ sites for each parity. As such, we consider one of such commutator in this case, and obtain its bound
\begin{align}
    &\quad|| [[\sum_{\{ \vec{n}_{i,p}\}}\frac{\hat{T}^{(1)}_{\vec{n}_{i,p}}+\hat{T}^{(4)}_{\vec{n}_{i,p}}}{2}, \sum_{\{ \vec{n}_{i,p}\}}\frac{\hat{T}^{(2)}_{\vec{n}_{i,p}}+\hat{T}^{(3)}_{\vec{n}_{i,p}}}{2}],\sum_{\{ \vec{n}_{i,p'}\}}\sum_{a=1}^{4}\frac{\hat{T}^{(a)}_{\vec{n}_{i,p'}}}{2}] || \nonumber \\
    &= || [\sum_{\{ \vec{n}_{i,p}\}}[\frac{\hat{T}^{(1)}_{\vec{n}_{i,p}}+\hat{T}^{(4)}_{\vec{n}_{i,p}}}{2}, \frac{\hat{T}^{(2)}_{\vec{n}_{i,p}}+\hat{T}^{(3)}_{\vec{n}_{i,p}}}{2}],\sum_{\{ \vec{n}_{i,p'}\}}\sum_{a=1}^{4}\frac{\hat{T}^{(a)}_{\vec{n}_{i,p'}}}{2}] || \nonumber \\
    &\leq 4 \frac{L^{d-2}}{2} ||\frac{\hat{T}^{(1)}_{\vec{n}_{i,p}}+\hat{T}^{(4)}_{\vec{n}_{i,p}}}{2} ||^2 || \sum_{a=1}^{4}\frac{\hat{T}^{(a)}_{\vec{n}_{i,p'}}}{2} || \nonumber \\
    &= \frac{L^{d-2}}{2a^3}.
    \label{eq:U1_ttt_e1}
\end{align}

Similarly, in the third case, $\hat{H}_i$ and $\hat{H}_k$ act on the same links, while $\hat{H}_j$ acts on links in a different direction of the same parity. There are $d^2-d$ of such instances, as in the second case, and each instance is bounded by $\frac{L^{d-2}}{2a^3}$, following similar arguments in (\ref{eq:U1_ttt_e1}). The fourth case we consider is when $\hat{H}_i$ and $\hat{H}_j$ act on links in a different direction of the same parity, and $\hat{H}_k = \hat{H}_j = \sum_{\{ \vec{n}_{j,p'}\}}\sum_{a=1}^{4}\frac{\hat{T}^{(a)}_{\vec{n}_{j,p'}}}{2}$. There are twice as many instances as the second case, i.e. $2d^2-2d$, because $\hat{H}_i$ can be of the kinds $(1,4)$ or $(2,3)$. Once again, each instance is bounded by $\frac{L^{d-2}}{2a^3}$, following similar arguments in (\ref{eq:U1_ttt_e1}). We now consider the fifth case, where $\hat{H}_i$, $\hat{H}_j$, and $\hat{H}_k$ all act on links of different directions, but the same parity. There are $\frac{2}{3}(d^3-3d^2+2d)$ such instances. This is so because if $\hat{H}_i$ acts on links in the $n$th direction, then $\hat{H}_j$ and $\hat{H}_k$ can act on links in $d-n$ and $d-n-1$ different directions. Thus, we obtain $2[(d-1)(d-2)+(d-2)(d-3)+...+2]=\frac{2}{3}(d^3-3d^2+2d)$ for the total number of instances, where the factor of two is due to the two parity degrees of freedom. Furthermore, since each edge is a $(d-1)-$dimensional surface, $\hat{H}_i$, $\hat{H}_j$, and $\hat{H}_k$ will collide on a $(d-3)$-dimensional surface, containing $\frac{L^{d-3}}{2}$ sites for each parity. As such, we consider one of such commutator in this case, and obtain its bound
\begin{align}
    &\quad|| [[\sum_{\{ \vec{n}_{i,p}\}}\frac{\hat{T}^{(a)}_{\vec{n}_{i,p}}+\hat{T}^{(b)}_{\vec{n}_{i,p}}}{2}, \sum_{\{ \vec{n}_{i,p'}\}}\sum_{c=1}^{4}\frac{\hat{T}^{(c)}_{\vec{n}_{i,p'}}}{2}],\sum_{\{ \vec{n}_{i,p''}\}}\sum_{d=1}^{4}\frac{\hat{T}^{(d)}_{\vec{n}_{i,p''}}}{2}] || \nonumber \\
    &\leq 4 \frac{L^{d-3}}{2} ||\frac{\hat{T}^{(a)}_{\vec{n}_{i,p}}+\hat{T}^{(b)}_{\vec{n}_{i,p}}}{2} ||\cdot || \sum_{c=1}^{4}\frac{\hat{T}^{(c)}_{\vec{n}_{i,p'}}}{2} ||^2 \nonumber \\
    &= \frac{L^{d-3}}{a^3}.
    \label{eq:U1_ttt_e2}
\end{align}
All together, the type-(ii) bound is given by
\begin{equation}
    \frac{N_E}{2a^3}2d+\frac{2L^{d-2}}{a^3}(d^2-d)+\frac{2L^{d-3}}{3a^3}(d^3-3d^2+2d).
\end{equation}

Now, we consider type-(iii) terms. We divide the terms into two cases. In the first case, the edge terms, $\hat{H}_i$ and $\hat{H}_j$, act on the same links. There are $4d^2$ of $[[\hat{H}_i,\hat{H}_j],\hat{H}_k]$ commutators, in this case, since there are $2d$ choices for $\hat{H}_i$ and $\hat{H}_j$ pairs and another factor of $2d$ for $\hat{H}_{k}$, but $2d$ of them are trivially zero. This is so since for a given parity and direction for $\hat{H}_i$ and $\hat{H}_j$, there is one $\hat{H}_k$ term, of the type $\sum_{a=1}^{4} \frac{\hat{T}^{(a)}}{2}$, with the same parity and direction as $\hat{H}_i$ and $\hat{H}_j$, that commute with $\hat{H}_i$ and $\hat{H}_j$. Therefore, there are $4d^2-2d$ non-vanishing commutators in this case, and each is bounded by $\frac{2N_E}{a^3}$, where we have replaced $N_B$ with the smaller $N_E$ in (\ref{eq:U1_ttt1_comm}). In the second case, $\hat{H}_i$ and $\hat{H}_j$ act on different links. There are $4d^2-2d$ such pairs of $\hat{H}_i$ and $\hat{H}_j$. Since, there are $2d$ choices of $\hat{H}_k$, there are $8d^3-4d^2$ terms in this case, and each of which is bounded by $\frac{4N_E}{a^3}$, using (\ref{eq:U1_ttt2_comm}). Therefore, we obtain the bound for the sum of all type-(iii) terms to be
\begin{equation}
    \frac{N_E}{a^3}(32d^3-8d^2-4d).
\end{equation}
Note for type-(iv) terms, the same bound can be obtained along similar lines of reasoning, with the only difference being $\hat{H}_j$ is in the bulk, and $\hat{H}_k$ is in the edge. 

Lastly, type-(v) terms can be separated into two cases. The first case is when the bulk terms $\hat{H}_j = \hat{H}_k = \sum_{\{ \vec{n}_{j,p'}\}}\sum_{a=1}^{4}\frac{\hat{T}^{(a)}_{\vec{n}_{j,p'}}}{2}$ are acting on the same links. For each choice of $\hat{H}_i$, there are $(2d-1)$ choices of $\hat{H}_j$ that yields a non-zero commutator, with the subtraction by one arising from the same parity and direction. Since there are $4d$ choices of $\hat{H}_i$, there are then $4d(2d-1) = 8d^2-4d$ instances in this first case. The bound of each instance is obtained as,
\begin{align}
    &|| [[\sum_{\{ \vec{n}_{i,p}\}}\frac{\hat{T}^{(a)}_{\vec{n}_{i,p}}+\hat{T}^{(b)}_{\vec{n}_{i,p}}}{2}, \sum_{\{ \vec{n}_{j,p'}\}}\sum_{c=1}^{4}\frac{\hat{T}^{(c)}_{\vec{n}_{j,p'}}}{2}],\sum_{\{ \vec{n}_{j,p'}\}}\sum_{c=1}^{4}\frac{\hat{T}^{(c)}_{\vec{n}_{j,p'}}}{2}] || \nonumber \\
    &= || \sum_{\{ \vec{n}_{j,p'}\}}[[\sum_{\{ \vec{n}_{i,p}\}}\frac{\hat{T}^{(a)}_{\vec{n}_{i,p}}+\hat{T}^{(b)}_{\vec{n}_{i,p}}}{2}, \sum_{c=1}^{4}\frac{\hat{T}^{(c)}_{\vec{n}_{j,p'}}}{2}],\sum_{c=1}^{4}\frac{\hat{T}^{(c)}_{\vec{n}_{j,p'}}}{2}] || \nonumber \\
    &\leq 4N_E ||\frac{\hat{T}^{(a)}_{\vec{n}_{i,p}}+\hat{T}^{(b)}_{\vec{n}_{i,p}}}{2} ||\cdot ||2 \sum_{c=1}^{4}\frac{\hat{T}^{(c)}_{\vec{n}_{j,p'}}}{2} ||^2 \nonumber \\
    &= \frac{8N_E}{a^3}.
\end{align}
In the second case, $\hat{H}_j$ and $\hat{H}_k$ act on different links. Now, for each non-trivial pair of $\hat{H}_i$ and $\hat{H}_j$, there are $2d-2$ choices of $\hat{H}_k$. Thus, there are $4d(2d-1)(2d-2)$ such terms, each of which is bounded by $\frac{4N_E}{a^3}$, using (\ref{eq:U1_ttt2_comm}). Therefore, type-(v) terms are bounded by
\begin{align}
    \frac{N_E}{a^3}(64d^3-32d^2).
\end{align}
As such, $C_{2,8}$ is bounded by
\begin{align}
    \frac{L^d}{a^3}(\frac{32}{3} d^3 - 4 d^2 + \frac{11}{6} d) + \frac{L^{d-1}}{a^3}(\frac{160}{3}d^3 - 20 d^2 - \frac{16}{3}d)+\frac{L^{d-2}}{a^3}(2d^2-2d)+\frac{2L^{d-3}}{3a^3}(d^3-3d^2+2d).
\end{align}

Now we consider the $C_{2,9}$ term, where $\hat{H}_i$ and $\hat{H}_j$ are kinetic terms, and $\hat{H}_k$ is a magnetic term. The terms can be separated into three types. Type (i): the terms where $\hat{H}_i$ and $\hat{H}_j$ are both of an edge kind. Type (ii): the terms where $\hat{H}_i$ and $\hat{H}_j$ are both of a bulk kind. The terms where $\hat{H}_i$ is in the edge and $\hat{H}_j$ is in the bulk belong to type (iii). We further divide type (i) into three cases. The first case is when $\hat{H}_i$ and $\hat{H}_j$ act on the same link. There are $2d$ such pairs of $\hat{H}_i$ and $\hat{H}_j$, and for each pair, there are $2(d-1)$ magnetic terms that act on links in the same direction. As such, there are $4d^2-4d$ commutators in this case, each of which is bounded by
\begin{align}
    &\quad|| [[\sum_{\{ \vec{n}_{i,p}\}} \frac{\hat{T}^{(1)}_{\vec{n}_{i,p}}+\hat{T}^{(4)}_{\vec{n}_{i,p}}}{2}, \sum_{\{ \vec{n}_{i,p}\}}\frac{\hat{T}^{(2)}_{\vec{n}_{i,p}}+\hat{T}^{(3)}_{\vec{n}_{i,p}}}{2}],\sum_{\vec{n}}\hat{L}_{\vec{n}}^{(B)}] || \nonumber \\
    &\leq ||\sum_{\{ \vec{n}_{i,p}\}} [[ \frac{\hat{T}^{(1)}_{\vec{n}_{i,p}}+\hat{T}^{(4)}_{\vec{n}_{i,p}}}{2}, \frac{\hat{T}^{(2)}_{\vec{n}_{i,p}}+\hat{T}^{(3)}_{\vec{n}_{i,p}}}{2}],\hat{L}_{\vec{n}_{i,\Box}}^{(B)}] || \nonumber \\
    &\leq 4N_E || \frac{\hat{T}^{(1)}_{\vec{n}_{i,p}}+\hat{T}^{(4)}_{\vec{n}_{i,p}}}{2} ||^2 \cdot || \frac{-1}{2g^2 a^{4-d}}(\hat{U}\hat{U}\hat{U}^\dag \hat{U}^\dag + h.c.) ||\nonumber \\
    &= \frac{N_E}{g^2 a^{6-d}},
\end{align}
where in step two, we have used the fact that only kinetic and magnetic operators that act on the same links yield non-zero commutators. The bound for all the terms in this case is then
\begin{equation}
    \frac{4N_E}{g^2 a^{6-d}}(d^2-d).
\end{equation}
The second case is when $\hat{H}_i$ and $\hat{H}_j$ act on different links, but in the same direction. $\hat{H}_i$ and $\hat{H}_j$ have to be even and odd, respectively, due to the ordering. Since the links in the edge for any given direction are not connected, $\hat{H}_i$ and $\hat{H}_j$ must commute. Therefore, the bound for this case is zero.

In the third case, $\hat{H}_i$ and $\hat{H}_j$ act on links in different directions. There are $4d^2-4d$ pairs in this case. This is so because if $\hat{H}_i$ acts on links in the $n$th direction, then $\hat{H}_j$ can act on links in $d-n$ different directions. Thus, we obtain $8[(d-1)+(d-2)+...+1]=4d^2-4d$, where the factor of 8 is due to the fact that $\hat{H}_i$ can be of the kinds $(1,4)$ or $(2,3)$, and each of $\hat{H}_i$ and $\hat{H}_j$ can be of two parities. For each pair of $\hat{H}_i$ and $\hat{H}_j$, there are $2$ magnetic terms that do not commute with both, since the two directions from the pair form a plane, and for each plane the magnetic terms can be of different parities. Each inner commutator acts on three links, due to the fact that there are two $\hat{H}_j$ terms acting on the same sites as an $\hat{H}_i$ term. For each of such triple-link configuration, there are $2(d-2)$ magnetic terms acting on each single link, where the factor of two comes from the two parities, and the factor of $(d-2)$ is due to the fact that each magnetic term do not act on both links at the same time. Therefore, for each pair of $\hat{H}_i$ and $\hat{H}_j$, the commutator is bounded by
\begin{align}
    &\quad|| [[\sum_{\{ \vec{n}_{i,p}\}} \frac{\hat{T}^{(a)}_{\vec{n}_{i,p}}+\hat{T}^{(b)}_{\vec{n}_{i,p}}}{2}, \sum_{\{ \vec{n}_{j,p}\}}\sum_{c=1}^{4} \frac{\hat{T}^{(c)}_{\vec{n}_{j,p'}}}{2}],\sum_{\vec{n}}\hat{L}_{\vec{n}}^{(B)}] || \nonumber \\
    &\leq 4N_E || \frac{\hat{T}^{(a)}_{\vec{n}_{i,p}}+\hat{T}^{(b)}_{\vec{n}_{i,p}}}{2} || \cdot|| 2 \sum_{c=1}^{4} \frac{\hat{T}^{(c)}_{\vec{n}_{j,p'}}}{2} || \cdot || \frac{-(2+3\cdot 2(d-2))}{2g^2 a^{4-d}}(\hat{U}\hat{U}\hat{U}^\dag \hat{U}^\dag + h.c.) ||\nonumber \\
    &= \frac{(24d-40)N_E}{g^2 a^{6-d}}.
    \label{eq:U1_ttl}
\end{align}
Therefore, the third case is bounded by
\begin{equation}
    \frac{N_E}{g^2 a^{6-d}}(24d-40)(4d^2-4d).
\end{equation}
Finally, the bound for type-(i) terms is then
\begin{equation}
    \frac{N_E}{g^2 a^{6-d}}(96 d^3 - 252 d^2 + 156 d).
\end{equation}

Similarly, we separate type-(ii) terms into three cases, and obtain the bounds for case (i) and (iii) as
\begin{equation}
    \frac{4N_B}{g^2 a^{6-d}}(d^2-d)
\end{equation}
and
\begin{equation}
    \frac{N_B}{g^2 a^{6-d}}(24d-40)(4d^2-4d),
\end{equation}
respectively. In the second case, $\hat{H}_i$ and $\hat{H}_j$ act on different links, but in the same direction. There are $2d$ such pairs of $\hat{H}_i$ and $\hat{H}_j$ because $\hat{H}_i$ and $\hat{H}_j$ have to be even and odd, respectively, due to the ordering, and $\hat{H}_i$ can be either $(1,4)$ or $(2,3)$, while $\hat{H}_j$ is of the form $\sum_{a=1}^{4}\frac{\hat{T}^{(a)}}{2}$. For each pairs of $\hat{H}_i$ and $\hat{H}_j$, the commutator is bounded by
\begin{align}
\label{eq:U1_ttt3_comm}
    &\quad|| [[\sum_{\{ \vec{n}_{e,p}\}} \frac{\hat{T}^{(a)}_{\vec{n}_{e,p}}+\hat{T}^{(b)}_{\vec{n}_{e,p}}}{2}, \sum_{\{ \vec{n}_{o,p}\}}\sum_{c=1}^{4} \frac{\hat{T}^{(c)}_{\vec{n}_{o,p}}}{2}],\sum_{\vec{n}}\hat{L}_{\vec{n}}^{(B)}] || \nonumber \\
    &\leq 4N_B || \frac{\hat{T}^{(a)}_{\vec{n}_{e,p}}+\hat{T}^{(b)}_{\vec{n}_{e,p}}}{2} || \cdot|| 2 \sum_{c=1}^{4} \frac{\hat{T}^{(c)}_{\vec{n}_{o,p}}}{2} || \cdot || \frac{-3\cdot 2(d-1)}{2g^2 a^{4-d}}(\hat{U}\hat{U}\hat{U}^\dag \hat{U}^\dag + h.c.) ||\nonumber \\
    &= \frac{24(d-1)N_B}{g^2 a^{6-d}},
\end{align}
where the factor two in the second norm term of the first inequality is because of the fact that there are two elements $\vec{n}_{o,p}$ that result in non-vanishing inner commutator due to collision of sites for a given $\vec{n}_{e,p}$. Further, for each inner commutator, which acts on three links, there are $3\cdot 2(d-1)$ magnetic terms that overlap on these links. As such, the second case is bounded by
\begin{equation}
    \frac{48N_B}{g^2 a^{6-d}}(d^2-d).
\end{equation}
Thus, the bound of type-(ii) terms is
\begin{equation}
    \frac{N_B}{g^2 a^{6-d}}(96 d^3 - 204 d^2 + 108 d).
\end{equation}

Now we consider type (iii), where $\hat{H}_i$ and $\hat{H}_j$ are from the edge and bulk, respectively. Once again, we divide the terms up into three cases. In the first case, $\hat{H}_i$ and $\hat{H}_j$ act on links in the same direction and sites of the same parity. There are $4d$ such pairs, and they commute with each other. The second case is where $\hat{H}_i$ and $\hat{H}_j$ act on links in the same direction, but sites of different parities. There are $4d$ such pairs, and there are $3(2d-2)$ magnetic terms that do not commute with each pair. Therefore, for each pair of $\hat{H}_i$ and $\hat{H}_j$, the bound for the commutator is $\frac{(24d-24)N_E}{g^2 a^{6-d}}$, using similar arguments as (\ref{eq:U1_ttt3_comm}). The third case contains commutators, in which $\hat{H}_i$ and $\hat{H}_j$ act on links in different directions. There are $8d^2-8d$ such pairs. This is so because there are $4d$ $\hat{H}_i$ and for each $\hat{H}_i$, $\hat{H}_j$ can be of $(d-1)$ directions and 2 parities. Further, there are $6d-10$ magnetic terms that do not commute with each pair, following similar arguments in (\ref{eq:U1_ttl}). Once again, for each pair of $\hat{H}_i$ and $\hat{H}_j$, the bound for the commutator is $\frac{(24d-40)N_E}{g^2 a^{6-d}}$, using similar arguments as (\ref{eq:U1_ttl}). As such the bound for type-(iii) terms is given by
\begin{equation}
    \frac{N_E}{g^2 a^{6-d}}(192d^3-416d^2+224d).
\end{equation}
Summing up the bounds for type-(i), -(ii), and -(iii) terms, we obtain the bound for $C_{2,9}$ as
\begin{equation}
    \frac{L^d}{g^2 a^{6-d}}(48d^3-102d^2+54d) + \frac{L^{d-1}}{g^2 a^{6-d}}(96d^3-232d^2+136d).
\end{equation}

Now for $C_{2,10}$, we can use a similar approach. The only difference is that in each case, we only need to consider the magnetic terms, $\hat{H}_j$, that do not commute with the first kinetic term, $\hat{H}_i$. This is because if $[\hat{H}_i, \hat{H}_j]=0$, $[[\hat{H}_i,\hat{H}_j],\hat{H}_k]=0$. If we divide up the terms into cases based on the kinetic terms $\hat{H}_i$ and $\hat{H}_k$ as we did for $C_{2,9}$, and use the fact that there are $2(d-1)$ magnetic terms $\hat{H}_j$ that do not commute with each $\hat{H}_i$, we can compute the bound for each case. For case (i) of type (i) and (ii), we obtain the bound
\begin{equation}
    \frac{N_E+N_B}{g^2a^{6-d}}(2d)2(d-1).
\end{equation}
For case (ii) of type (i), there are $2d$ pairs of $\hat{H}_i$ and $\hat{H}_k$ because there $d$ directions and $\hat{H}_i$ can be of type $(1,4)$ or $(2,3)$. For each pair, the commutator is bounded by
\begin{align}
    &\quad|| [[\sum_{\{ \vec{n}_{e,p}\}} \frac{\hat{T}^{(a)}_{\vec{n}_{e,p}}+\hat{T}^{(b)}_{\vec{n}_{e,p}}}{2},\sum_{\vec{n}_{k,\Box}}\hat{L}_{\vec{n}_{k,\Box}}^{(B)}], \sum_{\{ \vec{n}_{o,p}\}}\sum_{c=1}^{4} \frac{\hat{T}^{(c)}_{\vec{n}_{o,p}}}{2}] || \nonumber \\
    &\leq 4N_E || \frac{\hat{T}^{(a)}_{\vec{n}_{e,p}}+\hat{T}^{(b)}_{\vec{n}_{e,p}}}{2} ||  \cdot || \frac{-2(d-1)}{2g^2 a^{4-d}}(\hat{U}\hat{U}\hat{U}^\dag \hat{U}^\dag + h.c.) ||\cdot|| 2 \sum_{c=1}^{4} \frac{\hat{T}^{(c)}_{\vec{n}_{o,p}}}{2} ||\nonumber \\
    &= \frac{8(d-1)N_E}{g^2 a^{6-d}},
    \label{eq:U1_c210_21}
\end{align}
where in the second line, the factor of $2$ in the third norm term is because of the fact that for a given $\vec{n}_{e,p}$, there are two odd kinetic terms, of the type $\vec{n}_{o,p}$, that collide on two links with the magnetic term. For case (ii) of type (ii), once again, there are $2d$ pairs of $\hat{H}_i$ and $\hat{H}_k$. For each pair, the commutator is bounded by
\begin{align}
    &\quad|| [[\sum_{\{ \vec{n}_{e,p}\}} \frac{\hat{T}^{(a)}_{\vec{n}_{e,p}}+\hat{T}^{(b)}_{\vec{n}_{e,p}}}{2},\sum_{\vec{n}_{k,\Box}}\hat{L}_{\vec{n}_{k,\Box}}^{(B)}], \sum_{\{ \vec{n}_{o,p}\}}\sum_{c=1}^{4} \frac{\hat{T}^{(c)}_{\vec{n}_{o,p}}}{2}] || \nonumber \\
    &\leq 4N_E || \frac{\hat{T}^{(a)}_{\vec{n}_{e,p}}+\hat{T}^{(b)}_{\vec{n}_{e,p}}}{2} ||  \cdot || \frac{-2(d-1)}{2g^2 a^{4-d}}(\hat{U}\hat{U}\hat{U}^\dag \hat{U}^\dag + h.c.) ||\cdot|| 4 \sum_{c=1}^{4} \frac{\hat{T}^{(c)}_{\vec{n}_{o,p}}}{2} ||\nonumber \\
    &= \frac{16(d-1)N_B}{g^2 a^{6-d}},
\end{align}
where in the second line, the factor of $4$ in the third norm term is because of the fact that each even kinetic term of the type $\vec{n}_{e,p}$ collides with two odd kinetic terms of the type $\vec{n}_{o,p}$, and each magnetic term collides with two odd kinetic terms. Thus, for case (ii) of type (i) and (ii), we obtain the bound
\begin{equation}
    \frac{4N_E+8N_B}{g^2 a^{6-d}}2d\cdot 2(d-1).
\end{equation}
For case (iii) of type (i) and (ii), we obtain the bound
\begin{equation}
    \frac{4N_E+4N_B}{g^2 a^{6-d}}2(d-1)(4d^2-4d),
\end{equation}
by replacing the number of non-commuting magnetic terms. Next, there are $4d$ pairs of $\hat{H}_i$ and $\hat{H}_k$ in case (ii) of type (iii), and the commutator bound for each pair has the same bound as (\ref{eq:U1_c210_21}), $\frac{8(d-1)N_E}{g^2 a^{6-d}}$. For case (iii) of type (iii), there are $(8d^2-8d)$ pairs of $\hat{H}_i$ and $\hat{H}_k$ in case (ii) of type (iii), and the commutator bound for each pair has the same bound as (\ref{eq:U1_c210_21}), $\frac{8(d-1)N_E}{g^2 a^{6-d}}$. Thus, for type (iii), we obtain the bound 
\begin{equation}
    \frac{4N_E}{g^2a^{6-d}}(8d^2-4d)2(d-1),
\end{equation}
by replacing the number of non-commuting magnetic terms. By summing up the bounds for all cases, we obtain the bound for $C_{2,10}$,
\begin{align}
    &\quad\frac{N_E+N_B}{g^2a^{6-d}}[(2d)(2d-2) + 8d\cdot 2(d-1) + 8(d-1)(4d^2-4d)] \nonumber \\
    &+ \frac{N_E}{g^2 a^{6-d}}(8d^2-4d)(8d-8) + \frac{N_B}{g^2 a^{6-d}}8d(2d-2)\nonumber \\
    &= \frac{L^d}{g^2a^{6-d}}(16 d^3 - 10 d^2 -6 d) + \frac{L^{d-1}}{g^2a^{6-d}}(32 d^3 - 56 d^2+24d).
\end{align}

Lastly, we consider $C_{2,11}$, where $\hat{H}_i$ is a kinetic term, and $\hat{H}_j$ and $\hat{H}_k$ are magnetic terms. We evaluate its bound as follows:
\begin{align}
    &\quad \sum_{i=e}^{o} \sum_{p =1}^{2d} \sum_{(a,b)=(1,4)}^{(2,3)}|| [[\sum_{\{ \vec{n}_{i,p}\}} \frac{\hat{T}^{(a)}_{\vec{n}_{i,p}}+\hat{T}^{(b)}_{\vec{n}_{i,p}}}{2},\sum_{\vec{n}}\hat{L}_{\vec{n}}^{(B)}],\sum_{\vec{n}'}\hat{L}_{\vec{n}'}^{(B)}]] || \nonumber \\
    &\leq 8d (N_E+N_B)\cdot 4 ||\frac{\hat{T}^{(a)}_{\vec{n}_{i,p}}+\hat{T}^{(b)}_{\vec{n}_{i,p}}}{2} || \cdot || \frac{-2(d-1)}{2a^{4-d}g^2}(\hat{U}\hat{U}\hat{U}^\dag \hat{U}^\dag + h.c.)||\cdot || \frac{-(14d-20)}{2a^{4-d}g^2}(\hat{U}\hat{U}\hat{U}^\dag \hat{U}^\dag + h.c.)|| \nonumber \\
    &= \frac{L^d(224 d^3-544 d^2 + 320 d)}{a^{9-2d} g^4},
    \label{eq:U1_KBB}
\end{align}
where in the second line, the factor of $2(d-1)$ in the second norm term is the number of magnetic terms that act on the same link as the kinetic term, since there are $(d-1)$ planes that share a direction with the kinetic term, and on each plane, there is a pair of magnetic terms of two different parities. For each pair of such magnetic terms, there are 8 colliding magnetic terms on the same plane, and $7\cdot 2(d-2)$ colliding magnetic terms on different planes, where 7 is the number of links acted on by the pair and $2(d-2)$ is the number of colliding magnetic terms on each of the 7 links. This explains the factor of $8+7\cdot 2(d-2) = 14d-20$.

As a final step, we include the second-order Trotter error for the magnetic term, of which the ordering is given by (\ref{eq:U1_mag_trot}). We use (\ref{eq:comm_bound}) to bound both the commutators $[[\hat{H}_i,\sum_{j>i}\hat{H}_j],\hat{H}_i]$ and $[[\hat{H}_i,\sum_{j>i}\hat{H}_j],\sum_{k>i}\hat{H}_k]$. We remind the readers that each $\hat{H}_i$ is of the form 
\begin{equation}
    \frac{-1}{2a^{4-d}g^2}\hat{U}^{(1)^\dag \alpha}\hat{U}^{(2)\dag \beta} \hat{U}^{(3)\dag \gamma} \hat{U}^{(4)^\dag \delta} \hat{R}_{\Box}\hat{U}^{(1) \alpha} \hat{U}^{(2) \beta} \hat{U}^{(3) \gamma}\hat{U}^{(4) \delta},
\end{equation}
where $(\alpha, \beta, \gamma, \delta) \in S_{GC} \equiv \{GC(0),GC(1),...,GC(15)\}$. Excluding the prefactor $\frac{-1}{2a^{4-d}g^2}$, the norm of the operator is two. Furthermore, $\hat{H}_i$ operators with different values of $\alpha,\beta,\gamma,\delta$ are submatrices, which act on disjoint sets of states and have no overlapping elements, of
\begin{equation}
    \frac{-1}{2a^{4-d}g^2}(\hat{U}\hat{U}\hat{U}^\dag \hat{U}^\dag + h.c.),
\end{equation}
of which the norm is two, excluding the prefactor $\frac{-1}{2a^{4-d}g^2}$. Therefore, we obtain the inequality
\begin{equation}
    ||\sum_{(\alpha, \beta, \gamma, \delta) \in b; b\subseteq S_{GC}} \frac{-1}{2a^{4-d}g^2}\hat{U}^{(1)^\dag \alpha}\hat{U}^{(2)\dag \beta} \hat{U}^{(3)\dag \gamma} \hat{U}^{(4)^\dag \delta} \hat{R}_{\Box}\hat{U}^{(1) \alpha} \hat{U}^{(2) \beta} \hat{U}^{(3) \gamma}\hat{U}^{(4) \delta}|| \leq  \frac{1}{a^{4-d}g^2}.
\end{equation}
Using this relation, we evaluate the bound for the latter type of commutator
\begin{align}
    [[\hat{H}_i,\sum_{j>i}\hat{H}_j],\sum_{k>i}\hat{H}_k] \leq 4||\frac{1}{a^{4-d}g^2}||^3 = \frac{4}{a^{12-3d} g^6}.
\end{align}
As such, in accordance to (\ref{eq:U1_trotter_err}) and a straightforward counting argument, the error of the magnetic term is given by
\begin{equation}
    \frac{1}{12}\sum_{i}||[[\hat{H}_i,\sum_{j>i}\hat{H}_j],\hat{H}_i]|| + \frac{1}{24} \sum_{i}||[[\hat{H}_i,\sum_{j>i}\hat{H}_j],\sum_{k>i}\hat{H}_k] || \leq L^d\frac{d(d-1)}{2}\frac{8}{a^{12-3d} g^6},
\end{equation}
where $L^d$ is the number of sites, $\frac{d(d-1)}{2}$ is the number of plaquettes per site, and $\frac{1}{a^{12-3d} g^6}$ is the Trotter error per plaquette.

\subsubsection{Synthesis errors}
\label{subsubsec:Synth_U1}

Here, we compute the synthesis errors for $R_z$ gates required for
each of the four terms, i.e., $
e^{i\hat{D}_{\vec{n}}^{(M)}t},
e^{i\hat{D}_{\vec{n}}^{(E)}t},
e^{i\hat{T}_{\vec{n}}^{(K)}t},
e^{i\hat{L}_{\vec{n}}^{(B)}t},
$ that were described in detail in Sec.~\ref{sec:SimCircSynth}.
To start, we consider the mass term.
In this term, we have $\lfloor\log(L^d)+1 \rfloor$ $R_z$ gates to implement.
Therefore, we incur for each mass term $\lfloor\log(L^d)+1 \rfloor \cdot \epsilon(R_z)$ amount 
of error, where $\epsilon(R_z)$ denotes the error per $R_z$ gate.

Next, we consider the electric term, which has $2(\eta + 1) \lfloor\log(dL^d)+1 \rfloor$ $R_z$ gates. Therefore, each electric term incurs $2(\eta + 1)\lfloor\log(dL^d)+1 \rfloor \cdot \epsilon(R_z)$ amount of error. If we instead use the phase gradient operation, once the gadget state $\ket{\psi_M}$ in (\ref{eq:phgradstate}) is prepared, each quantum adder call to implement the operation does not incur any synthesis error. We come back to the error incurred in preparing the gadget state itself in the next section.

For the kinetic term, there are $4d (\lfloor \log(L^d-L^{d-1}) + 1\rfloor + \lfloor \log(L^{d-1}) + 1\rfloor)$ $R_z$ gates to apply in total. Therefore, the total error per kinetic term is $4d (\lfloor \log(L^d-L^{d-1}) + 1\rfloor + \lfloor \log(L^{d-1}) + 1\rfloor) \cdot \epsilon(R_z)$.

For the magnetic term, there are $16d(d-1)\lfloor\log(L^d)+1\rfloor$ $R_z$ gates to apply. Hence, the amount of error per magnetic term is $16d(d-1)\cdot\lfloor\log(L^d)+1\rfloor\cdot \epsilon(R_z)$.

Lastly, we compute the total synthesis error $\epsilon_{\rm synthesis}$. Note that each term appears twice per Trotter step in the second-order product formula in (\ref{eq:2nd_PF}). However, the implementation of the diagonal terms can be optimized. In particular, the diagonal mass and electric terms applied at the beginning (end) of each Trotter step can be applied together with the terms at the end (beginning) of the previous (next) Trotter step, unless the terms are at the very beginning or end of the simulation. As such, there are $r+1$ diagonal mass and electric terms, and $2r$ off-diagonal kinetic and magnetic terms to implement in total. Thus, $\epsilon_{\rm synthesis}$ is given by
\begin{align}
    \epsilon_{\rm synthesis} &= \{ (r+1) \cdot [\lfloor\log(L^d)+1\rfloor+2(\eta + 1)\lfloor\log(dL^d)+1 \rfloor] + 2r\cdot [(16d^2-16d)\cdot\lfloor\log(L^d)+1\rfloor \nonumber \\
    &\quad +4d (\lfloor \log(L^d-L^{d-1}) + 1\rfloor
    + \lfloor \log(L^{d-1}) + 1\rfloor)] \}\cdot \epsilon(R_z),
    \label{eq:U1_synth_err}
\end{align}
where $r$, to reiterate for the convenience of the readers, is the total number of Trotter steps.

\subsubsection{Complexity analysis}
\label{subsubsec:Analysis_U1}

Having computed both the Trotter and synthesis errors in the two previous sections,
we are now ready to perform the complexity analysis for the U(1) LGT.

The total error is given by
\begin{equation}
    \epsilon_{\rm total} = \epsilon_{\rm synthesis} + \epsilon_{\rm Trotter}.
\end{equation}
Here, we evenly distribute the total error between the synthesis and Trotter errors. Focusing on the Trotter error, we obtain the number of Trotter steps by
\begin{equation}
  \epsilon_{\rm Trotter} = \frac{\epsilon_{\rm total}}{2} \:\implies\: r = \lceil \frac{T^{3/2}2^{1/2}\rho^{1/2}}{\epsilon_{\rm total}^{1/2}} \rceil.   
\end{equation}
As such, we can compute the error each $R_z$ gate can incur by
\begin{align}
    &\epsilon_{\rm synthesis} = \frac{\epsilon_{\rm total}}{2} \implies \nonumber \\
    &\epsilon(R_z) = \frac{\epsilon_{\rm total}}{2} \{ (\lceil \frac{T^{3/2}2^{1/2}\rho^{1/2}}{\epsilon_{\rm total}^{1/2}} \rceil+1) \cdot [\lfloor\log(L^d)+1\rfloor+2(\eta + 1)\lfloor\log(dL^d)+1 \rfloor] \nonumber \\
    &\quad + 2\lceil \frac{T^{3/2}2^{1/2}\rho^{1/2}}{\epsilon_{\rm total}^{1/2}} \rceil\cdot [(16d^2-16d)\cdot\lfloor\log(L^d)+1\rfloor +4d (\lfloor \log(L^d-L^{d-1}) + 1\rfloor
    + \lfloor \log(L^{d-1}) + 1\rfloor)] \}^{-1}.
\end{align}
With this, we obtain the number of T gates required to synthesize each $R_z$ gate using RUS circuit \cite{bocharov2015efficient},
\begin{equation}
    \text{Cost}(R_z)=1.15\log(\frac{1}{\epsilon(R_z)}).
\end{equation}
Combining the T gates required for implementation of $R_z$ gates and the T gates used elsewhere in the circuit, we obtain the total number of T gates for the entire circuit as
\begin{align}
   &\quad  \{ (\lceil \frac{T^{3/2}2^{1/2}\rho^{1/2}}{\epsilon_{\rm total}^{1/2}} \rceil+1) \cdot [\lfloor\log(L^d)+1\rfloor+2(\eta + 1)\lfloor\log(dL^d)+1 \rfloor]+ 2\lceil \frac{T^{3/2}2^{1/2}\rho^{1/2}}{\epsilon_{\rm total}^{1/2}} \rceil\cdot [(16d^2-16d)\cdot\lfloor\log(L^d)+1\rfloor \nonumber \\
   &+4d (\lfloor \log(L^d-L^{d-1}) + 1\rfloor
    + \lfloor \log(L^{d-1}) + 1\rfloor)] \}\cdot \text{Cost}(R_z)
   + (\lceil \frac{T^{3/2}2^{1/2}\rho^{1/2}}{\epsilon_{\rm total}^{1/2}} \rceil+1) [4(L^d-\text{Weight}(L^d))\nonumber \\
   &+ 8d L^d(\eta-2) + 8d L^d\eta(12\eta -3\lfloor \log(\eta+1)\rfloor - 2)+8(\eta + 1)(dL^d - \text{Weight}(dL^d))]
    \nonumber \\
   &+ 2\lceil \frac{T^{3/2}2^{1/2}\rho^{1/2}}{\epsilon_{\rm total}^{1/2}} \rceil\cdot[16d(2L^d - \text{Weight}(L^d-L^{d-1}) - \text{Weight}(L^{d-1}) + L^d(\eta - 2)) + 16d(d-1)(L^d[8+2\eta] \nonumber\\
   &- 4\text{Weight}(L^d) )] .
\end{align}
The size of the ancilla register is given by the ancilla qubits required by the electric Hamiltonian, since it requires the most out of all circuit elements. Taking this into account, we obtain the total number of qubits required for the simulation by summing up those in the ancilla, fermionic and gauge-field registers, which is given by
\begin{equation}
L^d + \eta dL^d + 3(\eta + 1)dL^d
+dL^d - \text{Weight}(dL^d) = [4d(\eta + 1)+1]L^d-\text{Weight}(dL^d).
\end{equation}

Note that in the case where the electric term is implemented using phase gradient operation, the T-gate count changes by 
\begin{align}
    &\quad (\lceil \frac{T^{3/2}2^{1/2}\rho^{1/2}}{\epsilon_{total}^{1/2}} \rceil+1)\cdot [4dL^d\log(\frac{2\pi a^{d-2}}{g^2 t})+O(dL^d) -2(\eta+1)\lfloor\log(dL^d)+1 \rfloor\cdot \text{Cost}(R_z)\nonumber \\
    &-8(\eta+1)(dL^d - \text{Weight}(dL^d))]+\text{Cost}(\ket{\psi_M}),
\end{align} 
where the Cost$(R_z)$ needs to be modified, since $\epsilon(R_z)$ has changed to
\begin{align}
    &\epsilon(R_z) = \frac{\epsilon_{\rm total}}{2} \{ (\lceil \frac{T^{3/2}2^{1/2}\rho^{1/2}}{\epsilon_{\rm total}^{1/2}} \rceil+1) \cdot \lfloor\log(L^d)+1\rfloor + 2\lceil \frac{T^{3/2}2^{1/2}\rho^{1/2}}{\epsilon_{\rm total}^{1/2}} \rceil\cdot [(16d^2-16d)\cdot\lfloor\log(L^d)+1\rfloor \nonumber \\
    &\quad +4d (\lfloor \log(L^d-L^{d-1}) + 1\rfloor
    + \lfloor \log(L^{d-1}) + 1\rfloor)] \}^{-1}.
\end{align}
Further, $\text{Cost}(\ket{\psi_M})$, which denotes the one-time synthesis costs of the phase gradient gadget state. Here, we choose to use the synthesis method delineated in \cite{nam2020approximate}. Briefly, we apply Hadamard gates to the register $\ket{00...0}$, and then apply gates $Z, Z^{-1/2},...,Z^{-1/2^{M-1}}$. Each $Z^\alpha$ gates are synthesized using RUS circuits \cite{bocharov2015efficient}. Let $\delta$ be the error of preparing the gadget state $\ket{\psi_M}$. Then, each gate can incur at most $M/\delta$ error, and thus, costs $1.15\log(M/\delta)$, using RUS circuits \cite{bocharov2015efficient}. Thus, the gadget state preparation costs $1.15M\log(M/\delta)$.

Finally, in this case, the ancilla-qubit count is given by that of the magnetic term and the phase gradient state. As such, the total qubit count is given by
\begin{equation}
    L^d+dL^d\eta+(L^d - \text{Weight}(L^d)) +\eta + \log(\frac{2\pi a^{d-2}}{g^2 t}) 
    = L^d(2+d\eta) +\eta - \text{Weight}(L^d) + \log(\frac{2\pi a^{d-2}}{g^2 t}).
\end{equation}

\section{SU(2) Lattice Gauge Theory}
\label{sec:SU2}
In this section, we introduce the non-Abelian SU(2) lattice gauge theory. We follow the same format as in the U(1) case to guide the readers.

\subsection{Preliminaries}
\label{sec:SU2_prelim}
Once again, we aim to simulate our system, governed by four types of Hamiltonian, i.e., the electric Hamiltonian $H_E$, magnetic Hamiltonian $H_B$, mass Hamiltonian $H_M$ and kinetic Hamiltonian $H_K$. Just as in the U(1) case, $H_E$ and $H_B$ act on the links that connect two fermionic sites, $H_M$ acts on the fermions themselves, and $H_K$ acts on nearest pairs of fermionic sites and the links that connect the pairs. Thus, we consider two different types of qubit registers, one for the fields ($H_E$, $H_B$, $H_K$) and the other for the fermions ($H_M$, $H_K$).

To simulate this system, we need to choose a good basis for each register, as in the U(1) case. For the fermionic register, we consider an occupation basis. Note however though, in the current case of SU(2), the fermions may assume two different colors in the fundamental representation. As such, the mass Hamiltonian is now of the form
\begin{equation}
    \hat{H}_M = m\sum_{\vec{n}} \sum_{\alpha=1}^{2} (-1)^{\vec{n}} \hat{\psi}_{\alpha}^{\dag}(\vec{n})\hat{\psi}_{\alpha}(\vec{n}),
\end{equation}
where $\alpha \in \{1,2\}$ denotes the color. This means that we have two subregisters, each for the two different colors, that comprise the full fermion register.
For concreteness and simplicity, we use the JW transformation~\cite{wigner1928paulische} for the rest of this section to map the fermion operators to the qubit operators.

For the link register, just as in the U(1) theory, we need to write down Gauss' law.
For SU(2), we have
\begin{equation}
    \hat{G}^{a}(\vec{n})=\sum_{k}(\hat{E}^{a}_{L}({\vec{n},k}) + \hat{E}^{a}_{R}({\vec{n},k})) + \hat{Q}^{a}({\vec{n}}),
\end{equation}
where $\hat{G}^{a}$ is the Gauss operator for the projection axis $a$. 
The charge operator for the projection axis $a$ is
\begin{equation}
    \hat{Q}^{a}(\vec{n}) = \sum_{\alpha,\beta=1}^{2}\hat{\psi}_{\alpha}^\dag \frac{1}{2}\sigma^{a}_{\alpha \beta} \hat{\psi}_{\beta},
\end{equation}
and it satisfies the SU(2) algebra
\begin{equation}
    [\hat{Q}^{a},\hat{\psi}_{\alpha}]=-\frac{1}{2}\sum_{\beta=1}^{2}\sigma^{a}_{\alpha \beta}\hat{\psi}_{\beta}.
\end{equation}
$\hat{E}^{a}_{L}({\vec{n},k})$ and $\hat{E}^{a}_{R}({\vec{n},k})$ 
are the left and right chromoelectric field operators for the projection
axis $a$, whose mathematical
properties will be discussed later.
Here, $\sigma^{a}$ are Pauli-$a$ matrices, $a \in \{1,2,3\}$ for the three different kinds, which generate the fundamental representation of SU(2). $\hat{\psi}_{\alpha}^\dag$ and $\hat{\psi}_{\alpha}$
are the fermion creation and annihilation operators of different colors $\alpha$.
Similar to the U(1) lattice gauge theory,
the SU(2) Hamiltonian commutes with all of the
Gauss operators $\hat{G}^a(\vec{n})$,
implied by gauge invariance.

The physical, gauge-invariant Hilbert space $\mathcal{H}_{G}$ is defined through the eigenstates of the Gauss operator:
\begin{equation}
\label{su2gauss}
    \mathcal{H}_{G} = \{ \ket{\Psi} \in \mathcal{H}_G \: |\:   \hat{G}^{a}(\vec{n}) \ket{\Psi} = 0, \: \forall \vec{n}, a \}.
\end{equation}
Based on the U(1) case, one may be tempted to
write the field terms in the eigenbasis of the chromoelectric field operators.  
However, as can be straightforwardly checked, the chromoelectric field and Gauss 
operators do not all commute. 
Indeed, the complete set of commuting observables on a link is conventionally taken to be
$\{ \hat{E}^{2},\hat{E}^{3}_{L},\hat{E}^{3}_{R} \}$, where
$\hat{E}^{2}$ is the total electric field squared, known as the Casimir operator~\cite{georgi1982lie}, and
$\hat{E}^{3}_{L}$ and $\hat{E}^{3}_{R}$ are the third component of the left and right
chromoelectric fields, respectively.
Specifically, on each link, the left and right electric fields each forms SU(2) Lie algebras, and they obey the commutation relations
\begin{align}
     [\hat{E}^{a}_{L},\hat{E}^{b}_{L}] &= i \sum_{c=1}^{3}\epsilon^{abc}\hat{E}^{c}_{L}, \\
     [\hat{E}^{a}_{R},\hat{E}^{b}_{R}] &= i \sum_{c=1}^{3}\epsilon^{abc}\hat{E}^{c}_{R}, \\
     [\hat{E}^{a}_{L},\hat{E}^{b}_{R}] &= 0,
\end{align}
where $\epsilon^{abc}$ is the Levi-Civita symbol. While the left and right field operators commute, we note that they are not independent, and are related by a unitary change of reference frame \cite{kogut1975hamiltonian}. Consequently, the squared fields on either side are equal:
\begin{equation}
    \hat{E}^{2} \equiv \sum_{a=1}^{3}\hat{E}^{a}_{L}\hat{E}^{a}_{L} = \sum_{a=1}^{3} \hat{E}^{a}_{R}\hat{E}^{a}_{R}.
\end{equation}
The eigenbasis for this set of operators is an angular momentum basis that describes the quantum state of a link by its irreducible representation or total angular momentum, $j$, and associated third-component projections at the left ($m^{L}$) and right ($m^{R}$) end of the link, $\ket{j,m^{L},m^{R}}$. 
In this basis, the eigenvalues are given by \cite{robson1982gauge}
\begin{align}
    &\hat{E}^{2} \ket{j,m^{L},m^{R}}  = j(j+1) \ket{j,m^{L},m^{R}},\: j=n/2,\: n\in\mathbbm{N}, \label{eq:SU2_eig1} \\
    &\hat{E}^{3}_{L/R} \ket{j,m^{L},m^{R}}  = m^{L/R}\ket{j,m^{L},m^{R}},\: m^{L/R}=-j,-j+1,...,j. \label{eq:SU2_eig2}
\end{align}

Lastly, we express the SU(2) parallel transporters, $\hat{U}_{\alpha \beta}$, in the above angular momentum basis. Dropping the link position index for notational brevity, they are defined according to 
\begin{align}
    \hat{U}_{\alpha \beta}\ket{j,m^{L},m^{R}} &= \sum_{J = |j-1/2|}^{j+1/2}\sqrt{\frac{2j+1}{2J+1}} \braket{J,M_L}{j,m^{L};\frac{1}{2},\alpha'}\braket{J,M_R}{j,m^{R};\frac{1}{2},\beta'} \nonumber \\
    &\quad \times  \ket{J,M_L=m^{L}+\alpha',M_R=m^{R}+\beta'},
\end{align}
where $\braket{J,M}{j,m,1/2,\Delta m}$ are the Clebsch–Gordan coefficients for SU(2) in the fundamental representation, and $\alpha'/ \beta'=\frac{1}{2} (-\frac{1}{2})$ for $\alpha / \beta = 1 (2)$. Note that the Clebsch-Gordan coefficients $c_{\alpha \beta}$ can be evualated using the formulas provided in Table \ref{tb:CG_SU2}. The above definition of $\hat{U}_{\alpha \beta}$ operators can be used to directly verify the proper commutation relations
\begin{align}
    [\hat{E}^{a}_{L},\hat{U}_{\alpha \beta}] &= \sum_{\gamma=1}^{2}-\frac{1}{2}\sigma^{a}_{\alpha \gamma} \hat{U}_{\gamma \beta}, \\
    [\hat{E}^{a}_{R},\hat{U}_{\alpha \beta}] &= \sum_{\gamma=1}^{2}\frac{1}{2}\hat{U}_{\alpha \gamma }\sigma^{a}_{ \gamma \beta},
\end{align}
where $\alpha$, $\beta$, $\gamma$, $\delta \in \{1,2 \}$, required by the SU(2) lattice gauge theory.
These operators follow the commutation relations of
\begin{gather}
    [\hat{U}_{\alpha \beta},\hat{U}_{\gamma \delta}] = [\hat{U}_{\alpha \beta},\hat{U}^{\dag}_{\gamma \delta}] = 0,\\
    \hat{U}_{22} = \hat{U}^{\dag}_{11}, \:\: \hat{U}_{21} = -\hat{U}^{\dag}_{12},
    \label{eq:SU2_Uhc}
\end{gather}
where $\alpha \beta\neq \gamma \delta$.

\begin{table}[htp]
\centering
 \begin{tabular}{||c c c c||} 
 \hline
 Coefficient & $\Delta j$ & $\Delta m$ & Formula \\ [1.5ex] 
 \hline
 $c_{11}$ & $1/2$ & $1/2$ & $\sqrt{\frac{j+m +1}{2j+1}}$ \\ 
 $c_{12}$ & $1/2$ & $-1/2$ & $\sqrt{\frac{j-m +1}{2j+1}}$ \\
 $c_{21}$ & $-1/2$ & $1/2$ & $-\sqrt{\frac{j-m}{2j+1}}$ \\
 $c_{22}$ & $-1/2$ & $-1/2$ & $\sqrt{\frac{j+m}{2j+1}}$ \\ [1.5ex] 
 \hline
 \end{tabular}
\caption{The formulas for Clebsch-Gordan coefficients $\braket{J=j+\Delta j,M=m + \Delta m}{j,m;\frac{1}{2},\Delta m}$. They can be efficiently computed classically \cite{biedenharn1968pattern,louck1970recent}.}
\label{tb:CG_SU2}
\end{table}

\subsection{Simulation circuit synthesis}
\label{sec:SimCircSynth_SU2}

Following the encoding described in \cite{byrnes2006simulating}, the infinite-dimensional gauge-field register consists of three subregisters, $\ket{j}, \ket{m^{L}}, \ket{m^{R}}$, each representing a quantum number. As in the U(1) case, we impose a cutoff on the electric field. In particular, for a given link, $j \in \{0,\frac{1}{2},...,\Lambda\}$ and $m^{L}, m^{R} \in \{-\Lambda, -\Lambda + \frac{1}{2}, ..., \Lambda\}$. In this basis, we import the definitions of the following useful operators from \cite{byrnes2006simulating} and slightly modify them for our discussion:
\begin{align}
    \hat{J}^{\pm} &= \sum_{j=0}^{\Lambda}\ketbra{j \pm \frac{1}{2}}{j}, \\
    \hat{N}_{\alpha} &= \begin{cases} \sum_{j=0}^{\Lambda}\sqrt{\frac{2j+1}{2j+2}}\ketbra{j}{j}, & \alpha = 1, \\ \sum_{j=0}^{\Lambda}\sqrt{\frac{2j+1}{2j}} \ketbra{j}{j}, & \alpha = 2, \end{cases} \label{eq:SU2_norm_op}\\
    \hat{M}^{L/R}_{\alpha}  &= \begin{cases} \sum_{m^{L/R}=-\Lambda}^{\Lambda} \ketbra{m^{L/R} + \frac{1}{2}}{m^{L/R}}, & \alpha = 1, \\ \sum_{m^{L/R}=-\Lambda}^{\Lambda} \ketbra{m^{L/R} - \frac{1}{2}}{m^{L/R}}, & \alpha = 2, \end{cases} \\
    \hat{c}^{L/R}_{\alpha \beta} &= \sum_{j=0}^{\Lambda} \sum_{m^{L/R}=-\Lambda}^{\Lambda} c_{\alpha \beta}(j,m^{L/R}) \ketbra{j}{j} \otimes \ketbra{m^{L/R}}{m^{L/R}}, \label{eq:SU2_CG_op}
\end{align}
where the quantum numbers $j, m^L$ and $m^R$ are incremented and decremented by $\frac{1}{2}$ at a time, $c_{\alpha \beta}$ are Clebsch-Gordon coefficients, as provided in Table \ref{tb:CG_SU2}, $\hat{N}_{\alpha}$ is the normalization operator. The formulas in Table \ref{tb:CG_SU2} could evaluate to complex numbers, if the quantum numbers are outside the ranges given in (\ref{eq:SU2_eig1}) and (\ref{eq:SU2_eig2}), i.e., $\ket{j\leq \Lambda,-j\leq m^L \leq j,-j \leq m^R \leq j}$. However, the coefficients are real if the quantum numbers are within the allowed ranges. Thus, we set the elements, which correspond to the disallowed states, of the diagonal Clebsch-Gordan operator in (\ref{eq:SU2_CG_op}) to zeros, thereby ensuring its Hermiticity. Using the operators defined above, we can express the $\hat{U}_{\alpha \beta}$ operators in the new basis as
\begin{equation}
\label{eq:SU2_U_enc}
    \hat{U}_{\alpha \beta} \equiv \hat{M}^{L}_{\alpha}\hat{M}^{R}_{\beta}[\hat{J}^+  \hat{c}^{L}_{1 \alpha}  \hat{c}^{R}_{1 \beta} \hat{N}_{1} + \hat{J}^-  \hat{c}^{L}_{2 \alpha}  \hat{c}^{R}_{2 \beta} \hat{N}_{2}],
\end{equation} 
which can straightforwardly be shown to satisfy (\ref{eq:SU2_Uhc}). Note that the encoded $\hat{U}_{\alpha \beta}$ operator in (\ref{eq:SU2_U_enc}) only maps states with the allowed quantum numbers to each other, and the disallowed states are in the operator's null space. In order to conveniently represent them on a quantum computer, we map $j,m^L,m^R$ to positive full integers. Thus, $j \in \{0,1,...,2\Lambda\}$ and $m^{L}, m^{R} \in \{0,1, ..., 4\Lambda\}$. Using the binary computational basis, the number of qubits on the subregister $\ket{j}$ is $\eta = \log(2\Lambda+1)$, and that on the subregisters $\ket{m^L}$ and $\ket{m^R}$ is $\eta+1$. Then, the operators in the qubit basis are written as
\begin{align}
    \hat{J}^{\pm} &\rightarrow \sum_{j=0}^{2\Lambda}\ketbra{j \pm 1}{j}, \label{eq:SU2_j_op}\\
    \hat{N}_{\alpha} &\rightarrow \begin{cases} \sum_{j=0}^{2\Lambda}\sqrt{\frac{j+1}{j+2}}\ketbra{j}{j}, & \alpha = 1, \\ \sum_{j=0}^{2\Lambda}\sqrt{\frac{j+1}{j}} \ketbra{j}{j}, & \alpha = 2, \end{cases} \\
    \hat{M}^{L/R}_{\alpha}  &\rightarrow \begin{cases} \sum_{m^{L/R}=0}^{4\Lambda} \ketbra{m^{L/R} + 1}{m^{L/R}}, & \alpha = 1, \\ \sum_{m^{L/R}=0}^{4\Lambda} \ketbra{m^{L/R} - 1}{m^{L/R}}, & \alpha = 2, \end{cases} \label{eq:SU2_m_op}\\
    \hat{c}^{L/R}_{\alpha \beta} &\rightarrow \sum_{j=0}^{2\Lambda} \sum_{m^{L/R}=0}^{4\Lambda} c_{\alpha \beta}(\frac{j}{2},\frac{m^{L/R}}{2}-\Lambda)\ketbra{j}{j} \otimes \ketbra{m^{L/R}}{m^{L/R}}.
    \label{eq:SU2_CG_op_shift}
\end{align}
The slight modification from the original definitions in \cite{byrnes2006simulating} is that we have introduced a periodic-wrapping term to each of $\hat{J}^{+}, \hat{M}^{L/R}_{1}$, which are now numerically equivalent to the $\hat{U}$ operators from the U(1) model, and can be represented as binary incrementers. Similarly, $\hat{J}^{-}, \hat{M}^{L/R}_{2}$, like the $\hat{U}^\dag$ operators from the U(1) model, and can be represented as binary decrementers. The undesirable effect of the periodic-wrapping terms can be removed by applying circuits that include multiply controlled gates, similar to the U(1) case. On the other hand, the fermionic register is different from the U(1) case in that we now have two different colors of fermions or anti-fermions to consider for each site. We thus allocate two qubits per site, each qubit corresponding to a different color. 

Equipped with all the necessary operator definitions and qubit register structure, we now decompose the Hamiltonian into separate parts. Just as in U(1), we have for SU(2) LGT
\begin{equation}
\hat{H} = \sum_{\vec{n}}\left[\hat{D}^{(M)}_{\vec{n}} + \hat{D}^{(E)}_{\vec{n}} + \hat{T}^{(K)}_{\vec{n}} + \hat{L}^{(B)}_{\vec{n}} \right],
\end{equation}
where
\begin{align}
    \hat{D}_{\vec{n}}^{(M)} &= \frac{m}{2}(-1)^{\vec{n}} (\hat{Z}_1(\vec{n})+\hat{Z}_2({\vec{n}})), \\
    \hat{D}_{\vec{n}}^{(E)} &=\frac{g^2}{2a^{d-2}} \sum_{l=1}^{d}\hat{E}^2(\vec{n},l)
\end{align}
are diagonal operators, where $\hat{Z}_i(\vec{n})$, with $i\in\{1,2\}$, are Pauli-$z$ operators that act on the fermion of color $i$ at site $\vec{n}$, and $(-1)^{\vec{n}}$, is either $+1$ or $-1$ depending on whether $\vec{n}$ is a fermion or anti-fermion site, respectively, reflective of the use of staggered fermions \cite{kogut1975hamiltonian}, and
\begin{equation}
\hat{T}_{\vec{n}}^{(K)} = \frac{1}{2a}\sum_{l=1}^{d}\sum_{\alpha,\beta = 1}^{2}[\hat{U}_{\alpha\beta}(\vec{n},l) \hat{\sigma}^{-}_{\alpha}(\vec{n})\hat{\sigma}^{+}_{\beta}(\vec{n}+\hat{l})\hat{\zeta}_{\alpha \beta,\vec{n},l}+h.c.]
\end{equation}
is an off-diagonal operator, which corresponds to kinetic Hamiltonian. The operators $\hat{\sigma}_{\alpha}^{\pm}(\vec{n})$ are Pauli raising and lowering operators on the fermion of color $\alpha$ at site $\vec{n}$. Further, the operators $\hat{\zeta}_{\alpha \beta,\vec{n},l}$ tensor products of $\hat{Z}$, which arise from the JW transformation and have an additional color-dependence when compared to the U(1) case. If we consider a $d$-dimensional $L^d-$site lattice, the length of each $\hat{\zeta}_{\alpha \beta,\vec{n},l}$ is $O((2L)^{d-1})$. For brevity, we suppress the $\hat{\zeta}_{\alpha \beta,\vec{n},l}$ operators in the remaining part of the section. The second off-diagonal operator that corresponds to the magnetic Hamiltonian is given by
\begin{align}
\hat{L}^{(B)}_{\vec{n}} &= -\frac{1}{2a^{4-d}g^2}\sum_{i=1}^{d} \sum_{j\neq i; j=1}^{d} \sum_{\alpha, \beta, \delta, \gamma =1}^{2} (\hat{U}_{\alpha \beta}(\vec{n},i)\hat{U}_{ \beta \gamma}(\vec{n}+\hat{i},j)\hat{U}_{\gamma \delta}^{\dag}(\vec{n}+\hat{j},i)\hat{U}_{\delta \alpha}^{\dag}(\vec{n},j) + h.c.).
\end{align}

Assuming we use Suzuki-Trotter formula \cite{suzuki1991general} as our simulation method, each Trotter terms to be implemented then are of the form
$
e^{i\hat{D}_{\vec{n}}^{(M)}t},
e^{i\hat{D}_{\vec{n}}^{(E)}t},
e^{i\hat{T}_{\vec{n}}^{(K)}t},
e^{i\hat{L}_{\vec{n}}^{(B)}t},
$
where $t$ is a sufficiently small number to ensure the Trotter error incurred is within a pre-specified tolerance. In the remaining part of this subsection, we discuss circuit syntheses for each of the four Trotter terms.

\subsubsection{Mass term \texorpdfstring{$e^{i\hat{D}_{\vec{n}}^{(M)}t}$}{}}
\label{sec:SU2_mass}

The implementation of this term is straightforward. Two single-qubit $R_{z}(\theta) = \exp(-i\theta\hat{Z}/2)$ gates, where $\theta = -m(-1)^{\vec{n}}t$, applied to the two qubits, which correspond to the two fermions at site $\vec{n}$, in the site register suffice. As in the implementation of U(1) mass term, we once again use the weight-sum trick \cite{gidney2018halving,nam2019low}, except for SU($2$), the number of same angle $R_z$ gates increased by a factor of two. Again, briefly, if the original subcircuit applies the same angle $R_z$ gates on $p$ qubits simultaneously, we can reduce the number of $R_z$ gates to $\lfloor\log(p)+1 \rfloor$, while incurring $p-{\rm Weight}(p)$ ancilla qubits and at most $4(p - {\rm Weight}(p))$ T gates. For a $d$-dimensional lattice with $L^d$ lattice sites, $p = 2L^d$.

\subsubsection{Electric term \texorpdfstring{$e^{i\hat{D}_{\vec{n}}^{(E)}t}$}{}}
\label{sec:SU2_electric}

Here, we present a method to implement the electric term. $\hat{D}_{\vec{n}}^{(E)}$ is a sum of $d$ commuting terms, and thus, its evolution can be implemented exactly as a product of $d$ sub-evolutions,
\begin{equation}
    e^{i\hat{D}_{\vec{n}}^{(E)}t} = \prod_{l=1}^{d}e^{i\frac{g^2 t}{2a^{d-2}} \hat{E}^2(\vec{n},l)}.
\end{equation}
Without loss of generality, we only discuss one sub-evolution. For notational convenience, we drop the link location index. Since the eigenvalue equation 
\begin{align}
\hat{E}^2 \ket{j,m^{L},m^{R}} &= \frac{j}{2}(\frac{j}{2}+1) \ket{j,m^{L},m^{R}}\nonumber \\
&= \frac{1}{4}[(j+1)^2-1] \ket{j,m^{L},m^{R}}
\label{eq:SU2_elec_eig}
\end{align}
only depends on the subregister $\ket{j}$, and not on the subregisters $\ket{m^{L}}$ and $\ket{m^{R}}$, we will implement the operator $e^{it \frac{g^2 t}{2a^{d-2}} \hat{E}^2}$ according to the eigenvalue equation,
\begin{equation}
    e^{i \frac{g^2 t}{2a^{d-2}} \hat{E}^2} \ket{j} = e^{i \frac{g^2 t}{8a^{d-2}} (j+1)^2} \ket{j},
\end{equation}
where the global phase of $-\frac{g^2 t}{8a^{d-2}}$ has been neglected. As in the U(1) electric term, we implement the term by first computing $(j+1)^2$ into an ancilla register, then impart the phase on all links in parallel, using the weight-sum trick, and finally, uncomputing the ancilla register. The T-gate and ancilla-qubit counts for the arithmetic operations are the same as that for the U(1) electric term. Thus, we will simply state the arithmetic costs: $8dL^d[(\eta-2)+\eta (12\eta - 3\lfloor \log(\eta+1)\rfloor - 2)]$ T gates, $3(\eta + 1)dL^d$ ancilla qubits to store $\ket{j+1}$ and $\ket{(j+1)^2}$ for all links, and $3(\eta + 1) - \lfloor \log(\eta+1)\rfloor-1$ reusable workspace ancilla qubits. For more details on the implementation, we refer the readers to Sec. \ref{sec:U1_electric}. We now discuss the phase induction. The correct phase can be induced by applying $R_z(2^k \theta)$, where $\theta = \frac{g^2 t}{8a^{d-2}}$, on the $k$th qubit of the $2(\eta + 1)-$bit ancilla state, $\ket{(j+1)^2}$. Hence, there are $2(\eta+1)$ sets of $dL^d$ same-angle $R_z$ rotations to implement, where each set can be effected using the weight-sum trick. Once again, we first compute Weight$(dL^d)$ into the ancilla register, incurring $4(dL^d - \text{Weight}(dL^d))$ T gates and $dL^d - \text{Weight}(dL^d)$ ancilla qubits, and then, applying $\lfloor \log(dL^d) + 1 \rfloor$ $R_z$ gates to the ancilla register to induce the right phase. 

Alternatively, for simulations with a fixed $g^2$, $d$ and $t$, where $a$ can be chosen such that $\frac{g^2 t}{8a^{d-2}} = \frac{\pi}{2^M}$ with $M>1$. Then, the electric evolution can be implemented as
\begin{equation}
    \ket{j} \mapsto e^{i \frac{\pi}{2^M} (j+1)^2} \ket{j}.
\end{equation}
Once again, as in U(1), we implement this by first computing $(j+1)^2$ into the ancilla register, and then, impart the phase one link at a time, using the phase gradient operation.

Note that the operations required for the electric term evolution here are almost the same as that in the U($1$) case. The only difference is the size of the adders required for the phase gradient operations. The number of T gates needed for the phase gradient adders is $4 dL^d \log \left(\frac{8a^{d-2}\pi}{g^2 t}\right)+O(dL^d)$. We refer the readers to \ref{sec:U1_electric} for a detailed discussion on the synthesis costs of the phase-gradient operation.

\subsubsection{Kinetic term \texorpdfstring{$e^{i\hat{T}_{\vec{n}}^{(K)}t}$}{}}
\label{sec:SU2_kin}

Here we present a method to implement the kinetic term. This method takes advantage of the fact that the $\hat{U}_{\alpha \beta}$ operators are block-diagonal, just as in method 1 for the U(1) kinetic and magnetic term. For brevity, we drop the site location indices. We rewrite the $\hat{U}_{\alpha \beta}$ operators as follows:
\begin{align}
    \hat{U}_{11} &= \sum_{j=0}^{2^{\eta}-1} \sum_{m^L,m^R=0}^{2^{\eta+1}-1} \sum_{\Delta j = -1}^{1} f_{11}(j,\Delta j,m^L,m^R) \ketbra{j+\Delta j}{j}\otimes\ketbra{m^L +1}{m^L}\otimes\ketbra{m^R +1}{m^R}, \\
    \hat{U}_{12} &= \sum_{j=0}^{2^{\eta}-1} \sum_{m^L,m^R=0}^{2^{\eta+1}-1} \sum_{\Delta j = -1}^{1} f_{12}(j,\Delta j,m^L,m^R)  \ketbra{j+\Delta j}{j}\otimes\ketbra{m^L +1}{m^L}\otimes\ketbra{m^R -1}{m^R}, \\
    \hat{U}_{21} &= \sum_{j=0}^{2^{\eta}-1} \sum_{m^L,m^R=0}^{2^{\eta+1}-1} \sum_{\Delta j = -1}^{1} f_{21}(j,\Delta j,m^L,m^R)  \ketbra{j+\Delta j}{j}\otimes\ketbra{m^L -1}{m^L}\otimes\ketbra{m^R +1}{m^R}, \\
    \hat{U}_{22} &= \sum_{j=0}^{2^{\eta}-1} \sum_{m^L,m^R=0}^{2^{\eta+1}-1} \sum_{\Delta j = -1}^{1} f_{22}(j,\Delta j,m^L,m^R)  \ketbra{j+\Delta j}{j}\otimes\ketbra{m^L -1}{m^L}\otimes\ketbra{m^R -1}{m^R},
\end{align}
where 
\begin{align}
f_{11}(j,1,m^L,m^R)&=\sqrt{\frac{(\frac{j+m^{L}}{2}-\Lambda+1)(\frac{j+m^{R}}{2}-\Lambda+1)}{(j+1)(j+2)}}, \nonumber \\
f_{11}(j,-1,m^L,m^R)&=\sqrt{\frac{(\frac{j-m^{L}}{2}+\Lambda)(\frac{j-m^{R}}{2}+\Lambda)}{j(j+1)}}, \nonumber \\
f_{12}(j,1,m^L,m^R)&=\sqrt{\frac{(\frac{j+m^{L}}{2}-\Lambda+1)(\frac{j-m^{R}}{2}+\Lambda+1)}{(j+1)(j+2)}},\nonumber \\
f_{12}(j,-1,m^L,m^R)&=-\sqrt{\frac{(\frac{j-m^{L}}{2}+\Lambda)(\frac{j+m^{R}}{2}-\Lambda)}{j(j+1)}}, \nonumber \\
f_{21}(j,1,m^L,m^R)&=\sqrt{\frac{(\frac{j-m^{L}}{2}+\Lambda+1)(\frac{j+m^{R}}{2}-\Lambda+1)}{(j+1)(j+2)}}, \nonumber \\
f_{21}(j,-1,m^L,m^R)&=-\sqrt{\frac{(\frac{j+m^{L}}{2}-\Lambda)(\frac{j-m^{R}}{2}+\Lambda)}{j(j+1)}}, \nonumber \\
f_{22}(j,1,m^L,m^R)&=\sqrt{\frac{(\frac{j-m^{L}}{2}+\Lambda+1)(\frac{j-m^{R}}{2}+\Lambda+1)}{(j+1)(j+2)}},\nonumber \\
f_{22}(j,-1,m^L,m^R)&=\sqrt{\frac{(\frac{j+m^{L}}{2}-\Lambda)(\frac{j+m^{R}}{2}-\Lambda)}{j(j+1)}}.
\label{eq:SU2_fab_def}
\end{align}
The operators $\hat{U}_{\alpha\beta}$ are similar to the U(1) plaquette operator $\hat{U}\hat{U}\hat{U}^\dag \hat{U}^\dag$ in that both of them raise and lower multiple registers simultaneously. The main difference between them is that the matrix elements of $\hat{U}_{\alpha\beta}$ are not all ones. However, we can still employ the trick of splitting the sums over all $j,m^L,m^R$ into series of even and odd $j,m^L,m^R$, i.e.,
\begin{equation}
    \sum_{j=0}^{2^{\eta}-1} \sum_{m^L,m^R=0}^{2^{\eta+1}-1} = \sum_{j,m^L,m^R \in \{ even, odd\}},
\end{equation}
where $j=0,1,...,2^{\eta}-1$ and $m^{L/R} = 0,1,...,2^{\eta+1}-1$. Without loss of generality, we consider the all-even series of $\hat{U}_{11}$ with $\Delta j = 1$, i.e.,
\begin{align}
\label{eq:SU2_kin_even}
    &\sum_{j,m^L,m^R\: even} f_{11}(j,1,m^L,m^R)  \ketbra{j+1}{j}\otimes\ketbra{m^L +1}{m^L}\otimes\ketbra{m^R +1}{m^R} \nonumber \\
    &= \sum_{j,m^L,m^R\: even} f_{11}(j,1,m^L,m^R) \ketbra{j_{\eta-1}...j_1}{j_{\eta-1}...j_1}\otimes \ketbra{j_0=1}{j_0=0}\otimes \ketbra{m^L_{\eta}...m^L_1}{m^L_{\eta}...m^L_1}\nonumber \\
    &\otimes \ketbra{m^L_0=1}{m^L_0=0} \otimes \ketbra{m^R_{\eta}...m^R_1}{m^R_{\eta}...m^R_1}\otimes \ketbra{m^R_0=1}{m^R_0=0} \nonumber \\
    &\equiv \hat{D}_{11}(\Delta j = 1) \otimes \hat{\sigma}^{+}\hat{\sigma}^{+}\hat{\sigma}^{+},
\end{align}
where $j_i$ denotes the $i$th binary digit of $j$, likewise for $m^L$ and $m^R$, and the diagonal part $\hat{D}_{11}(\Delta j = 1)$ is defined via the operator
\begin{align}
    \hat{D}_{\alpha \beta}(\Delta j) &= \sum_{\substack{j_{\eta-1}...j_1 \\ m^L_\eta ... m^L_1 \\ m^R_\eta...m^R_1} = 0}^1  f_{\alpha \beta}(j,\Delta j,m^L,m^R) 
    \ketbra{j_{\eta-1}...j_1}{j_{\eta-1}...j_1}
    \otimes \ketbra{m^L_\eta ... m^L_1}{m^L_\eta ... m^L_1}  \nonumber \\
    &\quad \otimes \ketbra{m^R_\eta...m^R_1}{m^R_\eta...m^R_1},
\end{align}
where $\alpha, \beta \in \{1,2\}$, $\Delta j \in \{-1,1 \}$, and the zeroth digits of $j, m^L, m^R$ are classically known. The remaining digits can be gleaned from the qubits, and thus, with an abuse of notation for j, $f_{\alpha \beta}(j,\Delta j,m^L,m^R)$ can be evaluated on a quantum computer. By conjugating (\ref{eq:SU2_kin_even}) with binary decrementers and incrementers, the three sub-registers of different parities can be addressed, as shown by the following expressions for $\hat{U}_{11}, \hat{U}_{12}, \hat{U}_{21}$ and $\hat{U}_{22}$:
\begin{align}
    \hat{U}_{11} &= \sum_{a,b,c=0}^{1} (\hat{J}^{+})^a (\hat{M}_{1}^{L})^b(\hat{M}_{1}^{R})^c[\hat{D}_{11}(\Delta j = 1) \otimes \hat{\sigma}^{+}\hat{\sigma}^{+}\hat{\sigma}^{+} \nonumber \\
    &\quad +\hat{D}_{11}(\Delta j = -1) \otimes \hat{\sigma}^{-}\hat{\sigma}^{+}\hat{\sigma}^{+}](\hat{J}^{-})^a (\hat{M}_{2}^{L})^b (\hat{M}_{2}^{R})^c \\
    \hat{U}_{12} &= \sum_{a,b,c=0}^{1} (\hat{J}^{+})^a (\hat{M}_{1}^{L})^b(\hat{M}_{1}^{R})^c[\hat{D}_{12}(\Delta j = 1) \otimes \hat{\sigma}^{+}\hat{\sigma}^{+}\hat{\sigma}^{-} \nonumber \\
    &\quad+ \hat{D}_{12}(\Delta j = -1) \otimes \hat{\sigma}^{-}\hat{\sigma}^{+}\hat{\sigma}^{-}](\hat{J}^{-})^a (\hat{M}_{2}^{L})^b (\hat{M}_{2}^{R})^c \\
    \hat{U}_{21} &= \sum_{a,b,c=0}^{1} (\hat{J}^{+})^a (\hat{M}_{1}^{L})^b(\hat{M}_{1}^{R})^c[\hat{D}_{21}(\Delta j = 1) \otimes \hat{\sigma}^{+}\hat{\sigma}^{-}\hat{\sigma}^{+} \nonumber \\
    &\quad +\hat{D}_{21}(\Delta j = -1) \otimes \hat{\sigma}^{-}\hat{\sigma}^{-}\hat{\sigma}^{+}](\hat{J}^{-})^a (\hat{M}_{2}^{L})^b (\hat{M}_{2}^{R})^c \\
    \hat{U}_{22} &= \sum_{a,b,c=0}^{1} (\hat{J}^{+})^a (\hat{M}_{1}^{L})^b(\hat{M}_{1}^{R})^c[\hat{D}_{22}(\Delta j = 1) \otimes \hat{\sigma}^{+}\hat{\sigma}^{-}\hat{\sigma}^{-} \nonumber \\
    &\quad+ \hat{D}_{22}(\Delta j = -1) \otimes \hat{\sigma}^{-}\hat{\sigma}^{-}\hat{\sigma}^{-}](\hat{J}^{-})^a (\hat{M}_{2}^{L})^b (\hat{M}_{2}^{R})^c,
\end{align}
where, as defined in (\ref{eq:SU2_j_op}) and (\ref{eq:SU2_m_op}), $(\hat{J}^{+})^\dag = \hat{J}^{-}$ and $(\hat{M}_{1}^{L/R})^\dag = \hat{M}_{2}^{L/R}$. We simplify the notation and express $\hat{U}_{\alpha \beta}$ as
\begin{align}
    \hat{U}_{\alpha \beta}+h.c. &= \sum_{a,b,c=0}^{1} (\hat{J}^{+})^a (\hat{M}_{1}^{L})^b(\hat{M}_{1}^{R})^c [\sum_{\Delta j = -1}^{1}\hat{D}_{\alpha \beta}(\Delta j) \otimes (\hat{P}_{\alpha \beta}(\Delta j)+h.c.)](\hat{J}^{-})^a (\hat{M}_{2}^{L})^b (\hat{M}_{2}^{R})^c,
    \label{eq:SU2_kin_par}
\end{align}
where $\hat{P}_{\alpha \beta}(\Delta j)$ are Pauli operators of the form $\otimes_{i=0}^{2}\hat{\sigma}_{i}^{\pm}$ that depend on $\alpha$, $\beta$, and $\Delta j$.

Using $\alpha, r$ and $\beta, r+1$ to denote two colors $\alpha, \beta$ at two sites $r, r+1$, without loss of generality, we write
\begin{align}
    \hat{U}_{\alpha \beta}\hat{\sigma}_{\alpha,r}^{-}\hat{\sigma}_{\beta,r+1}^{+} + h.c. &= \sum_{a,b,c=0}^{1} (\hat{J}^{+})^a (\hat{M}_{1}^{L})^b(\hat{M}_{1}^{R})^c [\sum_{\Delta j = -1}^{1}\hat{D}_{\alpha \beta}(\Delta j) \otimes \hat{F}_{\alpha \beta}(\Delta j)](\hat{J}^{-})^a (\hat{M}_{2}^{L})^b (\hat{M}_{2}^{R})^c,
    \label{eq:SU2_U_decomp}
\end{align}
where $\hat{F}_{\alpha \beta}(\Delta j)=\hat{P}_{\alpha \beta}(\Delta j)\hat{\sigma}_{\alpha,r}^{-}\hat{\sigma}_{\beta,r+1}^{+} + h.c.$ acts on both the gauge field and fermionic registers. We may now approximate the kinetic evolution $ e^{it \hat{T}_r^{(K)} }$ as
\begin{align}
    e^{it \hat{T}_r^{(K)} } \approx \prod_{(c,b,a)=GC(0)}^{GC(7)} (\hat{J}^{+})^a (\hat{M}_{1}^{L})^b(\hat{M}_{1}^{R})^c  \left[\prod_{\alpha, \beta=0}^{1} \prod_{\Delta j = -1}^{1} e^{i\frac{t}{2a}\hat{D}_{\alpha \beta}(\Delta j)\hat{F}_{\alpha \beta}(\Delta j)}\right] (\hat{J}^{-})^a (\hat{M}_{2}^{L})^b (\hat{M}_{2}^{R})^c,
\end{align}
where we have minimized the number of binary incrementers and decrementers using the Gray code. In particular, the number of binary incrementers or decrementers is reduced to eight, i.e., four acting on the $\eta-$bit $\ket{j}$ register and four acting on the $(\eta+1)-$bit $\ket{m^L}$ and $\ket{m^R}$ registers, which cost $4\cdot 4(\eta-2) + 4 \cdot 4(\eta + 1 -2) = 32\eta - 48$ T gates and $\eta+1$ reuseable ancilla qubits in total \cite{shaw2020quantum}. 

The operators $e^{i\frac{t}{2a}\hat{D}_{\alpha \beta}\hat{F}_{\alpha \beta}(\Delta j)}$ can be implemented by first diagonalizing $\hat{F}_{\alpha \beta}(\Delta j)$ with CNOT gates and two Hadamard gates, just as in the U(1) case. Let us denotes $\hat{\mathcal{D}}_{\alpha \beta}(\Delta)$ as the diagonalized $\hat{F}_{\alpha \beta}(\Delta j)$. In order to understand the action of the diagonalized evolution $e^{i\frac{t}{2a} \hat{D}_{\alpha \beta} \hat{\mathcal{D}}_{\alpha \beta}(\Delta)}$, it is instrumental to first consider a hypothetical scenario where the operator $\hat{D}_{\alpha \beta}$ is an identity operator. Then, in the exponent of $e^{i\frac{t}{2a}\hat{D}_{\alpha \beta}\hat{F}_{\alpha \beta}(\Delta j)}$, we are left with $\hat{F}_{\alpha \beta}(\Delta j)$, i.e. a string of Pauli ladder operators of the kind $\otimes_{i=0}^{4} \sigma^{\pm}_i + h.c.$, which can be reduced to a quadruply-controlled $R_x$ gate with $8$ CNOT gates. When we diagonalize the quadruply-controlled $R_x$ gate with a pair of Hadamard gates on the target qubit, the resulting quadruply-controlled $R_z$ gate imparts phases to the states $\ket{11110}$ and $\ket{11111}$, assuming the last qubit is the target. Now, we return to the original case with the operator $\hat{D}_{\alpha \beta}$ restored. Here, instead of the multiply-controlled $R_z$ gate, we have a multiply-controlled diagonal gate. We can implement this via a phase oracle $\hat{O}_{\alpha \beta}^{(\Delta j)}$, which is defined according to
\begin{align}
    &\quad \hat{O}_{\alpha \beta}^{(\Delta j)} \ket{j^\prime}\ket{m^{L\prime}}\ket{m^{R\prime}}\ket{f_{\alpha,r}^\prime}\ket{f_{\beta,r+1}^\prime} \nonumber \\
    & \mapsto e^{i f_{\alpha \beta}(j^\prime,\Delta j, m^{L\prime}, m^{R\prime})[\frac{t}{2a}(-1)^{j_0^\prime}m^{L\prime}_0 m^{R\prime}_0 f_{\alpha,r}^\prime f_{\beta,r+1}^\prime]} \ket{j^\prime}\ket{m^{L\prime}}\ket{m^{R\prime}}\ket{f_{\alpha,r}^\prime}\ket{f_{\beta,r+1}^\prime},
    \label{eq:SU2_kin_oracle}
\end{align}
where ${j_0'}, m^{L\prime}_0, m^{R\prime}_0$ are the zeroth digits of ${j'}, m^{L\prime}, m^{R\prime}$, the first three registers of qubits are for a link, and the latter two registers of length one each denote the fermions of colors $\alpha$ and $\beta$ that sit at sites $r$ and $r+1$, respectively. This oracle can be implemented efficiently with qRAM \cite{giovannetti2008quantum} in polylogarithmic time. Alternatively, we choose to synthesize this oracle directly as a diagonal gate, which imparts the phase $(-1)^{j_0'}f_{\alpha \beta}(j, \Delta j, m^L, m^R)\frac{t}{2a}$, controlled by $m^{L\prime}_0, m^{R\prime}_0,f_{\alpha,r}^\prime f_{\beta,r+1}^\prime$. In particular, for each link, $f_{\alpha \beta}(j, \Delta j, m^L, m^R)$ can be computed efficiently using fixed point arithmetic circuits \cite{bhaskar2016quantum}, and the phases can be induced using $R_z$ gates. We refer readers to sec.~\ref{subsubsec:oracle_SU2} for the detailed implementation.

Hereafter, we assume in our resource analysis that the oracles with different values of $\alpha, \beta, \Delta j$ have the same cost $\mathcal{C}^{(K)}$ per query. Then, the implementation of the kinetic term for a pair of nearest-neightbor sites and the link joining them costs $32\eta -48$ T gates and $64 \mathcal{C}^{(K)}$. Note that the neglected $\hat{\zeta}$ multi-site Pauli-$z$ operators can be straightforwardly accommodated by conjugating the circuit with controlled-Z gates, as in the U(1) kinetic term. The $\hat{\zeta}$ operators arise from the JW transformation. In order to account for the two fermions per site in the JW transformation, we slightly modify the zigzagging JW path by assigning two nodes, instead of one node in the case of U($1$), per lattice site. In total, there are $dL^d$ links, and thus, the cost of implementing the kinetic term on all links is $dL^d(32\eta -48)$ T gates and $64dL^d \mathcal{C}^{(K)}$.

\subsubsection{Magnetic term \texorpdfstring{$e^{i\hat{L}_{\vec{n}}^{(B)}t}$}{}}
\label{sec:SU2_mag}

Here, we extend the block-diagonal decomposition method used for the kinetic term to implement the magnetic term. Once again, we drop the location indices for brevity. The magnetic term for a single plaquette is given by
\begin{align}
    \hat{L}_{r}^{(B)}=-\frac{1}{2a^{4-d}g^2}\sum_{\alpha \beta \delta \gamma =1}^{2} (\hat{U}_{\alpha \beta}\hat{U}_{ \beta \gamma}\hat{U}_{\gamma \delta}^{\dag}\hat{U}_{\delta \alpha}^{\dag} + h.c.).
    \label{eq:SU2_plaq}
\end{align}
Now, we rewrite one term in (\ref{eq:SU2_plaq}) as
\begin{align}
    &\quad \hat{U}_{\alpha \beta}\hat{U}_{ \beta \gamma}\hat{U}_{\gamma \delta}^{\dag}\hat{U}_{\delta \alpha}^{\dag} +h.c. \nonumber \\
    &= \sum_{q_1, q_2,...,q_{12}=0}^{1} (\hat{J}^{(1)+})^{q_1} (\hat{M}_{1}^{(1),L})^{q_2}(\hat{M}_{1}^{(1),R})^{q_3}...(\hat{J}^{(4)+})^{q_{10}} (\hat{M}_{1}^{(4),L})^{q_{11}}(\hat{M}_{1}^{(4),R})^{q_{12}} \nonumber \\
    &\quad[\sum_{\Delta j_1,...,\Delta j_4=-1}^{1}
    \hat{D}_{\alpha \beta}(\Delta j_1)\hat{P}_{\alpha \beta}(\Delta j_1) \hat{D}_{\beta \gamma}(\Delta j_2)\hat{P}_{\beta \gamma}(\Delta j_2)\hat{D}_{\gamma \delta}(\Delta j_3)\hat{P}^{\dag}_{\gamma \delta}(\Delta j_3)\hat{D}_{\delta \alpha}(\Delta j_4)\hat{P}^{\dag}_{\delta \alpha}(\Delta j_4)+h.c.]\nonumber \\
    &\quad (\hat{J}^{(1)-})^{q_1} (\hat{M}_{2}^{(1),L})^{q_2}(\hat{M}_{2}^{(1),R})^{q_3}...(\hat{J}^{(4)-})^{q_{10}} (\hat{M}_{2}^{(4),L})^{q_{11}}(\hat{M}_{2}^{(4),R})^{q_{12}} \nonumber \\
    &\equiv \sum_{q_1, q_2,...,q_{12}=0}^{1} (\hat{J}^{(1)+})^{q_1} (\hat{M}_{1}^{(1),L})^{q_2}...(\hat{M}_{1}^{(4),R})^{q_{12}}\nonumber\\
    &\quad[\sum_{\Delta j_1,...,\Delta j_4=-1}^{1} \hat{D}_{\alpha \beta \gamma \delta}(\Delta j_1, \Delta j_2, \Delta j_3, \Delta j_4)\hat{P}_{\alpha \beta \gamma \delta}(\Delta j_1, \Delta j_2, \Delta j_3, \Delta j_4)]\nonumber \\
    &\quad(\hat{J}^{(1)-})^{q_1}(\hat{M}_{2}^{(1),L})^{q_2}...(\hat{M}_{2}^{(4),R})^{q_{12}},
    \label{eq:SU2_mag_param}
\end{align}
where the superscript $(i)$ and subindex $i$ of $\Delta j_i$ are used to differentiate between the four links of a plaquette. Once again, in the implementation $e^{i\hat{L}_{r}^{(B)}t}$, we can minimize the number of binary incrementers and decrementers using the Gray code, i.e.,
\begin{align}
    &\quad e^{i\hat{L}_{r}^{(B)}t} \nonumber\\
    &\approx \prod_{(q_1,q_2,...q_{12})=GC(0)}^{GC(2^{12}-1)}(\hat{J}^{(1)+})^{q_1} (\hat{M}_{1}^{(1),L})^{q_2} ...(\hat{M}_{1}^{(4),R})^{q_{12}} \nonumber \\
    &\quad \left[\prod_{\alpha, \beta, \delta, \gamma = 1}^{2}\prod_{\Delta j_1,...,\Delta j_4=-1}^{1}e^{it\frac{-1}{2ag^2}\hat{D}_{\alpha \beta \gamma \delta}(\Delta j_1, \Delta j_2, \Delta j_3, \Delta j_4)\hat{P}_{\alpha \beta \gamma \delta}(\Delta j_1, \Delta j_2, \Delta j_3, \Delta j_4)}\right] \nonumber \\
    &\quad (\hat{J}^{(1)-})^{q_1} (\hat{M}_{2}^{(1),L})^{q_2}...(\hat{M}_{2}^{(4),R})^{q_{12}}.
\end{align}
Since $\hat{P}_{\alpha \beta \gamma \delta}$ is a twelve-qubit Pauli operator of the form $\otimes_{i=0}^{11}\hat{\sigma}_{i}^{\pm}+h.c.$, $e^{it\frac{-1}{2a^{4-d}g^2}\hat{D}_{\alpha \beta \gamma \delta}\hat{P}_{\alpha \beta \gamma \delta}}$ can be diagonalized using twenty-two CNOT and two Hadamard gates into an operator of the form
\begin{equation}
    e^{it\frac{-1}{2a^{4-d}g^2}\hat{D}_{\alpha \beta \gamma \delta}(\Delta j_1, \Delta j_2, \Delta j_3, \Delta j_4)\hat{\mathcal{D}}_{\alpha \beta \gamma \delta}(\Delta j_1, \Delta j_2, \Delta j_3, \Delta j_4)},
\end{equation}
where $\hat{\mathcal{D}}_{\alpha \beta \gamma \delta}$ is the diagonalized Pauli part of the operator. Before defining the oracle, we define some useful notations. We denote the state of a plaquette as $\otimes_{i=1}^{4} \ket{j_i}\ket{m^L_i}\ket{m^R_i}$, where the values of $i$ represent the four links of a plaquette. Next, the zeroth digits of $j_i, m^L_i, m^R_i$ are denoted as $j_{i,0}, m^L_{i,0}, m^R_{i,0}$, respectively. Further, we define
\begin{equation}
    f_{\alpha \beta \gamma \delta} = f_{\alpha \beta}(j_1, \Delta j_1, m^L_1, m^R_1) f_{\beta \gamma}(j_2, \Delta j_2, m^L_2, m^R_2)
    f_{\gamma \delta}(j_3, \Delta j_3, m^L_3, m^R_3) f_{\delta \alpha}(j_4, \Delta j_4, m^L_4, m^R_4).
    \label{eq:SU2_fabcd}
\end{equation}
As such, the phase oracle is defined as
\begin{align}
    &\quad \hat{O}_{\alpha \beta \gamma \delta}^{(\Delta j_1, \Delta j_2, \Delta j_3, \Delta j_4)} \otimes_{i=1}^{4} \ket{j^\prime_i}\ket{m^{L\prime}_i}\ket{m^{R\prime}_i}\nonumber \\
    &\mapsto e^{if_{\alpha \beta \gamma \delta}[(\frac{-1}{2a^{4-d}g^2}(-1)^{j^\prime_{1,0}}m_{1,0}^{L\prime} m_{1,0}^{R\prime} {j_{2,0}^\prime}m_{2,0}^{L\prime} m_{2,0}^{R\prime}
    {j_{3,0}^\prime}m_{3,0}^{L\prime} m_{3,0}^{R\prime} {j_{4,0}^\prime}m_{4,0}^{L\prime} m_{4,0}^{R\prime} )]}\otimes_{i=1}^{4} \ket{j_i^\prime}\ket{m^{L\prime}_i}\ket{m^{R\prime}_i},
    \label{eq:SU2_mag_oracle}
\end{align}
where the primed states are the states that have been acted on by the diagonalization circuit consisting of CNOT and Hadamard gates, and $\ket{j'_{1,0}}$ is the target qubit of the diagonalized Pauli part of the operator. As in the kinetic term implementation, we can implement this diagonal operator as a phase oracle using qRAM, or directly synthesize it using quantum arithmetic circuits. Once again, we compute $\pm f_{\alpha \beta \gamma \delta}$ using fixed point arithmetic circuits \cite{bhaskar2016quantum}, and induce the approximate phases using $R_z$ gates. See \ref{subsubsec:oracle_SU2} for the implementation details.

Assuming the cost of the oracle $\mathcal{C}^{(B)}$ is the same for all parameters choices, we calculate the implementation cost of $e^{i\hat{L}_{r}^{(B)}t}$. First, it incurs $2^{12}$ incrementers or decrementers, of which a significant portion act on the $\eta-$bit $\ket{j}$, but, without affecting complexity arguments, we assume that all the incrementers or decrementers are for the larger register of size $\eta+1$ qubits. As such, they cost $2^{12}\cdot 4(\eta-1)$ T gates, and $\eta+1$ reusable ancilla qubits. Second, it needs $2^{12}\cdot 16^2=1048576$ oracle queries to account for the different combinations of $\alpha$, $\beta$, $\gamma$, $\delta$, and $\Delta j_i$. In order to obtain the cost of the entire magnetic term, we need to multiply both the T-gate count and the number of oracle queries by the number of plaquettes, $\frac{d(d-1)}{2} L^d$.

\subsection{Resource requirement estimates}
\label{sec:SU2_res}

In this section, we analyze the algorithmic and synthesis errors for our simulations. In Sec.~\ref{subsubsec:Trotter_SU2} we compute the algorithmic error for the Suzuki-Trotter formula for our SU(2) Hamiltonian. Therein we show our result first, then show a full derivation of it for completeness. In, Sec.~\ref{subsubsec:oracle_SU2}, we calculate the error incurred by the arithmetic circuits used to synthesize the oracles. In Sec.~\ref{subsubsec:Synth_SU2} we compute the $R_z$ synthesis error. In Sec.~\ref{subsubsec:Analysis_SU2} we combine the two errors discussed in Secs.~\ref{subsubsec:Trotter_SU2} and~\ref{subsubsec:Synth_SU2} to report the gate and query complexity, and ancilla requirements.

\subsubsection{Trotter errors}
\label{subsubsec:Trotter_SU2}
As in the U(1) case, we choose to use the second-order PF as our simulation algorithm, and evaluate the commutator bound for the error given in (\ref{eq:U1_trotter_err}). The result is
\begin{align}
    || e^{-i\hat{H}T}-\hat{U}_2^r(t)||
    \leq r\left(\frac{T}{r}\right)^3 \rho
    \equiv \epsilon_{Trotter},
\end{align}
where
\begin{align}
    \rho &= \frac{1}{12} \Big[ 
    \frac{16 dL^d m^2}{a}+
    \frac{dL^d g^4(4\Lambda + 3)^2}{32a^{2d-3}}+
    \frac{2L^d d(d-1)g^2(4\Lambda + 3)^2}{a^d} +\frac{8192 d(d-1)L^d}{a^{6-d}g^2}\nonumber \\
    &\quad+
    \frac{(2048 d^2 - 48 d)L^d}{a^3}+
    \frac{L^d}{a^{12-3d}g^6}(134217728 d^3 - 301990160 d^2 + 167772432 d) \Big]
    \nonumber \\
     &\quad+ \frac{1}{24}\Big[
    \frac{mdL^d g^2(4\Lambda +3)}{a^{d-1}}+
    \frac{(128 d^2- 16 d)mL^d}{a^2}+
    \frac{256 mL^d (d^2-d)}{a^{5-d}g^2} +\frac{8 d^2 L^d g^2(4\Lambda + 3)}{a^d}\nonumber \\
    &\quad+
    \frac{32 d(d-1)L^d (4\Lambda + 3)}{a^3}+
    \frac{128 d(d-1)L^d (4\Lambda +3)}{a^3}  +\frac{L^d(4\Lambda + 3)}{a^{6-d}g^2}(1024 d^3 - 2432 d^2  \nonumber \\
    &\quad+ 1408 d) +(131072 \frac{d^3}{3} - 25728 d^2 + 52528 \frac{d}{3})\frac{L^d}{a^3}+
    \frac{L^d}{a^{6-d}g^2}(81920 d^3 - 148992 d^2 \nonumber\\
    &\quad+ 67072 d) +(32768 d^3 + 6656 d^2 - 39424 d)\frac{L^d}{a^{6-d}g^2}+
    \frac{L^d}{a^{9-2d}g^4}(1835008 d^3  \nonumber \\
    &\quad- 4456448 d^2+ 2621440 d) + ( 17179869184 \frac{d^5}{5} - 17179869184 \frac{d^4}{3} + 49392123904 \frac{d^3}{3} \nonumber \\
    &\quad - 322659435248 \frac{d^2}{3} +1207422766768\frac{d}{15}-8589934592 ) \frac{L^d}{a^{12-3d}g^6}
    \Big].
\end{align}
For completeness, we show below a full derivation of the results shown above. Readers interested in how the results compare with the size of the synthesis error and how, together, they affect our simulation gate complexity should proceed to Secs.~\ref{subsubsec:Synth_SU2} and~\ref{subsubsec:Analysis_SU2}.

We start our derivation by first ordering the terms in the Hamiltonian $\hat{H}$. As in the U(1) case, we implement the diagonal mass and electric terms first, then the off-diagonal kinetic and magnetic terms. We consider only the kinetic terms for the moment. As in the U(1) case, nearest-neighbor terms do not commute as they collide on fermionic sites. Therefore, we choose to group together the commuting terms, which originate from sites of the same parity, and act on links of the same direction. In particular, using $(p, l)$ to indicate the parity- and direction-dependence of each kinetic term, we order the kinetic terms according to the ordered list, i.e.,
\begin{equation}
    \{ (p, l) \} \equiv \{ ({e}, 1),({o}, 1), ({e}, 2), ({o}, 2),...,({e}, d), ({o}, d) \},
\end{equation}
where $p=e,o$ labels the even and odd sites, respectively. For a fixed $(p, l)$, the kinetic terms also depend on the parameters $a,b,c,\Delta j, \alpha, \beta$, which are defined in (\ref{eq:SU2_U_decomp}). For instance, $a,b,c$ are the exponents on the incrementers and decrementers, which are ordered according to the Gray code, i.e.,
\begin{equation}
    \{ a,b,c \} \equiv \{ GC(q) \}_{q=1}^{7}.
\end{equation}
Each of the remaining parameters, i.e., $\Delta j, \alpha, \beta$, can take on two values. Thus, we order them according to the standard binary code. Representing each kinetic term with the label $(p$,$l$,$a$,$b$,$c$,$\Delta j$, $\alpha$,$\beta)$, the ordering of the kinetic term is then given by the ordered list, i.e.,
\begin{equation}
    \mathbbm{T} = \{ (p, l) \} \times \{ a,b,c \} \times \{ \Delta j, \alpha, \beta \},
\end{equation}
where $\times$ is an order-preserving Cartesian product, such that $l$ and $\beta$ are the variables that vary the slowest and fastest in $\mathbbm{T}$, respectively. We denote an element in $\mathbbm{T}$ as $\hat{h}_{T}$, and an element that is implemented after $\hat{h}_{T}$ as $\hat{h}_{T'}$ with $T' > T$. Each $\hat{h}_{T}$ with a fixed set of parameters is a sum of $\frac{L^d}{2}$ all even or odd kinetic operators, each of which acts on a link in the fixed direction $l$ originated from an even or odd site $\vec{n}_p$, respectively, and the sites connected by the link, and is represented with $\hat{h}_{T}(\vec{n}_p,l)$. Moreover, the number of elements in $\mathbbm{T}$ is given by 
\begin{equation}
    |\mathbbm{T}| = |\{ (p, l) \}| \cdot |\{ a,b,c \}| \cdot |\{ \Delta j, \alpha, \beta \}| = 2d\cdot 2^3 \cdot 2^3 = 2^7 d.
\end{equation}
We proceed to describe the ordering for the magnetic terms. Magnetic terms acting on neighboring plaquettes do not commute in general. Therefore, we group together the terms that originate from sites of the same parity and act on the same two-dimensional plane. Further, we order the magnetic terms according to the ordered list,
\begin{equation}
    \{ (p, j, k) \} = \{(e,1,2),(o,1,2),(e,1,3),(o,1,3),...(o,1,d),(e,2,3),...,(e,d-1,d),(o,d-1,d)\},
\end{equation}
where $p=e,o$ labels the parity, the two directions of a given plane are represented by $j=1,2,...,d-1$ and $k=j+1,...,d$, and $p$ is the fastest-varying parameter in the list. We represent each magnetic term with a tuple of its parameters, $(p, j, k, q_1,q_2,...,q_{12},\Delta j_1,\Delta j_2,\Delta j_3,\Delta j_4, \alpha, \beta, \gamma, \delta)$, as defined in (\ref{eq:SU2_mag_param}), and order the magnetic terms according to the following ordered list,
\begin{equation}
    \mathbbm{L} = \{ (p, j, k) \} \times \{q_1, q_2, ..., q_{12} \} \times \{ \Delta j_1,\Delta j_2,\Delta j_3,\Delta j_4 \} \times \{ \alpha, \beta, \gamma, \delta \},
\end{equation}
where $q_1, q_2, ..., q_{12}$ are ordered using the Gray code, $\Delta j_1,\Delta j_2,\Delta j_3,\Delta j_4$ and $\alpha, \beta, \gamma, \delta$ are ordered according to the binary code. We denote an element in $\mathbbm{L}$ as $\hat{h}_{L}$, and an element that is implemented after $\hat{h}_{L}$ as $\hat{h}_{L'}$ with $L' > L$. Each $\hat{h}_{L}$ with a fixed set of parameters is a sum of $\frac{L^d}{2}$ all even or odd magnetic operators, each of which acts on a plaquette on a fixed plane $(j,k)$ originated from an even or odd site $\vec{n}_p$, respectively, and is represented with $\hat{h}_{L}(\vec{n}_p,j,k)$. Further, the number of elements in $\mathbbm{L}$ is given by
\begin{align}
    |\mathbbm{L}| &= |\{ (p, j, k) \}| \cdot |\{q_1, q_2, ..., q_{12} \}| \cdot |\{ \Delta j_1,\Delta j_2,\Delta j_3,\Delta j_4 \}| \cdot  |\{ \alpha, \beta, \gamma, \delta \}| \nonumber \\
    &= 2\frac{d(d-1)}{2} \cdot 2^{12} \cdot 2^4 \cdot 2^4 = 2^{20}d(d-1).
\end{align}
Finally, the ordering of the terms in the Hamiltonian $\hat{H}$ is given by the following ordered list:
\begin{equation}
    \{ \hat{H}_x \} = \{ \sum_{\vec{n}} \hat{D}_{\vec{n}}^{(M)}, \sum_{\vec{n}} \hat{D}_{\vec{n}}^{(E)} \} \cup \mathbbm{T} \cup \mathbbm{L},
\end{equation}
where $\cup$ is denotes the order-preserving union.

We now proceed to evaluate the Trotter error incurred by the second-order PF. First, we focus on the first sum in (\ref{eq:U1_trotter_err}), i.e.,
\begin{align}
\sum_{i} \lvert\lvert [[\hat{H}_i, \sum_{j>i}\hat{H}_j],\hat{H}_i] \rvert\rvert \leq \sum_{k=1}^{8} \lvert\lvert C_{1,k} \rvert\rvert,
\label{eq:SU2_trot1}
\end{align}
where
\begin{align}
C_{1,1} =& [[ \sum_{\vec{n}} \hat{D}_{\vec{n}}^{(M)} , \sum_{\vec{n}'} \hat{D}_{\vec{n}'}^{(E)} ]  , \sum_{\vec{n}} \hat{D}_{\vec{n}}^{(M)} ], \nonumber \\
C_{1,2} =& [[ \sum_{\vec{n}} \hat{D}_{\vec{n}}^{(M)} , \sum_{\vec{n}'} \hat{T}_{\vec{n}'}^{(K)} ]  , \sum_{\vec{n}} \hat{D}_{\vec{n}}^{(M)} ], \nonumber \\
C_{1,3} =& [[ \sum_{\vec{n}} \hat{D}_{\vec{n}}^{(M)} , \sum_{\vec{n}'} \hat{L}_{\vec{n}'}^{(B)} ]  , \sum_{\vec{n}} \hat{D}_{\vec{n}}^{(M)} ], \nonumber \\
C_{1,4} =& [[ \sum_{\vec{n}} \hat{D}_{\vec{n}}^{(E)} , \sum_{\vec{n}'} \hat{T}_{\vec{n}'}^{(K)} ]  , \sum_{\vec{n}} \hat{D}_{\vec{n}}^{(E)} ], \nonumber \\
C_{1,5} =& [[ \sum_{\vec{n}} \hat{D}_{\vec{n}}^{(E)} , \sum_{\vec{n}'} \hat{L}_{\vec{n}'}^{(B)} ]  , \sum_{\vec{n}} \hat{D}_{\vec{n}}^{(E)} ], \nonumber \\
C_{1,6} =& \sum_{\hat{h}_{T}\in \mathbbm{T}} [[ \hat{h}_{T} , \sum_{\vec{n}} \hat{L}_{\vec{n}}^{(B)} ]  , \hat{h}_{T} ], \nonumber \\
C_{1,7} =& \sum_{\hat{h}_{T}\in \mathbbm{T}} [[ \hat{h}_{T} , \sum_{\hat{h}_{T'}\in \mathbbm{T};T' > T} \hat{h}_{T'} ]  , \hat{h}_{T} ], \nonumber \\
C_{1,8} =& \sum_{\hat{h}_{L}\in \mathbbm{L}} [[ \hat{h}_{L} , \sum_{\hat{h}_{L'}\in \mathbbm{L};L' > L} \hat{h}_{L'} ]  , \hat{h}_{L} ].
\label{eq:SU2_trotter_err_1}
\end{align}
Note the following expressions will be useful in the foregoing evaluations of the terms:
\begin{equation}
    ||\hat{U}_{\alpha \beta}|| \leq 1,\label{eq:SU2_useful_k1}
\end{equation} 
which follows from the fact that in the original, untruncated theory, $\hat{U}_{\alpha \beta}$ are elements of a matrix in the considered unitary gauge group SU(2) in the so-called group element basis~\footnote{The basis used in this paper and the group element basis are related via a generalized, non-Abelian Fourier transform~\cite{zohar2015formulation}.}~\cite{creutz1977gauge,zohar2015formulation,tong2021provably}. \eqref{eq:SU2_useful_k1} still holds after the truncation because of two reasons. First, for all input state $\ket{\psi}$ such that the output state $\ket{\phi} = \hat{U}_{\alpha \beta}\ket{\psi}$ are within the truncated space, the truncated and untruncated $\hat{U}_{\alpha \beta}$ produce the same output. Second, if in the untruncated theory, $\ket{\phi}$ contains any basis state $\ket{j,m^L,m^R}$ that lies outside the truncated space, $\ket{j,m^L,m^R}$ will be replaced by 0 in the truncated theory. Thus, the norm of any output state of $\hat{U}_{\alpha \beta}$ in the truncated theory is upper-bounded by that in the untruncated theory.

Using (\ref{eq:SU2_useful_k1}), we straightforwardly obtain
\begin{gather}
    ||\hat{U}_{\alpha \beta}+\hat{U}_{\alpha \beta}^\dag|| \leq 2, \label{eq:SU2_useful_k2} \\
    ||\hat{U}_{\alpha \beta}\hat{\sigma}_{\alpha}^- \hat{\sigma}_{\beta}^+ +\hat{U}_{\alpha \beta}^\dag \hat{\sigma}_{\alpha}^+ \hat{\sigma}_{\beta}^-|| \leq 2 \label{eq:SU2_useful_k3}.
\end{gather}
We note that the norm of each term in the block-diagonal decomposition of $\hat{U}_{\alpha \beta}\hat{\sigma}_{\alpha}^- \hat{\sigma}_{\beta}^+ +\hat{U}_{\alpha \beta}^\dag \hat{\sigma}_{\alpha}^+ \hat{\sigma}_{\beta}^-$, as shown in (\ref{eq:SU2_U_decomp}), is upper-bounded by $2$. The reason is that, for each term in the decomposition, the norms of the incrementers, decrementers and Pauli ladder operators are bounded from above by one, and that of the diagonal part $\hat{D}_{\alpha \beta}$, of which the elements are defined via $f_{\alpha \beta}$, is also upper-bounded by one. This implies that 
\begin{equation}
    ||\hat{h}_{T}(\vec{n}_p,l)|| \leq \frac{1}{a} \label{eq:SU2_useful_k4},
\end{equation}
because $\hat{h}_{T}(\vec{n}_p,l)$ is, up to a multiplicative factor of $\frac{1}{2a}$, a term in the block-diagonal decomposition.

Furthermore, we consider the terms $\hat{h}_T(\vec{n},l)$, which act on a pair of nearest-neighbor sites and the link that connects them, i.e., $(\vec{n},l)$. We denote this set of terms as $\mathbbm{T}|_{(\vec{n},l)}$. Then,
\begin{equation}
    \sum_{\hat{h}_T(\vec{n},l)\in \mathbbm{T}|_{(\vec{n},l)}} \hat{h}_T(\vec{n},l) = \frac{1}{2a}\sum_{\alpha,\beta=1}^{2}\hat{U}_{\alpha \beta}(\vec{n},l)\hat{\sigma}_{\alpha}^-(\vec{n}) \hat{\sigma}_{\beta}^+(\vec{n}+\hat{l}) +h.c.,
    \label{eq:SU2_useful_k5}
\end{equation}
which we combine with (\ref{eq:SU2_useful_k3}) to obtain
\begin{equation}
    ||\sum_{\substack{\hat{h}_T(\vec{n},l)\in S;\\ S \subseteq \mathbbm{T}|_{(\vec{n},l)}}} \hat{h}_T(\vec{n},l)|| \leq || \frac{1}{2a}\sum_{\alpha,\beta=1}^{2}\hat{U}_{\alpha \beta}(\vec{n},l)\hat{\sigma}_{\alpha}^-(\vec{n}) \hat{\sigma}_{\beta}^+(\vec{n}+\hat{l}) +h.c.|| \leq \frac{4}{a}. \label{eq:SU2_useful_k6}
\end{equation}

Similarly, for the magnetic term, we obtain the following bound:
\begin{equation}
    ||\hat{U}_{\alpha\beta}\hat{U}_{\beta\gamma}\hat{U}_{\gamma\delta}^\dag\hat{U}_{\delta\alpha}^\dag + h.c.|| \leq 2||\hat{U}_{\alpha\beta}\hat{U}_{\beta\gamma}\hat{U}_{\gamma\delta}^\dag\hat{U}_{\delta\alpha}^\dag|| \leq 2||\hat{U}_{\alpha\beta}||^4 \leq 2, \label{eq:SU2_useful_m1}
\end{equation}
where we have used (\ref{eq:SU2_useful_k1}) for the last equality. As in the kinetic term, the norm of each term in the block-diagonal decomposition of $\hat{U}_{\alpha\beta}\hat{U}_{\beta\gamma}\hat{U}_{\gamma\delta}^\dag\hat{U}_{\delta\alpha}^\dag+h.c.$, as shown in (\ref{eq:SU2_mag_param}), is bounded by $2$. The reason is that the norms of the incrementers, decrementers and Pauli ladder operators are bounded from above by one, and that of the diagonal part $\hat{D}_{\alpha \beta \gamma \delta}$, of which the elements are defined via $f_{\alpha \beta \gamma \delta}$, as defined in (\ref{eq:SU2_fabcd}), is also upper-bounded by one. This implies that
\begin{equation}
    ||\hat{h}_L(\vec{n}_p,i,j)|| \leq \frac{1}{a^{4-d}g^2} \label{eq:SU2_useful_m2}
\end{equation}
because each $\hat{h}_L(\vec{n}_p,i,j)$ is a product between $\frac{-1}{2a^{4-d}g^2}$ and a term in (\ref{eq:SU2_mag_param}).

Lastly, we consider the terms $\hat{h}_L(\vec{n},i,j)$, which act on a single plaquette, denoted by $(\vec{n}, i,j)$. Let these terms form a set $\mathbbm{L}|_{(\vec{n},i, j)}$. Then,
\begin{equation}
    \sum_{\hat{h}_L(\vec{n},i,j) \in \mathbbm{L}|_{(\vec{n}, i,j)}}\hat{h}_L{(\vec{n},i,j)}=\frac{-1}{2a^{4-d}g^2}\sum_{\alpha,\beta,\gamma,\delta=1}^{2}\hat{U}_{\alpha \beta}(\vec{n},i)\hat{U}_{ \beta \gamma}(\vec{n}+\hat{i},j)\hat{U}_{\gamma \delta}^{\dag}(\vec{n}+\hat{j},i)\hat{U}_{\delta \alpha}^{\dag}(\vec{n},j) + h.c..
    \label{eq:SU2_useful_m3}
\end{equation}
Using the above relation and (\ref{eq:SU2_useful_m1}), we obtain
\begin{align}
    || \sum_{\substack{\hat{h}_L(\vec{n},i,j)\in S;\\ S\subseteq \mathbbm{L}|_{(\vec{n}, i,j)}}}\hat{h}_L{(\vec{n},i,j)}||&\leq ||\frac{-1}{2a^{4-d}g^2}\sum_{\alpha,\beta,\gamma,\delta=1}^{2}\hat{U}_{\alpha \beta}(\vec{n},i)\hat{U}_{ \beta \gamma}(\vec{n}+\hat{i},j)\hat{U}_{\gamma \delta}^{\dag}(\vec{n}+\hat{j},i)\hat{U}_{\delta \alpha}^{\dag}(\vec{n},j) + h.c.|| \nonumber \\
    &\leq \frac{16}{a^{4-d}g^2}.
    \label{eq:SU2_useful_m4}
\end{align}
Whenever these useful expressions (\ref{eq:SU2_useful_k1}-\ref{eq:SU2_useful_m4}) are used, we use them without explicit references for brevity.

We now evaluate in the following each term $C_{1,n}$.
First, $C_{1,1}$ and $C_{1,3}$ both evaluate to zero because the mass term commutes with both the electric and magnetic terms.
$C_{1,2}$ is bounded by
\begin{align}
    &\quad ||[[ \sum_{\vec{n}} \hat{D}_{\vec{n}}^{(M)} , \sum_{\vec{n}'} \hat{T}_{\vec{n}'}^{(K)} ]  , \sum_{\vec{n}} \hat{D}_{\vec{n}}^{(M)} ]|| \nonumber \\
    &\leq \sum_{\vec{n}} \sum_{l=1}^{d} || [[\hat{D}_{\vec{n}}^{(M)}+\hat{D}_{\vec{n}+\hat{l}}^{(M)},\frac{1}{2a}\sum_{\alpha, \beta=1}^{2}(\hat{U}_{\alpha \beta}(\vec{n},l)\hat{\sigma}_{\alpha}^{-}(\vec{n})\hat{\sigma}_{\beta}^{+}(\vec{n}+\hat{l}) + h.c.)], \hat{D}_{\vec{n}}^{(M)}+\hat{D}_{\vec{n}+\hat{l}}^{(M)}]|| \nonumber \\
    &\leq dL^d \sum_{\alpha, \beta=1}^{2}||[[ \frac{m}{2}((-1)^{\vec{n}} \hat{Z}_{\alpha}(\vec{n})+(-1)^{\vec{n}+\hat{l}} \hat{Z}_{\beta}(\vec{n}+\hat{l})) , \frac{1}{2a}(\hat{U}_{\alpha \beta}(\vec{n},l)\hat{\sigma}_{\alpha}^{-}(\vec{n})\hat{\sigma}_{\beta}^{+}(\vec{n}+\hat{l}) + h.c.) ]  \nonumber \\
    &\quad , \frac{m}{2}((-1)^{\vec{n}} \hat{Z}_{\alpha}(\vec{n})+(-1)^{\vec{n}+\hat{l}} \hat{Z}_{\beta}(\vec{n}+\hat{l})) ] || \nonumber \\
    &\leq 4dL^d \cdot 4 ||m||^2 ||\frac{1}{a}||=\frac{16dL^d m^2}{a},
\end{align}
where in the first two inequalities, we have used the fact that only the fermionic sites of colors $\alpha,\beta$ at $\vec{n},\vec{n}+\hat{l}$, respectively, are acted on by the kinetic term with color indices $\alpha\beta$.

Before evaluating $C_{1,4}$, we provide some useful properties about the kinetic operators $\frac{1}{2a}$ $(\hat{U}_{\alpha \beta}(\vec{n},l)\hat{\sigma}_{\alpha}^{-}(\vec{n})\hat{\sigma}_{\beta}^{+}(\vec{n}+\hat{l}) + h.c.)$. At the fermionic sites $\vec{n}$ and $\vec{n}+\hat{l}$ of colors $\alpha$ and $\beta$, respectively, the kinetic operator takes a computational basis state to another basis state. Acting on the gauge field on the link $(\vec{n},l)$, it maps the subregister $\ket{j} \mapsto \ket{j \pm 1}$, where $j\in [0,2\Lambda]$, up to a multiplicative constant. Therefore, if we evaluate the commutator between an electric and kinetic operator acting on the same link, we obtain
\begin{align}
    &\quad || [\frac{g^2}{2a^{d-2}}\hat{E}^2(\vec{n},l),\frac{1}{2a}(\hat{U}_{\alpha \beta}(\vec{n},l)\hat{\sigma}_{\alpha}^{-}(\vec{n})\hat{\sigma}_{\beta}^{+}(\vec{n}+\hat{l}) + h.c.)] ||  \nonumber \\
    &\mapsto  \frac{g^2}{2a^{d-2}} \frac{1}{4} [(j+1 \pm 1)^2-(j+1)^2]||\frac{1}{2a}(\hat{U}_{\alpha \beta}(\vec{n},l)\hat{\sigma}_{\alpha}^{-}(\vec{n})\hat{\sigma}_{\beta}^{+}(\vec{n}+\hat{l}) + h.c.)|| \nonumber \\
    &\leq \frac{g^2}{8a^{d-2}}(2j+3)\frac{1}{a} \leq \frac{g^2}{8a^{d-1}}(4\Lambda +3),
    \label{eq:SU2_EK}
\end{align}
where in the second line, we have used the eigenvalue equation in (\ref{eq:SU2_elec_eig}). Using this relation, we evaluate the bound for $C_{1,4}$, and obtain
\begin{align}
    &\quad || [[\sum_{\vec{n}}\hat{D}_{\vec{n}}^{(E)}, \sum_{ \vec{n} } \hat{T}_{\vec{n}'}^{(K)}] ,\sum_{\vec{n}}\hat{D}_{\vec{n}}^{(E)}] || \nonumber \\
    &\leq \sum_{\vec{n}}\sum_{l=1}^{d}\sum_{\alpha, \beta=1}^{2} ||[ [\frac{g^2}{2a^{d-2}}\hat{E}^2(\vec{n},l),\frac{1}{2a}(\hat{U}_{\alpha \beta}(\vec{n},l)\hat{\sigma}_{\alpha}^{-}(\vec{n})\hat{\sigma}_{\beta}^{+}(\vec{n}+\hat{l}) + h.c.)],\frac{g^2}{2a^{d-2}}\hat{E}^2(\vec{n},l)] ||\nonumber \\
    &\mapsto \sum_{\vec{n}}\sum_{l=1}^{d}\sum_{\alpha, \beta=1}^{2}  \frac{g^2}{8a^{d-2}} [(j+1 \pm 1)^2-(j+1)^2] \nonumber \\
    &\quad \cdot ||[\frac{g^2}{2a^{d-2}}\hat{E}^2(\vec{n},l),\frac{1}{2a}(\hat{U}_{\alpha \beta}(\vec{n},l)\hat{\sigma}_{\alpha}^{-}(\vec{n})\hat{\sigma}_{\beta}^{+}(\vec{n}+\hat{l}) + h.c.)]|| \nonumber \\
    &\leq 4dL^d \frac{g^2}{8a^{d-2}} (4\Lambda +3)||[\frac{g^2}{2a^{d-2}}\hat{E}^2(\vec{n},l),\frac{1}{2a}(\hat{U}_{\alpha \beta}(\vec{n},l)\hat{\sigma}_{\alpha}^{-}(\vec{n})\hat{\sigma}_{\beta}^{+}(\vec{n}+\hat{l}) + h.c.)]|| \nonumber \\
    &\leq 4dL^d \frac{g^2}{8a^{d-2}} (4\Lambda +3) \frac{g^2}{8a^{d-1}} (4\Lambda +3) = \frac{dL^d g^4(4\Lambda + 3)^2}{32a^{2d-3}}. \nonumber \\
\end{align}

Next, we evaluate the commutators between the electric and magnetic terms, which are trivial unless the operators act on the same link. Thus, we evaluate the bound for the commutator between the electric and magnetic operators acting on a single plaquette as follows:
\begin{align}
    &\quad || [ \frac{g^2}{2a^{d-2}} (\hat{E}^2(\vec{n},i)+\hat{E}^2(\vec{n}+\hat{i},j)+\hat{E}^2(\vec{n}+\hat{j},i)+\hat{E}^2(\vec{n},j) )  \nonumber \\
    &\quad, -\frac{1}{2a^{4-d}g^2} \sum_{\alpha, \beta, \delta, \gamma =1}^{2} (\hat{U}_{\alpha \beta}(\vec{n},i)\hat{U}_{ \beta \gamma}(\vec{n}+\hat{i},j)\hat{U}_{\gamma \delta}^{\dag}(\vec{n}+\hat{j},i)\hat{U}_{\delta \alpha}^{\dag}(\vec{n},j) + h.c.) ] || \nonumber \\
    &\leq 4 \sum_{\alpha, \beta, \delta, \gamma =1}^{2} || [ \frac{g^2}{2a^{d-2}} \hat{E}^2 ,-\frac{1}{2a^{4-d}g^2}( \hat{U}_{\alpha \beta}\hat{U}_{ \beta \gamma}\hat{U}_{\gamma \delta}^{\dag}\hat{U}_{\delta \alpha}^{\dag}+h.c.) ] || \nonumber \\
    &\leq 4 \cdot 2^4 \frac{g^2}{2a^{d-2}}\frac{1}{4} [ (j+1\pm 1)^2-(j+1)^2 ]|| \frac{1}{2a^{4-d}g^2} (\hat{U}_{\alpha \beta}\hat{U}_{ \beta \gamma}\hat{U}_{\gamma \delta}^{\dag}\hat{U}_{\delta \alpha}^{\dag}+h.c.) || \nonumber \\
    &\leq \frac{8}{a^2}(4\Lambda + 3),
    \label{eq:SU2_elec_mag}
\end{align}
where in the first inequality, we have dropped the location indices for brevity, since the commutator between any of the four electric terms and the magnetic terms shares the same bound, and in the last inequality, we have used (\ref{eq:SU2_useful_m1}). Now we use the above relation to compute the bound for $C_{1,5}$, and obtain
\begin{align}
    &\quad ||[[ \sum_{\vec{n}} \hat{D}_{\vec{n}}^{(E)} , \sum_{\vec{n}'} \hat{L}_{\vec{n}'}^{(B)} ]  , \sum_{\vec{n}} \hat{D}_{\vec{n}}^{(E)} ]|| \nonumber \\
    &\leq || \sum_{\vec{n}}\sum_{i=1}^{d}\sum_{j\neq i;j=1}^{d} [[\frac{g^2}{2a^{d-2}}(\hat{E}^2(\vec{n},i)+\hat{E}^2(\vec{n}+\hat{i},j)+\hat{E}^2(\vec{n}+\hat{j},i)+\hat{E}^2(\vec{n},j) ) \nonumber \\
    &\quad , -\frac{1}{2a^{4-d}g^2} \sum_{\alpha, \beta, \delta, \gamma =1}^{2} (\hat{U}_{\alpha \beta}(\vec{n},i)\hat{U}_{ \beta \gamma}(\vec{n}+\hat{i},j)\hat{U}_{\gamma \delta}^{\dag}(\vec{n}+\hat{j},i)\hat{U}_{\delta \alpha}^{\dag}(\vec{n},j) + h.c.) ] \nonumber \\
    &\quad , \frac{g^2}{2a^{d-2}}(\hat{E}^2(\vec{n},i)+\hat{E}^2(\vec{n}+\hat{i},j)+\hat{E}^2(\vec{n}+\hat{j},i)+\hat{E}^2(\vec{n},j) )] || \nonumber \\
    &\leq L^d\frac{d(d-1)}{2} 4|| \frac{g^2}{2a^{d-2}}\frac{1}{4}[(j+1\pm 1)^2 - (j+1)^2] [\frac{g^2}{2a^{d-2}}(\hat{E}^2(\vec{n},i)+\hat{E}^2(\vec{n}+\hat{i},j)+\hat{E}^2(\vec{n}+\hat{j},i) \nonumber \\
    &\quad +\hat{E}^2(\vec{n},j) ) , -\frac{1}{2a^{4-d}g^2} \sum_{\alpha, \beta, \delta, \gamma =1}^{2} (\hat{U}_{\alpha \beta}(\vec{n},i)\hat{U}_{ \beta \gamma}(\vec{n}+\hat{i},j)\hat{U}_{\gamma \delta}^{\dag}(\vec{n}+\hat{j},i)\hat{U}_{\delta \alpha}^{\dag}(\vec{n},j) + h.c.) ] || \nonumber \\
    &\leq L^d\frac{d(d-1)}{2}\frac{g^2}{2a^{d-2}}(4\Lambda + 3)\frac{8}{a^2}(4\Lambda + 3) = \frac{2L^d d(d-1)g^2(4\Lambda + 3)^2}{a^d},
\end{align}
where in the second equality, we have used the fact that there are $L^d\frac{d(d-1)}{2}$ plaquettes on the lattice, and in the last inequality, we have used (\ref{eq:SU2_elec_mag}). 

Now, we move on to $C_{1,6}$, which is bounded from above by
\begin{align}
    &\quad \sum_{\hat{h}_T \in \mathbbm{T}} ||  [[ \hat{h}_T , \sum_{\vec{n}}\hat{L}_{\vec{n}}^{(B)} ], \hat{h}_T ]  || \nonumber \\
    &\leq \sum_{\hat{h}_T \in \mathbbm{T}} ||  [[ \sum_{\vec{n}_p}\hat{h}_T(\vec{n}_p,l) , \sum_{\vec{n}}\hat{L}_{\vec{n}}^{(B)} ], \sum_{\vec{n}_p} \hat{h}_T(\vec{n}_p,l) ]  ||\nonumber \\
    &\leq 2^7 d\frac{L^d}{2} 4|| \hat{h}_{T}(\vec{n}_p,l)|| \cdot || \frac{2(d-1)}{2a^{4-d}g^2}\sum_{\alpha, \beta, \gamma, \delta=1}^{2} (\hat{U}_{\alpha\beta}\hat{U}_{\beta\gamma}\hat{U}_{\gamma\delta}^\dag\hat{U}_{\delta\alpha}^\dag + h.c.) || \cdot || \hat{h}_{T}(\vec{n}_p,l) || \nonumber \\
    &\leq 2^8 L^d d ||\frac{1}{a}||^2 || \frac{32(d-1)}{a^{4-d}g^2} || = \frac{8192 d(d-1)L^d}{a^{6-d}g^2},
\end{align}
where in the first inequality, the factor of $2^7d$ outside the norm expression is the cardinality of $\mathbbm{T}$, and in the second inequality, the factor $\frac{L^d}{2}$ is the number of even or odd sites $\vec{n}_p$, the factor of $2(d-1)$ is due to the fact that each link $(\vec{n}_p,l)$ is acted on concurrently by a kinetic term and $2(d-1)$ plaquette operators, and we have used (\ref{eq:comm_bound}).

Now we consider $C_{1,7}$, which consists of the commutators between kinetic terms. There are two types of commutators; those between (i) terms acting on the same link, and (ii) terms acting on neighboring links that are connected via the sites. We begin with the analysis of case (i). First, we denote the subset of $\mathbbm{T}$ that consists of elements with a fixed parity and direction by $\mathbbm{T}|_{(p,l)}$. The number of elements in $\mathbbm{T}|_{(p,l)}$ is $2^6$, which is the number of remaining free parameters, $a,b,c, \Delta j, \alpha, \beta$. We consider the bound of the commutator, where $\hat{h}_T$ is one of the first $2^6-4$ elements of $\mathbbm{T}|_{(p,l)}$, and obtain
\begin{align}
    &\quad ||  [[\hat{h}_T(\vec{n}_p,l),\sum_{\hat{h}_{T'}\in \mathbbm{T}|_{(p,l)}; T'>T}\hat{h}_{T'}],\hat{h}_T(\vec{n}_p,l)]  || \nonumber \\
    &\leq 4\cdot ||\hat{h}_T(\vec{n}_p,l)||^2 ||\sum_{\hat{h}_{T'}\in \mathbbm{T}|_{(p,l)}; T'>T}\hat{h}_{T'}|| \nonumber \\
    &\leq 4\cdot ||\hat{h}_T(\vec{n}_p,l)||^2 || \frac{1}{2a}\sum_{\alpha, \beta=1}^{2}(\hat{U}_{\alpha \beta}(\vec{n}_p,l)\hat{\sigma}_{\alpha}^{-}(\vec{n}_p)\hat{\sigma}_{\beta}^{+}(\vec{n}_p+\hat{l}) + h.c.) || \nonumber \\
    &\leq 4\cdot ||\frac{1}{a}||^2||\frac{4}{a}|| = \frac{16}{a^3},
\end{align}
where in the second inequality, we used (\ref{eq:SU2_useful_k6}) to obtain the second norm expression. The last $4$ elements of $\mathbbm{T}|_{(p,l)}$ have the same $a,b,c,\Delta j$. The terms with $\alpha \beta = 11,22$,$(12,21)$ commute with each other, but not with terms with $\alpha \beta = 12,21$,$(11,22)$. Therefore, for the last four $\hat{h}_T \in \mathbbm{T}|_{(p,l)}$, which are implemented in the order of $\alpha \beta = 11,12,21,22$, there are four non-trivial commutators, each of which is bounded by
\begin{equation}
    4||\frac{1}{a}||^3 = \frac{4}{a^3}
\end{equation}
Therefore, the bound for the commutators that belong to type (i) is given by
\begin{equation}
    2d\frac{L^d}{2}[(2^6-4)\frac{16}{a^3} + \frac{16}{a^3}] = \frac{976  dL^d}{a^3},
\end{equation}
where $2d$ is the number of different $(p,l)$, and $\frac{L^d}{2}$ is the number of sites of a given parity. Next, we analyze type (ii), which we further divide into two cases. Case (i) consists of commutators where $\hat{h}_{T}, \hat{h}_{T'}$ act on links in the same direction, but sites of different parities. Since we implement even terms before odd terms, there are $d$ commutators in case (i). In case (ii), $\hat{h}_{T}, \hat{h}_{T'}$ act on links in different directions. Let $\hat{h}_{T}$ and $\hat{h}_{T'}$ be labelled by $(p,l)$ and $(p',l')$. Then, there are $\sum_{p',p=e}^{o} \sum_{l=1}^{d} \sum_{l'>l,l'=1}^{d} = 4\frac{d(d-1)}{2}=2d^2-2d$ commutators in case (ii). In both types, $\hat{h}_{T}, \hat{h}_{T'}$ could collide on one or two sites, depending on their respective color indices. If $\hat{h}_{T}$ and $\hat{h}_{T'}$ are labelled by $\alpha \beta$ and $\beta \alpha$, respectively, then each $\hat{h}_{T}$ will collide with two $\hat{h}_{T'}$ on sites of both colors $\alpha$ and $\beta$. If $\hat{h}_{T}$ is labelled by $\alpha \beta$, and $\hat{h}_{T'}$ by $\beta \beta$ $(\alpha \alpha)$, then each $\hat{h}_{T}$ will collide with one $\hat{h}_{T'}$ on the site of color $\beta$ ($\alpha$). In the former scenario, where collisions occur at two sites, the bound for each commutator is
\begin{equation}
    4\cdot ||\hat{h}_T(\vec{n}_p,l)||^2 \cdot ||2 \cdot \frac{1}{2a}(\hat{U}_{\beta\alpha }\hat{\sigma}_{\beta}^{-}\hat{\sigma}_{\alpha}^{+} + h.c.) || \leq \frac{8}{a^3}.
\end{equation}
In the latter scenario, where there is a collision on one site, the bound for each commutator is
\begin{equation}
    4\cdot ||\hat{h}_T(\vec{n}_p,l)||^2 \cdot || \frac{1}{2a}\sum_{\alpha' \beta'=\beta \beta}^{\alpha \alpha}(\hat{U}_{\alpha' \beta' }\hat{\sigma}_{\alpha'}^{-}\hat{\sigma}_{\beta'}^{+} + h.c.) || \leq \frac{8}{a^3}.
\end{equation}
The bound for the commutators in type (ii) is thus
\begin{equation}
    \frac{(8+8)}{a^3}2^6\cdot(2d^2-2d + d) \frac{L^d}{2} = \frac{(2048 d^2 - 1024 d)L^d}{a^3}.
\end{equation}
As such, the bound for $C_{1,7}$ is
\begin{equation}
    \frac{(2048 d^2 - 48 d)L^d}{a^3}.
\end{equation}

Finally, we consider $C_{1,8}$, which consists of the commutators between magnetic terms. We separate the terms into two cases: cases (i) and (ii), respectively, consist of intra- and inter-plaquette commutators, where $\hat{h}_L$ and $\hat{h}_{L'}$ act on the same and different plaquettes, respectively. We first examine case (i). We denote the subset of $\mathbbm{L}$ that consists of elements with a fixed parity and two-dimensional plane by $\mathbbm{L}|_{(p,j,k)}$. As such, the number of elements in $\mathbbm{L}|_{(p,j,k)}$, $2^{20}$, is the number of remaining free parameters. We compute the bound of the commutator, in which $\hat{h}_{L}$ is amongst the first $2^{20}-16$ elements of $\mathbbm{L}|_{(p,j,k)}$, and obtain
\begin{align}
    &\quad||[[\hat{h}_{L}(p,j,k), \sum_{\hat{h}_{L'}\in \mathbbm{L}|_{(p,j,k)};L'>L}\hat{h}_{L'}],\hat{h}_{L}(p,j,k)]|| \nonumber \\
    &\leq 4 ||\hat{h}_{L}(\vec{n}_p,j,k)||^2 \cdot ||\frac{-1}{2a^{4-d}g^2}(\sum_{\alpha,\beta,\gamma,\delta=1}^{2}\hat{U}_{\alpha \beta}\hat{U}_{\beta \gamma}\hat{U}_{\gamma\delta}^\dag \hat{U}_{\delta \alpha}^\dag  + h.c.)|| = \frac{4\cdot 16}{a^{12-3d}g^6}= \frac{64}{a^{12-3d}g^6}.
\end{align}
Since the last sixteen terms all have different color indices $\alpha\beta\gamma\delta$, they do not commute with each other in general. Thus, the bound for the last $16$ terms is
\begin{equation}
    \sum_{q=15}^{1} 4\cdot ||\hat{h}_{L}(\vec{n}_p,j,k)||\cdot ||q\cdot\hat{h}_{L'}(\vec{n}_p,j,k)|| \cdot ||\hat{h}_{L}(\vec{n}_p,j,k)|| = 480 ||\hat{h}_{L}(\vec{n}_p,j,k)||^3 = \frac{480}{a^{12-3d}g^6}
\end{equation}
Therefore, the bound for case-(i) terms is
\begin{equation}
    L^d\frac{d(d-1)}{2}[(2^{20}-16)\frac{64}{a^{12-3d}g^6} + \frac{480}{a^{12-3d}g^6}] = \frac{33554160 L^d d(d-1)}{a^{12-3d}g^6},
\end{equation}
We divide case (ii) into two types; those where (i) $\hat{h}_L$ and $\hat{h}_{L'}$ act on neighboring plaquettes with different parities on the same two-dimensional plane, and those where (ii) $\hat{h}_L$ and $\hat{h}_{L'}$ act on plaquettes that share only one common dimension. In type (i), the number of pairs of $\hat{h}_L$ and $\hat{h}_{L'}$ is the number of two-dimensional planes, $\frac{d(d-1)}{2}$. Since we have chosen to implement even terms before odd terms, each commutator is bounded by
\begin{equation}
    4\cdot ||\hat{h}_L (\vec{n}_e,j,k)||^2 \cdot ||4\cdot 16 \hat{h}_{L'} (\vec{n}_o,j,k)|| = \frac{256}{a^{12-3d}g^6},
\end{equation}
where in the second norm expression, the factor of $4$ is due to the fact that there are four $\hat{h}_{L'} (\vec{n}_o)$ terms acting on the four links, which form the plaquette that $\hat{h}_L (\vec{n}_e)$ acts on, and the factor of $16$ is the number of different color indices of the magnetic operators. There are $2d^3-6d^2+4d$ pairs of $\hat{h}_L$ and $\hat{h}_{L'}$ in type (ii). See Table \ref{tb:SU2_plaq1} for the explanation. Each commutator is bounded by
\begin{equation}
    4\cdot ||\hat{h}_L (\vec{n}_p,j,k)||^2 \cdot ||2\cdot 16 \hat{h}_{L'} (\vec{n}_p',j',k')|| = \frac{128}{a^{12-3d}g^6},
\end{equation}
where in the second norm expression, the factor of $2$ is due to the fact that there are two $\hat{h}_{L'}$ that collide with each $\hat{h}_L$. Therefore, the bound for case-(ii) terms is given by
\begin{equation}
    2^{20}\frac{L^d}{2}[\frac{256}{a^{12-3d}g^6}\frac{d(d-1)}{2} + \frac{128}{a^{12-3d}g^6}(2d^3-6d^2+4d)] = \frac{67108864 L^d}{a^{12-3d}g^6}(2 d^3- 5 d^2+3 d)
\end{equation}
In total, $C_{1,8}$ is bounded by
\begin{equation}
    \frac{L^d}{a^{12-3d}g^6}(134217728 d^3 - 301990160 d^2 + 167772432 d).
\end{equation}

\begin{table}[t]
\centering
\begin{tabular}{|l|l|l|} 
\hline
$(p,k,l)$, $k<l$ & colliding tuples & $\#$ tuples  \\ 
\hline
$1$       & $(p,k,j)$, $j>l$       & $d-l$               \\ 
\hline
$2$       & $(p,l,j)$, $j>l$       & $d-l$               \\ 
\hline
$3$       & $(p,j,l)$, $k<j<l$      & $l-k-1$              \\ 
\hline
$4$       & $(p',k,j)$, $j>l$    & $d-l$               \\ 
\hline
$5$       & $(p',l,j)$, $j>l$      & $d-l$               \\ 
\hline
$6$       & $(p',j,l)$, $k<j<l$     & $l-k-1$             \\
\hline
\end{tabular}
\caption{The number of tuples that share a dimension with $(p,k,l)$ and are implemented after $(p,k,l)$ in the ordered list $\mathbbm{L}$. Each term $\hat{h}_{L}\in \mathbbm{L}$ partially labelled by $(p,k,l)$ collides on two links with two $\hat{h}_{L'}$ terms, respectively, labelled with tuples in the second column. Here, $p,p' \in \{even,odd\}$ and $p\neq p'$. Therefore, the total number colliding tuples is given by $\sum_{k=1}^{d-1} \sum_{l>k}^{d} 8d-4(l+k)-4= 2d^3-6d^2+4d$.}
\label{tb:SU2_plaq1}
\end{table}

Next, we analyze the second sum in (\ref{eq:U1_trotter_err}), which is given by
\begin{align}
\sum_{i}|| [[\hat{H}_i, \sum_{j>i}\hat{H}_j],\sum_{k>i}\hat{H}_k] || \leq
\sum_{n=1}^{12} || C_{2,n} ||,
\label{eq:SU2_trot2}
\end{align}
where
\begin{align}
C_{2,1} =& [[ \sum_{\vec{n}} \hat{D}_{\vec{n}}^{(M)} , \sum_{\vec{n}'} \hat{T}_{\vec{n}'}^{(K)} ]  , \sum_{\vec{n}''} \hat{D}_{\vec{n}''}^{(E)} ], \nonumber \\
C_{2,2} =& [[ \sum_{\vec{n}} \hat{D}_{\vec{n}}^{(M)} , \sum_{\vec{n}'} \hat{T}_{\vec{n}'}^{(K)} ]  , \sum_{\vec{n}''} \hat{T}_{\vec{n}''}^{(K)} ], \nonumber \\
C_{2,3} =& [[ \sum_{\vec{n}} \hat{D}_{\vec{n}}^{(M)} , \sum_{\vec{n}'} \hat{T}_{\vec{n}'}^{(K)} ]  , \sum_{\vec{n}''} \hat{L}_{\vec{n}''}^{(B)} ], \nonumber \\
C_{2,4} =& [[ \sum_{\vec{n}} \hat{D}_{\vec{n}}^{(E)} , \sum_{\vec{n}'} \hat{T}_{\vec{n}'}^{(K)} ]  , \sum_{\vec{n}''} \hat{T}_{\vec{n}''}^{(K)} ], \nonumber \\
C_{2,5} =& [[ \sum_{\vec{n}} \hat{D}_{\vec{n}}^{(E)} , \sum_{\vec{n}'} \hat{T}_{\vec{n}'}^{(K)} ]  , \sum_{\vec{n}''} \hat{L}_{\vec{n}''}^{(B)} ], \nonumber \\
C_{2,6} =& [[ \sum_{\vec{n}} \hat{D}_{\vec{n}}^{(E)} , \sum_{\vec{n}'} \hat{L}_{\vec{n}'}^{(B)} ]  , \sum_{\vec{n}''} \hat{T}_{\vec{n}''}^{(K)} ], \nonumber \\
C_{2,7} =& [[ \sum_{\vec{n}} \hat{D}_{\vec{n}}^{(E)} , \sum_{\vec{n}'} \hat{L}_{\vec{n}'}^{(B)} ]  , \sum_{\vec{n}''} \hat{L}_{\vec{n}''}^{(B)} ], \nonumber \\
C_{2,8} =& \sum_{\hat{h}_{T}\in \mathbbm{T}}[[ \hat{h}_T , \sum_{\hat{h}_{T'}\in \mathbbm{T};T'>T} \hat{h}_{T'} ]  , \sum_{\hat{h}_{T''}\in \mathbbm{T};T''>T} \hat{h}_{T''} ], \nonumber \\
C_{2,9} =& \sum_{\hat{h}_{T}\in \mathbbm{T}} [[ \hat{h}_T , \sum_{\hat{h}_{T'}\in \mathbbm{T};T'>T} \hat{h}_{T'} ]  , \sum_{\vec{n}} \hat{L}_{\vec{n}}^{(B)} ], \nonumber \\
C_{2,10} =& \sum_{\hat{h}_{T}\in \mathbbm{T}}[[ \hat{h}_T , \sum_{\vec{n}'} \hat{L}_{\vec{n}'}^{(B)} ]  , \sum_{\hat{h}_{T'}\in \mathbbm{T};T'>T} \hat{h}_{T'} ], \nonumber \\
C_{2,11} =& \sum_{\hat{h}_{T}\in \mathbbm{T}} [[ \hat{h}_T , \sum_{\vec{n}} \hat{L}_{\vec{n}}^{(B)} ]  , \sum_{\vec{n}'} \hat{L}_{\vec{n}'}^{(B)} ], \nonumber \\
C_{2,12} =& \sum_{\hat{h}_{L}\in \mathbbm{L}} [[ \hat{h}_L , \sum_{\hat{h}_{L'}\in \mathbbm{L};L'>L} \hat{h}_{L'} ]  , \sum_{\hat{h}_{L''}\in \mathbbm{L};L''>L} \hat{h}_{L''} ].
\label{eq:SU2_trotter_err_2}
\end{align}

For $C_{2,1}$, we obtain the bound
\begin{align}
    &\quad ||[[ \sum_{\vec{n}} \hat{D}_{\vec{n}}^{(M)} , \sum_{\vec{n}'} \hat{T}_{\vec{n}'}^{(K)} ]  , \sum_{\vec{n}''} \hat{D}_{\vec{n}''}^{(E)} ]|| \nonumber \\
    &\leq || \sum_{\vec{n}}\sum_{l=1}^{d}\sum_{\alpha, \beta=1}^{2}[[\frac{m}{2}((-1)^{\vec{n}}\hat{Z}_{\alpha}(\vec{n})+(-1)^{\vec{n}+\hat{l}}\hat{Z}_{\beta}(\vec{n}+\hat{l})),\frac{1}{2a}(\hat{U}_{\alpha \beta}(\vec{n},l)\hat{\sigma}_{\alpha}^-(\vec{n})\hat{\sigma}_{\beta}^+(\vec{n}) +h.c.)]\nonumber\\
    &\quad ,\frac{g^2}{2a^{d-2}}\hat{E}^2(\vec{n},l)]  ||\nonumber \\
    &\leq 4dL^d||\frac{g^2}{2a^{d-2}}\frac{1}{4}[(j+1\pm 1)^2-(j+1)^2][\frac{m}{2}((-1)^{\vec{n}}\hat{Z}_{\alpha}(\vec{n})+(-1)^{\vec{n}+\hat{l}}\hat{Z}_{\beta}(\vec{n}+\hat{l}))\nonumber \\
    &\quad ,\frac{1}{2a}(\hat{U}_{\alpha \beta}(\vec{n},l)\hat{\sigma}_{\alpha}^-(\vec{n})\hat{\sigma}_{\beta}^+(\vec{n}) +h.c.)]||\nonumber \\
    &\leq 4dL^d\frac{g^2}{8a^{d-2}}(4\Lambda +3)||[\frac{m}{2}((-1)^{\vec{n}}\hat{Z}_{\alpha}(\vec{n})+(-1)^{\vec{n}+\hat{l}}\hat{Z}_{\beta}(\vec{n}+\hat{l})),\frac{1}{2a}(\hat{U}_{\alpha \beta}(\vec{n},l)\hat{\sigma}_{\alpha}^-(\vec{n})\hat{\sigma}_{\beta}^+(\vec{n}) +h.c.)]||\nonumber \\
    &\leq \frac{dL^d g^2(4\Lambda +3)}{2a^{d-2}} 2||m|| \cdot ||\frac{1}{a}|| = \frac{mdL^d g^2(4\Lambda +3)}{a^{d-1}},
\end{align}
where, in the first inequality, we used the fact that the mass terms at $\vec{n},\vec{n}+\hat{l}$, of which the respective colors are not $\alpha,\beta$, commute with the kinetic term with color indices $\alpha\beta$.

For $C_{2,2}$, we divide the commutators up into two types; type (i) where the kinetic terms in the inner commutator $\hat{T}_{\vec{n}'}^{(K)}$, and outer commutator $\hat{T}_{\vec{n}''}^{(K)}$ act on the same links, and type (ii) where $\hat{T}_{\vec{n}'}^{(K)}$ and $\hat{T}_{\vec{n}''}^{(K)}$ act on different links. We evaluate the bound for type (i), and obtain
\begin{align}
    &\quad || \sum_{\vec{n}}\sum_{l=1}^{d}\sum_{\alpha,\beta=1}^{2} [[\frac{m}{2}((-1)^{\vec{n}}\hat{Z}_{\alpha}(\vec{n})+(-1)^{\vec{n}+\hat{l}}\hat{Z}_{\beta}(\vec{n}+\hat{l})),\frac{1}{2a}(\hat{U}_{\alpha \beta}(\vec{n},l)\hat{\sigma}_{\alpha}^-(\vec{n})\hat{\sigma}_{\beta}^+(\vec{n}+\hat{l}) +h.c.)] \nonumber \\
    &,\frac{1}{2a}\sum_{\alpha'\beta'}(\hat{U}_{\alpha', \beta'}(\vec{n},l)\hat{\sigma}_{\alpha'}^-(\vec{n})\hat{\sigma}_{\beta'}^+(\vec{n}+\hat{l}) +h.c.)] || \leq 4dL^d \cdot 4 ||m|| \cdot ||\frac{1}{a}|| \cdot ||3\frac{1}{a}|| = \frac{48 md L^d}{a^2},
\end{align}
where the factor of three in the third norm expression in the last line is due to the fact that one of the four kinetic terms commute with the inner commutator. In particular, if $\alpha' = \alpha$ or $\beta' = \beta$, then the fermionic part of the outer kinetic term does not commute with the mass term. However, if $\alpha'\beta'$ and $\alpha \beta$ do not share any common color, then the fermionic operators in the inner and outer commutators act on different registers representing fermionic sites of different colors, the two gauge field operators commute due to (\ref{eq:SU2_Uhc}), and thus the commutator vanishes. 

We split type (ii) into two cases. Suppose the color indices of $\hat{T}_{\vec{n}'}^{(K)}$ is $\alpha\beta$. Then, in case (i), $\hat{T}_{\vec{n}''}^{(K)}$ has color indices $\beta \alpha$, and in case (ii), $\hat{T}_{\vec{n}''}^{(K)}$ has color indices $\beta\beta$ or $\alpha \alpha$. The bound for case (i) is as follows:
\begin{align}
    &\quad || \sum_{\vec{n}}\sum_{l=1}^{d}\sum_{\alpha,\beta=1}^{2} [[\frac{m}{2}((-1)^{\vec{n}}\hat{Z}_{\alpha}(\vec{n})+(-1)^{\vec{n}+\hat{l}}\hat{Z}_{\beta}(\vec{n}+\hat{l})),\frac{1}{2a}(\hat{U}_{\alpha \beta}(\vec{n},l)\hat{\sigma}_{\alpha}^-(\vec{n})\hat{\sigma}_{\beta}^+(\vec{n}+\hat{l}) +h.c.)] \nonumber \\
    &,\frac{4d-2}{2a}(\hat{U}_{\beta\alpha}\hat{\sigma}_{\beta}^-\hat{\sigma}_{\alpha}^+ +h.c.)] || \leq 4dL^d \cdot 4 ||m|| \cdot ||\frac{1}{a}|| \cdot ||\frac{(4d-2)}{a}|| = \frac{(64 d^2-32 d)mL^d}{a^2},
\end{align}
where the factor of $4d-2$, in front of the third term of the commutator, is the number of $\hat{U}_{\beta \alpha}\hat{\sigma}_{\beta}^-\hat{\sigma}_{\alpha}^+ +h.c.$ that collide with each $\hat{U}_{ \alpha\beta}\hat{\sigma}_{\alpha}^-\hat{\sigma}_{\beta}^+ +h.c.$. The bound for case (ii) is
\begin{align}
    &\quad || \sum_{\vec{n}}\sum_{l=1}^{d}\sum_{\alpha,\beta=1}^{2} [[\frac{m}{2}((-1)^{\vec{n}}\hat{Z}_{\alpha}(\vec{n})+(-1)^{\vec{n}+\hat{l}}\hat{Z}_{\beta}(\vec{n}+\hat{l})),\frac{1}{2a}(\hat{U}_{\alpha \beta}(\vec{n},l)\hat{\sigma}_{\alpha}^-(\vec{n})\hat{\sigma}_{\beta}^+(\vec{n}+\hat{l}) +h.c.)] \nonumber \\
    &,\frac{2d-1}{2a}\sum_{\alpha'\beta'=\beta\beta}^{\alpha\alpha}(\hat{U}_{\alpha'\beta'}\hat{\sigma}_{\alpha'}^-\hat{\sigma}_{\beta'}^+ +h.c.)] || \leq 4dL^d \cdot 4 ||m|| \cdot ||\frac{1}{a}|| \cdot ||\frac{2(2d-1)}{a}|| = \frac{(64 d^2- 32 d)mL^d}{a^2},
\end{align}
where the factor of $2d-1$, in front of the third term of the commutator, is the number of $\hat{U}_{\alpha'\beta'}\hat{\sigma}_{\alpha'}^-\hat{\sigma}_{\beta'}^+ +h.c.$, with $\alpha'\beta'=\beta\beta$ or $\alpha\alpha$, that collide with each $\hat{U}_{ \alpha\beta}\hat{\sigma}_{\alpha}^-\hat{\sigma}_{\beta}^+ +h.c.$.
Adding up the bounds for type (i) and (ii), we obtain the bound for the $C_{2,2}$ as follows:
\begin{equation}
    \frac{(128 d^2- 16 d)mL^d}{a^2}.
\end{equation}

The bound for $C_{2,3}$ is given by
\begin{align}
    &\quad ||[[ \sum_{\vec{n}} \hat{D}_{\vec{n}}^{(M)} , \sum_{\vec{n}'} \hat{T}_{\vec{n}'}^{(K)} ]  , \sum_{\vec{n}''} \hat{L}_{\vec{n}''}^{(B)} ]|| \nonumber \\
    &\leq ||\sum_{\vec{n}}\sum_{l=1}^{d}\sum_{\alpha, \beta = 1}^{2} [[\frac{m}{2}((-1)^{\vec{n}}\hat{Z}_{\alpha}(\vec{n})+(-1)^{\vec{n}+\hat{l}}\hat{Z}_{\beta}(\vec{n}+\hat{l})),\frac{1}{2a}(\hat{U}_{\alpha \beta}(\vec{n},l)\hat{\sigma}_{\alpha}^-(\vec{n})\hat{\sigma}_{\beta}^+(\vec{n}+\hat{l}) +h.c.)]\nonumber \\
    &\quad , \sum_{\vec{n}''} \hat{L}_{\vec{n}''}^{(B)} ]|| \nonumber \\
    &\leq 4dL^d\cdot 4||m||\cdot ||\frac{1}{a}|| \cdot ||\frac{2(d-1)}{2a^{4-d}g^2} \sum_{\alpha,\beta,\gamma,\delta=1}^{2} (\hat{U}_{\alpha \beta}\hat{U}_{\beta \gamma}\hat{U}_{\gamma \delta}^\dag \hat{U}_{\delta \alpha}^\dag + h.c.) || \leq \frac{256 mL^d (d^2-d)}{a^{5-d}g^2},
\end{align}
where $2(d-1)$ is the number of plaquettes that consist of the link $(\vec{n},l)$.

Next, we consider $C_{2,4}$. We divide the commutators up into two cases the same way we did for $C_{2,2}$. The bound for case (i), where the kinetic terms act on the same links, is given by
\begin{align}
    &\quad || \sum_{\vec{n}}\sum_{l=1}^{d}\sum_{\alpha,\beta=1}^{2} [[\frac{g^2}{2a^{d-2}}\hat{E}^2(\vec{n},l),\frac{1}{2a}(\hat{U}_{\alpha \beta}(\vec{n},l)\hat{\sigma}_{\alpha}^-(\vec{n})\hat{\sigma}_{\beta}^+(\vec{n}+\hat{l}) +h.c.)] \nonumber \\
    &,\frac{1}{2a}\sum_{\alpha',\beta'=1}^{2}(\hat{U}_{\alpha' \beta'}(\vec{n},l)\hat{\sigma}_{\alpha'}^-(\vec{n})\hat{\sigma}_{\beta'}^+(\vec{n}+\hat{l}) +h.c.)] || \nonumber \\
    &\leq 4dL^d \cdot 2|| [\frac{g^2}{2a^{d-2}}\hat{E}^2(\vec{n},l),\frac{1}{2a}(\hat{U}_{\alpha \beta}(\vec{n},l)\hat{\sigma}_{\alpha}^-(\vec{n})\hat{\sigma}_{\beta}^+(\vec{n}+\hat{l}) +h.c.)]|| \nonumber \\
    &\quad \cdot || \frac{1}{2a}\sum_{\alpha',\beta'=1}^{2}(\hat{U}_{\alpha' \beta'}(\vec{n},l)\hat{\sigma}_{\alpha'}^-(\vec{n})\hat{\sigma}_{\beta'}^+(\vec{n}+\hat{l}) +h.c.) || \nonumber \\
    &\leq 8dL^d \frac{g^2}{8a^{d-1}}(4\Lambda + 3) \cdot \frac{4}{a} = \frac{4g^2 dL^d (4\Lambda + 3)}{a^d},
\end{align}
where we have used (\ref{eq:SU2_EK}) to evaluate the first norm expression in the first inequality. As in $C_{2,2}$, we separate case (ii), where the kinetic terms act on different links, into two types. The bound for type (i), where the color indices for the outer kinetic term is $\beta \alpha$, is given by
\begin{align}
    &\quad || \sum_{\vec{n}}\sum_{l=1}^{d}\sum_{\alpha,\beta=1}^{2} [[\frac{g^2}{2a^{d-2}}\hat{E}^2(\vec{n},l),\frac{1}{2a}(\hat{U}_{\alpha \beta}(\vec{n},l)\hat{\sigma}_{\alpha}^-(\vec{n})\hat{\sigma}_{\beta}^+(\vec{n}+\hat{l}) +h.c.)] \nonumber \\
    &,\frac{4d-2}{2a} (\hat{U}_{\beta \alpha}\hat{\sigma}_{\beta}^-\hat{\sigma}_{\alpha}^+ +h.c.)] || \nonumber \\
    &\leq 4dL^d \cdot 2 \frac{g^2}{8a^{d-1}}(4\Lambda + 3) \cdot \frac{4d-2}{a} = \frac{(4d^2-2d)L^d g^2(4\Lambda + 3)}{a^d}.
\end{align}
For type (ii), where the color indices for the outer kinetic term is $\beta \beta$ or $\alpha \alpha$, we obtain the following bound:
\begin{align}
    &\quad || \sum_{\vec{n}}\sum_{l=1}^{d}\sum_{\alpha,\beta=1}^{2} [[\frac{g^2}{2a^{d-2}}\hat{E}^2(\vec{n},l),\frac{1}{2a}(\hat{U}_{\alpha \beta}(\vec{n},l)\hat{\sigma}_{\alpha}^-(\vec{n})\hat{\sigma}_{\beta}^+(\vec{n}+\hat{l}) +h.c.)] \nonumber \\
    &,\frac{2d-1}{2a}\sum_{\alpha'\beta'=\beta\beta}^{\alpha\alpha}(\hat{U}_{\alpha'\beta'}\hat{\sigma}_{\alpha'}^-\hat{\sigma}_{\beta'}^+ +h.c.)] || \nonumber \\
    &\leq 4dL^d \cdot 2 \frac{g^2}{8a^{d-1}}(4\Lambda + 3) \cdot \frac{2(2d-1)}{a} = \frac{(4d^2-2d)L^d g^2(4\Lambda + 3)}{a^d}.
\end{align}
Therefore, the bound for $C_{2,4}$ is
\begin{equation}
    \frac{8 d^2 L^d g^2(4\Lambda + 3)}{a^d}.
\end{equation}

We proceed to evaluate the bound for $C_{2,5}$ as follows:
\begin{align}
    &\quad [[ \sum_{\vec{n}} \hat{D}_{\vec{n}}^{(E)} , \sum_{\vec{n}'} \hat{T}_{\vec{n}'}^{(K)} ]  , \sum_{\vec{n}''} \hat{L}_{\vec{n}''}^{(B)} ] \nonumber \\
    &\leq ||\sum_{\vec{n}}\sum_{l=1}^{d}\sum_{\alpha, \beta = 1}^{2} [[\frac{g^2}{2a^{d-2}}\hat{E}^2(\vec{n},l),\frac{1}{2a}(\hat{U}_{\alpha \beta}(\vec{n},l)\hat{\sigma}_{\alpha}^-(\vec{n})\hat{\sigma}_{\beta}^+(\vec{n}+\hat{l}) +h.c.)], \sum_{\vec{n}''} \hat{L}_{\vec{n}''}^{(B)}] || \nonumber \\
    &\leq 4dL^d\cdot 2 ||[\frac{g^2}{2a^{d-2}}\hat{E}^2(\vec{n},l),\frac{1}{2a}(\hat{U}_{\alpha \beta}(\vec{n},l)\hat{\sigma}_{\alpha}^-(\vec{n})\hat{\sigma}_{\beta}^+(\vec{n}+\hat{l}) +h.c.)]|| \cdot ||\frac{2(d-1)}{2a^{4-d}g^2}\sum_{\alpha,\beta,\gamma,\delta=1}^{2}(\hat{U}_{\alpha \beta}\hat{U}_{\beta \gamma}\hat{U}_{\gamma \delta}^\dag \hat{U}_{\delta \alpha}^\dag + h.c. )||\nonumber \\
    &\leq 4dL^d\cdot 2 \frac{g^2}{8a^{d-1}}(4\Lambda + 3) \cdot \frac{32(d-1)}{a^{4-d}g^2} = \frac{32 d(d-1)L^d (4\Lambda + 3)}{a^3},
\end{align}
where in the second inequality, we have used (\ref{eq:SU2_EK}) and the fact that there are $2(d-1)$ plaquettes consisting of the link $(\vec{n},l)$. The bound for the $C_{2,6}$ is given by
\begin{align}
    &\quad ||[[ \sum_{\vec{n}} \hat{D}_{\vec{n}}^{(E)} , \sum_{\vec{n}'} \hat{L}_{\vec{n}'}^{(B)} ]  , \sum_{\vec{n}''} \hat{T}_{\vec{n}''}^{(K)} ]|| \nonumber \\
    &= ||  [\sum_{\vec{n}} \sum_{i=1}^{d}\sum_{j\neq i;j=1}^{d} [ \frac{g^2}{2a^{d-2}} (\hat{E}^2(\vec{n},i) + \hat{E}^2(\vec{n}+\hat{i},j)+\hat{E}^2(\vec{n}+\hat{j},i) +\hat{E}^2(\vec{n},j) ) \nonumber \\
    &\quad , \frac{-1}{2a^{4-d}g^2} \sum_{\alpha,\beta,\gamma,\delta=1}^{2} (\hat{U}_{\alpha \beta}(\vec{n},i)\hat{U}_{ \beta \gamma}(\vec{n}+\hat{i},j)\hat{U}_{\gamma \delta}^{\dag}(\vec{n}+\hat{j},i)\hat{U}_{\delta \alpha}^{\dag}(\vec{n},j) + h.c.) ]\nonumber \\
    &\quad , \frac{1}{2a} \sum_{\alpha',\beta'=1}^{2}[\hat{U}_{\alpha'\beta'}(\vec{n},i) \hat{\sigma}^{-}_{\alpha'}(\vec{n})\hat{\sigma}^{+}_{\beta'}(\vec{n}+\hat{i}) +\hat{U}_{\alpha'\beta'}(\vec{n}+\hat{i},j) \hat{\sigma}^{-}_{\alpha'}(\vec{n}+\hat{i})\hat{\sigma}^{+}_{\beta'}(\vec{n}+\hat{i}+\hat{j}) \nonumber \\
    &\quad  + \hat{U}_{\alpha'\beta'}(\vec{n}+\hat{j},i) \hat{\sigma}^{-}_{\alpha'}(\vec{n}+\hat{j})\hat{\sigma}^{+}_{\beta'}(\vec{n}+\hat{j}+\hat{i}) +\hat{U}_{\alpha'\beta'}(\vec{n},j) \hat{\sigma}^{-}_{\alpha'}(\vec{n})\hat{\sigma}^{+}_{\beta'}(\vec{n}+\hat{j})
    +h.c.] ]  || \nonumber \\
    &\leq L^d \frac{d(d-1)}{2} \cdot 2 \cdot || \frac{8 (4\Lambda + 3)}{a^2} ||  \cdot 4|| \frac{1}{2a}\sum_{\alpha',\beta'=1}^{2} (\hat{U}_{\alpha'\beta'}\hat{\sigma}^{-}_{\alpha'}\hat{\sigma}^{+}_{\beta'} +h.c.) || \nonumber \\
    &\leq L^d \frac{d(d-1)}{2} \cdot 2 \cdot || \frac{8 (4\Lambda + 3)}{a^2} ||  \cdot 4|| \frac{4}{a}  || = \frac{128 d(d-1)L^d (4\Lambda +3)}{a^3},
\end{align}
where we have used (\ref{eq:SU2_elec_mag}) to evaluate the first norm expression of the first inequality. The bound for $C_{2,7}$ is given by
\begin{align}
    &\quad ||[[ \sum_{\vec{n}} \hat{D}_{\vec{n}}^{(E)} , \sum_{\vec{n}'} \hat{L}_{\vec{n}'}^{(B)} ]  , \sum_{\vec{n}''} \hat{L}_{\vec{n}''}^{(B)} ]|| \nonumber \\
    &\leq ||  [\sum_{\vec{n}} \sum_{i=1}^{d}\sum_{j\neq i;j=1}^{d} [ \frac{g^2}{2a^{d-2}} (\hat{E}^2(\vec{n},i) + \hat{E}^2(\vec{n}+\hat{i},j)+\hat{E}^2(\vec{n}+\hat{j},i) +\hat{E}^2(\vec{n},j) ) \nonumber \\
    &\quad , \frac{-1}{2a^{4-d}g^2}\sum_{\alpha,\beta,\gamma,\delta=1}^{2} (\hat{U}_{\alpha \beta}(\vec{n},i)\hat{U}_{ \beta \gamma}(\vec{n}+\hat{i},j)\hat{U}_{\gamma \delta}^{\dag}(\vec{n}+\hat{j},i)\hat{U}_{\delta \alpha}^{\dag}(\vec{n},j) + h.c.) ], \sum_{\vec{n}''} \hat{L}_{\vec{n}''}^{(B)} ]||\nonumber \\
    &\leq L^d\frac{d(d-1)}{2} \cdot 2 ||\frac{8 (4\Lambda + 3)}{a^2}||\cdot || \frac{8d-11}{2a^{4-d}g^2} \sum_{\alpha,\beta,\gamma,\delta=1}^{2} (\hat{U}_{\alpha \beta}\hat{U}_{\beta \gamma}\hat{U}_{\gamma \delta}^\dag \hat{U}_{\delta \alpha}^\dag +h.c.) ||\nonumber\\
    &\leq L^d\frac{d(d-1)8(4\Lambda + 3)}{a^2} \cdot \frac{(8d-11)16}{a^{4-d}g^2} = \frac{(1024 d^3 - 2432 d^2 + 1408 d)L^d(4\Lambda + 3)}{a^{6-d}g^2},
\end{align}
where in the second inequality, we have used (\ref{eq:SU2_elec_mag}) to evaluate the first norm expression, and the factor of $8d-11$ in the second norm term is the number of plaquettes that collide on links with the plaquette acted on by the magnetic operators in the inner commutator, as explained in the paragraph below (\ref{eq:U1_EBB}).

Next we consider $C_{2,8}$, and divide the commutators up into five cases: (i) $\hat{h}_{T}$, $\hat{h}_{T'}$ and $\hat{h}_{T''}$ all act on the same links, (ii) $\hat{h}_{T}$ and $\hat{h}_{T'}$ act on the same links, while $\hat{h}_{T''}$ acts on neighboring links that are connected via the fermionic sites, (iii) $\hat{h}_{T}$ and $\hat{h}_{T''}$ act on the same links, while $\hat{h}_{T'}$ acts on neighboring links that are connected via the fermionic sites, (iv) $\hat{h}_{T'}$ and $\hat{h}_{T''}$ act on the same links, while $\hat{h}_{T}$ acts on neighboring links that are connected via the fermionic sites, and (v) $\hat{h}_{T}$, $\hat{h}_{T'}$ and $\hat{h}_{T''}$ all act on different but connected links. We begin with case (i) by considering $\hat{h}_{T}$, $\hat{h}_{T'}$ and $\hat{h}_{T''}$ that act on one link $(\vec{n}_p,l)$ only. There are $2^6$ such $\hat{h}_{T}$ terms, each with different parameters $a,b,c,\Delta j, \alpha, \beta$ in $\mathbbm{T}$. Since $\hat{h}_{T'}$ and $\hat{h}_{T''}$ are terms that succeed $\hat{h}_T$ in $\mathbbm{T}$, they can have different or same color indices $\alpha, \beta$ as $\hat{h}_T$, if $\hat{h}_T$ is among the first $2^6-4$ terms of $\mathbbm{T}$, in which case the bound is given by
\begin{align}
    &\quad (2^6-4) \sum_{p=e}^{o}\sum_{l=1}^{d}||[[ \sum_{\vec{n}_p}\hat{h}_T(\vec{n}_p,l), \sum_{\hat{h}_{T'}\in \mathbbm{T};T'>T} \sum_{\vec{n}_p} \hat{h}_{T'}(\vec{n}_p,l) ]  , \sum_{\hat{h}_{T''}\in \mathbbm{T};T''>T} \sum_{\vec{n}_p} \hat{h}_{T''}(\vec{n}_p,l) ]|| \nonumber \\
    &\leq  (2^6-4) 2d \sum_{\vec{n}_p} 4|| \hat{h}_T(\vec{n}_p,l) || \cdot ||\sum_{\hat{h}_{T'}\in \mathbbm{T};T'>T} \hat{h}_{T'}(\vec{n}_p,l)|| \cdot || \sum_{\hat{h}_{T''}\in \mathbbm{T};T''>T} \hat{h}_{T''}(\vec{n}_p,l) ]|| \nonumber \\
    &\leq 240dL^d ||\frac{1}{a}|| \cdot ||\frac{1}{2a}\sum_{\alpha,\beta = 1}^{2}(\hat{U}_{\alpha \beta}(\vec{n}_p,l)\hat{\sigma}_{\alpha}^-(\vec{n}_p)\hat{\sigma}_{\beta}^+(\vec{n}_p+\hat{l}) +h.c.)||^2 \nonumber \\
    &\leq \frac{240dL^d}{a}(\frac{4}{a})^2 = \frac{3840 dL^d}{a^3}.
\end{align}
If $\hat{h}_T$ is one of the last four terms of $\mathbbm{T}$, then $\hat{h}_{T'}$ and $\hat{h}_{T''}$ must have different color indices. As a result of (\ref{eq:SU2_Uhc}), if $\hat{h}_T$ is the fourth-to-last term, which has $\alpha \beta = 11$, then the commutator $[[\hat{h}_{T}(\vec{n}_p,l),\hat{h}_{T'}(\vec{n}_p,l)],\hat{h}_{T''}(\vec{n}_p,l)]$ is non-zero if both $\hat{h}_{T'},\hat{h}_{T''}$ have color indices equal to either $12$ or $21$, in which case the bound is given by
\begin{equation}
    dL^d \cdot 4 ||\frac{1}{a}||\cdot ||2\frac{1}{a}||^2,
\end{equation}
where the factor of two outside the fraction in the second norm expression reflects the fact that $\hat{h}_{T'},\hat{h}_{T''}$ can have $\alpha \beta = 12,21$. If $\hat{h}_T$ is the second- and third-to-last terms, which have $\alpha \beta = 12,21$, respectively, then the commutator $[[\hat{h}_{T}(\vec{n}_p,l),\hat{h}_{T'}(\vec{n}_p,l)],\hat{h}_{T''}(\vec{n}_p,l)]$ is non-zero if $\hat{h}_{T'},\hat{h}_{T''}$ have color indices equal to $22$, in which each case the bound is given by
\begin{equation}
    dL^d \cdot 4 ||\frac{1}{a}||^3.
\end{equation}
As such, the bound for case (i) is
\begin{equation}
    \frac{3920 dL^d}{a^3}.
\end{equation}
Similarly, for case (ii), we analyze the first $2^6-4$ and last $4$ $\hat{h}_T \in \mathbbm{T}$ separately. The bound for the first $2^6-4$ terms is
\begin{align}
    &\quad (2^6-4) \sum_{p=e}^{o}\sum_{l=1}^{d}||[[ \sum_{\vec{n}_p}\hat{h}_T(\vec{n}_p,l), \sum_{\hat{h}_{T'}\in \mathbbm{T};T'>T} \sum_{\vec{n}_p} \hat{h}_{T'}(\vec{n}_p,l) ]  , \sum_{\hat{h}_{T''}\in \mathbbm{T};T''>T} \hat{h}_{T''} ]|| \nonumber \\
    &\leq  (2^6-4) 2d \sum_{\vec{n}_p} 4|| \hat{h}_T(\vec{n}_p,l) || \cdot ||\sum_{\hat{h}_{T'}\in \mathbbm{T};T'>T} \hat{h}_{T'}(\vec{n}_p,l)|| \cdot || \sum_{\hat{h}_{T''}\in \mathbbm{T};T''>T} \hat{h}_{T''} || \nonumber \\
    &\leq 240 dL^d ||\frac{1}{a}|| \cdot ||\frac{1}{2a}\sum_{\alpha,\beta = 1}^{2}(\hat{U}_{\alpha \beta}(\vec{n}_p,l)\hat{\sigma}_{\alpha}^-(\vec{n}_p)\hat{\sigma}_{\beta}^+(\vec{n}_p+\hat{l}) +h.c.)|| \cdot ||\frac{4d-1}{2a}\sum_{\alpha,\beta = 1}^{2}(\hat{U}_{\alpha \beta}\hat{\sigma}_{\alpha}^-\hat{\sigma}_{\beta}^+ +h.c.)|| \nonumber \\
    &\leq \frac{240 dL^d}{a}\frac{4}{a}\frac{4(4d-1)}{a} = \frac{(15360 d^2 - 3840 d)L^d}{a^3},
    \label{eq:SU2_c28_2}
\end{align}
where the factor of $4d-1$ in the third norm expression of the second inequality is the number of links connected to $(\vec{n}_p,l)$, via the fermionic sites on both its ends. We admit that the constants can be improved slightly by optimizing over the color indices. For instance, if the inner commutator consists of only terms with $\alpha \beta = 11$, then it commutes with $\hat{h}_{T''}$ with $\alpha \beta = 22$, as they do act on fermions of different colors on the ends of $(\vec{n}_p,l)$, and it commutes with $\hat{h}_{T''}$ with $\alpha \beta = 12, 21$ on one of the fermionic sites on the ends of $(\vec{n}_p,l)$. Hereafter, we neglect this type of optimization, unless specified. We now consider the fourth-to-last $\hat{h}_T \in \mathbbm{T}$, with $\alpha \beta = 11$, and compute its bound in the following:
\begin{align}
    4dL^d || \frac{1}{a} || \cdot ||\frac{2}{a}|| \cdot || \frac{2d-1}{a}8|| = \frac{(128 d^2 - 64 d)L^d}{a^3},
\end{align}
where the $2$ in the numerator of the second norm expression is due to the fact that $\hat{h}_T$ does not commute with two of the remaining terms, with $\alpha \beta = 12,21$. For the the second- and third-to-last $\hat{h}_T \in \mathbbm{T}$, with $\alpha \beta = 12,21$, we compute the bound for each term as follows:
\begin{align}
    4dL^d || \frac{1}{a} || \cdot ||\frac{1}{a}|| \cdot || \frac{2d-1}{a}8|| = \frac{(64 d^2- 32 d)L^d}{a^3},
\end{align}
where the second norm expression is due to the fact that $\hat{h}_T$ does not commute with the remaining term with $\alpha \beta = 22$. Therefore, the case (ii) is bounded by
\begin{equation}
    \frac{(15616 d^2 - 3968 d)L^d}{a^3}.
\end{equation}

We separate case (iii) into two types: type (i) consists of commutators where $\hat{h}_{T'}$ acts on links in the same direction, but of different parity, as $\hat{h}_{T},\hat{h}_{T''}$; type (ii) consists of commutators where $\hat{h}_{T'}$ acts on links in different directions from $\hat{h}_{T},\hat{h}_{T''}$. Consider type (i), since we implement the even terms before the odd terms, the commutator bound for the first $2^6-4$ $\hat{h}_{T}$ is
\begin{align}
    &\quad (2^6-4)\sum_{l=1}^{d}||[[ \sum_{\vec{n}_e}\hat{h}_T(\vec{n}_e,l), \sum_{\hat{h}_{T'}\in \mathbbm{T};T'>T} \sum_{\vec{n}_o} \hat{h}_{T'}(\vec{n}_o,l) ]  , \sum_{\hat{h}_{T''}\in \mathbbm{T};T''>T} \sum_{\vec{n}_e} \hat{h}_{T''}(\vec{n}_e,l) ]|| \nonumber \\
    &\leq  (2^6-4) d \sum_{\vec{n}_e} 4|| \hat{h}_T(\vec{n}_e,l) || \cdot ||2 \frac{1}{2a}\sum_{\alpha,\beta = 1}^{2}(\hat{U}_{\alpha \beta}(\vec{n}_o,l)\hat{\sigma}_{\alpha}^-(\vec{n}_o)\hat{\sigma}_{\beta}^+(\vec{n}_o+\hat{l}) +h.c.) || \nonumber \\
    &\quad \cdot || 3 \frac{1}{2a}\sum_{\alpha,\beta = 1}^{2}(\hat{U}_{\alpha \beta}(\vec{n}_e,l)\hat{\sigma}_{\alpha}^-(\vec{n}_e)\hat{\sigma}_{\beta}^+(\vec{n}_e+\hat{l}) +h.c.) ]|| \nonumber \\
    &\leq 240d\frac{L^d}{2} ||\frac{1}{a}|| \cdot ||\frac{8}{a}||\cdot || \frac{12}{a}|| = \frac{11520 dL^d}{a^3},
\end{align}
where in the first inequality, the factor of $2$ in the second norm expression is the number of odd links that are connected to each $(\vec{n}_e,l)$, and the factor of $3$ in the third norm expression is the number of links that collide with the original link, and are connected with the two odd links connected to the original link. There are four non-trivial commutators for the last $4$ $\hat{h}_{T}$; if $\hat{h}_{T}$ has $\alpha \beta = 11$, then $\hat{h}_{T'}$ must have $\alpha \beta = 12,21$; if $\hat{h}_{T}$ has $\alpha \beta = 12,21$, then $\hat{h}_{T'}$ must have $\alpha \beta = 22$. Thus, each of the last four commutators is bounded by
\begin{align}
    &\quad \sum_{l=1}^{d} \sum_{\vec{n}_e} 4||\hat{h}_T(\vec{n}_e,l)||\cdot ||2 \frac{1}{2a}(\hat{U}_{\alpha \beta}(\vec{n}_o,l)\hat{\sigma}_{\alpha}^-(\vec{n}_o)\hat{\sigma}_{\beta}^+(\vec{n}_o+\hat{l}) +h.c.) || \nonumber \\
    &\quad \cdot || 3 \frac{1}{2a}\sum_{\alpha,\beta = 1}^{2}(\hat{U}_{\alpha \beta}(\vec{n}_e,l)\hat{\sigma}_{\alpha}^-(\vec{n}_e)\hat{\sigma}_{\beta}^+(\vec{n}_e+\hat{l}) +h.c.) ]|| = \frac{48 dL^d}{a^3},
\end{align}
and thus, type (i) is bounded by
\begin{equation}
    \frac{11712 dL^d}{a^3}.
\end{equation}

Similarly, we obtain bound for the first $2^6-4$ terms in type (ii) as follows
\begin{align}
    &\quad (2^6-4) \sum_{p,p'=e}^{o} \sum_{l'>l} \sum_{l=1}^{d}||[[ \sum_{\vec{n}_p}\hat{h}_T(\vec{n}_p,l), \sum_{\hat{h}_{T'}\in \mathbbm{T};T'>T} \sum_{\vec{n}_{p'}} \hat{h}_{T'}(\vec{n}_{p'},l') ]  , \sum_{\hat{h}_{T''}\in \mathbbm{T};T''>T} \sum_{\vec{n}_p} \hat{h}_{T''}(\vec{n}_p,l) ]|| \nonumber \\
    &\leq  (2^6-4) 4\frac{d(d-1)}{2} \sum_{\vec{n}_p} 4|| \hat{h}_T(\vec{n}_p,l) || \cdot ||2 \frac{1}{2a}\sum_{\alpha,\beta = 1}^{2}(\hat{U}_{\alpha \beta}(\vec{n}_p',l')\hat{\sigma}_{\alpha}^-(\vec{n}_p')\hat{\sigma}_{\beta}^+(\vec{n}_p'+\hat{l}') +h.c.) || \nonumber \\
    &\quad \cdot || \frac{1}{2a}\sum_{\alpha,\beta = 1}^{2}(\hat{U}_{\alpha \beta}(\vec{n}_p,l)\hat{\sigma}_{\alpha}^-(\vec{n}_p)\hat{\sigma}_{\beta}^+(\vec{n}_p+\hat{l}) +h.c.) ]|| \nonumber \\
    &\leq 240d(d-1)L^d ||\frac{1}{a}|| \cdot ||\frac{8}{a}||\cdot || \frac{4}{a}|| = \frac{7680 (d^2-d)L^d}{a^3},
\end{align}
where in the first inequality, the factor of $2$ in the second norm expression is the number of links $(\vec{n}_{p'},l)$ that are connected to each $(\vec{n}_p,l)$. The bound for each of the last four commutators is bounded by
\begin{align}
    &\quad \sum_{p,p'=e}^{o} \sum_{l'>l} \sum_{l=1}^{d}||[[ \sum_{\vec{n}_p}\hat{h}_T(\vec{n}_p,l), \sum_{\hat{h}_{T'}\in \mathbbm{T};T'>T} \sum_{\vec{n}_{p'}} \hat{h}_{T'}(\vec{n}_{p'},l') ]  , \sum_{\hat{h}_{T''}\in \mathbbm{T};T''>T} \sum_{\vec{n}_p} \hat{h}_{T''}(\vec{n}_p,l) ]|| \nonumber \\
    &\leq  4\frac{d(d-1)}{2} \sum_{\vec{n}_p} 4|| \hat{h}_T(\vec{n}_p,l) || \cdot ||2 \frac{1}{2a}(\hat{U}_{\alpha \beta}(\vec{n}_p',l')\hat{\sigma}_{\alpha}^-(\vec{n}_p')\hat{\sigma}_{\beta}^+(\vec{n}_p'+\hat{l}') +h.c.) || \nonumber \\
    &\quad \cdot || \frac{1}{2a}\sum_{\alpha,\beta = 1}^{2}(\hat{U}_{\alpha \beta}(\vec{n}_p,l)\hat{\sigma}_{\alpha}^-(\vec{n}_p)\hat{\sigma}_{\beta}^+(\vec{n}_p+\hat{l}) +h.c.) ]|| \nonumber \\
    &\leq 4d(d-1)L^d ||\frac{1}{a}|| \cdot ||\frac{2}{a}||\cdot || \frac{4}{a}|| = \frac{32 (d^2-d)L^d}{a^3}.
\end{align}
Therefore, the bound for case (iii) is given by
\begin{equation}
    \frac{L^d}{a^3}(7808 d^2 + 3904 d).
\end{equation}
We divide case (iv) into two types: type-(i) commutators are those where $\hat{h}_{T}$, $\hat{h}_{T'}$ and $\hat{h}_{T''}$ act on links in the same direction; and type-(ii) commutators are those where $\hat{h}_{T}$, $\hat{h}_{T'}$ and $\hat{h}_{T''}$ act on links in different directions. We consider type (i) first. Since even terms are implemented before odd terms, we obtain its bound as follows
\begin{align}
    &\quad 2^6 \sum_{l=1}^{d}||[[ \sum_{\vec{n}_e}\hat{h}_T(\vec{n}_e,l), \sum_{\hat{h}_{T'}\in \mathbbm{T};T'>T} \sum_{(\vec{n}_o,l)} \hat{h}_{T'}(\vec{n}_o,l) ]  , \sum_{\hat{h}_{T''}\in \mathbbm{T};T''>T} \sum_{(\vec{n}_o,l)} \hat{h}_{T''}(\vec{n}_o,l) ]|| \nonumber \\
    &\leq 2^6 d \cdot \sum_{\vec{n}_e}4 || \hat{h}_T(\vec{n}_e,l) || \cdot || \frac{2}{2a}\sum_{\alpha, \beta=1}^{2} (\hat{U}_{\alpha \beta}(\vec{n}_o,l)\hat{\sigma}_{\alpha}^-(\vec{n}_o)\hat{\sigma}_{\beta}^+(\vec{n}_o+\hat{l}) +h.c.)||^2 \nonumber \\
    &\leq \frac{2^6 dL^d}{2} 4||\frac{1}{a}||\cdot ||\frac{8}{a}||^2 = \frac{8192 dL^d}{a^2},
\end{align}
where the factor of $2$ in the numerator of the second norm expression of the second line is due to the fact that $(\vec{n}_e,l)$ is connected to two $(\vec{n}_o,l)$. Next, we evaluate the bound for type (ii) as follows:
\begin{align}
    &\quad 2^6 \sum_{p,p'=e}^{o}\sum_{l=1}^{d}\sum_{l'>l}||[[ \sum_{\vec{n}_p}\hat{h}_T(\vec{n}_p,l), \sum_{\hat{h}_{T'}\in \mathbbm{T};T'>T} \sum_{(\vec{n}_{p'},l')} \hat{h}_{T'}(\vec{n}_{p'},l') ]  , \sum_{\hat{h}_{T''}\in \mathbbm{T};T''>T} \sum_{(\vec{n}_{p'},l')} \hat{h}_{T''}(\vec{n}_{p'},l') ]|| \nonumber \\
    &\leq 2^6\cdot 4\frac{d(d-1)}{2} \cdot 4 \sum_{\vec{n}_p}|| \hat{h}_{T}(\vec{n}_p,l) ||\cdot || \frac{2}{2a}\sum_{\alpha \beta} (\hat{U}_{\alpha \beta}(\vec{n}_{p'},l')\hat{\sigma}_{\alpha}^-(\vec{n}_{p'})\hat{\sigma}_{\beta}^+(\vec{n}_{p'}+\hat{l}') +h.c.)||^2 \nonumber \\
    &\leq 2^7d(d-1)\frac{L^d}{2} 4 ||\frac{1}{a}|| \cdot ||\frac{8}{a}||^2 = \frac{16384(d^2-d)L^d}{a^3},
\end{align}
where the factor of $2$ in the numerator of the second norm expression of the second line is due to the fact that the link $(\vec{n}_p,l)$, acted on by $\hat{h}_T$, is connected to two links $(\vec{n}_{p'},l')$, acted on by $\hat{h}_{T'}$ and $\hat{h}_{T''}$. Therefore, case (iv) is bounded by
\begin{equation}
    \frac{(16384 d^2 - 8192 d)L^d}{a^3}.
\end{equation}
Lastly, we obtain the bound for case (v)
\begin{align}
    &\quad 2^6 \sum_{(p,l)}\sum_{\substack{(p',l');\\(p',l')>(p,l)}}\sum_{\substack{(p'',l'');\\(p'',l'')>(p',l')}}||[[ \sum_{\vec{n}_p}\hat{h}_T(\vec{n}_p,l), \sum_{\substack{\hat{h}_{T'}\in \mathbbm{T};\\T'>T}} \sum_{(\vec{n}_{p'},l')} \hat{h}_{T'}(\vec{n}_{p'},l') ]  , \sum_{\substack{\hat{h}_{T''}\in \mathbbm{T};\\T''>T}} \sum_{(\vec{n}_{p''},l'')} \hat{h}_{T''}(\vec{n}_{p''},l'') ]|| \nonumber \\
    &\leq 2^6 \frac{4}{3}(2d^3-3d^2+d) \cdot 4 \sum_{\vec{n}_p} || \hat{h}_T(\vec{n}_p,l)|| \cdot || \frac{2}{2a}\sum_{\alpha \beta} (\hat{U}_{\alpha \beta}(\vec{n}_{p'},l')\hat{\sigma}_{\alpha}^-(\vec{n}_{p'})\hat{\sigma}_{\beta}^+(\vec{n}_{p'}+\hat{l}') +h.c.)|| \nonumber \\
    &\quad \cdot || \frac{4}{2a}\sum_{\alpha \beta} (\hat{U}_{\alpha \beta}(\vec{n}_{p''},l'')\hat{\sigma}_{\alpha}^-(\vec{n}_{p''})\hat{\sigma}_{\beta}^+(\vec{n}_{p''}+\hat{l}'') +h.c.)||\nonumber \\
    &= \frac{2^{9} L^d}{3a}(2d^3-3d^2+d) ||\frac{8}{a}||\cdot ||\frac{16}{a}|| = \frac{L^d}{3a^3}(131072 d^3 - 196608 d^2 + 65536 d),
\end{align}
where $(p',l')>(p,l)$ means that $(p',l')$ appears after $(p,l)$ in $\mathbbm{T}$, and since there are $2d$ $(p,l)$, the triple sum outside the norm expression in the first line evaluates to
\begin{equation}
    \sum_{q=1}^{2d} (2d-q)(2d-q-1) = \frac{4}{3}(2d^3-3d^2+d).
\end{equation}
In the inequality, the factor of $2$ in the numerator of the second norm expression is because of the fact that $(\vec{n}_p,l)$ is connected to at most $2$ $(\vec{n}_{p'},l')$, and thus, each of the inner commutators acts on at most three links and four sites. Further, each of these four sites are connected to at most one $(\vec{n}_{p''},l'')$, hence the factor of $4$ in the numerator of the third norm expression. Note that the constants can be further tightened by considering the color indices of the fermionic operators. Finally, adding the bounds for all cases, we arrive at the bound for $C_{2,8}$
\begin{equation}
    (131072 \frac{d^3}{3} - 25728 d^2 + 52528 \frac{d}{3})\frac{L^d}{a^3}.
\end{equation}

Now for $C_{2,9}$, we separate the commutators into three cases. Case-(i) commutators consists of $\hat{h}_{T}$ and $\hat{h}_{T'}$ that act on the same links. Case-(ii) commutators consists of $\hat{h}_{T}$ and $\hat{h}_{T'}$ that act on links in the same directions, but of different parities. Case-(iii) commutators consists of $\hat{h}_{T}$ and $\hat{h}_{T'}$ that act on links in different directions, but connected via fermionic sites. For case (i), once again, we consider the first $2^6-4$ and last $4$ $\hat{h}_T$ terms in $\mathbbm{T}$ separately. The bound for the first $2^6-4$ $\hat{h}_T$ terms is given by
\begin{align}
    &\quad (2^6-4) \sum_{p=e}^{o} \sum_{l=1}^{d} [[ \sum_{\vec{n}_p}\hat{h}_T(\vec{n}_p,l) , \sum_{\hat{h}_{T'}\in \mathbbm{T};T'>T} \sum_{\vec{n}_p} \hat{h}_{T'}(\vec{n}_p,l) ]  , \sum_{\vec{n}} \hat{L}_{\vec{n}}^{(B)} ]\nonumber \\
    &\leq (2^6-4) 2d \sum_{\vec{n}_p} 4||\hat{h}_T(\vec{n}_p,l)|| \cdot || \frac{1}{2a}\sum_{\alpha \beta} (\hat{U}_{\alpha \beta}(\vec{n}_{p},l)\hat{\sigma}_{\alpha}^-(\vec{n}_{p})\hat{\sigma}_{\beta}^+(\vec{n}_{p}+\hat{l}) +h.c.) || \nonumber \\
    &\quad \cdot || \frac{2(d-1)}{2a^{4-d}g^2}\sum_{\alpha,\beta,\gamma,\delta=1}^{2} (\hat{U}_{\alpha \beta}\hat{U}_{\beta \gamma}\hat{U}_{\gamma \delta}^\dag \hat{U}_{\delta \alpha}^\dag +h.c.) ||\nonumber \\
    &\leq 60dL^d \cdot 4||\frac{1}{a}||\cdot ||\frac{4}{a}||\cdot ||\frac{32(d-1)}{a^{4-d}g^2} || = \frac{(30720 d^2 - 30720 d)L^d}{a^{6-d}g^2},
\end{align}
where in the first inequality, the numerator $2(d-1)$ in the third norm term is the number of magnetic operators that act on link $(\vec{n}_p, l)$. The last four $\hat{h}_{T}$ terms contribute four commutators, each of which is bounded by
\begin{align}
    &\quad 4dL^d||\hat{h}_T(\vec{n}_p,l)|| \cdot || \frac{1}{2a} (\hat{U}_{\alpha \beta}(\vec{n}_{p},l)\hat{\sigma}_{\alpha}^-(\vec{n}_{p})\hat{\sigma}_{\beta}^+(\vec{n}_{p}+\hat{l}) +h.c.) || \nonumber \\
    &\quad \cdot || \frac{2(d-1)}{2a^{4-d}g^2}\sum_{\alpha,\beta,\gamma,\delta=1}^{2} (\hat{U}_{\alpha \beta}\hat{U}_{\beta \gamma}\hat{U}_{\gamma \delta}^\dag \hat{U}_{\delta \alpha}^\dag +h.c.) ||\nonumber \\
    &\leq \frac{128 (d^2-d)L^d}{a^{6-d}g^2}.
\end{align}
As such, the bound for case-(i) commutators is
\begin{align}
    \frac{31232( d^2 -  d)L^d}{a^{6-d}g^2}.
\end{align}

For case (ii), there are two types of commutators; those where (i) $\hat{h}_T$ and $\hat{h}_{T'}$ have color indices $\alpha \beta$ and $\beta \alpha$, respectively, and thus, collide on two fermionic sites, and where (ii) $\hat{h}_T$ and $\hat{h}_{T'}$ have color indices $\alpha \beta$ and $\beta \beta$ or $\alpha \alpha$, respectively, and thus, collide on one fermionic sites. Thus, considering type (i), the inner commutators act on three links, which collide with $3\cdot 2(d-1)$ magnetic operators. Since we implement the even terms before the odd terms, we obtain the bound for the type-(i) commutators as follows:
\begin{align}
    &\quad 2^6 \sum_{l=1}^{d} 4 \sum_{\vec{n}_e} ||\hat{h}_T(\vec{n}_e,l)|| \cdot || \frac{2}{2a} (\hat{U}_{ \beta\alpha}(\vec{n}_{o},l)\hat{\sigma}_{\beta}^-(\vec{n}_{o})\hat{\sigma}_{\alpha}^+(\vec{n}_{o}+\hat{l}) +h.c.) || \nonumber \\
    &\quad \cdot || \frac{6(d-1)}{2a^{4-d}g^2} \sum_{\alpha,\beta,\gamma,\delta=1}^{2} (\hat{U}_{\alpha \beta}\hat{U}_{\beta \gamma}\hat{U}_{\gamma \delta}^\dag \hat{U}_{\delta \alpha}^\dag +h.c.) || \nonumber \\
    &\leq 2^7 dL^d ||\frac{1}{a}|| \cdot ||\frac{2}{a}||\cdot ||\frac{96 (d-1)}{a^{4-d}g^2}|| = \frac{24576 (d^2-d)L^d}{a^{6-d}g^2}.
\end{align}
For type (ii), the inner commutators act on two links, and thus collide with $2\cdot 2(d-1)$ magnetic operators. We obtain the bound for the type-(ii) commutators as follows:
\begin{align}
    &\quad 2^6 \sum_{l=1}^{d} 4 \sum_{\vec{n}_e} ||\hat{h}_T(\vec{n}_e,l)|| \cdot || \frac{1}{2a} \sum_{\alpha' \beta' = \beta \beta}^{\alpha \alpha}(\hat{U}_{ \alpha' \beta'}(\vec{n}_{o},l)\hat{\sigma}_{\alpha'}^-(\vec{n}_{o})\hat{\sigma}_{\beta'}^+(\vec{n}_{o}+\hat{l}) +h.c.) || \nonumber \\
    &\quad \cdot || \frac{4(d-1)}{2a^{4-d}g^2} \sum_{\alpha,\beta,\gamma,\delta=1}^{2} (\hat{U}_{\alpha \beta}\hat{U}_{\beta \gamma}\hat{U}_{\gamma \delta}^\dag \hat{U}_{\delta \alpha}^\dag +h.c.) || \nonumber \\
    &\leq 2^7 dL^d ||\frac{1}{a}|| \cdot ||\frac{2}{a}||\cdot ||\frac{64 (d-1)}{a^{4-d}g^2}|| = \frac{16384 (d^2-d)L^d}{a^{6-d}g^2}.
\end{align}
Thus, for case (ii), we obtain the bound
\begin{equation}
    \frac{40960(d^2-d)L^d}{a^{6-d}g^2}.
\end{equation}
In the third case, $\hat{h}_{T}$ and $\hat{h}_{T'}$ act on links in different directions. As in the second case, we divide up case (iii) based on the color indices of $\hat{h}_{T}$ and $\hat{h}_{T'}$. Focusing on the first type, the inner commutators act on three links, which collide with $2+3\cdot2(d-2)=6d-10$ magnetic operators, where two of them act on all three links, and there are $2(d-2)$ magnetic operators acting on each one link, but not the other two links. Thus, we obtain the bound for type (i)
commutators as follows:
\begin{align}
    &\quad 2^6 \sum_{p,p'=e}^{o}\sum_{l=1}^{d}\sum_{l'>l} 4 \sum_{\vec{n}_p} ||\hat{h}_T(\vec{n}_p,l)|| \cdot || \frac{2}{2a} (\hat{U}_{ \beta\alpha}(\vec{n}_{p'},l')\hat{\sigma}_{\beta}^-(\vec{n}_{p'})\hat{\sigma}_{\alpha}^+(\vec{n}_{p'}+\hat{l}') +h.c.) || \nonumber \\
    &\quad \cdot || \frac{6d-10}{2a^{4-d}g^2} \sum_{\alpha,\beta,\gamma,\delta=1}^{2} (\hat{U}_{\alpha \beta}\hat{U}_{\beta \gamma}\hat{U}_{\gamma \delta}^\dag \hat{U}_{\delta \alpha}^\dag +h.c.) || \nonumber \\
    &\leq 2^6 \cdot 4 \frac{d(d-1)}{2} \cdot 4\frac{L^d}{2} ||\frac{1}{a}|| \cdot ||\frac{2}{a}||\cdot ||\frac{32 (3d-5)}{a^{4-d}g^2}|| =(49152 d^3 - 131072 d^2 + 81920 d) \frac{L^d}{a^{6-d}g^2}.
\end{align}
Moving onto the second type, the inner commutators act on two links, which collide with $1+2\cdot2(d-2)=4d-7$ magnetic operators, where one of them acts on both links, and there are $2(d-2)$ magnetic operators acting on each one link, but not the other. Hence, we evaluate the bound for type (ii), and obtain,
\begin{align}
    &\quad 2^6 \sum_{p,p'=e}^{o}\sum_{l=1}^{d}\sum_{l'>l} 4 \sum_{\vec{n}_p} ||\hat{h}_T(\vec{n}_p,l)|| \cdot || \frac{1}{2a} \sum_{\alpha' \beta' = \beta \gamma}^{\gamma \alpha}(\hat{U}_{ \alpha' \beta'}(\vec{n}_{p'},l)\hat{\sigma}_{\alpha'}^-(\vec{n}_{p'})\hat{\sigma}_{\beta'}^+(\vec{n}_{p'}+\hat{l}) +h.c.) || \nonumber \\
    &\quad \cdot || \frac{4d-7}{2a^{4-d}g^2} \sum_{\alpha,\beta,\gamma,\delta=1}^{2} (\hat{U}_{\alpha \beta}\hat{U}_{\beta \gamma}\hat{U}_{\gamma \delta}^\dag \hat{U}_{\delta \alpha}^\dag +h.c.) || \nonumber \\
    &\leq 2^6 \cdot 4 \frac{d(d-1)}{2} \cdot 4\frac{L^d}{2} ||\frac{1}{a}|| \cdot ||\frac{2}{a}||\cdot ||\frac{16(4d-7)}{a^{4-d}g^2}|| =(32768 d^3 - 90112 d^2 + 57344 d) \frac{L^d}{a^{6-d}g^2}.
\end{align}
Therefore, the bound for case (iii) is
\begin{equation}
    (81920 d^3 - 221184 d^2 + 139264 d) \frac{L^d}{a^{6-d}g^2}.
\end{equation}
Summing up the bounds for all three cases, we obtain the bound for $C_{2,9}$
\begin{equation}
    (81920 d^3 - 148992 d^2 + 67072 d)\frac{L^d}{a^{6-d}g^2}.
\end{equation}

Now for $C_{2,10}$, we divide the commutators up into cases and types, as we have done for $C_{2,9}$. The bound for case (i) of both the $C_{2,9}$ and $C_{2,10}$ is the same, and is given by
\begin{equation}
    \frac{31232(d^2-d)L^d}{a^{6-d}g^2}.
\end{equation}
The bound for case (ii) can be obtained from the case-(ii) bounds for $C_{2,9}$ after some slight modifications. First, since the kinetic operator $\hat{h}_T$ only acts on one link, there are only $2(d-1)$ plaquette operators in the inner commutator that do not commute with each $\hat{h}_T$ because the plaquettes may lie on $d-1$ two-dimensional planes and can be of two different parities. Second, the kinetic operators $\hat{h}_{T'}$ not only collide with $\hat{h}_{T}$ via fermionic sites, but also with the magnetic operators on links. Thus, we obtain the bound for type (i) of case (ii) 
\begin{align}
    &\quad 2^6 \sum_{l=1}^{d} 4 \sum_{\vec{n}_e} ||\hat{h}_T(\vec{n}_e,l)||  \cdot || \frac{2 (d-1)}{2a^{4-d}g^2} \sum_{\alpha,\beta,\gamma,\delta=1}^{2} (\hat{U}_{\alpha \beta}\hat{U}_{\beta \gamma}\hat{U}_{\gamma \delta}^\dag \hat{U}_{\delta \alpha}^\dag +h.c.) ||\nonumber \\
    &\quad \cdot || \frac{4}{2a} (\hat{U}_{ \beta\alpha}(\vec{n}_{o},l)\hat{\sigma}_{\beta}^-(\vec{n}_{o})\hat{\sigma}_{\alpha}^+(\vec{n}_{o}+\hat{l}) +h.c.) || \nonumber \\
    &\leq 2^7 d L^d ||\frac{1}{a}|| \cdot ||\frac{32(d-1)}{a^{4-d}g^2}||\cdot ||\frac{4}{a}|| = \frac{16384(d^2-d)L^d}{a^{6-d}g^2},
\end{align}
where a factor of $4$ in the numerator of the third norm expression is due to the fact that two $\hat{h}_{T'}$, of color indices $\beta \alpha$, collide with each of $\hat{h}_{T}$, of color indices $\alpha \beta$, and the pair of plaquette operators that lie on the same plane. The bound for the second type of case (ii) is
\begin{align}
    &\quad 2^6  \sum_{l=1}^{d} 4 \sum_{\vec{n}_e} ||\hat{h}_T(\vec{n}_e,l)||   \cdot || \frac{2 (d-1)}{2a^{4-d}g^2} \sum_{\alpha,\beta,\gamma,\delta=1}^{2} (\hat{U}_{\alpha \beta}\hat{U}_{\beta \gamma}\hat{U}_{\gamma \delta}^\dag \hat{U}_{\delta \alpha}^\dag +h.c.) ||\nonumber \\
    &\quad \cdot || \frac{3}{2a} \sum_{\alpha' \beta' = \beta \beta}^{\alpha \alpha}(\hat{U}_{ \alpha' \beta'}(\vec{n}_{o},l)\hat{\sigma}_{\alpha'}^-(\vec{n}_{o})\hat{\sigma}_{\beta'}^+(\vec{n}_{o}+\hat{l}) +h.c.) ||\nonumber \\
    &\leq 2^7 dL^d ||\frac{1}{a}|| \cdot ||\frac{32(d-1)}{a^{4-d}g^2}||\cdot ||\frac{6}{a}|| = \frac{24576 (d^2-d)L^d}{a^{6-d}g^2},
\end{align}
where the numerator $3$ in the third norm expression is due to the fact that each $\hat{h}_{T'}$, of color indices $\alpha \alpha$ or $\beta \beta$, collides with $\hat{h}_{T}$, of color indices $\alpha \beta$, and two $\hat{h}_{T'}$ collide with the pair of magnetic operators that lie on the same plane. Thus, for case (ii), the bound is given by
\begin{equation}
    \frac{40960(d^2-d)L^d}{a^{6-d}g^2}.
\end{equation}
Now we consider the third case. Once again, we modify the case-(iii) bounds of $C_{2,9}$. Thus, we obtain the respective bounds for type (i) and (ii) commutators as follows:
\begin{align}
    &\quad 2^6 \sum_{p,p'=e}^{o}\sum_{l=1}^{d}\sum_{l'>l} 4 \sum_{\vec{n}_p} ||\hat{h}_T(\vec{n}_p,l)|| \cdot || \frac{2(d-1)}{2a^{4-d}g^2} \sum_{\alpha,\beta,\gamma,\delta=1}^{2} (\hat{U}_{\alpha \beta}\hat{U}_{\beta \gamma}\hat{U}_{\gamma \delta}^\dag \hat{U}_{\delta \alpha}^\dag +h.c.) || \nonumber \\
    &\quad \cdot || \frac{2}{2a} (\hat{U}_{ \beta\alpha}(\vec{n}_{p'},l')\hat{\sigma}_{\beta}^-(\vec{n}_{p'})\hat{\sigma}_{\alpha}^+(\vec{n}_{p'}+\hat{l}') +h.c.) || \nonumber \\
    &\leq 2^6 \cdot 4 \frac{d(d-1)}{2} \cdot 4\frac{L^d}{2} ||\frac{1}{a}||\cdot ||\frac{32(d-1)}{a^{4-d}g^2}||  \cdot ||\frac{2}{a}||=
    (16384 d^3 - 32768 d^2 + 16384 d)\frac{L^d}{a^{6-d}g^2},
\end{align}
and
\begin{align}
    &\quad 2^6 \sum_{p,p'=e}^{o}\sum_{l=1}^{d}\sum_{l'>l} 4 \sum_{\vec{n}_p} ||\hat{h}_T(\vec{n}_p,l)|| \cdot || \frac{2(d-1)}{2a^{4-d}g^2} \sum_{\alpha,\beta,\gamma,\delta=1}^{2} (\hat{U}_{\alpha \beta}\hat{U}_{\beta \gamma}\hat{U}_{\gamma \delta}^\dag \hat{U}_{\delta \alpha}^\dag +h.c.) || \nonumber \\
    &\quad \cdot || \frac{1}{2a} \sum_{\alpha' \beta' = \beta \gamma}^{\gamma \alpha}(\hat{U}_{ \alpha' \beta'}(\vec{n}_{p'},l)\hat{\sigma}_{\alpha'}^-(\vec{n}_{p'})\hat{\sigma}_{\beta'}^+(\vec{n}_{p'}+\hat{l}) +h.c.) ||  \nonumber \\
    &\leq 2^6 \cdot 4 \frac{d(d-1)}{2} \cdot 4\frac{L^d}{2} ||\frac{1}{a}||\cdot ||\frac{32(d-1)}{a^{4-d}g^2}||\cdot ||\frac{2}{a}|| =(16384 d^3 - 32768 d^2 + 16384 d) \frac{L^d}{a^{6-d}g^2}.
\end{align}
Therefore, the bound for case (iii) is
\begin{equation}
    (32768 d^3 - 65536 d^2 + 32768 d) \frac{L^d}{a^{6-d}g^2}.
\end{equation}
Summing up the bounds for all cases, we obtain the bound for $C_{2,10}$
\begin{equation}
    (32768 d^3 + 6656 d^2 - 39424 d)\frac{L^d}{a^{6-d}g^2}.
\end{equation}

We compute the bound for $C_{2,11}$, and obtain
\begin{align}
    &\quad \sum_{\hat{h}_{T}\in \mathbbm{T}} ||[[ \hat{h}_T , \sum_{\vec{n}} \hat{L}_{\vec{n}}^{(B)} ]  , \sum_{\vec{n}'} \hat{L}_{\vec{n}'}^{(B)} ]|| \nonumber \\
    &\leq 2^7 d\frac{L^d}{2}\cdot 4 ||\hat{h}_T(\vec{n}_p,l)||\cdot ||\frac{2(d-1)}{2a^{4-d}g^2} \sum_{\alpha,\beta,\gamma,\delta=1}^{2} (\hat{U}_{\alpha \beta}\hat{U}_{\beta \gamma}\hat{U}_{\gamma \delta}^\dag \hat{U}_{\delta \alpha}^\dag +h.c.) || \nonumber \\
    &\quad \cdot ||\frac{14d-20}{2a^{4-d}g^2}\sum_{\alpha,\beta,\gamma,\delta=1}^{2} (\hat{U}_{\alpha \beta}\hat{U}_{\beta \gamma}\hat{U}_{\gamma \delta}^\dag \hat{U}_{\delta \alpha}^\dag +h.c.)|| \nonumber \\
    &\leq \frac{2^8d{L^d}}{a}\frac{32(d-1)}{a^{4-d}g^2}\frac{16(14d-20)}{a^{4-d}g^2} = \frac{(1835008 d^3 - 4456448 d^2 + 2621440 d)L^d}{a^{9-2d}g^4},
\end{align}
where in the first inequality, the factors of $2(d-1)$ and $14d-20$ are explained in the paragraph below (\ref{eq:U1_KBB}).

Lastly, we consider $C_{2,12}$, which consists of commutators between only magnetic operators. The commutators are either \textit{intra-plaquette} or \textit{inter-plaquette}, where $\hat{h}_{L}$, $\hat{h}_{L'}$ and $\hat{h}_{L''}$ act on the same or different plaquettes, respectively. We consider intra-plaquette terms first. We remind the readers that there are $2^{20}$ $\hat{h}_{L}(\vec{n}_p,j,k)$ terms acting on each plaquette $(\vec{n}_p,j,k)$. We analyze the commutators, where $\hat{h}_{T}(\vec{n}_p,j,k)$ is among the first $2^{20}-16$, and last $16$ terms, separately. When $\hat{h}_{T}(\vec{n}_p,j,k)$ is among the first $2^{20}-16$ terms, the bound is given by
\begin{align}
    &\quad (2^{20}-16)\sum_{k\neq j; k=1}^{d} \sum_{j=1}^{d}\sum_{\vec{n}_p}\sum_{p=e}^o||[[ \hat{h}_L(\vec{n}_p,j,k) , \sum_{\hat{h}_{L'}\in \mathbbm{L};L'>L} \hat{h}_{L'}(\vec{n}_p,j,k) ]  , \sum_{\hat{h}_{L''}\in \mathbbm{L};L''>L} \hat{h}_{L''}(\vec{n}_p,j,k) ]|| \nonumber \\
    &\leq \frac{(2^{20}-16)d(d-1)L^d}{2} 4||\hat{h}_L(\vec{n}_p,j,k)|| \cdot ||\frac{1}{2a^{4-d}g^2}\sum_{\alpha,\beta,\gamma,\delta=1}^{2} (\hat{U}_{\alpha \beta}\hat{U}_{\beta \gamma}\hat{U}_{\gamma \delta}^\dag \hat{U}_{\delta \alpha}^\dag +h.c.)||^2 \nonumber \\
    &\leq (2^{21}-32)d(d-1)L^d ||\frac{1}{a^{4-d}g^2}||\cdot ||\frac{16}{a^{4-d}g^2}||^2 = \frac{536862720 (d^2-d) L^d}{a^{12-3d}g^6}.
\end{align}
When $\hat{h}_{L}(\vec{n}_p,j,k)$ is among the last $16$ terms, the bound is given by
\begin{equation}
    \frac{d(d-1)L^d}{2}4 ||\frac{1}{a^{4-d}g^2}||\cdot \sum_{q=15}^{1}||\frac{16q}{a^{4-d}g^2}||^2 = \frac{2480 (d^2-d) L^d}{a^{12-3d}g^6}.
\end{equation}
Thus, the bound for the intra-plaquette commutators is
\begin{equation}
    \frac{536865200 (d^2-d) L^d}{a^{12-3d}g^6}.
\end{equation}

We now proceed to analyze the inter-plaquette commutators. Since each $\hat{h}_{L''}$ operator is non-zero, the inner commutator $[ \hat{h}_L , \sum_{\hat{h}_{L'}\in \mathbbm{L};L'>L} \hat{h}_{L'} ]$ must be non-zero to guarantee a non-trivial triple commutator
\begin{equation}
    [[ \hat{h}_L , \sum_{\hat{h}_{L'}\in \mathbbm{L};L'>L} \hat{h}_{L'} ]  , \sum_{\hat{h}_{L''}\in \mathbbm{L};L''>L} \hat{h}_{L''} ].\nonumber
\end{equation}
Given a non-zero inner commutator, we further divide the inter-plaquette commutators into three types. Type (i) commutators satisfy
\begin{equation}
    [ \hat{h}_L , \sum_{\hat{h}_{L'}\in \mathbbm{L};L'>L} \hat{h}_{L'} ] \neq 0, \: [ \hat{h}_L , \sum_{\hat{h}_{L''}\in \mathbbm{L};L''>L} \hat{h}_{L''} ] \neq 0,\: [ \sum_{\hat{h}_{L'}\in \mathbbm{L};L'>L} \hat{h}_{L'} , \sum_{\hat{h}_{L''}\in \mathbbm{L};L''>L} \hat{h}_{L''} ]\neq 0.
    \label{eq:SU_BBB1}
\end{equation}
Type (ii) commutators satisfy
\begin{equation}
    [ \hat{h}_L , \sum_{\hat{h}_{L'}\in \mathbbm{L};L'>L} \hat{h}_{L'} ] \neq 0, \: [ \hat{h}_L , \sum_{\hat{h}_{L''}\in \mathbbm{L};L''>L} \hat{h}_{L''} ] \neq 0,\: [ \sum_{\hat{h}_{L'}\in \mathbbm{L};L'>L} \hat{h}_{L'} , \sum_{\hat{h}_{L''}\in \mathbbm{L};L''>L} \hat{h}_{L''} ]= 0.
    \label{eq:SU_BBB2}
\end{equation}
Type (iii) commutators satisfy
\begin{equation}
    [ \hat{h}_L , \sum_{\hat{h}_{L'}\in \mathbbm{L};L'>L} \hat{h}_{L'} ] \neq 0, \: [ \hat{h}_L , \sum_{\hat{h}_{L''}\in \mathbbm{L};L''>L} \hat{h}_{L''} ] = 0,\: [ \sum_{\hat{h}_{L'}\in \mathbbm{L};L'>L} \hat{h}_{L'} , \sum_{\hat{h}_{L''}\in \mathbbm{L};L''>L} \hat{h}_{L''} ]\neq 0.
    \label{eq:SU_BBB3}
\end{equation}
In order to facilitate the counting of the commutators in each case, we provide a geometric interpretation to the conditions for each type. For instance, the condition satisfied by all three cases $[ \hat{h}_L , \sum_{\hat{h}_{L'}\in \mathbbm{L};L'>L} \hat{h}_{L'} ] \neq 0$ implies that, some or all of the links of each plaquette, which is acted on by $\hat{h}_L$, must also be acted on by each $\hat{h}_{L'}$. As such, we can further infer that the plaquettes acted on by $\hat{h}_L$ and $\hat{h}_{L'}$ share at least one dimension. We extend this geometric interpretation to compute the number of non-trivial commutators in each type.

For type (i), the plaquettes acted on by $\hat{h}_L$, collide with those acted on by $\hat{h}_{L'}$ and $\hat{h}_{L''}$, and those acted on by $\hat{h}_{L'}$ also collide with those acted on by $\hat{h}_{L''}$. Suppose $\hat{h}_L$ is labelled by $(p,k,l)$. The possible parity-location tuples that label $\hat{h}_{L'}$ and $\hat{h}_{L''}$ are given in Table \ref{tb:SU2_bbb_1} and \ref{tb:SU2_bbb_2} for $p=$ $even$ and $odd$, respectively. Consider first the case where $p=even$, and $\hat{h}_{L}$ and $\hat{h}_{L'}$ are labelled by items $1-4$, $6$ and $7$ in Table \ref{tb:SU2_bbb_1}. Then, each plaquette acted on by $\hat{h}_{L}$ is acted on by two $\hat{h}_{L'}$ as they share one dimension. Further, the plaquettes acted on by $\hat{h}_{L}$ and $\hat{h}_{L'}$ are acted on by either $2,4,6$ or $8$ $\hat{h}_{L''}$. In particular, if the plaquettes acted on by $\hat{h}_{L''}$ (i) share one common dimension with $\hat{h}_{L}$, and two common dimensions and parity with $\hat{h}_{L'}$, or (ii) share one common dimension with $\hat{h}_{L}$ and a different dimension with $\hat{h}_{L'}$, then $\hat{h}_{L}$ and $\hat{h}_{L'}$ collide with two $\hat{h}_{L''}$. We compute the number of combinations of parity-location labels that satisfy these conditions using Table \ref{tb:SU2_bbb_1}, and obtain 
\begin{equation}
    \sum_{l>k}\sum_{k=1}^{d-1} 12(d-l)+2(l-k-1) = \frac{7}{3}(d^3-3d^2+2d).
\end{equation}
The bound in this case is given by
\begin{align}
    &\quad 2^{20} ||[[ \sum_{\vec{n}_e}\hat{h}_L(e,k,l) ,\sum_{\hat{h}_{L'}\in \mathbbm{L};L'>L} \hat{h}_{L'} ]  , \sum_{\hat{h}_{L''}\in \mathbbm{L};L''>L} \hat{h}_{L''} ]|| \nonumber \\
    &\leq \frac{2^{20} \cdot 7}{3}(d^3-3d^2+2d) \sum_{\vec{n}_e} 4\cdot || \hat{h}_L(e,k,l) || \cdot ||\frac{2}{2a^{4-d}g^2}\sum_{\alpha,\beta,\gamma,\delta=1}^{2} (\hat{U}_{\alpha \beta}\hat{U}_{\beta \gamma}\hat{U}_{\gamma \delta}^\dag \hat{U}_{\delta \alpha}^\dag +h.c.)|| \nonumber \\
    &\quad  \cdot || \frac{2}{2a^{4-d}g^2}\sum_{\alpha,\beta,\gamma,\delta=1}^{2} (\hat{U}_{\alpha \beta}\hat{U}_{\beta \gamma}\hat{U}_{\gamma \delta}^\dag \hat{U}_{\delta \alpha}^\dag +h.c.)||\nonumber \\
    &= \frac{L^d}{3a^{12-3d}g^6}15032385536 (d^3 - 3 d^2 + 2 d),
\end{align}
where $2^{20}$ is the number of $\hat{h}_L$ terms per plaquette.

If the plaquettes acted on by $\hat{h}_{L''}$ share only one dimension with both $\hat{h}_{L}$ and $\hat{h}_{L'}$, then $\hat{h}_{L}$ and $\hat{h}_{L'}$ collide with four $\hat{h}_{L''}$. Using Table \ref{tb:SU2_bbb_1}, we find the number of combinations of parity-location labels that satisfy this condition, i.e.
\begin{align}
    &\quad \sum_{l>k}\sum_{k=1}^{d-1} (d-l)[8(d-l-1)+4(l-k-1)] + (l-k-1)[4(d-l)+4(l-k-2)] \nonumber \\
    &= \frac{4}{3}(d^4-6d^3+11d^2-6d).
\end{align}
The bound in this case is given by
\begin{align}
    &\quad 2^{20} ||[[ \sum_{\vec{n}_e}\hat{h}_L(e,k,l) ,\sum_{\hat{h}_{L'}\in \mathbbm{L};L'>L} \hat{h}_{L'} ]  , \sum_{\hat{h}_{L''}\in \mathbbm{L};L''>L} \hat{h}_{L''} ]|| \nonumber \\
    &\leq \frac{2^{20} \cdot 4}{3}(d^4-6d^3+11d^2-6d) \sum_{\vec{n}_e} 4\cdot || \hat{h}_L(e,k,l) || \cdot ||\frac{2}{2a^{4-d}g^2}\sum_{\alpha,\beta,\gamma,\delta=1}^{2} (\hat{U}_{\alpha \beta}\hat{U}_{\beta \gamma}\hat{U}_{\gamma \delta}^\dag \hat{U}_{\delta \alpha}^\dag +h.c.)|| \nonumber \\
    &\quad  \cdot || \frac{4}{2a^{4-d}g^2}\sum_{\alpha,\beta,\gamma,\delta=1}^{2} (\hat{U}_{\alpha \beta}\hat{U}_{\beta \gamma}\hat{U}_{\gamma \delta}^\dag \hat{U}_{\delta \alpha}^\dag +h.c.)||\nonumber \\
    &= \frac{L^d}{3a^{12-3d}g^6}17179869184(d^4-6d^3+11d^2-6d).
\end{align}

If the plaquettes acted on by $\hat{h}_{L''}$ share only one dimension with $\hat{h}_{L'}$, and share both dimensions, but not the parity, with $\hat{h}_{L}$, then $\hat{h}_{L}$ and $\hat{h}_{L'}$ collide with six $\hat{h}_{L''}$. Once again, we use Table \ref{tb:SU2_bbb_1} to obtain the number of combinations of parity-location labels that satisfy this condition, i.e.
\begin{equation}
    \sum_{l>k}\sum_{k=1}^{d-1} 4(d-l)+2(l-k-1) = (d^3-3d^2+2d).
\end{equation}
The bound in this case is given by
\begin{align}
    &\quad 2^{20} ||[[ \sum_{\vec{n}_e}\hat{h}_L(e,k,l) ,\sum_{\hat{h}_{L'}\in \mathbbm{L};L'>L} \hat{h}_{L'} ]  , \sum_{\hat{h}_{L''}\in \mathbbm{L};L''>L} \hat{h}_{L''} ]|| \nonumber \\
    &\leq 2^{20} (d^3-3d^2+2d) \sum_{\vec{n}_e} 4\cdot || \hat{h}_L(e,k,l) || \cdot ||\frac{2}{2a^{4-d}g^2}\sum_{\alpha,\beta,\gamma,\delta=1}^{2} (\hat{U}_{\alpha \beta}\hat{U}_{\beta \gamma}\hat{U}_{\gamma \delta}^\dag \hat{U}_{\delta \alpha}^\dag +h.c.)|| \nonumber \\
    &\quad  \cdot || \frac{6}{2a^{4-d}g^2}\sum_{\alpha,\beta,\gamma,\delta=1}^{2} (\hat{U}_{\alpha \beta}\hat{U}_{\beta \gamma}\hat{U}_{\gamma \delta}^\dag \hat{U}_{\delta \alpha}^\dag +h.c.)||\nonumber \\
    &= \frac{L^d}{a^{12-3d}g^6}6442450944(d^3-3d^2+2d).
\end{align}

If the plaquettes acted on by $\hat{h}_{L''}$ share only one dimension with $\hat{h}_{L}$, and share both dimensions, but not the parity, with $\hat{h}_{L'}$, then $\hat{h}_{L}$ and $\hat{h}_{L'}$ collide with eight $\hat{h}_{L''}$. Once again, we use Table \ref{tb:SU2_bbb_1} to obtain the number of combinations of parity-location labels that satisfy this condition, i.e.
\begin{equation}
    \sum_{l>k}\sum_{k=1}^{d-1} 4(d-l)+2(l-k-1) = (d^3-3d^2+2d).
\end{equation}
The bound in this case is given by
\begin{align}
    &\quad 2^{20} ||[[ \sum_{\vec{n}_e}\hat{h}_L(e,k,l) ,\sum_{\hat{h}_{L'}\in \mathbbm{L};L'>L} \hat{h}_{L'} ]  , \sum_{\hat{h}_{L''}\in \mathbbm{L};L''>L} \hat{h}_{L''} ]|| \nonumber \\
    &\leq 2^{20} (d^3-3d^2+2d) \sum_{\vec{n}_e} 4\cdot || \hat{h}_L(e,k,l) || \cdot ||\frac{2}{2a^{4-d}g^2}\sum_{\alpha,\beta,\gamma,\delta=1}^{2} (\hat{U}_{\alpha \beta}\hat{U}_{\beta \gamma}\hat{U}_{\gamma \delta}^\dag \hat{U}_{\delta \alpha}^\dag +h.c.)|| \nonumber \\
    &\quad  \cdot || \frac{8}{2a^{4-d}g^2}\sum_{\alpha,\beta,\gamma,\delta=1}^{2} (\hat{U}_{\alpha \beta}\hat{U}_{\beta \gamma}\hat{U}_{\gamma \delta}^\dag \hat{U}_{\delta \alpha}^\dag +h.c.)||\nonumber \\
    &= \frac{L^d}{a^{12-3d}g^6}8589934592(d^3-3d^2+2d).
\end{align}

If $\hat{h}_{L''}$ acts on plaquettes that share only one dimension with $\hat{h}_{L}$ and $\hat{h}_{L'}$, then $\hat{h}_{L}$ and $\hat{h}_{L'}$ collide with eight $\hat{h}_{L''}$. There are
\begin{equation}
    \sum_{l>k}\sum_{k=1}^{d-1} 4(d-l)+2(l-k-1) = d^3-3d^2+2d
\end{equation}
combinations of parity-location labels that satisfy this condition. The bound in this case is given by
\begin{align}
    &\quad 2^{20} ||[[ \sum_{\vec{n}_e}\hat{h}_L(e,k,l) ,\sum_{\hat{h}_{L'}\in \mathbbm{L};L'>L} \hat{h}_{L'} ]  , \sum_{\hat{h}_{L''}\in \mathbbm{L};L''>L} \hat{h}_{L''} ]|| \nonumber \\
    &\leq 2^{20} (d-1) \sum_{\vec{n}_e} 4\cdot || \hat{h}_L(e,k,l) || \cdot ||\frac{4}{2a^{4-d}g^2}\sum_{\alpha,\beta,\gamma,\delta=1}^{2} (\hat{U}_{\alpha \beta}\hat{U}_{\beta \gamma}\hat{U}_{\gamma \delta}^\dag \hat{U}_{\delta \alpha}^\dag +h.c.)|| \nonumber \\
    &\quad  \cdot || \frac{8}{2a^{4-d}g^2}\sum_{\alpha,\beta,\gamma,\delta=1}^{2} (\hat{U}_{\alpha \beta}\hat{U}_{\beta \gamma}\hat{U}_{\gamma \delta}^\dag \hat{U}_{\delta \alpha}^\dag +h.c.)||\nonumber \\
    &= \frac{L^d}{a^{12-3d}g^6}17179869184(d^3-3d^2+2d).
\end{align}

Consider now the case where $\hat{h}_{L}$ and $\hat{h}_{L'}$ collide on two dimensions, but have different parities, i.e., item 5 in Table \ref{tb:SU2_bbb_1}. Since we implement even terms before odd ones, the parities of $\hat{h}_{L}$ and $\hat{h}_{L'}$ are even and odd, respectively. Moreover, $\hat{h}_{L}$ acting on a plaquette collides with four $\hat{h}_{L'}$ on the four links. If $\hat{h}_{L''}$ act on plaquettes that share one dimension with $\hat{h}_{L}$, and both dimensions and the parity with those acted on by $\hat{h}_{L'}$, then $\hat{h}_{L}$ and $\hat{h}_{L'}$ collide with four $\hat{h}_{L''}$. There are
\begin{equation}
    \sum_{k=1}^{d-1} 1 = d-1
\end{equation}
combinations of parity-location labels that satisfy this condition. The bound in this case is given by
\begin{align}
    &\quad 2^{20} ||[[ \sum_{\vec{n}_e}\hat{h}_L(e,k,l) ,\sum_{\hat{h}_{L'}\in \mathbbm{L};L'>L} \hat{h}_{L'} ]  , \sum_{\hat{h}_{L''}\in \mathbbm{L};L''>L} \hat{h}_{L''} ]|| \nonumber \\
    &\leq 2^{20} (d-1) \sum_{\vec{n}_e} 4\cdot || \hat{h}_L(e,k,l) || \cdot ||\frac{4}{2a^{4-d}g^2}\sum_{\alpha,\beta,\gamma,\delta=1}^{2} (\hat{U}_{\alpha \beta}\hat{U}_{\beta \gamma}\hat{U}_{\gamma \delta}^\dag \hat{U}_{\delta \alpha}^\dag +h.c.)|| \nonumber \\
    &\quad  \cdot || \frac{4}{2a^{4-d}g^2}\sum_{\alpha,\beta,\gamma,\delta=1}^{2} (\hat{U}_{\alpha \beta}\hat{U}_{\beta \gamma}\hat{U}_{\gamma \delta}^\dag \hat{U}_{\delta \alpha}^\dag +h.c.)||\nonumber \\
    &= \frac{L^d}{a^{12-3d}g^6}8589934592(d-1).
\end{align}

Therefore, type-(i) commutators, where $\hat{h}_{L}$ acts on even plaquettes are bounded by
\begin{equation}
    (17179869184 \frac{d^4}{3}+8589934592 \frac{d^3}{3}  - 146028888064 \frac{d^2}{3} + 146028888064 \frac{d}{3} -8589934592)\frac{L^d}{a^{12-3d}g^6}.
\end{equation}
Similarly, we obtain the bound for the commutators where $\hat{h}_{L}$ acts on odd plaquettes, i.e.,
\begin{equation}
    ( 17179869184 \frac{d^4}{3}- 10737418240\frac{d^3}{3}- 88046829568 \frac{d^2}{3}+81604378624 \frac{d}{3})\frac{L^d}{a^{12-3d}g^6},
\end{equation}
by considering separately the cases, in which $\hat{h}_L$ and $\hat{h}_{L'}$ collide with 2, 4, or 8 $\hat{h}_{L''}$, listed in Table \ref{tb:SU2_bbb_2}. Thus, the bound for all type-(i) commutators is
\begin{equation}
    (34359738368 \frac{d^4}{3} - 2147483648 \frac{d^3}{3} - 234075717632 \frac{d^2}{3}+ 227633266688 \frac{d}{3}-8589934592)\frac{L^d}{a^{12-3d}g^6}.
\end{equation}

We proceed to analyze type-(ii) commutators. By definition, $\hat{h}_{L}$ does not commute with both $\hat{h}_{L'}$ and $\hat{h}_{L''}$, but $\hat{h}_{L'}$ and $\hat{h}_{L''}$ commute with each other. On the lattice, this implies that $\hat{h}_{L}$ shares one common dimension each with $\hat{h}_{L'}$ and $\hat{h}_{L''}$, but $\hat{h}_{L'}$ and $\hat{h}_{L''}$ share no common dimension. Thus, each plaquette acted on by $\hat{h}_{L}$ is also acted on by two $\hat{h}_{L'}$ and $\hat{h}_{L''}$. Using table \ref{tb:SU2_bbb_3}, we obtain the number of combinations of parity-location labels that satisfy this condition as follows
\begin{equation}
    \sum_{l>k}\sum_{k=1}^{d-1} 2(d-l) [4(l-k-1)+8(d-l-1)] + 8(l-k-1)(d-l) = 2d^4 -12d^3+11d^2-6d.
\end{equation}
Hence, the bound for all type-(ii) commutators is
\begin{align}
    &\quad 2^{20} \sum_{p=e}^{o} ||[[ \sum_{\vec{n}_p}\hat{h}_L(p,k,l) ,\sum_{\hat{h}_{L'}\in \mathbbm{L};L'>L} \hat{h}_{L'} ]  , \sum_{\hat{h}_{L''}\in \mathbbm{L};L''>L} \hat{h}_{L''} ]|| \nonumber \\
    &\leq 2^{21}(2d^4 -12d^3+11d^2-6d)  \sum_{\vec{n}_p} 4\cdot || \hat{h}_L(p,k,l) || \cdot ||\frac{2}{2a^{4-d}g^2}\sum_{\alpha,\beta,\gamma,\delta=1}^{2} (\hat{U}_{\alpha \beta}\hat{U}_{\beta \gamma}\hat{U}_{\gamma \delta}^\dag \hat{U}_{\delta \alpha}^\dag +h.c.)|| \nonumber \\
    &\quad  \cdot || \frac{2}{2a^{4-d}g^2}\sum_{\alpha,\beta,\gamma,\delta=1}^{2} (\hat{U}_{\alpha \beta}\hat{U}_{\beta \gamma}\hat{U}_{\gamma \delta}^\dag \hat{U}_{\delta \alpha}^\dag +h.c.)||\nonumber \\
    &= \frac{L^d}{a^{12-3d}g^6}4294967296 (2 d^4 - 12 d^3 + 11 d^2 - 6 d).
\end{align}

Last but not least, for type (iii) commutators, $\hat{h}_{L'}$ does not commute with both $\hat{h}_{L}$ and $\hat{h}_{L''}$, but $\hat{h}_{L}$ and $\hat{h}_{L''}$ commute with each other. On the lattice, this implies that $\hat{h}_{L'}$ share one common dimension each with $\hat{h}_{L}$ and $\hat{h}_{L''}$, but $\hat{h}_{L}$ and $\hat{h}_{L''}$ share no common dimension. Thus, each plaquette acted on by $\hat{h}_{L}$ is also acted on by two $\hat{h}_{L'}$, and each plaquette acted on by $\hat{h}_{L'}$ is in turn acted on by two $\hat{h}_{L''}$. Using table \ref{tb:SU2_bbb_4}, we evaluate the number of combinations of parity-location labels that satisfy this condition, and obtain
\begin{equation}
    \sum_{\substack{j>l;\\ l>k}}\sum_{k=1}^{d-1} 16(d-l)(d-j+l-k-1) + 8(l-k-1)[(l-j-1)+(j-k-1)] = \frac{2}{5}(2d^5 - 15d^4 + 40d^3 - 45 d^2 + 18d).
\end{equation}
Hence, the bound for all type-(iii) commutators is
\begin{align}
    &\quad 2^{20} \sum_{p=e}^{o} ||[[ \sum_{\vec{n}_p}\hat{h}_L(p,k,l) ,\sum_{\hat{h}_{L'}\in \mathbbm{L};L'>L} \hat{h}_{L'} ]  , \sum_{\hat{h}_{L''}\in \mathbbm{L};L''>L} \hat{h}_{L''} ]|| \nonumber \\
    &\leq \frac{2^{22}}{5}(2d^5 - 15d^4 + 40d^3 - 45 d^2 + 18d) \sum_{\vec{n}_p} 4\cdot || \hat{h}_L(p,k,l) || \cdot  \nonumber \\
    &\quad  \cdot ||\frac{2}{2a^{4-d}g^2}\sum_{\alpha,\beta,\gamma,\delta=1}^{2} (\hat{U}_{\alpha \beta}\hat{U}_{\beta \gamma}\hat{U}_{\gamma \delta}^\dag \hat{U}_{\delta \alpha}^\dag +h.c.)||\cdot || \frac{2}{2a^{4-d}g^2}\sum_{\alpha,\beta,\gamma,\delta=1}^{2} (\hat{U}_{\alpha \beta}\hat{U}_{\beta \gamma}\hat{U}_{\gamma \delta}^\dag \hat{U}_{\delta \alpha}^\dag +h.c.)||\nonumber \\
    &= \frac{L^d}{a^{12-3d}g^6}\frac{8589934592}{5}(2d^5 - 15d^4 + 40d^3 - 45 d^2 + 18d).
\end{align}
Finally, summing up the bounds for all three types of commutators, we obtain the bound for $C_{2,12}$,
\begin{equation}
 ( 17179869184 \frac{d^5}{5} - 17179869184 \frac{d^4}{3} + 49392123904 \frac{d^3}{3} - 322659435248 \frac{d^2}{3} +1207422766768\frac{d}{15}-8589934592 ) \frac{L^d}{a^{12-3d}g^6}.
\end{equation}

\setlength{\LTcapwidth}{\linewidth}
\begin{longtable}{|l|l|l|l|l|l|l|l|l|}
\caption{The number of tuples, which label $\hat{h}_{L'}$ and $\hat{h}_{L''}$, such that (\ref{eq:SU_BBB1}) holds and $\hat{h}_{L}$ is labelled by $(even,k,l)$ with $k<l$. The tuples that label $\hat{h}_{L'}$ are given in the numerically labelled rows. The alphabetical rows, which immediately follow each numerically labelled row, but precede the next one, list the tuples that label $\hat{h}_{L''}$. The number of tuples that label $\hat{h}_{L'}$ for each $\hat{h}_{L}$, satisfying $L'>L$, and that label $\hat{h}_{L''}$ for each pair of $\hat{h}_{L}$ and $\hat{h}_{L'}$, satisfying $L''>L$, are given in the fifth column. The sixth column indicates the number of common dimensions $\hat{h}_{L'}$ or $\hat{h}_{L''}$ at each row share with $\hat{h}_{L}$. The seventh column denotes whether $\hat{h}_{L'}$ or $\hat{h}_{L''}$ at each row acts on plaquettes of the same parity as $\hat{h}_{L}$ do. The eighth column indicates the number of common dimensions $\hat{h}_{L''}$ at each row shares with $\hat{h}_{L'}$. The seventh column denotes whether $\hat{h}_{L''}$ at each row acts on plaquettes of the same parity as $\hat{h}_{L'}$ does. Combining the information in the sixth to ninth columns, one can straightforwardly compute, for each combination of tuples that label $\hat{h}_{L}$, $\hat{h}_{L'}$ and $\hat{h}_{L''}$, the number of links $\hat{h}_{L}$, $\hat{h}_{L'}$ and $\hat{h}_{L''}$ collide on.}\\ \hline
\label{tb:SU2_bbb_1}
$\hat{h}_{L}$ & \multicolumn{3}{l|}{$(even,k,l), k<l$}                          & $\#$ tuples & $\#$ common dimensions & same parity & \multicolumn{2}{l|}{}                \\ \hline
\endfirsthead

\multicolumn{9}{c}%
{{\bfseries \tablename\ \thetable{} -- continued from previous page}} \\
\hline
$\hat{h}_{L}$ & \multicolumn{3}{l|}{$(even,k,l), k<l$}                          & $\#$ tuples & $\#$ common dimensions & same parity & \multicolumn{2}{l|}{}                \\ \hline
\endhead

\hline \multicolumn{9}{|r|}{{Continued on next page}} \\ \hline
\endfoot

\hline \hline
\endlastfoot

$1$           & $\hat{h}_{L'}$ & \multicolumn{2}{l|}{$(even,k,j), j>l$}         & $d-l$       & $1$                    & Y           & $\#$ common dimensions & same parity \\ \hline
              & a              & $\hat{h}_{L''}$ & $(even,k,i), i=j$            & $1$         & $1$                    & Y           & $2$                    & Y           \\ \hline
              & b              &                 & $(even,k,i), i>l, i\neq j$   & $d-l-1$     & $1$                    & Y           & $1$                    & Y           \\ \hline
              & c              &                 & $(even,l,i), i=j$            & $1$         & $1$                    & Y           & $1$                    & Y           \\ \hline
              & d              &                 & $(odd,k,i), i=j$             & $1$         & $1$                    & N           & $2$                    & N           \\ \hline
              & e              &                 & $(odd,k,i), i=l$             & $1$         & $2$                    & N           & $1$                    & N           \\ \hline
              & f              &                 & $(odd,k,i), i>l, i\neq j$    & $d-l-1$     & $1$                    & N           & $1$                    & N           \\ \hline
              & g              &                 & $(odd,l,i), i=j$             & $1$         & $1$                    & N           & $1$                    & N           \\ \hline
$2$           & $\hat{h}_{L'}$ & \multicolumn{2}{l|}{$(even,l,j), j>l$}         & $d-l$       & $1$                    & Y           & \multicolumn{2}{l|}{}                \\ \hline
              & a              & $\hat{h}_{L''}$ & $(even,k,i), i=j$            & $1$         & $1$                    & Y           & $1$                    & Y           \\ \hline
              & b              &                 & $(even,i,l), k<i<l$          & $l-k-1$     & $1$                    & Y           & $1$                    & Y           \\ \hline
              & c              &                 & $(even,l,i), i=j$            & $1$         & $1$                    & Y           & $2$                    & Y           \\ \hline
              & d              &                 & $(even,l,i), i>l, i\neq j$   & $d-l-1$     & $1$                    & Y           & $1$                    & Y           \\ \hline
              & e              &                 & $(odd,k,i), i=l$             & $1$         & $2$                    & N           & $1$                    & N           \\ \hline
              & f              &                 & $(odd,k,i), i=j$             & $1$         & $1$                    & N           & $1$                    & N           \\ \hline
              & g              &                 & $(odd,l,i), i=j$             & $1$         & $1$                    & N           & $2$                    & N           \\ \hline
              & h              &                 & $(odd,l,i), i>l, i\neq j$    & $d-l-1$     & $1$                    & N           & $1$                    & N           \\ \hline
              & i              &                 & $(odd,i,l), k<i<l$           & $l-k-1$     & $1$                    & N           & $1$                    & N           \\ \hline
$3$           & $\hat{h}_{L'}$ & \multicolumn{2}{l|}{$(even,j,l), k<j<l$}       & $l-k-1$     & $1$                    & Y           & \multicolumn{2}{l|}{}                \\ \hline
              & a              & $\hat{h}_{L''}$ & $(even,l,i), i>l$            & $d-l$       & $1$                    & Y           & $1$                    & Y           \\ \hline
              & b              &                 & $(even,i,l), i=j$            & $1$         & $1$                    & Y           & $2$                    & Y           \\ \hline
              & c              &                 & $(even,i,l), k<i<l, i\neq j$ & $l-k-2$     & $1$                    & Y           & $1$                    & Y           \\ \hline
              & d              &                 & $(odd,l,i), i>l$             & $d-l$       & $1$                    & Y           & $1$                    & Y           \\ \hline
              & e              &                 & $(odd,k,i), i=l$             & $1$         & $2$                    & N           & $1$                    & N           \\ \hline
              & f              &                 & $(odd,i,l), i=j$             & $1$         & $1$                    & N           & $2$                    & N           \\ \hline
              & g              &                 & $(odd,i,l), k<i<l, i\neq j$  & $l-k-2$     & $1$                    & N           & $1$                    & N           \\ \hline
$4$           & $\hat{h}_{L'}$ & \multicolumn{2}{l|}{$(odd,k,j), j>l$}          & $d-l$       & $1$                    & N           & \multicolumn{2}{l|}{}                \\ \hline
              & a              & $\hat{h}_{L''}$ & $(even,k,i), i=j$            & $1$         & $1$                    & Y           & $2$                    & N           \\ \hline
              & b              &                 & $(even,k,i), i>l, i\neq j$   & $d-l-1$     & $1$                    & Y           & $1$                    & N           \\ \hline
              & c              &                 & $(even,l,i), i=j$            & $1$         & $1$                    & Y           & $1$                    & N           \\ \hline
              & d              &                 & $(odd,k,i), i=j$             & $1$         & $1$                    & N           & $2$                    & Y           \\ \hline
              & e              &                 & $(odd,k,i), i=l$             & $1$         & $2$                    & N           & $1$                    & Y           \\ \hline
              & f              &                 & $(odd,k,i), i>l, i\neq j$    & $d-l-1$     & $1$                    & N           & $1$                    & Y           \\ \hline
              & g              &                 & $(odd,l,i),i=j$              & $1$         & $1$                    & N           & $1$                    & Y           \\ \hline
$5$           & $\hat{h}_{L'}$ & \multicolumn{2}{l|}{$(odd,k,j), j=l$}          & $1$         & $2$                    & N           & \multicolumn{2}{l|}{}                \\ \hline
              & a              & $\hat{h}_{L''}$ & $(even,k,i), i>l$            & $d-l$       & $1$                    & Y           & $1$                    & N           \\ \hline
              & b              &                 & $(even,l,i),  i>l$           & $d-l$       & $1$                    & Y           & $1$                    & N           \\ \hline
              & c              &                 & $(even,i,l), k<i<l$          & $l-k-1$     & $1$                    & Y           & $1$                    & N           \\ \hline
              & d              &                 & $(odd,k,i), i=l$             & $1$         & $2$                    & N           & $2$                    & Y           \\ \hline
              & e              &                 & $(odd,k,i), i>l$             & $d-l$       & $1$                    & N           & $1$                    & Y           \\ \hline
              & f              &                 & $(odd,l,i), i>l$             & $d-l$       & $1$                    & N           & $1$                    & Y           \\ \hline
              & g              &                 & $(odd,i,l), k<i<l$           & $l-k-1$     & $1$                    & N           & $1$                    & Y           \\ \hline
$6$           & $\hat{h}_{L'}$ & \multicolumn{2}{l|}{$(odd,l,j), j>l$}          & $d-l$       & $1$                    & N           & \multicolumn{2}{l|}{}                \\ \hline
              & a              & $\hat{h}_{L''}$ & $(even,k,i), i=j$            & $1$         & $1$                    & Y           & $1$                    & N           \\ \hline
              & b              &                 & $(even,i,l), k<i<l$          & $l-k-1$     & $1$                    & Y           & $1$                    & N           \\ \hline
              & c              &                 & $(even,l,i), i=j$            & $1$         & $1$                    & Y           & $2$                    & N           \\ \hline
              & d              &                 & $(even,l,i), i>l, i\neq j$   & $d-l-1$     & $1$                    & Y           & $1$                    & N           \\ \hline
              & e              &                 & $(odd,k,i), i=l$             & $1$         & $2$                    & N           & $1$                    & Y           \\ \hline
              & f              &                 & $(odd,k,i), i=j$             & $1$         & $1$                    & N           & $1$                    & Y           \\ \hline
              & g              &                 & $(odd,l,i), i=j$             & $1$         & $1$                    & N           & $2$                    & Y           \\ \hline
              & h              &                 & $(odd,l,i), i>l, i\neq j$    & $d-l-1$     & $1$                    & N           & $1$                    & Y           \\ \hline
              & i              &                 & $(odd,i,l), k<i<l$           & $l-k-1$     & $1$                    & N           & $1$                    & Y           \\ \hline
$7$           & $\hat{h}_{L'}$ & \multicolumn{2}{l|}{$(odd,j,l), k<j<l$}        & $l-k-1$     & $1$                    & N           & \multicolumn{2}{l|}{}                \\ \hline
              & a              & $\hat{h}_{L''}$ & $(even,l,i), i>l$            & $d-l$       & $1$                    & Y           & $1$                    & N           \\ \hline
              & b              &                 & $(even,i,l), i=j$            & $1$         & $1$                    & Y           & $2$                    & N           \\ \hline
              & c              &                 & $(even,i,l), k<i<l,i\neq j$  & $l-k-2$     & $1$                    & Y           & $1$                    & N           \\ \hline
              & d              &                 & $(odd,l,i), i>l$             & $d-l$       & $1$                    & N           & $1$                    & Y           \\ \hline
              & e              &                 & $(odd,k,i), i=l$             & $1$         & $2$                    & N           & $1$                    & Y           \\ \hline
              & f              &                 & $(odd,i,l), i=j$             & $1$         & $1$                    & N           & $2$                    & Y           \\ \hline
              & g              &                 & $(odd,i,l), k<i<l, i\neq j$  & $l-k-2$     & $1$                    & N           & $1$                    & Y           \\ \hline
\end{longtable}

\begin{longtable}{|l|l|l|l|l|l|l|l|l|}
\caption{The number of tuples, which label $\hat{h}_{L'}$ and $\hat{h}_{L''}$, such that (\ref{eq:SU_BBB1}) holds and $\hat{h}_{L}$ is labelled by $(odd,k,l)$ with $k<l$.}\\ \hline
\label{tb:SU2_bbb_2}

$\hat{h}_{L}$ & \multicolumn{3}{l|}{$(odd,k,l), k<l$}                          & $\#$ tuples & $\#$ common dimensions & same parity & \multicolumn{2}{l|}{}                \\ \hline
\endfirsthead

\multicolumn{9}{c}%
{{\bfseries \tablename\ \thetable{} -- continued from previous page}} \\
\hline
$\hat{h}_{L}$ & \multicolumn{3}{l|}{$(odd,k,l), k<l$}                          & $\#$ tuples & $\#$ common dimensions & same parity & \multicolumn{2}{l|}{}                \\ \hline
\endhead

\hline \multicolumn{9}{|r|}{{Continued on next page}} \\ \hline
\endfoot

\hline \hline
\endlastfoot

$1$           & $\hat{h}_{L'}$ & \multicolumn{2}{l|}{$(even,k,j), j>l$}         & $d-l$       & $1$                    & N           & $\#$ common dimensions & same parity \\ \hline
              & a              & $\hat{h}_{L''}$ & $(even,k,i), i=j$            & $1$         & $1$                    & N           & $2$                    & Y           \\ \hline
              & b              &                 & $(even,k,i), i>l, i\neq j$   & $d-l-1$     & $1$                    & N           & $1$                    & Y           \\ \hline
              & c              &                 & $(even,l,i), i=j$            & $1$         & $1$                    & N           & $1$                    & Y           \\ \hline
              & d              &                 & $(odd,k,i), i=j$             & $1$         & $1$                    & Y           & $2$                    & N           \\ \hline
              & e              &                 & $(odd,k,i), i>l, i\neq j$    & $d-l-1$     & $1$                    & Y           & $1$                    & N           \\ \hline
              & f              &                 & $(odd,l,i), i=j$             & $1$         & $1$                    & Y           & $1$                    & N           \\ \hline
$2$           & $\hat{h}_{L'}$ & \multicolumn{2}{l|}{$(even,l,j), j>l$}         & $d-l$       & $1$                    & N           & \multicolumn{2}{l|}{}                \\ \hline
              & a              & $\hat{h}_{L''}$ & $(even,k,i), i=j$            & $1$         & $1$                    & N           & $1$                    & Y           \\ \hline
              & b              &                 & $(even,i,l), k<i<l$          & $l-k-1$     & $1$                    & N           & $1$                    & Y           \\ \hline
              & c              &                 & $(even,l,i), i=j$            & $1$         & $1$                    & N           & $2$                    & Y           \\ \hline
              & d              &                 & $(even,l,i), i>l, i\neq j$   & $d-l-1$     & $1$                    & N           & $1$                    & Y           \\ \hline
              & e              &                 & $(odd,k,i), i=j$             & $1$         & $1$                    & Y           & $1$                    & N           \\ \hline
              & f              &                 & $(odd,l,i), i=j$             & $1$         & $1$                    & Y           & $2$                    & N           \\ \hline
              & g              &                 & $(odd,l,i), i>l, i\neq j$    & $d-l-1$     & $1$                    & Y           & $1$                    & N           \\ \hline
              & h              &                 & $(odd,i,l), k<i<l$           & $l-k-1$     & $1$                    & Y           & $1$                    & N           \\ \hline
$3$           & $\hat{h}_{L'}$ & \multicolumn{2}{l|}{$(even,j,l), k<j<l$}       & $l-k-1$     & $1$                    & N           & \multicolumn{2}{l|}{}                \\ \hline
              & a              & $\hat{h}_{L''}$ & $(even,l,i), i>l$            & $d-l$       & $1$                    & N           & $1$                    & Y           \\ \hline
              & b              &                 & $(even,i,l), i=j$            & $1$         & $1$                    & N           & $2$                    & Y           \\ \hline
              & c              &                 & $(even,i,l), k<i<l, i\neq j$ & $l-k-2$     & $1$                    & N           & $1$                    & Y           \\ \hline
              & d              &                 & $(odd,l,i), i>l$             & $d-l$       & $1$                    & Y           & $1$                    & N           \\ \hline
              & e              &                 & $(odd,i,l), i=j$             & $1$         & $1$                    & Y           & $2$                    & N           \\ \hline
              & f              &                 & $(odd,i,l), k<i<l, i\neq j$  & $l-k-2$     & $1$                    & Y           & $1$                    & N           \\ \hline
$4$           & $\hat{h}_{L'}$ & \multicolumn{2}{l|}{$(odd,k,j), j>l$}          & $d-l$       & $1$                    & Y           & \multicolumn{2}{l|}{}                \\ \hline
              & a              & $\hat{h}_{L''}$ & $(even,k,i), i=j$            & $1$         & $1$                    & N           & $2$                    & N           \\ \hline
              & b              &                 & $(even,k,i), i>l, i\neq j$   & $d-l-1$     & $1$                    & N           & $1$                    & N           \\ \hline
              & c              &                 & $(even,l,i), i=j$            & $1$         & $1$                    & N           & $1$                    & N           \\ \hline
              & d              &                 & $(odd,k,i), i=j$             & $1$         & $1$                    & Y           & $2$                    & Y           \\ \hline
              & e              &                 & $(odd,k,i), i>l, i\neq j$    & $d-l-1$     & $1$                    & Y           & $1$                    & Y           \\ \hline
              & f              &                 & $(odd,l,i),i=j$              & $1$         & $1$                    & Y           & $1$                    & Y           \\ \hline
$5$           & $\hat{h}_{L'}$ & \multicolumn{2}{l|}{$(odd,l,j), j>l$}          & $d-l$       & $1$                    & Y           & \multicolumn{2}{l|}{}                \\ \hline
              & a              & $\hat{h}_{L''}$ & $(even,k,i), i=j$            & $1$         & $1$                    & N           & $1$                    & N           \\ \hline
              & b              &                 & $(even,i,l), k<i<l$          & $l-k-1$     & $1$                    & N           & $1$                    & N           \\ \hline
              & c              &                 & $(even,l,i), i=j$            & $1$         & $1$                    & N           & $2$                    & N           \\ \hline
              & d              &                 & $(even,l,i), i>l, i\neq j$   & $d-l-1$     & $1$                    & N           & $1$                    & N           \\ \hline
              & e              &                 & $(odd,k,i), i=j$             & $1$         & $1$                    & Y           & $1$                    & Y           \\ \hline
              & f              &                 & $(odd,l,i), i=j$             & $1$         & $1$                    & Y           & $2$                    & Y           \\ \hline
              & g              &                 & $(odd,l,i), i>l, i\neq j$    & $d-l-1$     & $1$                    & Y           & $1$                    & Y           \\ \hline
              & h              &                 & $(odd,i,l), k<i<l$           & $l-k-1$     & $1$                    & Y           & $1$                    & Y           \\ \hline
$6$           & $\hat{h}_{L'}$ & \multicolumn{2}{l|}{$(odd,j,l), k<j<l$}        & $l-k-1$     & $1$                    & Y           & \multicolumn{2}{l|}{}                \\ \hline
              & a              & $\hat{h}_{L''}$ & $(even,l,i), i>l$            & $d-l$       & $1$                    & N           & $1$                    & N           \\ \hline
              & b              &                 & $(even,i,l), i=j$            & $1$         & $1$                    & N           & $2$                    & N           \\ \hline
              & c              &                 & $(even,i,l), k<i<l,i\neq j$  & $l-k-2$     & $1$                    & N           & $1$                    & N           \\ \hline
              & d              &                 & $(odd,l,i), i>l$             & $d-l$       & $1$                    & Y           & $1$                    & Y           \\ \hline
              & e              &                 & $(odd,i,l), i=j$             & $1$         & $1$                    & Y           & $2$                    & Y           \\ \hline
              & f              &                 & $(odd,i,l), k<i<l, i\neq j$  & $l-k-2$     & $1$                    & Y           & $1$                    & Y           \\ \hline
\end{longtable}

\begin{table*}[ht]
\centering
\begin{tabular}{|l|l|l|l|l|l|l|}
\hline
$\hat{h}_{L}$ & \multicolumn{3}{l|}{$(p,k,l), k<l$}                         & $\#$ tuples \\ \hline
$1$           & $\hat{h}_{L'}$ & \multicolumn{2}{l|}{$(p',k,j), j>l$}       & $d-l$       \\ \hline
              & a              & $\hat{h}_{L''}$ & $(p,i,l), k<i<l$         & $l-k-1$     \\ \hline
              & b              &                 & $(p,l,i), i>l, i\neq j$  & $d-l-1$     \\ \hline
              & c              &                 & $(p',i,l), k<i<l$        & $l-k-1$     \\ \hline
              & d              &                 & $(p',l,i), i>l, i\neq j$ & $d-l-1$     \\ \hline
$2$           & $\hat{h}_{L'}$ & \multicolumn{2}{l|}{$(p',l,j), j>l$}       & $d-l$       \\ \hline
              & a              & $\hat{h}_{L''}$ & $(p',k,i), i>l, i\neq j$ & $d-l-1$     \\ \hline
              & b              &                 & $(p,k,i), i>l, i\neq j$  & $d-l-1$     \\ \hline
$3$           & $\hat{h}_{L'}$ & \multicolumn{2}{l|}{$(p',j,l), k<j<l$}     & $l-k-1$     \\ \hline
              & a              & $\hat{h}_{L''}$ & $(p',k,i), i>l$          & $d-l$       \\ \hline
              & b              &                 & $(p,k,i), i>l$           & $d-l$       \\ \hline
$4$           & $\hat{h}_{L'}$ & \multicolumn{2}{l|}{$(p,k,j), j>l$}        & $d-l$       \\ \hline
              & a              & $\hat{h}_{L''}$ & $(p,i,l), k<i<l$         & $l-k-1$     \\ \hline
              & b              &                 & $(p,l,i), i>l, i\neq j$  & $d-l-1$     \\ \hline
              & c              &                 & $(p',i,l), k<i<l$        & $l-k-1$     \\ \hline
              & d              &                 & $(p',l,i), i>l, i\neq j$ & $d-l-1$     \\ \hline
$5$           & $\hat{h}_{L'}$ & $(p,l,j), j>l$  &                          & $d-l$       \\ \hline
              & a              & $\hat{h}_{L''}$ & $(p',k,i), i>l, i\neq j$ & $d-l-1$     \\ \hline
              & b              &                 & $(p,k,i), i>l, i\neq j$  & $d-l-1$     \\ \hline
$6$           & $\hat{h}_{L'}$ & \multicolumn{2}{l|}{$(p,j,l), k<j<l$}      & $l-k-1$     \\ \hline
              & a              & $\hat{h}_{L''}$ & $(p',k,i), i>l$          & $d-l$       \\ \hline
              & b              &                 & $(p,k,i), i>l$           & $d-l$       \\ \hline
\end{tabular}
\caption{The number of tuples, which label $\hat{h}_{L'}$ and $\hat{h}_{L''}$, such that (\ref{eq:SU_BBB2}) holds and $\hat{h}_{L}$ is labelled by $(p,k,l)$ with $k<l$. Here, $p \neq p'$.}
\label{tb:SU2_bbb_3}
\end{table*}

\begin{table*}[ht]
\centering
\begin{tabular}{|l|l|l|l|l|}
\hline
$\hat{h}_{L}$ & \multicolumn{3}{l|}{$(p,k,l), k<l$}                    & $\#$ tuples \\ \hline
$1$           & $\hat{h}_{L'}$ & \multicolumn{2}{l|}{$(p',k,j), j>l$}  & $d-l$       \\ \hline
              & a              & $\hat{h}_{L''}$   & $(p,j,i), j<i$    & $d-j$       \\ \hline
              & b              &                   & $(p',j,i),j<i$    & $d-j$       \\ \hline
              & c              &                   & $(p,i,j), k<i<l$  & $l-k-1$     \\ \hline
              & d              &                   & $(p',i,j), k<i<l$ & $l-k-1$     \\ \hline
$2$           & $\hat{h}_{L'}$ & \multicolumn{2}{l|}{$(p',l,j), j>l$}  & $d-l$       \\ \hline
              & a              & $\hat{h}_{L''}$   & $(p,j,i), j<i$    & $d-j$       \\ \hline
              & b              &                   & $(p',j,i),j<i$    & $d-j$       \\ \hline
              & c              &                   & $(p,i,j), k<i<l$  & $l-k-1$     \\ \hline
              & d              &                   & $(p',i,j), k<i<l$ & $l-k-1$     \\ \hline
$3$           & $\hat{h}_{L'}$ & $(p',j,l), k<j<l$ &                   & $l-k-1$     \\ \hline
              & a              & $\hat{h}_{L''}$   & $(p,j,i), j<i<l$  & $l-j-1$     \\ \hline
              & b              &                   & $(p',j,i),j<i<l$  & $l-j-1$     \\ \hline
              & c              &                   & $(p,i,j), k<i<j$  & $j-k-1$     \\ \hline
              & d              &                   & $(p',i,j), k<i<j$ & $j-k-1$     \\ \hline
$4$           & $\hat{h}_{L'}$ & $(p,k,j), j>l$    &                   & $d-l$       \\ \hline
              & a              & $\hat{h}_{L''}$   & $(p,j,i), j<i$    & $d-j$       \\ \hline
              & b              &                   & $(p',j,i),j<i$    & $d-j$       \\ \hline
              & c              &                   & $(p,i,j), k<i<l$  & $l-k-1$     \\ \hline
              & d              &                   & $(p',i,j), k<i<l$ & $l-k-1$     \\ \hline
$5$           & $\hat{h}_{L'}$ & \multicolumn{2}{l|}{$(p,l,j), j>l$}   & $d-l$       \\ \hline
              & a              & $\hat{h}_{L''}$   & $(p,j,i), j<i$    & $d-j$       \\ \hline
              & b              &                   & $(p',j,i),j<i$    & $d-j$       \\ \hline
              & c              &                   & $(p,i,j), k<i<l$  & $l-k-1$     \\ \hline
              & d              &                   & $(p',i,j), k<i<l$ & $l-k-1$     \\ \hline
$6$           & $\hat{h}_{L'}$ & $(p,j,l), k<j<l$  &                   & $l-k-1$     \\ \hline
              & a              & $\hat{h}_{L''}$   & $(p,j,i), j<i<l$  & $l-j-1$     \\ \hline
              & b              &                   & $(p',j,i),j<i<l$  & $l-j-1$     \\ \hline
              & c              &                   & $(p,i,j), k<i<j$  & $j-k-1$     \\ \hline
              & d              &                   & $(p',i,j), k<i<j$ & $j-k-1$     \\ \hline
\end{tabular}
\caption{The number of tuples, which label $\hat{h}_{L'}$ and $\hat{h}_{L''}$, such that (\ref{eq:SU_BBB3}) holds and $\hat{h}_{L}$ is labelled by $(p,k,l)$ with $k<l$. Here, $p \neq p'$.}
\label{tb:SU2_bbb_4}
\end{table*}

\subsubsection{Oracle errors}
\label{subsubsec:oracle_SU2}

Here, we describe the direct syntheses of the kinetic and magnetic oracles, and compute the errors incurred by the fixed point arithmetic circuits.

{\bf Syntheses of the kinetic oracles:}
The kinetic oracle, defined in (\ref{eq:SU2_kin_oracle}), can be directly synthesized as two controlled-diagonal gates, which impart the phases $(-1)^{j_0'}f_{\alpha \beta}(j, \Delta j, m^L, m^R)\frac{t}{2a}$, where $f_{\alpha \beta}(\cdot)$ is defined in (\ref{eq:SU2_fab_def}), for $j_0'=0,1$, if the control bits $m^{L\prime}_0, m^{R\prime}_0,f_{\alpha,r}^\prime$, and $f_{\beta,r+1}^\prime$ are all ones. Without loss of generality, we consider the case where $j_0'=0$. The implementation of the diagonal phase gate is similar to that of the electric term. Before providing the details, we provide a high-level description of the implementation. We begin by computing $f_{\alpha \beta}(j, \Delta j, m^L, m^R)$ into an ancilla register, conditioned upon the values of the control bits. Then, by applying $R_z$ gates to the ancilla state $\ket{f_{\alpha \beta}(j, \Delta j, m^L, m^R)}$, we induce the correct phase. Finally, we uncompute $\ket{f_{\alpha \beta}(j, \Delta j, m^L, m^R)}$. The computation of $f_{\alpha \beta}(j, \Delta j, m^L, m^R)$, each of which is the square root of a fraction as defined in (\ref{eq:SU2_fab_def}), can be broken down into five steps. In the first and second steps, we compute the numerator and denominator, respectively. In the third step, we approximate the inverse of the denominator, using the circuits in \cite{bhaskar2016quantum}. In the fourth step, we approximate the argument of the square-root by multiplying together the numerator and the inverse of the denominator. Lastly, we approximate the square-root, using the circuits in \cite{bhaskar2016quantum}. Only the third and last steps incur approximation errors. Hereafter, we use the logarithmic depth out-of-place adder developed in \cite{draper2006logarithmic}, unless one of the inputs is classically known in which case we use the adder proposed in \cite{gidney2018halving}, and the multiplier proposed in \cite{shaw2020quantum}.

We consider the computation of the numerator. First, we perform two additions or subtractions between $j$ and $m^L$, and $j$ and $m^R$, which requires two $(\eta+1)$-bit adders. They cost $40(\eta+1)-24\lfloor \log(\eta+1)\rfloor - 8$ T gates. Next, we divide the outputs by $2$, which is accomplished by shifting the decimal point. Then, for each output, we add or subtract an $\eta$-bit classically known number, costing $4(\eta-1)$ T gates. At this point, we have two numbers with at most $(\eta+3)$ bits, which we need to multiply together. We need to guarantee that both numbers are positive, which implies the realness of the Clebsch-Gordan coefficients, and that the quantum numbers are within the allowed ranges given in (\ref{eq:SU2_eig1}) and (\ref{eq:SU2_eig2}), i.e., $j\leq \Lambda,-j\leq m^L \leq j,-j \leq m^R \leq j$. This can be accomplished by applying a Toffoli to an ancilla bit, controlled by the bits representing the signs of the two numbers, and then, apply the multiplication, conditioned upon the value of the ancilla bit. The singly-controlled multiplication operation is implemented in two stages. First, we copy one of the numbers into an ancilla register, controlled by the ancilla bit, which costs $(\eta+3)$ Toffoli gates and ancilla qubits. The Toffoli gates can be synthesized with $4\eta+16$ T gates \cite{jones2013low}. Second, we apply a multiplier to the register that holds the controlled-copy, and the uncopied number. This requires one $(\eta+3)$-bit multiplier, which costs $4(\eta+2)(12\eta -3\lfloor \log(\eta+3)\rfloor+23)+4$ T gates. The number of ancilla qubits required to store the outputs is $5\eta + 15$, and that required for the workspace is $3(\eta+3)-\lfloor \log(\eta+3)\rfloor-1$. For the computation of the denominator, we have to add a classical number to $j$ with at most $2$ bits twice, costing $8(\eta-2)$ in total and resulting in two $(\eta+1)$-bit numbers. Then, we multiply them together, which costs $4\eta(12\eta - 3\lfloor \log(\eta+1)\rfloor)+4$ T gates. The number of ancilla qubits required to store the outputs is $4\eta + 4$, and that required for the workspace is $3(\eta+1)-\lfloor \log(\eta+1)\rfloor-1$.

Next, we consider the third step, where we compute the inverse of the denominator using the algorithm in \cite{bhaskar2016quantum}. Briefly, the algorithm takes a fixed precision $n$-bit binary number $w \geq 1$, with the first $m$ bits representing the integer part, and approximates $\frac{1}{w}$ by applying Newton's root-finding algorithm to $f(x)=\frac{1}{w}-x$. This produces a sequence of estimates, according to the recurrence equation
\begin{equation}
    x_i = -w \tilde{x}_{i-1}^2 + 2\tilde{x}_{i-1}, \: i=1,2,...,s,
\end{equation}
where $\tilde{x}_{i-1}$ is obtained by truncating $x_{i-1}$ to $b \geq n$ bits after the decimal point. If the initial estimate $\tilde{x}_0 = 2^{-p}$ with $p\in \mathbbm{N}$ and $2^p>w\geq 2^{p-1}$, and $s=\lceil \log(b) \rceil$, then the approximation error is
\begin{equation}
    |\tilde{x}_s - \frac{1}{w} | \leq \frac{2+ \log(b)}{2^b}.
\end{equation}
Since the input $w$, in this case, is the denominator, at most an $(2\eta+2)-$bit integer, the algorithmic parameters $n=m=2\eta+2$. Thus, the preparation of the initial estimate requires $2\eta+1$ Toffoli gates, which cost $4$ T gates each \cite{jones2013low}, and one triply-controlled Toffoli gate, which costs $15$ T gates \cite{maslov2016advantages}. The evaluation of each iteration can be split into four steps: (i) squaring $\tilde{x}_{i-1}$; (ii) multiplying $w$ and $\tilde{x}_{i-1}^2$; (iii) appending a $0$ to $\tilde{x}_{i-1}$ to obtain $2\tilde{x}_{i-1}$; and (iv) adding $2\tilde{x}_{i-1}$ to the negated $w\tilde{x}_{i-1}^2$. Using the fact that each estimate $\tilde{x}_i \leq \frac{1}{w} \leq 1$ is at most a $b-$bit number, we obtain the T-gate count for inverting the denominator, i.e., $\lceil \log(b)\rceil \cdot [48b^2+96 b(2\eta+2) -12 (b-1) \lfloor \log(b)\rfloor-12(2\eta+1)\lfloor \log(b)\rfloor -152b - 32 (2\eta+2)-12\lfloor \log(2b+2\eta+2)\rfloor+104]+8\eta+19$. Furthermore, the number of storage and workspace ancilla qubits required are $\lceil\log(b)\rceil\cdot(4\eta+7b+6)$ and $6b-\lfloor\log(6b)\rfloor-1$, respectively.

In the fourth step, we multiply the numerator and the inverse of the denominator, which are at most $(2\eta+6)-$bit and $b-$bit numbers, respectively, to obtain the fraction. Assuming $b\geq 2\eta+6$, this costs $4b+4(2\eta+5)(12b-3\lfloor \log(b)\rfloor-13)$ T gates, $b+2\eta+6$ storage ancilla qubits, and $3b-\lfloor \log(b)\rfloor-1$ workspace ancilla qubits. In the fifth step, we compute the square-root of the fraction using the algorithm in \cite{bhaskar2016quantum}. Briefly, given an $n-$bit input $w\geq 1$ with an $m$-bit integer part, the algorithm first calls the inverse algorithm to obtain $\tilde{x}_s \approx \frac{1}{w}$. Then, it applies Newton's root-finding algorithm to $f(y) = \frac{1}{y^2}-\frac{1}{w}$ to approximate $\sqrt{w}$ via the recursive relation
\begin{equation}
    y_j = \frac{1}{2}(3\tilde{y}_{j-1}-\tilde{x}_s \tilde{y}_{j-1}^3), \: j = 1,2,...,s,
\end{equation}
where $\tilde{y}_{j-1}$ is obtained by truncating ${y}_{j-1}$ to $c\geq \max\{2m,4\}$ bits after the decimal point. If $\tilde{y}_0=2^{\lfloor (q-1)/2 \rfloor}$ with $q\in \mathbbm{N}$ and $2^{1-q}> \tilde{x}_s \geq 2^{-q}$, and $s=\lceil \log(c)\rceil$, then the approximation error is \begin{equation}
    |\tilde{y}_s - \sqrt{w}| \leq \left(\frac{3}{4}\right)^{c-2m}(2+c+\log(c)).
\end{equation}
The output of the fourth step is at most a $(b+4\eta+6)-$bit fraction that is bounded above and below by $1$ and $0$, respectively. Since the square-root algorithm requires an input $w\geq 1$, we shift the decimal point of the fraction $(b+4\eta+6)$ times to obtain a $(b+4\eta+6)-$bit integer $w \geq 1$, and shift it back $(b+4\eta+6)/2$ times once we obtain the root. We assume without loss of generality that we have chosen an even $b$ in the third step. We divide the fifth step into three stages: calling the inverse algorithm, preparing $\ket{\tilde{y}_0}$, and Newton's iteration for $\ket{\tilde{y}_s}$. The costs of the inverse algorithm has been discussed above. The second stage requires $\lceil \frac{3c}{2}\rceil$ storage ancilla qubits, $\lceil \frac{c}{2}\rceil$ Toffoli gates, $\lceil \frac{c}{2}\rceil-1$ triply-controlled Toffoli gates, and one quadruply-controlled Toffoli gates. We further divide the third stage into four steps: (i) multiplying $\tilde{x}_s$ and $\tilde{y}_0$; (ii) multiplying $\tilde{x}_s \tilde{y}_0$ and $\tilde{y}_0$; (iii) add $3$ to the negated $\tilde{x}_s \tilde{y}_0^2$; and (iv) multiplying $3-\tilde{x}_s \tilde{y}_0^3$ by $\tilde{y}_0$ and then, $\frac{1}{2}$. We use the fact that each estimate $\tilde{y}_i$ is at most a $\lceil \frac{3c}{2} \rceil-$bit number to obtain the total T-gate count, i.e.,
$\lceil\log(c)\rceil\cdot[528c^2+96bc+384c\eta-c(12\lfloor \log(c)\rfloor+12\lfloor \log(\lceil \frac{3c}{2}\rceil)\rfloor+12\lfloor \log(\lceil \frac{5c}{2}\rceil)\rfloor+18\lfloor \log(1+4c)\rfloor)-12b\lfloor \log(2c)\rfloor-48\eta\lfloor \log(2c)\rfloor+12(\lfloor \log(c)\rfloor+\lfloor \log(\lceil \frac{3c}{2}\rceil)\rfloor)-60\lfloor \log(2c)\rfloor+12\lfloor \log(\lceil \frac{5c}{2}\rceil)\rfloor+12\lfloor \log(1+4c)\rfloor-12\lfloor \log(2c+b+4\eta+6)\rfloor+50c-32b-128\eta-28]+\lceil \frac{19c}{2}\rceil+4b+16\eta+43$. The number of storage and workspace ancilla qubits required are $\lceil \log(c)\rceil \cdot (\lceil \frac{51c}{2}\rceil+8\eta+2b+17)$ and $12c-\lfloor \log(4c+2)\rfloor + 5$, respectively.

We impart the phase by applying $R_z(2^{k-c} \theta)$, where $c$ is the number of digits after the decimal point in the output of the square-root functions, and $\theta = \frac{t}{2a}$, to the $k$th qubit of the ancilla state $\ket{f_{\alpha \beta}(j, \Delta j, m^L, m^R)}$. In order to implement the controlled version of this phase gate, we control each $R_z$ gate by the four control bits and the ancilla bit that checks the realness of the Clebsch-Gordan coefficients. Each quintuply-controlled $R_z$ gate requires two quintuply-controlled Toffoli gates, which cost 31 T gates each \cite{maslov2016advantages}, two $R_z$ gates and one ancilla qubit~\cite{wang2020resource}. Since the state has at most $\lceil \frac{3c}{2}\rceil$ qubits and there are $dL^d$ links on the lattice, there are $\lceil \frac{3c}{2}\rceil dL^d$ multi-controlled $R_z$ gates to be applied.

We sum up the T-gate requirements for all the steps, and multiply the outcome by two to account for the uncomputation costs. Conjugating each controlled $R_z$ gate with a pair of CNOT gates, where $j_0'$ is the control bit, addresses both $j_0'=0,1$. As such, the T-gate count for one oracle call, i.e.,
\begin{align}
    \mathcal{T}^{(K)}&=2\lceil\log(c)\rceil\cdot[528c^2+96bc+384c\eta-c(12\lfloor \log(c)\rfloor+12\lfloor \log(\lceil \frac{3c}{2}\rceil)\rfloor+12\lfloor \log(\lceil \frac{5c}{2}\rceil)\rfloor\nonumber\\
    &\quad +18\lfloor \log(1+4c)\rfloor)-12b\lfloor \log(2c)\rfloor-48\eta\lfloor \log(2c)\rfloor+12(\lfloor \log(c)\rfloor+\lfloor \log(\lceil \frac{3c}{2}\rceil)\rfloor)\nonumber\\
    &\quad -60\lfloor \log(2c)\rfloor+12\lfloor \log(\lceil \frac{5c}{2}\rceil)\rfloor+12\lfloor \log(1+4c)\rfloor-12\lfloor \log(2c+b+4\eta+6)\rfloor+50c\nonumber\\
    &\quad -32b-128\eta-28]+19c+8b+32\eta+86+2\lceil \log(b)\rceil \cdot [48b^2+96 b(2\eta+2)\nonumber \\
    &\quad  -12 (b-1) \lfloor \log(b)\rfloor-12(2\eta+1)\lfloor \log(b)\rfloor -152b- 32 (2\eta+2)-12\lfloor \log(2b+2\eta+2)\rfloor \nonumber\\
    &\quad +104]+16\eta+38+8b+8(2\eta+5)(12b-3\lfloor \log(b)\rfloor-13)+88\eta-48\lfloor \log(\eta+1)\rfloor\nonumber\\
    &\quad +(8\eta+16)(12\eta -3\lfloor \log(\eta+3)\rfloor+23)+8\eta(12\eta - 3\lfloor \log(\eta+1)\rfloor)+74+186c+32\eta-48.
    \label{eq:SU2_kin_oracle_T}
\end{align}
The number of storage ancilla qubits required is
\begin{align}
    \lceil \log(c)\rceil \cdot (\lceil \frac{49c}{2}\rceil+2b+8\eta+17)+\lceil\log(b)\rceil\cdot(4\eta+7b+6)+b+9\eta+19,
\end{align}
and the number of workspace ancilla qubits required is
\begin{equation}
    12c-\lfloor\log(4c+1)\rfloor+5.
\end{equation}
There are two types of syntheses errors, i.e., arithmetic approximation errors and $R_z$ syntheses errors. The latter will be analyzed in Sec. \ref{subsubsec:Synth_SU2}. The arithmetic approximation errors per step is given by
\begin{equation}
 \frac{2+\log(b)}{2^b}+\left(\frac{3}{4}\right)^{c-( 2b+8\eta+12)}(2+c+\log(c)).
\end{equation}
In total, the approximation errors are 
\begin{equation}
    \epsilon^{(K)}=r\cdot 64dL^d[\frac{2+\log(b)}{2^b}+\left(\frac{3}{4}\right)^{c-( 2b+8\eta+12)}(2+c+\log(c))],
\end{equation}
where $r$ is the number of Trotter steps, and $64dL^d$ is the number of oracle calls. We divide the approximation errors evenly such that
\begin{equation}
    \epsilon^{(K)}_b=\frac{\epsilon^{(K)}}{2r\cdot 64dL^d},\:\epsilon^{(K)}_c=\frac{\epsilon^{(K)}}{2r\cdot 64dL^d},
\end{equation}
and
\begin{gather}
    \epsilon^{(K)}_b\geq \frac{2+\log(b)}{2^b}, \label{eq:SU2_eps_kinb}\\
    \epsilon^{(K)}_c\geq \left(\frac{3}{4}\right)^{c-( 2b+8\eta+12)}(2+c+\log(c)).\label{eq:SU2_eps_kinc}
\end{gather}
Note that this is not the optimal division of approximation errors. We let $b = \log(\frac{8}{\epsilon^{(K)}_b})$. Then, (\ref{eq:SU2_eps_kinb}) is always satisfied for $0<\epsilon^{(K)}_b<1$. We proceed to compute the upper bound for $c$. Since $c\geq 2b+8\eta+12 \geq 12\eta+24$, and $\eta\geq 1$, $\frac{44}{36}c \geq 2+c+\log(c)$. Inserting this relation and our choice of $b$ into (\ref{eq:SU2_eps_kinc}), we obtain
\begin{align}
    \epsilon^{(K)}_c\geq \left(\frac{3}{4}\right)^{c-( 2\log({8}/{\epsilon^{(K)}_b})+8\eta+12)}\frac{44}{36}c.
\end{align}
Let
\begin{equation}
    \tilde{\epsilon}^{(K)}_c = \epsilon^{(K)}_c\cdot \left(\frac{3}{4}\right)^{2\log({8}/{\epsilon^{(K)}_b})+8\eta+12}\frac{36}{44}.
\end{equation}
Then, we want to find a $c$ such that 
\begin{align}
    c\left(\frac{3}{4}\right)^c \leq \tilde{\epsilon}^{(K)}_c.
    \label{eq:SU2_kin_c}
\end{align}
Assuming that $0<\tilde{\epsilon}^{(K)}_c<1$, the choice
\begin{equation}
    c=\log_{\frac{3}{4}}\left(\frac{\frac{\tilde{\epsilon}^{(K)}_c}{2.28}}{\log_{\frac{3}{4}}(\frac{\tilde{\epsilon}^{(K)}_c}{2.28})}\right)
    \label{eq:SU2_chosen_kin_c}
\end{equation}
satisfies (\ref{eq:SU2_kin_c}). We have verified numerically with Mathematica that for $0<\tilde{\epsilon}_{\rm oracle}<1$, $\tilde{\epsilon}_{\rm oracle}-c\left(\frac{3}{4}\right)^c$ never exceeds $0.0431937$, given our choice of $c$.

We mention in passing that the phase-inducing step can be parallelized. As in U(1), we first divide the kinetic terms up into bulk and edge terms. Then, for each direction of the bulk or edge terms, we implement $O(L^{d}-L^{d-1})$ or $O(L^{d-1})$ $R_z(2^{k-c} \theta)$ in parallel, using the weight-sum trick, and thus, exponentially reducing the number of $R_z$ gates required to $O(c d \log(L^d))$. In this work, we focus on the resource analysis of the serial implementation, and will leave that of the parallel implementation for future work.

{\bf Syntheses of the magnetic oracles:} Similar to the implementation of the kinetic oracle, we directly synthesize the magnetic oracle, defined in (\ref{eq:SU2_mag_oracle}), as two controlled-diagonal gates, which impart the phases $\frac{-f_{\alpha \beta \gamma \delta}}{2a^{4-d}g^2}(-1)^{j^\prime_{1,0}}$ for ${j^\prime_{1,0}}=0,1$, if the control bits $m_{1,0}^{L\prime}$, $m_{1,0}^{R\prime}$, ${j_{2,0}^\prime}$, $m_{2,0}^{L\prime}$, $m_{2,0}^{R\prime}$, ${j_{3,0}^\prime}$, $m_{3,0}^{L\prime}$, $m_{3,0}^{R\prime}$, ${j_{4,0}^\prime}$, $m_{4,0}^{L\prime}$, and $m_{4,0}^{R\prime}$ are all ones. The function $f_{\alpha \beta \gamma \delta}$ is defined in (\ref{eq:SU2_fabcd}) as a product of four functions, $f_{\alpha \beta}(j_1, \Delta j_1, m^L_1, m^R_1)$, $f_{\beta \gamma}(j_2, \Delta j_2, m^L_2, m^R_2)$, $f_{\gamma \delta}(j_3, \Delta j_3, m^L_3, m^R_3)$, and $f_{\delta \alpha}(j_4, \Delta j_4, m^L_4, m^R_4)$, where $\ket{j_i,m^L_i,m^R_i}$ represents the state for the $i$th link on a plaquette. Without loss of generality, we consider the case where ${j^\prime_{1,0}}=0$. We begin by computing $f_{\alpha \beta \gamma \delta}$ into an ancilla register, conditioned upon the values of the control bits. Then, we induce the correct phase by applying $R_z$ gates to the ancilla state $\ket{f_{\alpha \beta \gamma \delta}}$. Finally, we uncompute $\ket{f_{\alpha \beta \gamma \delta}}$. The computation of $f_{\alpha \beta \gamma \delta}$ can be broken down into five steps. In the first and second steps, we compute the numerator and denominator, respectively. In the third step, we approximate the inverse of the denominator, using the circuits in \cite{bhaskar2016quantum}. In the fourth step, we approximate the argument of the square-root by multiplying together the numerator and the inverse of the denominator. Lastly, we approximate the square-root, using the circuits in \cite{bhaskar2016quantum}. Only the third and last steps incur approximation errors. Hereafter, we use the logarithmic depth out-of-place adder developed in \cite{draper2006logarithmic}, unless one of the inputs is classically known in which case we use the adder proposed in \cite{gidney2018halving}, and the multiplier proposed in \cite{shaw2020quantum}.

First, we consider the computation of the numerator. We start with computing the numerators in the four functions that constitute $f_{\alpha \beta \gamma \delta}$. This costs four times as many T gates and storage ancilla qubits as those required for the computation of the numerator in a kinetic oracle, i.e., $16(\eta+2)(12\eta-3\lfloor\log(\eta+3)\rfloor+23)+240\eta-96\lfloor\log(\eta+1)\rfloor+208$ and $20\eta+60$, respectively, while the reusable workspace ancilla qubit requirement remains the same, i.e., $3(\eta+3)-\lfloor \log(\eta+3)\rfloor - 1$. The multiplication of the four numerators costs two $(2\eta+6)$-bit multipliers, and one $(4\eta+12)-$bit multiplier. In total, the multipliers cost $(16\eta+44)(48\eta - 3\lfloor \log(4\eta+12)\rfloor+131)+(8\eta+20)(24\eta - 3\lfloor \log(2\eta+6)\rfloor+59)+(32\eta+96)$ T gates, $16\eta+48$ storage ancilla qubits, and $12\eta-\lfloor \log(4\eta+12)\rfloor+35$ workspace ancilla qubits. Next, we compute the denominator. First, we perform two additions of a classical number to an $\eta-$bit $j_i$, for each $i\in \{1,2,3,4\}$, which costs $32(\eta-2)$T gates, and then, multiply together eight $(\eta+1)-$bit numbers, which requires four $(\eta+1)-$bit multipliers, two $(2\eta+2)-$bit multipliers, and one $(4\eta+4)-$bit multiplier. This costs $(16\eta+12)(48\eta - 3\lfloor \log(4\eta+4)\rfloor+35)+(16\eta+8)(24\eta - 3\lfloor \log(2\eta+2)\rfloor+11)+16\eta(12\eta - 3\lfloor \log(\eta+1)\rfloor-1)+(48\eta+48)$ T gates, $32\eta+32$ storage ancilla qubits, and $12\eta-\lfloor \log(4\eta+4)\rfloor+11$ workspace ancilla qubits.

We now proceed to compute the inverse of the denominator using the algorithm in \cite{bhaskar2016quantum}. We refer readers to the kinetic oracle implementation for a detailed overview of the algorithm. Here, we simply state the algorithmic parameters and costs. The input $w$ is at most an $(8\eta+8)$-bit integer. Truncating at $b\geq 8\eta+8$ bits after the decimal point, the approximation error is bounded from above by $\frac{2+ \log(b)}{2^{b}}$. The T-gate count is $\lceil\log(b)\rceil\cdot (48b^2+768b\eta-(12b+96\eta-12)\lfloor\log(b)\rfloor+616b-256\eta -12\lfloor\log(2b+8\eta+8)\rfloor-84\lfloor\log(2b)\rfloor-152) + 32\eta + 43$. The required number of storage ancilla qubits is $\lceil\log(b)\rceil\cdot(7b+16\eta+18)$, and that of workspace ancilla qubits is $6b-\lfloor\log(2b)\rfloor-1$. Next, we multiply the numerator, an $8\eta+24$-bit number, and the inverse of the denominator, a $b$-bit number to obtain the fraction. Assuming that $b \geq 8\eta+24$, this costs $384b\eta-96\eta\lfloor\log(b)\rfloor+1108b-416\eta-276\lfloor\log(b)\rfloor-1196$ T gates, $b+8\eta+24$ storage ancilla qubits, and $3b-\lfloor\log(b)\rfloor-1$ workspace ancilla qubits. Lastly, we compute the square-root function. The output of the previous step is a $(b+8\eta+24)$-bit fraction, bounded above and below by 1 and 0, respectively. As in the kinetic oracle implementation, we shift it by $b+8\eta+24$ bits to obtain an integer input $w$, and shift it back by $(b+8\eta+24)/2$ bits once we obtain the root, assuming without loss of generality that $b$ is even. Let $c\geq 2b+16\eta+48$. Then, we approximate $\sqrt{w}$, up to $\left(\frac{3}{4}\right)^{c-2m}(2+c+\log(c))$ error. This costs $\lceil\log(c)\rceil\cdot[528c^2+96bc+768c\eta-c(12\lfloor \log(c)\rfloor+12\lfloor \log(\lceil \frac{3c}{2}\rceil)\rfloor+12\lfloor \log(\lceil \frac{5c}{2}\rceil)\rfloor+18\lfloor \log(1+4c)\rfloor)-12b\lfloor \log(2c)\rfloor-96n\lfloor \log(2c)\rfloor+12\lfloor \log(c)\rfloor+12\lfloor \log(\lceil \frac{3c}{2}\rceil)\rfloor-276\lfloor \log(2c)\rfloor+12\lfloor \log(\lceil \frac{5c}{2}\rceil)\rfloor+12\lfloor \log(1+4c)\rfloor-12\lfloor \log(2c+b+8n+24)\rfloor+1778c-32b-256\eta-604]+\lceil \frac{19c}{2}\rceil+4b+32\eta+115$ T gates, $\lceil\log(c)\rceil\cdot(\lceil \frac{51c}{2}\rceil+2b+16\eta+53)$ storage ancilla qubits, and $12c-\lfloor \log(4c+1)\rfloor+5$ workspace ancilla qubits.

We impart the phase by applying $R_z(2^{k-c}\theta)$, where $c$ is the number of digits after the decimal in the output of the square-root function, and $\theta = \frac{-1}{2a^{4-d}g^2}$, to the $k$th qubit of the ancilla state $\ket{f_{\alpha\beta\gamma\delta}}$. In order to implement the controlled version of this phase gate, we control each $R_z$ gate by the eleven control bits and the four ancilla bits that check the realness of the Clebsch-Gordan coefficients. Each multi-controlled $R_z$ gate requires two Toffoli gates with fifteen controls, which cost 119 T gates each \cite{maslov2016advantages}, two $R_z$ gates and one ancilla qubit~\cite{wang2020resource}. Since the state has at most $\lceil \frac{3c}{2}\rceil$ qubits and there are $L^d\frac{d(d-1)}{2}$ plaquettes on the lattice, there are $3c L^d\frac{d(d-1)}{2}$ multi-controlled $R_z$ gates to be applied.

We sum up the T-gate requirements for all the steps, and multiply the outcome by two to account for the uncomputation costs. Conjugating each controlled $R_z$ gate with a pair of CNOT gates, where $j_0'$ is the control bit, addresses both $j_0'=0,1$. As such, the T-gate count for one oracle call, i.e.,
\begin{align}
    \mathcal{T}^{(B)}&=1248\eta^2-12\eta(\lceil\log(2\eta+6)\rceil-2\lceil\log(2\eta+2)\rceil-2\lceil\log(\eta+3)\rceil+2\lceil\log(\eta)\rceil\nonumber \\
    &\quad+2\lceil\log(4\eta+12)\rceil+\lceil\log(4\eta+4)\rceil)+3844\eta-6(2\lceil\log(2\eta+2)\rceil+8\lceil\log(\eta+1)\rceil\nonumber \\
    &\quad+8\lceil\log(\eta+3)\rceil+5\lceil\log(2\eta+6)\rceil+3\lceil\log(4\eta+4)\rceil+11\lceil\log(4\eta+12)\rceil)+4270\nonumber \\
    &\quad+2\lceil\log(b)\rceil\cdot (48b^2+768b\eta-(12b+96\eta-12)\lfloor\log(b)\rfloor+616b-256\eta \nonumber \\
    &\quad-12\lfloor\log(2b+8\eta+8)\rfloor-84\lfloor\log(2b)\rfloor-152) + 64\eta + 86 \nonumber \\
    &\quad+768b\eta-192\eta\lfloor\log(b)\rfloor+2216b-832\eta-552\lfloor\log(b)\rfloor-3624\nonumber \\
    &\quad+2\lceil\log(c)\rceil\cdot[528c^2+96bc+768c\eta-c(12\lfloor \log(c)\rfloor+12\lfloor \log(\lceil \frac{3c}{2}\rceil)\rfloor+12\lfloor \log(\lceil \frac{5c}{2}\rceil)\rfloor\nonumber\\
    &\quad +18\lfloor \log(2+4c)\rfloor)-12b\lfloor \log(2c)\rfloor-96\eta\lfloor \log(2c)\rfloor+12\lfloor \log(c)\rfloor+12\lfloor \log(\lceil \frac{3c}{2}\rceil)\rfloor\nonumber\\
    &\quad -276\lfloor \log(2c)\rfloor+12\lfloor \log(\lceil \frac{5c}{2}\rceil)\rfloor+12\lfloor \log(2+4c)\rfloor-12\lfloor \log(2c+b+8\eta+24)\rfloor\nonumber\\
    &\quad +1762c-32b-256\eta-596]+19c+8b+64\eta+96+714c+192\eta-320.
    \label{eq:SU2_mag_oracle_T}
\end{align}
The number of storage ancilla qubits required is
\begin{equation}
    \lceil\log(c)\rceil\cdot(\lceil \frac{49c}{2}\rceil+4b+32\eta+101)+\lceil\log(b)\rceil\cdot(7b+16\eta+18)+b+76\eta+164,
\end{equation}
and the number of workspace ancilla qubits required is
\begin{equation}
    12c-\lfloor \log(4c+1)\rfloor+5.
\end{equation}
There are two types of syntheses errors, i.e., arithmetic approximation errors and $R_z$ syntheses errors. The latter will be analyzed in Sec. \ref{subsubsec:Synth_SU2}. The arithmetic approximation errors per step is given by
\begin{equation}
 \frac{2+\log(b)}{2^b}+\left(\frac{3}{4}\right)^{c-(2b+16\eta+48)}(2+c+\log(c)).
\end{equation}
In total, the approximation errors are 
\begin{equation}
    \epsilon^{(B)}= r\cdot 1048576\cdot \frac{d(d-1)}{2}L^d(\frac{2+\log(b)}{2^{b}}+\left(\frac{3}{4}\right)^{c-(2b+16\eta+48)}(2+c+\log(c))),
\end{equation}
where $r$ is the number of Trotter steps, and $1048576\cdot \frac{d(d-1)}{2}L^d$ is the number of oracle calls. We divide the approximation errors evenly such that
\begin{equation}
    \epsilon^{(B)}_b=\frac{\epsilon^{(B)}}{r\cdot 1048576 d(d-1)L^d},\:\epsilon^{(B)}_c=\frac{\epsilon^{(B)}}{r\cdot 1048576 d(d-1)L^d},
\end{equation}
and
\begin{gather}
    \epsilon^{(B)}_b\geq \frac{2+\log(b)}{2^b}, \label{eq:SU2_eps_magb}\\
    \epsilon^{(B)}_c\geq \left(\frac{3}{4}\right)^{c-( 2b+16\eta+48)}(2+c+\log(c)).\label{eq:SU2_eps_magc}
\end{gather}
Note that this is not the optimal division of approximation errors. We let $b = \log(\frac{8}{\epsilon^{(B)}_b})$. Then, (\ref{eq:SU2_eps_magb}) is always satisfied for $0<\epsilon^{(B)}_b<1$. We proceed to compute the upper bound for $c$. Since $c\geq 2b+16\eta+48 \geq 32\eta+96$, and $\eta\geq 1$, $\frac{137}{128}c \geq 2+c+\log(c)$. Inserting this relation and our choice of $b$ into (\ref{eq:SU2_eps_magc}), we obtain
\begin{align}
    \epsilon^{(B)}_c\geq \left(\frac{3}{4}\right)^{c-(2\log({8}/{\epsilon^{(B)}_b})+16\eta+48)}\frac{137}{128}c.
\end{align}
Let
\begin{equation}
    \tilde{\epsilon}^{(B)}_c = \epsilon^{(B)}_c\cdot \left(\frac{3}{4}\right)^{ (2\log({8}/{\epsilon^{(B)}_b})+16\eta+48)}\frac{128}{137}.
\end{equation}
Then, we want to find a $c$ such that 
\begin{align}
    c\left(\frac{3}{4}\right)^c \leq \tilde{\epsilon}^{(B)}_c.
    \label{eq:SU2_mag_c}
\end{align}
Assuming that $0<\tilde{\epsilon}^{(B)}_c<1$, the choice
\begin{equation}
    c=\log_{\frac{3}{4}}\left(\frac{\frac{\tilde{\epsilon}^{(B)}_c}{2.28}}{\log_{\frac{3}{4}}(\frac{\tilde{\epsilon}^{(B)}_c}{2.28})}\right)
    \label{eq:SU2_chosen_mag_c}
\end{equation}
satisfies (\ref{eq:SU2_mag_c}).

We mention in passing that the phase-inducing step can be parallelized. As in U(1), we can implement the magnetic terms acting on the odd and even plaquettes on a given two dimensional plane in parallel. In particular, we effect $O(L^d)$ same-angle $R_z$ gates, using the weight-sum trick, for each bit of the ancilla state $\ket{\pm f_{\alpha \beta \gamma \delta} }$. Thus, the parallel implementation exponentially reduces the $R_z$-gate count to $O(c d^2 \log(L^d) )$. In this work, we focus on the resource analysis of the serial implementation, and will leave that of the parallel implementation for future work.

We divide the oracle error $\epsilon_{oracle}$ evenly between the 
kinetic and magnetic oracles, i.e.,
\begin{equation}
    \epsilon^{(K)} = \frac{ \epsilon_{\rm oracle}}{2}, \:\epsilon^{(B)} = \frac{ \epsilon_{\rm oracle}}{2} \: \implies \: \epsilon_{\rm oracle}=\epsilon^{(K)}+\epsilon^{(B)}.
    \label{eq:SU2_oracle_err}
\end{equation}

\subsubsection{Synthesis errors}
\label{subsubsec:Synth_SU2}

Here, we compute the synthesis errors for $R_z$ gates required for the mass and electric term, separately from those for the oracles required for the kinetic and magnetic terms. To start, we consider the mass term. In this term, we have $\lfloor\log(2L^d)+1 \rfloor$ $R_z$ gates to implement. Therefore, we incur for each mass term $\lfloor\log(2L^d)+1 \rfloor \cdot \epsilon(R_z)$ amount of error, where $\epsilon(R_z)$ denotes the error per $R_z$ gate.

Next, we consider the electric term, which has $2(\eta + 1) \lfloor\log(dL^d)+1 \rfloor$ $R_z$ gates. Therefore, each electric term incurs $2(\eta + 1)\lfloor\log(dL^d)+1 \rfloor \cdot \epsilon(R_z)$ amount of error. If we instead use the phase gradient operation, once the gadget state $\ket{\psi_M}$ in (\ref{eq:phgradstate}) is prepared, each quantum adder call to implement the operation does not incur any synthesis error. We come back to the error incurred in preparing the gadget state itself in the next section.

The errors per Trotter step due to the $R_z$ gates for the kinetic and magnetic terms are $64 dL^d\cdot \lceil \frac{3c^{(K)}}{2} \rceil \cdot \epsilon(R_z)$ and $1048576\cdot \frac{d(d-1)}{2}L^d \cdot \lceil \frac{3c^{(B)}}{2}\rceil \cdot \epsilon(R_z)$, where we have and will continue to denote the approximation parameter $c$ for the kinetic and magnetic oracles as  $c^{(K)}$ and $c^{(B)}$, respectively.

We add the error incurred for the four terms to obtain the synthesis error $\epsilon_{\rm synthesis}$. Note that, as in the U(1) case, the implementation of the diagonal mass and electric terms can be optimized. As in the U(1) case, there are $r+1$ diagonal mass and electric terms, and $2r$ off-diagonal kinetic and magnetic terms to implement in total. Thus, $\epsilon_{\rm synthesis}$ is given by
\begin{equation}
    \epsilon_{\rm synthesis} = \{(r+1) \cdot [\lfloor\log(2L^d)+1 \rfloor + 2(\eta + 1)\lfloor\log(dL^d)+1 \rfloor] + 2r\cdot [96 c^{(K)}dL^d + 786432 c^{(B)}d(d-1)L^d ]\}\cdot \epsilon(R_z),
    \label{eq:SU2_synth_err}
\end{equation}
where $r$, to reiterate for the convenience of the readers, is the total number of Trotter steps.

\subsubsection{Complexity analysis}
\label{subsubsec:Analysis_SU2}

Having computed the Trotter, oracle and synthesis errors, we proceed to perform the complexity analysis for the SU(2) LGT.

The total error is given by
\begin{equation}
    \epsilon_{\rm total}=\epsilon_{\rm Trotter}+\epsilon_{\rm oracle}+\epsilon_{\rm synthesis}.
\end{equation}
We choose to evenly distribute the total error between the Trotter, oracle, and synthesis errors. Focusing on the Trotter error, we obtain the number of Trotter steps by
\begin{equation}
  \epsilon_{\rm Trotter} = \frac{\epsilon_{\rm total}}{3} \:\implies\: r = \lceil \frac{T^{3/2}3^{1/2}\rho^{1/2}}{\epsilon_{\rm total}^{1/2}} \rceil.
\end{equation}

Next, we use the above relation and $\epsilon_{\rm oracle} = \frac{\epsilon_{\rm total}}{3}$ to obtain the expressions for $c^{(K)}$ and $c^{(B)}$. $c^{(K)}$ is given by (\ref{eq:SU2_chosen_kin_c}), where
\begin{align}
    \tilde{\epsilon}_c^{(K)} = \frac{\epsilon_{\rm total}}{12r\cdot 64 dL^d}\left(\frac{3}{4}\right)^{ 2\log(8\frac{12r\cdot 64 dL^d}{\epsilon_{\rm total}})+8\eta+12}\frac{36}{44}.
\end{align}
$c^{(B)}$ is given by (\ref{eq:SU2_chosen_mag_c}), where
\begin{align}
    \tilde{\epsilon}_c^{(B)} = \frac{\epsilon_{\rm total}}{6r\cdot 1048576 d(d-1)L^d}\left(\frac{3}{4}\right)^{ 2 \log(8\frac{6r\cdot 1048576 d(d-1)L^d}{\epsilon_{\rm total}})+16\eta+48}\frac{128}{137}.
\end{align}

Finally, we compute the error each $R_z$ gate can incur by
\begin{align}
    &\epsilon_{\rm synthesis} = \frac{\epsilon_{\rm total}}{3} \implies \nonumber \\
    &\epsilon(R_z) = \frac{\epsilon_{\rm total}}{3} \{(\lceil \frac{T^{3/2}3^{1/2}\rho^{1/2}}{\epsilon_{\rm total}^{1/2}} \rceil+1) \cdot [\lfloor\log(2L^d)+1 \rfloor + 2(\eta + 1)\lfloor\log(dL^d)+1 \rfloor] \nonumber \\
    &\quad + 2\lceil \frac{T^{3/2}3^{1/2}\rho^{1/2}}{\epsilon_{\rm total}^{1/2}} \rceil\cdot [96 c^{(K)}dL^d + 786432 c^{(B)}d(d-1)L^d ]\}^{-1}.
\end{align}
With this, we obtain the number of T gates required to synthesize each $R_z$ gate using RUS circuit \cite{bocharov2015efficient},
\begin{equation}
    \text{Cost}(R_z)=1.15\log(\frac{1}{\epsilon(R_z)}).
\end{equation}
Combining the T gates required for implementation of $R_z$ gates and the T gates used elsewhere in the circuit, we obtain the total number of T gates for the entire circuit as
\begin{align}
    &\quad \{ (\lceil \frac{T^{3/2}3^{1/2}\rho^{1/2}}{\epsilon_{\rm total}^{1/2}} \rceil+1)\cdot [\lfloor\log(2L^d)+1 \rfloor + 2(\eta + 1)\lfloor\log(dL^d)+1 \rfloor] + 2\lceil \frac{T^{3/2}3^{1/2}\rho^{1/2}}{\epsilon_{\rm total}^{1/2}} \rceil\cdot [96 c^{(K)}dL^d \nonumber\\
    &  + 786432 c^{(B)}d(d-1)L^d ] \} 
    \text{Cost}(R_z)+(\lceil \frac{T^{3/2}3^{1/2}\rho^{1/2}}{\epsilon_{\rm total}^{1/2}} \rceil+1)\cdot [4(2L^d-\text{Weight}(2L^d))+8dL^d(\eta-2)\nonumber \\
    &+8dL^d\eta(12\eta-3\lfloor\log(\eta+1)\rfloor-2)
    +(8\eta+8)(dL^d-\text{Weight}(dL^d))+dL^d(32\eta-48)]+2\lceil \frac{T^{3/2}3^{1/2}\rho^{1/2}}{\epsilon_{\rm total}^{1/2}} \rceil \nonumber \\
    &\cdot [64dL^d \mathcal{T}^{(K)} +2^{13}d(d-1)L^d+524288d(d-1)L^d \mathcal{T}^{(B)}],
\end{align}
where $\mathcal{T}^{(K)}, \mathcal{T}^{(B)}$ are given in (\ref{eq:SU2_kin_oracle_T},\ref{eq:SU2_mag_oracle_T}), respectively. The size of the ancilla register is given by the maximum between the ancilla qubits required by the electric and magnetic Hamiltonian, i.e.,
\begin{multline}
    Q_{\max}=\max \{(3\eta+1)dL^d+3(\eta+1)-\lfloor\log(\eta+1)\rfloor-1+dL^d-\text{Weight}(dL^d),\\ \lceil\log(c)\rceil\cdot(\lceil \frac{49c}{2}\rceil+4b+32\eta+101)+\lceil\log(b)\rceil\cdot(7b+16\eta+18)+b+76\eta+12c-\lfloor \log(4c+1)\rfloor+169 \}.
\end{multline}
Taking this into account, we obtain the total number of qubits required for the simulation by summing up those in the ancilla, fermionic and gauge-field registers, which is given by
\begin{equation}
    2L^d+(3\eta+2)dL^d+Q_{\max}.
\end{equation}

Note that in the case where the electric term is implemented using phase gradient operation, the T-gate count changes by 
\begin{align}
    &\quad (\lceil \frac{T^{3/2}3^{1/2}\rho^{1/2}}{\epsilon_{total}^{1/2}} \rceil+1)\cdot [4dL^d\log(\frac{8\pi a^{d-2}}{g^2 t})+O(dL^d) -2(\eta+1)\lfloor\log(dL^d)+1 \rfloor\cdot \text{Cost}(R_z)\nonumber \\
    &-(8\eta+8)(dL^d - \text{Weight}(dL^d))]+\text{Cost}(\ket{\psi_M}),
\end{align} 
where the Cost$(R_z)$ needs to be modified, since $\epsilon(R_z)$ has changed to
\begin{align}
    &\epsilon(R_z) = \frac{\epsilon_{\rm total}}{3} \{(\lceil \frac{T^{3/2}3^{1/2}\rho^{1/2}}{\epsilon_{\rm total}^{1/2}} \rceil+1) \cdot \lfloor\log(2L^d)+1 \rfloor \nonumber \\
    &\quad + 2\lceil \frac{T^{3/2}3^{1/2}\rho^{1/2}}{\epsilon_{\rm total}^{1/2}} \rceil\cdot [192c^{(K)}dL^d + 1572864c^{(B)}d(d-1)L^d ]\}^{-1}.
\end{align}
Further, $\text{Cost}(\ket{\psi_M})$, which denotes the one-time synthesis costs of the phase gradient gadget state. Here, we choose to use the synthesis method delineated in \cite{nam2020approximate}. Briefly, we apply Hadamard gates to the register $\ket{00...0}$, and then apply gates $Z, Z^{-1/2},...,Z^{-1/2^{M-1}}$. Each $Z^\alpha$ gates are synthesized using RUS circuits \cite{bocharov2015efficient}. Let $\delta$ be the error of preparing the gadget state $\ket{\psi_M}$. Then, each gate can incur at most $M/\delta$ error, and thus, costs $1.15\log(M/\delta)$, using RUS circuits \cite{bocharov2015efficient}. Thus, the gadget state preparation costs $1.15M\log(M/\delta)$.

Finally, in this case, the ancilla-qubit count is given by that of the maximum between the mass term and the phase gradient state, and the magnetic term, i.e.,
\begin{multline}
    Q_{\max}=\max \{ (2L^d - \text{Weight}(2L^d)) + \log(\frac{8\pi a^{d-2}}{g^2 t}), \lceil\log(c)\rceil\cdot(\lceil \frac{49c}{2}\rceil+4b+32\eta+101)\\ +\lceil\log(b)\rceil\cdot(7b+16\eta+18)+b+76\eta+12c-\lfloor \log(4c+1)\rfloor+169 \}.
\end{multline}
As such, the total qubit count is given by
\begin{equation}
    2L^d+dL^d(3\eta+2)+Q_{\max}.
\end{equation}

\section{SU(3) Lattice Gauge Theory}
\label{sec:SU3}
In this section, we introduce the non-Abelian SU(3) lattice gauge theory, or lattice QCD. We follow the same format as in the SU(2) case to guide the readers.
\subsection{Preliminaries}
\label{sec:SU3_prelim}
As in the cases of U(1) and SU(2), we aim to simulate our system, governed by four types of Hamiltonian, i.e., the electric Hamiltonian $H_E$, magnetic Hamiltonian $H_B$, mass Hamiltonian $H_M$ and kinetic Hamiltonian $H_K$. Once again, $H_E$ and $H_B$ act on the links that connect two fermionic sites, $H_M$ acts on the fermions themselves, and $H_K$ act on nearest pairs of fermionic sites and the links that connect the pairs. Thus, we consider two different types of qubit registers, one for the fields ($H_E$, $H_B$, $H_K$) and the other for the fermions ($H_M$, $H_K$).

To simulate this system, we again need to choose a good basis for each register, as in the SU($2$) case. For the fermionic register, we consider an occupation basis. Note however though, in the current case of SU($3$), the fermions may assume three different colors in the fundamental representation. For instance, the mass Hamiltonian is now of the form
\begin{equation}
    \hat{H}_M = m\sum_{\vec{n}} \sum_{\alpha=1}^{3} (-1)^{\vec{n}} \hat{\psi}_{\alpha}^{\dag}(\vec{n})\hat{\psi}_{\alpha}(\vec{n}),
\end{equation}
where $\alpha \in \{1,2,3\}$ denotes the color. This means that we have three subregisters, each for the three different colors, that comprise the full fermion register. For concreteness and simplicity, we use the JW transformation~\cite{wigner1928paulische} for the rest of this section to map the fermion operators to the qubit operators.

For the link register, we once again write down the Gauss' law,
\begin{equation}
    \hat{G}^{a}(\vec{n})=\sum_{k}(\hat{E}^{a}_{L}({\vec{n},k}) + \hat{E}^{a}_{R}({\vec{n},k})) + \hat{Q}^{a}({\vec{n}}),
\end{equation}
which, up to the charge operator, has the same form as the one for SU($2$). The SU($3$) charge operator for the projection axis $a$ is given by
\begin{equation}
    \hat{Q}^a(\vec{n}) = \sum_{\alpha,\beta=1}^{3} \hat{\psi}^\dag_\alpha \tau_{\alpha \beta}^a  \hat{\psi}_\beta,
\end{equation}
where $\tau^a$, $a=1,...,8$, are the eight generators of the fundamental representation of SU(3), which satisfy
\begin{equation}
    [\tau^a,\tau^b]=i\sum_{c=1}^{8}f^{abc}\tau^c, \: \text{Tr}[\tau^a \tau^b]=\frac{1}{2}\delta_{ab},
\end{equation}
where $f^{abc}$ are the group structure constants of SU(3), the $\hat{\psi}^\dag_\alpha$ and $\hat{\psi}_\alpha$ are the fermion creation and annihilation operators of color $\alpha$. Similar to the SU($2$) case, the charge operators at each vertex satisfy the SU($3$) algebra
\begin{equation}
    [\hat{Q}^a, \hat{\psi}_{\alpha}] = \sum_{\beta=1}^{3} -\tau^a_{\alpha \beta} \hat{\psi}_\beta.
\end{equation}
On each link, the left and right electric fields each also forms SU(3) Lie algebras, and they commute with each other, according to
\begin{align}
     [\hat{E}^{a}_{L},\hat{E}^{b}_{L}] &= i \sum_{c=1}^{8}f^{abc}\hat{E}^{c}_{L}, \\
     [\hat{E}^{a}_{R},\hat{E}^{b}_{R}] &= i \sum_{c=1}^{8}f^{abc}\hat{E}^{c}_{R}, \\
     [\hat{E}^{a}_{L},\hat{E}^{b}_{R}] &= 0.
\end{align}
Once again, due to gauge-invariance, the SU(3) Hamiltonian commutes with all of the Gauss operators $\hat{G}^a(\vec{n})$. 

The physical, gauge-invariant Hilbert space $\mathcal{H}_{G}$ is defined through the eigenstates of the Gauss operator:
\begin{equation}
\label{su3gauss}
    \mathcal{H}_{G} = \{ \ket{\Psi} \in \mathcal{H}_G \: |\:   \hat{G}^{a}(\vec{n}) \ket{\Psi} = 0, \: \forall \vec{n}, a \}.
\end{equation}
Due to the non-Abelian nature of SU(3), the electric fields and Gauss operators do not all commute. Conventionally, the complete set of commuting observables on a link is given by $\{\hat{E}^2,\hat{T}_L,\hat{T}_L^z,\hat{Y}_L,\hat{T}_R,\hat{T}_R^z,\hat{Y}_R\}$, where $\hat{T}_i,\hat{T}_i^z,\hat{Y}_i$, with $i=L,R$ for left and right, are physical quantities, known as the isospin, $z$-component of the isospin, and hypercharge~\cite{de1963octet,gasiorowicz1966elementary}, and $\hat{E}^2$ is the Casimir operator defined by
\begin{equation}
\label{eq:SU3_basis}
    \hat{E}^2 \equiv \sum_{a = 1}^8 \hat{E}^a_L \hat{E}^a_L = \sum_{a = 1}^8 \hat{E}^a_R \hat{E}^a_R.
\end{equation}
In this basis, the states are labelled by eight quantum numbers
\begin{equation}
    \ket{p,q,{T}_L,{T}_L^z,{Y}_L,{T}_R,{T}_R^z, {Y}_R},
\end{equation}
where $p$ and $q$ label the representation~\cite{gasiorowicz1966elementary}. The eigenvalues of the Casimir operator are a function of $p$ and $q$ given by~\cite{carruthers1966introduction}
\begin{equation}
\label{eq:SU3_elec}
    \hat{E}^2 \ket{p,q} = \frac{1}{3}[p^2+q^2+pq+3(p+q)]\ket{p,q};\: p,q\in \mathbbm{N},
\end{equation}
where the isospin and hypercharge labels are omitted for brevity.

Finally, we can define the SU(3) parallel transporters, $\hat{U}_{\alpha \beta}$, in the basis defined above. Dropping the link position index for notational convenience, they are given by
\begin{align}
    &\quad \hat{U}_{\alpha \beta} \ket{p,q,{T}_L,{T}_L^z,{Y}_L,{T}_R,{T}_R^z, {Y}_R} \nonumber \\
    &= \sum_{(p',q')}\sum_{T_{L}'=|T_L - t_L|}^{T_L + t_L}\sum_{T_{R}'=|T_R - t_R|}^{T_R + t_R} \sqrt{\frac{\text{dim}(p,q)}{\text{dim}(p',q')}} \braket{p',q',T_L',{T_L^z}',Y_L'}{p,q,T_L,T_L^z,Y_L;1,0,t_L,t_L^z,y_L} \nonumber \\
    & \quad \times \braket{p',q',T_R',{T_R^z}',Y_R'}{p,q,T_R,T_R^z,Y_R;1,0,t_R,t_R^z,y_R} \ket{p',q',{T}_L',{T_L^z} ',{Y}_L',{T}_R',{T_R^z} ', {Y}_R'},
    \label{eq:SU3_U_CG}
\end{align}
where $(p',q')\in\{ (p+1,q),(p-1,q+1),(p,q-1) \}$, $\braket{p',q',{T_i}',{T^z_i}',{Y}_i'}{p,q,T_i,T^z_i,Y_i;1,0,t_i,t^z_i,y_i}$, with $i=L,R$, are the Clebsch-Gordan coefficients for SU(3) in the fundamental representation, i.e., $(p=1,q=0)$, as provided in Table \ref{tb:CG_SU3}, the dimension of the representation $(p,q)$ is given by
\begin{equation}
\label{eq:SU3_dim}
    \text{dim}(p,q) = (1+p)(1+q)(1+\frac{p+q}{2}),
\end{equation}
and the left and right isospin and hypercharge values $(t_i,t^z_i,y_i)_{i=L,R}$ depend on $\alpha,\beta$, respectively, as follows:
\begin{equation}
    (t_{L/R},t^z_{L/R},y_{L/R}) =
    \begin{cases}
    (\frac{1}{2},\frac{1}{2},\frac{1}{3}), \alpha / \beta = 1, \\
    (\frac{1}{2},-\frac{1}{2},\frac{1}{3}), \alpha / \beta = 2, \\
    (0,0,-\frac{2}{3}), \alpha / \beta = 3.
    \end{cases}
\end{equation}
Further, the hypercharge $Y_i'$ and z-component isospin ${T_i^z}'$ values are obtained by
\begin{align}
    Y_i'&= Y_i + y_i,\\
    {T_i^z}' &= T_i^z + t_i^z,
\end{align}
for $i = L, R$. Lastly, the ranges of values for the quantum numbers are given by
\begin{gather}
    p = n,\: q = m; \:n,m \in \mathbbm{N}, \label{eq:SU3_qnum1}\\
    T_i = 0, \frac{1}{2},...,\frac{1}{2}(p+q),\label{eq:SU3_qnum2} \\
    T_i^z = -\frac{1}{2}(p+q), -\frac{1}{2}(p+q)+\frac{1}{2},...,\frac{1}{2}(p+q), \label{eq:SU3_qnum3}\\
    Y_i = -\frac{1}{3}(q+2p),-\frac{1}{3}(q+2p)+\frac{1}{3},...,\frac{1}{3}(p+2q),\label{eq:SU3_qnum4}
\end{gather}
where $i=L,R$. 

The above definition of $\hat{U}_{\alpha \beta}$ operators can be used to directly verify the proper commutation relations
\begin{align}
    [\hat{E}^{a}_{L},\hat{U}_{\alpha \beta}] &= -\sum_{\gamma=1}^{3}\tau^{a}_{\alpha \gamma} \hat{U}_{\gamma \beta}, \\
    [\hat{E}^{a}_{R},\hat{U}_{\alpha \beta}] &= \sum_{\gamma=1}^{3}\hat{U}_{\alpha \gamma }\tau^{a}_{ \gamma \beta},
\end{align}
where $\alpha$, $\beta$, $\gamma$, $\delta \in \{1,2,3 \}$, required by the SU(3) lattice gauge theory. Note that $\hat{U}_{\alpha \beta}$ are elements of the unitary operator-valued matrix $\hat{\bf{U}}$. Denoting the $\alpha \beta$th element of $\hat{\bf{U}}$ as $[\hat{\bf{U}}]_{\alpha \beta}$, the operators $\hat{U}_{\alpha \beta}$ are related to $\hat{\bf{U}}$ by $[\hat{\bf{U}}]_{\alpha \beta} = \hat{U}_{\alpha \beta}$ and thus, $[\hat{\bf{U}}^\dag]_{\alpha \beta} = \hat{U}_{\beta \alpha}^{\dag}$. It can be shown that in the fundamental representation of SU(3), the parallel transporters satisfy the relation \cite{carruthers1966introduction, kogut1976quantitative}
\begin{equation}
    \bra{0}[\hat{\bf{U}}]_{\alpha \beta}[\hat{\bf{U}}^\dag]_{\gamma \delta} \ket{0} =  \bra{0}\hat{U}_{\alpha \beta}\hat{U}^\dag_{\delta\gamma} \ket{0} = \frac{1}{3}\delta_{\alpha \gamma}\delta_{\beta \delta},
\label{eq:SU3_Uhc}
\end{equation}
where $\ket{0}$ here is used to denote the basis state in (\ref{eq:SU3_basis}) with all eight quantum numbers being zeros.

\begin{table}[htp]
\centering
\begin{tabular}{||c c||}
\hline
 Coefficient & Formula  \\[1.5ex] 
\hline
 $I_{11}$&$\sqrt{\frac{\Omega_1^+(\Omega_2^+ + 6)}{\Gamma_1 \Upsilon_1}}$  \\
 $I_{12}$&$-\sqrt{\frac{-\Omega_1^-(\Omega_2^- + 6)(\Omega_3^- + 6)}{\Gamma_1 \Upsilon_2}}$   \\
 $I_{13}$&$\sqrt{\frac{\Omega_1^- \Omega_1^+}{\Gamma_1 \Upsilon_3}}$   \\
 $I_{21}$&$-\sqrt{\frac{(6-\Omega_1^-)\Omega^-_2(\Omega_3^+ +6)}{\Gamma_2 \Upsilon_1}}$   \\
 $I_{22}$&$-\sqrt{\frac{(\Omega_1^+ - 6)\Omega_2^+ (\Omega_3^- + 6)}{\Gamma_2 \Upsilon_2}}$   \\
 $I_{23}$&$\sqrt{\frac{-\Omega_2^- \Omega_2^+}{\Gamma_2 \Upsilon_3}}$   \\
 $I_{31}$&$-\sqrt{\frac{(\Omega_1^- - 6) (\Omega_2^+ + 6) \Omega_3^-}{\Gamma_3 \Upsilon_1}}$   \\
 $I_{32}$&$\sqrt{\frac{-(\Omega_1^+ - 6) (\Omega_2^- + 6)\Omega_3^+}{\Gamma_3 \Upsilon_2}}$   \\
 $I_{33}$&$\sqrt{\frac{\Omega_3^+ \Omega_3^-}{\Gamma_3 \Upsilon_3}}$   \\[1.5ex] 
\hline
\end{tabular}
\caption{The formulas for the Isoscalar factors $I_{\alpha \beta}$ in the fundamental SU($3$) representation, where $\Omega_1^{\pm} = 4p+2q\pm6T_i - 3Y_i+9\pm3$, $\Omega_2^{\pm}=2p-2q\pm 6T_i + 3Y_i - 3 \pm 3$, $\Omega_3^{\pm}=2p+4q\pm6T_i+3Y_i+3\pm3$, $\Gamma_1=(1+p)(2+p+q)$, $\Gamma_2 = (1+p)(1+q)$, $\Gamma_3=(1+q)(2+p+q)$, $\Upsilon_1 = 432(1+T_i)$, $\Upsilon_2=432T_i$, and $\Upsilon_3=36$, taken from \cite{byrnes2006simulating}, whose authors obtained these formulas efficiently classically with methods developed in \cite{biedenharn1968pattern, louck1970recent}.}
\label{tb:iso}
\end{table}

\begin{table*}[t]
\centering
\begin{tabular}{||c c c c c c c||}
\hline
 Coefficient & $\Delta p$ & $\Delta q$ & $\Delta T$ & $\Delta T^z$ & $\Delta Y$ & Formula \\[1.5ex] 
\hline
 $C_{11}^{(a)}$& 1 & 0 & $\frac{1}{2}$ & $\frac{1}{2}$ & $\frac{1}{3}$ & $I_{11}c_{11}$  \\
 $C_{11}^{(b)}$& 1 & 0 & $-\frac{1}{2}$ & $\frac{1}{2}$ & $\frac{1}{3}$ & $I_{12}c_{21}$  \\
 $C_{12}^{(a)}$& 1 & 0 & $\frac{1}{2}$ & $-\frac{1}{2}$ & $\frac{1}{3}$ & $I_{11}c_{12}$  \\
 $C_{12}^{(b)}$& 1 & 0 & $-\frac{1}{2}$ & $-\frac{1}{2}$ & $\frac{1}{3}$ & $I_{12}c_{22}$  \\
 $C_{13}$& 1  & 0 & 0 & 0 & $-\frac{2}{3}$ & $I_{13}$\\
 $C_{21}^{(a)}$& -1 & 1 & $\frac{1}{2}$ & $\frac{1}{2}$ & $\frac{1}{3}$ & $I_{21}c_{11}$  \\
 $C_{21}^{(b)}$& -1 & 1 & $-\frac{1}{2}$ & $\frac{1}{2}$ & $\frac{1}{3}$ & $I_{22}c_{21}$  \\
 $C_{22}^{(a)}$& -1 & 1 & $\frac{1}{2}$ & $-\frac{1}{2}$ & $\frac{1}{3}$ & $I_{21}c_{12}$  \\
 $C_{22}^{(b)}$& -1 & 1 & $-\frac{1}{2}$ & $-\frac{1}{2}$ & $\frac{1}{3}$ & $I_{22}c_{22}$  \\
 $C_{23}$& -1 & 1 & 0 & 0 & $-\frac{2}{3}$ &$I_{23}$   \\
 $C_{31}^{(a)}$& 0 & -1 &$\frac{1}{2}$ & $\frac{1}{2}$ & $\frac{1}{3}$ & $I_{31}c_{11}$  \\
 $C_{31}^{(b)}$& 0 & -1 & $-\frac{1}{2}$ & $\frac{1}{2}$ & $\frac{1}{3}$ & $I_{32}c_{21}$  \\
 $C_{32}^{(a)}$& 0 & -1 & $\frac{1}{2}$ & $-\frac{1}{2}$ & $\frac{1}{3}$ & $I_{31}c_{12}$  \\
 $C_{32}^{(b)}$& 0 & -1 & $-\frac{1}{2}$ & $-\frac{1}{2}$ & $\frac{1}{3}$ & $I_{32}c_{22}$ \\
 $C_{33}$& 0 & -1 & 0 & 0 & $-\frac{2}{3}$ & $I_{33}$ \\ [1.5ex] 
\hline
\end{tabular}
\caption{The formulas for Clebsch-Gordan coefficients $\langle p+\Delta p,q+\Delta q,T_i+\Delta T,T_i^z+\Delta T^z,Y_i+\Delta Y|$ $p,q,T_i,$ $T^z_i,Y_i;1,0,\Delta T, \Delta T^z, \Delta Y\rangle$ may be obtained by combining the SU($2$) Clebsch-Gordan coefficients $c_{\alpha \beta}$ and Isoscalar factors $I_{\alpha \beta}$ from Tables \ref{tb:CG_SU2} and \ref{tb:iso}, respectively. When evaluating $c_{\alpha \beta}$, $\Delta T, \Delta T^z$ are inserted in place of $\Delta j, \Delta m$, respectively. See \cite{byrnes2006simulating} for details, where the authors computed these formulas classically efficiently with the methods developed in \cite{biedenharn1968pattern, louck1970recent}.}
\label{tb:CG_SU3}
\end{table*}

\subsection{Simulation circuit synthesis}
\label{sec:SimCircSynth_SU3}

The infinite-dimensional gauge-field register consists of eight subregisters, each representing a quantum number. We import the encodings for $\ket{p}$, $\ket{q}$, $\ket{{T}_L}$, $\ket{{T}_L^z}$, $\ket{{Y}_L}$, $\ket{{T}_R}$, $\ket{{T}_R^z}$ and $\ket{{Y}_R}$ from \cite{byrnes2006simulating}. In order to represent them on a finite quantum computer, we impose a cutoff on the electric field. In particular, for a given link, we truncate both $p$ and $q$ at $\Lambda$, i.e.,
\begin{gather}
    p, q \in \{0,1,...,\Lambda \}, \: T_i \in \{0,\frac{1}{2},...,\Lambda \}, \nonumber \\
    T_i^z \in \{-\Lambda, -\Lambda+\frac{1}{2},...,\Lambda \}, \: Y_i = \{-\Lambda, -\Lambda + \frac{1}{3},...,\Lambda  \},
\label{eq:SU3_range}
\end{gather}
where $i=L,R$. In this basis, in order to import the encoding of the parallel transporters $\hat{U}_{\alpha \beta}$ from \cite{byrnes2006simulating}, we import and will slightly modify the definitions of the following useful operators from \cite{byrnes2006simulating}:
\begin{gather}
    \hat{P}^{\pm} = \sum_{p=0}^{\Lambda} \ketbra{p\pm 1}{p},\: \hat{Q}^{\pm} = \sum_{q=0}^{\Lambda} \ketbra{q\pm 1}{q},\nonumber \\
    \hat{T}_i^{\pm} = \sum_{T_i = 0}^{\Lambda} \ketbra{T_i \pm \frac{1}{2}}{T_i},\: \hat{T}_i^{z \pm} = \sum_{T_i^z = -\Lambda}^{\Lambda} \ketbra{T_i^z \pm \frac{1}{2}}{T_i^z}, \nonumber \\
    \hat{Y}_i^{\pm}=\sum_{Y_i = -\Lambda}^{\Lambda}\ketbra{Y_i \pm \frac{1}{3}}{Y_i},
\label{eq:SU3_ladder}
\end{gather}
where $i=L,R$, and the quantum numbers $p$ and $q$, $T_i$ and $T_i^z$, and $Y_i$ are incremented and decremented by $1$, $\frac{1}{2}$ and $\frac{1}{3}$ at a time, respectively. The slight modification from the original definitions in \cite{byrnes2006simulating} is that we have introduced a periodic-wrapping term to the above ladder operators, such that they can be implemented as binary incrementers and decrementers on a quantum computer. The undesirable effect of the periodic-wrapping terms can be removed by applying circuits that include multiply controlled gates, similar to the U(1) case. We further import the following diagonal Clebsch-Gordan operators,
\begin{align}
    \hat{C}^{i(a)}_{\alpha \beta} &= \sum_{p,q,T_i,T_i^z,Y_i} C^{(a)}_{\alpha \beta}(p,q,T_i,T_i^z,Y_i,\Delta p,\Delta q,\Delta T_i,\Delta T_i^z,\Delta Y_i) \ketbra{p,q,T_i,T_i^z,Y_i}{p,q,T_i,T_i^z,Y_i},\nonumber \\
    &\quad
    \text{ for }\beta = 1,2, \nonumber \\
    \hat{C}^{i(b)}_{\alpha \beta} &= \sum_{p,q,T_i,T_i^z,Y_i} C^{(b)}_{\alpha \beta}(p,q,T_i,T_i^z,Y_i,\Delta p,\Delta q,\Delta T_i,\Delta T_i^z,\Delta Y_i) \ketbra{p,q,T_i,T_i^z,Y_i}{p,q,T_i,T_i^z,Y_i},\nonumber \\
    &\quad
    \text{ for }\beta = 1,2, \nonumber \\
    \hat{C}^{i}_{\alpha \beta} &= \sum_{p,q,T_i,T_i^z,Y_i} C_{\alpha \beta}(p,q,T_i,T_i^z,Y_i,\Delta p,\Delta q,\Delta T_i,\Delta T_i^z,\Delta Y_i) \ketbra{p,q,T_i,T_i^z,Y_i}{p,q,T_i,T_i^z,Y_i},\nonumber \\
    &\quad
    \text{ for }\beta =3,
    \label{eq:SU3_CG_op}
\end{align}
where $C_{\alpha \beta}^{i(a)}$, $C_{\alpha \beta}^{i(b)}$, and $C_{\alpha \beta}^i$ are the SU(3) Clebsch-Gordan coefficients evaluated using the formulas in Table \ref{tb:CG_SU3}, $\alpha \in \{1,2,3 \}$, $i=L,R$, the sums of $p$, $q$, $T_i$, $T_i^z$, $Y_i$ are over the ranges defined in (\ref{eq:SU3_range}), and $\ket{p,q,T_i,T_i^z,Y_i}$ is a tensor product of $\ket{p}$, $\ket{q}$, $\ket{T_i}$, $\ket{T_i^z}$, $\ket{Y_i}$. Further, the SU(3) Clebsch-Gordan coefficients shown in Table \ref{tb:CG_SU3} are real \cite{de1963octet,kaeding1995tables}. However, since we are summing over the ranges in (\ref{eq:SU3_range}), which include unphysical quantum numbers that violate (\ref{eq:SU3_qnum1}-\ref{eq:SU3_qnum4}), the formulas in Table \ref{tb:CG_SU3} could result in complex numbers. Thus, we set the complex elements of the diagonal Clebsch-Gordan operator in (\ref{eq:SU3_CG_op}) to zeros, thereby ensuring its Hermiticity. Next, using (\ref{eq:SU3_ladder}) and (\ref{eq:SU3_CG_op}), the following operators can be defined:
\begin{align}
    &\hat{\mathcal{C}}^{i}_{\alpha \beta} = \begin{cases}
        \hat{T}_{i}^+ \hat{C}_{\alpha \beta}^{i(a)} + \hat{T}_{i}^- \hat{C}_{\alpha \beta}^{i(b)}, \text{ for }\beta = 1,2, \\
        \hat{C}_{\alpha \beta}^{i},\text{ for }\beta = 3,
    \end{cases} \\
    &\hat{M}^{i}_{\alpha} =
    \begin{cases}
        \hat{T}_i^{z+}\hat{Y}_i^+,\text{ for }\alpha = 1, \\
        \hat{T}_i^{z-}\hat{Y}_i^+,\text{ for }\alpha = 2, \\
        (\hat{Y}_i^-)^2,\text{ for }\alpha = 3,
    \end{cases}
\end{align}
where $i=L,R$. Moreover, the diagonal normalization operators are defined as
\begin{equation}
\label{eq:SU3_N}
    \hat{N}_{\alpha} =
    \begin{cases}
        \sum_{p,q} \sqrt{\frac{\text{dim}(p,q)}{\text{dim}(p+1,q)}}\ketbra{p,q}{p,q},\text{ for }\alpha = 1,\\
        \sum_{p,q}\sqrt{\frac{\text{dim}(p,q)}{\text{dim}(p-1,q+1)}}\ketbra{p,q}{p,q} ,\text{ for }\alpha = 2,\\
        \sum_{p,q}\sqrt{\frac{\text{dim}(p,q)}{\text{dim}(p,q-1)}}\ketbra{p,q}{p,q} ,\text{ for }\alpha = 3,
    \end{cases}
\end{equation}
where $\text{dim}(p,q)$ are given in (\ref{eq:SU3_dim}). Finally, using (\ref{eq:SU3_ladder}), (\ref{eq:SU3_CG_op}) and (\ref{eq:SU3_N}), we import the encoding for $\hat{U}_{\alpha \beta}$ from \cite{byrnes2006simulating},
\begin{equation}
    \hat{U}_{\alpha \beta} = \hat{M}_{\alpha}^{L}\hat{M}_{\beta}^{R}[ \hat{P}^+ \hat{\mathcal{C}}^{L}_{1\alpha}\hat{\mathcal{C}}^{R}
    _{1\beta}\hat{N}_{1} + \hat{P}^- \hat{Q}^+ \hat{\mathcal{C}}^{L}_{2\alpha}\hat{\mathcal{C}}^{R}
    _{2\beta}\hat{N}_{2}
    + \hat{Q}^- \hat{\mathcal{C}}^{L}_{3\alpha}\hat{\mathcal{C}}^{R}
    _{3\beta}\hat{N}_{3} ],
\label{eq:SU3_U_enc}
\end{equation}
which can be straightforwardly shown to satisfy (\ref{eq:SU3_U_CG}).

In order to implement these operators on a quantum computer, we need to map the non-integer quantum numbers $T_i, T_i^z, Y_i$ with $i=L,R$ to positive full integers:
\begin{align}
\label{eq:SU3_shift}
    T_i \mapsto 2T_i,\: T_i^z \mapsto 2(T_i^z + \Lambda),\: Y_i \mapsto 3(Y_i + \Lambda),
\end{align}
such that
\begin{equation}
\label{eq:SU3_range_shift}
    T_i \in \{0,1,...,2\Lambda \},\:
    T_i^z \in \{0, 1,...,4\Lambda \}, \: Y_i = \{0,1,...,6\Lambda  \},
\end{equation}
and
\begin{equation}
    \hat{T}_i^{\pm} \mapsto \sum_{T_i = 0}^{2\Lambda} \ketbra{T_i \pm 1}{T_i},
    \hat{T}_i^{z \pm} \mapsto \sum_{T_i^z = 0}^{4\Lambda} \ketbra{T_i^z \pm 1}{T_i^z}, 
    \hat{Y}_i^{\pm} \mapsto \sum_{Y_i = 0}^{6\Lambda}\ketbra{Y_i \pm 1}{Y_i}.
    \label{eq:SU3_inc_dec}
\end{equation}
Thus, if the number of qubits required for each of $\ket{p}$ and $\ket{q}$ is $\eta = \log(\Lambda+1)$, then that for $\ket{T_i}$, $\ket{T_i^z}$ and $\ket{Y_i}$ are $\eta + 1$, $\eta + 2$ and $\lceil \log(6 \Lambda+1) \rceil = \eta + 3$. Since the quantum numbers have been scaled, we have to apply the inverse of (\ref{eq:SU3_shift}), i.e.,
\begin{align}
\label{eq:SU3_inv_shift}
    T_i \mapsto \frac{T_i}{2},\: T_i^z \mapsto \frac{T_i^z}{2}-\Lambda,\: Y_i \mapsto \frac{Y_i}{3}-\Lambda,
\end{align}
before evaluating the Clebsch-Gordan coefficients in (\ref{eq:SU3_CG_op}).

Equipped with all the necessary operator definitions and qubit register structure, we now decompose the Hamiltonian into separate parts. Specifically, just as in SU(2), we have for SU(3) LGT
\begin{equation}
\hat{H} = \sum_{\vec{n}}\left[\hat{D}^{(M)}_{\vec{n}} + \hat{D}^{(E)}_{\vec{n}} + \hat{T}^{(K)}_{\vec{n}} + \hat{L}^{(B)}_{\vec{n}} \right],
\end{equation}
where
\begin{align}
    \hat{D}_{\vec{n}}^{(M)} &= \frac{m}{2}(-1)^{\vec{n}} (\hat{Z}_1(\vec{n})+\hat{Z}_2({\vec{n}})+\hat{Z}_3({\vec{n}})), \\
    \hat{D}_{\vec{n}}^{(E)} &=\frac{g^2}{2a^{d-2}} \sum_{l=1}^{d}\hat{E}^2(\vec{n},l)
\end{align}
are diagonal operators, where $\hat{Z}_i(\vec{n})$, with $i\in\{1,2,3\}$, are Pauli-$z$ operators that act on the fermion of color $i$ at site $\vec{n}$, and $(-1)^{\vec{n}}$, is either $+1$ or $-1$ depending on whether $\vec{n}$ is a fermion or anti-fermion site, respectively, reflective of the use of staggered fermions \cite{kogut1975hamiltonian}, and
\begin{equation}
\hat{T}_{\vec{n}}^{(K)} = \frac{1}{2a}\sum_{l=1}^{d}\sum_{\alpha,\beta = 1}^{3}[\hat{U}_{\alpha\beta}(\vec{n},l) \hat{\sigma}^{-}_{\alpha}(\vec{n})\hat{\sigma}^{+}_{\beta}(\vec{n}+\hat{l})\hat{\zeta}_{\alpha \beta,\vec{n},l}+h.c.]
\end{equation}
is an off-diagonal operator, which corresponds to kinetic Hamiltonian. The operators $\hat{\sigma}_{\alpha}^{\pm}(\vec{n})$ are Pauli raising and lowering operators on the fermion of color $\alpha$ at site $\vec{n}$. Further, the operators $\hat{\zeta}_{\alpha \beta,\vec{n},l}$ are tensor products of $\hat{Z}$, which arise from the JW transformation and have an additional color-dependence when compared to the U(1) case. If we consider a $d$-dimensional $L^d-$site lattice, the length of each $\hat{\zeta}_{\alpha \beta,\vec{n},l}$ is $O((3L)^{d-1})$. For brevity, we suppress the $\hat{\zeta}_{\alpha \beta,\vec{n},l}$ operators in the remaining part of the section. The second off-diagonal operator that corresponds to the magnetic Hamiltonian is given by
\begin{align}
\hat{L}^{(B)}_{\vec{n}} &= -\frac{1}{2a^{4-d}g^2}\sum_{i=1}^{d} \sum_{j\neq i; j=1}^{d} \sum_{\alpha, \beta, \delta, \gamma =1}^{3} (\hat{U}_{\alpha \beta}(\vec{n},i)\hat{U}_{ \beta \gamma}(\vec{n}+\hat{i},j)\hat{U}_{\gamma \delta}^{\dag}(\vec{n}+\hat{j},i)\hat{U}_{\delta \alpha}^{\dag}(\vec{n},j) + h.c.).
\end{align}

Assuming we use Suzuki-Trotter formula \cite{suzuki1991general} as our simulation method, each Trotter terms to be implemented then are of the form
$
e^{i\hat{D}_{\vec{n}}^{(M)}t},
e^{i\hat{D}_{\vec{n}}^{(E)}t},
e^{i\hat{T}_{\vec{n}}^{(K)}t},
e^{i\hat{L}_{\vec{n}}^{(B)}t},
$
where $t$ is a sufficiently small number to ensure the Trotter error incurred is within a pre-specified tolerance.
In the remaining part of this subsection, we discuss circuit synthesis for each of the four Trotter terms.

\subsubsection{Mass term \texorpdfstring{$e^{i\hat{D}_{\vec{n}}^{(M)}t}$}{}}
\label{sec:SU3_mass}

The implementation of this term is straightforward. Three single-qubit $R_{z}(\theta) = \exp(-i\theta\hat{Z}/2)$ gate, where $\theta = -m(-1)^{\vec{n}}t$, applied to the three qubits, which correspond to the three fermions at site $\vec{n}$, in the site register suffices. As in the implementation of U(1) mass term, we once again use the weight-sum trick \cite{gidney2018halving,nam2019low}, except for SU($3$), the number of same-angle $R_z$ gates increased by a factor of three. Again, briefly, if the original subcircuit applies the same angle $R_z$ gates on $p$ qubits simultaneously, we can reduce the number of $R_z$ gates to $\lfloor\log(p)+1 \rfloor$, while incurring $p-{\rm Weight}(p)$ ancilla qubits and at most $4(p - {\rm Weight}(p))$ T gates. For a $d$-dimensional lattice with $L^d$ lattice sites, $p = 3L^d$.

\subsubsection{Electric term \texorpdfstring{$e^{i\hat{D}_{\vec{n}}^{(E)}t}$}{}}
\label{sec:SU3_electric}

Here, we present a method to implement the electric term, $\hat{D}_{\vec{n}}^{(E)}$. It is a sum of $d$ commuting terms, and thus, its evolution can be implemented exactly as a product of $d$ sub-evolutions,
\begin{equation}
    e^{i\hat{D}_{\vec{n}}^{(E)}t} = \prod_{l=1}^{d}e^{i \frac{g^2 t}{2a^{d-2}} \hat{E}^2(\vec{n},l)}.
\end{equation}
Without loss of generality, we restrict our discussion to one sub-evolution. For notational convenience, we drop the link location index. Since the eigenvalue equation in (\ref{eq:SU3_elec}) only depends on the subregisters $\ket{p}$ and $\ket{q}$, we implement the operator $e^{i \frac{g^2 t}{2a^{d-2}} \hat{E}^2}$ according to the eigenvalue equation,
\begin{align}
    e^{i\frac{g^2 t}{2a^{d-2}} \hat{E}^2} \ket{p}\ket{q} &= e^{i\frac{g^2 t}{6a^{d-2}}[p^2+q^2+pq+3(p+q)]} \ket{p}\ket{q} \nonumber \\
    &= e^{i\frac{g^2 t}{6a^{d-2}}[(p+q)(p+q+3)-pq]} \ket{p}\ket{q}.
    \label{eq:SU3_elec_eig}
\end{align}
Similar to the SU(2) electric term implementation, we first compute $(p+q)(p+q+3)-pq$ into an ancilla register, then induce the phase on all links in parallel, using the weight-sum trick, and finally uncomputing the ancilla register. The initial arithmetic operations create $\ket{p+q}$ and $\ket{-pq}$ with an $\eta-$bit adder and multiplier, respectively, incurring $4(5\eta-3\lfloor \log(\eta)\rfloor -4)$ and $4(\eta -1)(12\eta-3\lfloor \log(\eta)\rfloor-12)+4$ T gates, respectively \cite{draper2006logarithmic,shaw2020quantum}. Next, we compute $\ket{p+q+3}$ into the ancilla register by copying $\ket{p+q}$ and then adding $3$ with a $2\eta$-bit adder, incurring $4(2\eta-2)$ T gates. Finally, we compute $(p+q)(p+q+3)-pq$ with a multiplier and $(2\eta + 3)$-bit adder, incurring $4\eta(12\eta - 3\lfloor \log(\eta+2)\rfloor + 10)+4$ and $4(10\eta -3\lfloor \log(2\eta+3)\rfloor + 11)$ T gates, respectively. In total, the number of T gates required for computing and uncomputing of $(p+q)(p+q+3)-pq$ on all the links is $8dL^d[(8\eta - 8)+(5\eta - 3\lfloor\log(\eta)\rfloor-4)+(\eta - 1)(12\eta - 3\lfloor\log(\eta)\rfloor - 12)+1+\eta(12\eta-3\lfloor\log(\eta+2)\rfloor+10)+1+(10\eta-3\lfloor\log(2\eta+3)\rfloor+11)]$. The number of ancilla qubits used to store $\ket{p+q}$, $\ket{-pq}$, $\ket{p+q+3}$, $\ket{(p+q)(p+q+3)}$ and $\ket{(p+q)(p+q+3)-pq}$ for all the links is $dL^d(8\eta + 10)$. The number of reusable workspace ancilla qubits is $3(\eta+2)-\lfloor \log(\eta+2)\rfloor$, as required by the most expensive arithmetic step, the last multiplier.

We now discuss the phase induction. The correct phase can be induced by applying $R_z(2^k \theta)$, where $\theta = \frac{g^2 t}{6a^{d-2}}$, on the $k$th qubit of the $(2\eta + 4)-$bit ancilla state, $\ket{(p+q)(p+q+3)-pq}$. Hence, there are $(2\eta+4)$ sets of $dL^d$ same-angle $R_z$ rotations to implement, where each set can be effected using the weight sum trick. Once again, we first compute Weight$(dL^d)$ into the ancilla register, incurring $4(dL^d - \text{Weight}(dL^d))$ T gates and $dL^d - \text{Weight}(dL^d)$ ancilla qubits, and then, applying $\lfloor \log(dL^d) + 1 \rfloor$ $R_z$ gates to the ancilla register to induce the right phase. Thus, the number of $R_z$ gates required to impart the phase is $(2\eta+4)\lfloor\log(dL^d)+1\rfloor$.

As in the cases of U($1$) and SU($2$) electric terms, for simulations with a fixed $t$, $d$ and $g^2$, where $a$ can be chosen such that $\frac{g^2 t}{6a^{d-2}} = \frac{\pi}{2^M}$ with a sufficiently large $M>1$. Then, the electric evolution can be implemented by
\begin{equation}
    \ket{p}\ket{q} \mapsto e^{-i \frac{\pi}{2^M}[(p+q)(p+q+3)-pq]} \ket{p}\ket{q}.
\end{equation}
First, we compute $[(p+q)(p+q+3)-pq]$ into an ancilla register, then impart the phase by a phase gradient operation, consisting of an $M-$bit addition on the phase gradient state in (\ref{eq:phgradstate}), on the ancilla register, and finally uncompute the ancilla register. The arithmetic operations are carried out one link at a time, and hence, the number of storage ancilla qubits is $8\eta + 10$. The phase gradient operations on all links cost $4dL^d\log(\frac{6a^{d-2}\pi}{g^2 t})+O(dL^d)$ T gates in total.

\subsubsection{Kinetic term \texorpdfstring{$e^{i\hat{T}_{\vec{n}}^{(K)}t}$}{}}
\label{sec:SU3_kin}

In order to facilitate the circuit syntheses of terms involving the $\hat{U}_{\alpha \beta}$ operators, we rewrite them as
\begin{align}
    \hat{U}_{11} &= \sum_{p,q=0}^{\Lambda} \sum_{T_L,T_R=0}^{2\Lambda}\sum_{T_L^z,T_R^z=0}^{4\Lambda}\sum_{Y_L,Y_R=0}^{6\Lambda}\sum_{\Delta T_L, \Delta T_R \in \{-1,1 \} } \sum_{(\Delta p, \Delta q)} \nonumber \\
    &\quad  f_{11}(p,q,T_L,T_L^z,Y_L,T_R,T_R^z,Y_R,\Delta p,\Delta q,\Delta T_L,\Delta T_R) \nonumber \\
    &\quad |p+\Delta p, q+\Delta q, T_L + \Delta T_L, T_L^z+1, Y_L+1,\nonumber \\
    &\quad T_R+\Delta T_R, T_R^z+1,Y_R+1\rangle
    \bra{p,q,T_L,T_L^z,Y_L,T_R,T_R^z,Y_R} \\
    \hat{U}_{12} &= \sum_{p,q=0}^{\Lambda} \sum_{T_L,T_R=0}^{2\Lambda}\sum_{T_L^z,T_R^z=0}^{4\Lambda}\sum_{Y_L,Y_R=0}^{6\Lambda}\sum_{\Delta T_L, \Delta T_R \in \{-1,1 \} } \sum_{(\Delta p, \Delta q)} \nonumber \\
    &\quad  f_{12}(p,q,T_L,T_L^z,Y_L,T_R,T_R^z,Y_R,\Delta p,\Delta q,\Delta T_L,\Delta T_R) \nonumber \\
    &\quad  |p+\Delta p, q+\Delta q, T_L + \Delta T_L, T_L^z+1, Y_L+1,\nonumber \\
    &\quad T_R+\Delta T_R, T_R^z-1,Y_R+1\rangle
    \bra{p,q,T_L,T_L^z,Y_L,T_R,T_R^z,Y_R} \\
    \hat{U}_{13} &= \sum_{p,q=0}^{\Lambda} \sum_{T_L,T_R=0}^{2\Lambda}\sum_{T_L^z,T_R^z=0}^{4\Lambda}\sum_{Y_L,Y_R=0}^{6\Lambda}\sum_{\Delta T_L \in \{-1,1 \} } \sum_{(\Delta p, \Delta q)} \nonumber \\
    &\quad  f_{13}(p,q,T_L,T_L^z,Y_L,T_R,T_R^z,Y_R,\Delta p,\Delta q,\Delta T_L,\Delta T_R=0) \nonumber \\
    &\quad  |p+\Delta p, q+\Delta q, T_L + \Delta T_L, T_L^z+1, Y_L+1,\nonumber \\
    &\quad T_R, T_R^z,Y_R-2\rangle
    \bra{p,q,T_L,T_L^z,Y_L,T_R,T_R^z,Y_R} \\
    \hat{U}_{21} &= \sum_{p,q=0}^{\Lambda} \sum_{T_L,T_R=0}^{2\Lambda}\sum_{T_L^z,T_R^z=0}^{4\Lambda}\sum_{Y_L,Y_R=0}^{6\Lambda}\sum_{\Delta T_L, \Delta T_R \in \{-1,1 \} } \sum_{(\Delta p, \Delta q)} \nonumber \\
    &\quad  f_{21}(p,q,T_L,T_L^z,Y_L,T_R,T_R^z,Y_R,\Delta p,\Delta q,\Delta T_L,\Delta T_R) \nonumber \\
    &\quad  |p+\Delta p, q+\Delta q, T_L + \Delta T_L, T_L^z-1, Y_L+1,\nonumber \\
    &\quad T_R+\Delta T_R, T_R^z+1,Y_R+1\rangle
    \bra{p,q,T_L,T_L^z,Y_L,T_R,T_R^z,Y_R} \\
    \hat{U}_{22} &= \sum_{p,q=0}^{\Lambda} \sum_{T_L,T_R=0}^{2\Lambda}\sum_{T_L^z,T_R^z=0}^{4\Lambda}\sum_{Y_L,Y_R=0}^{6\Lambda}\sum_{\Delta T_L, \Delta T_R \in \{-1,1 \} } \sum_{(\Delta p, \Delta q)} \nonumber \\
    &\quad  f_{22}(p,q,T_L,T_L^z,Y_L,T_R,T_R^z,Y_R,\Delta p,\Delta q,\Delta T_L,\Delta T_R) \nonumber \\
    &\quad  |p+\Delta p, q+\Delta q, T_L + \Delta T_L, T_L^z-1, Y_L+1,\nonumber \\
    &\quad T_R+\Delta T_R, T_R^z-1,Y_R+1\rangle
    \bra{p,q,T_L,T_L^z,Y_L,T_R,T_R^z,Y_R} \\
    \hat{U}_{23} &= \sum_{p,q=0}^{\Lambda} \sum_{T_L,T_R=0}^{2\Lambda}\sum_{T_L^z,T_R^z=0}^{4\Lambda}\sum_{Y_L,Y_R=0}^{6\Lambda}\sum_{\Delta T_L \in \{-1,1 \} } \sum_{(\Delta p, \Delta q)} \nonumber \\
    &\quad  f_{23}(p,q,T_L,T_L^z,Y_L,T_R,T_R^z,Y_R,\Delta p,\Delta q,\Delta T_L,\Delta T_R=0) \nonumber \\
    &\quad  |p+\Delta p, q+\Delta q, T_L + \Delta T_L, T_L^z-1, Y_L+1,\nonumber \\
    &\quad T_R, T_R^z,Y_R-2\rangle
    \bra{p,q,T_L,T_L^z,Y_L,T_R,T_R^z,Y_R}\\
    \hat{U}_{31} &= \sum_{p,q=0}^{\Lambda} \sum_{T_L,T_R=0}^{2\Lambda}\sum_{T_L^z,T_R^z=0}^{4\Lambda}\sum_{Y_L,Y_R=0}^{6\Lambda}\sum_{\Delta T_R \in \{-1,1 \} } \sum_{(\Delta p, \Delta q)} \nonumber \\
    &\quad  f_{31}(p,q,T_L,T_L^z,Y_L,T_R,T_R^z,Y_R,\Delta p,\Delta q,\Delta T_L=0,\Delta T_R) \nonumber \\
    &\quad  |p+\Delta p, q+\Delta q, T_L, T_L^z, Y_L-2,T_R+\Delta T_R, \nonumber \\
    &\quad T_R^z+1,Y_R+1\rangle
    \bra{p,q,T_L,T_L^z,Y_L,T_R,T_R^z,Y_R} \\
    \hat{U}_{32} &= \sum_{p,q=0}^{\Lambda} \sum_{T_L,T_R=0}^{2\Lambda}\sum_{T_L^z,T_R^z=0}^{4\Lambda}\sum_{Y_L,Y_R=0}^{6\Lambda}\sum_{\Delta T_R \in \{-1,1 \} } \sum_{(\Delta p, \Delta q)} \nonumber \\
    &\quad  f_{32}(p,q,T_L,T_L^z,Y_L,T_R,T_R^z,Y_R,\Delta p,\Delta q,\Delta T_L=0,\Delta T_R) \nonumber \\
    &\quad  |p+\Delta p, q+\Delta q, T_L, T_L^z, Y_L-2,T_R+\Delta T_R, \nonumber \\
    &\quad T_R^z-1,Y_R+1\rangle \bra{p,q,T_L,T_L^z,Y_L,T_R,T_R^z,Y_R}\\
    \hat{U}_{33} &= \sum_{p,q=0}^{\Lambda} \sum_{T_L,T_R=0}^{2\Lambda}\sum_{T_L^z,T_R^z=0}^{4\Lambda}\sum_{Y_L,Y_R=0}^{6\Lambda} \sum_{(\Delta p, \Delta q)} \nonumber \\
    &\quad  f_{33}(p,q,T_L,T_L^z,Y_L,T_R,T_R^z,Y_R,\Delta p,\Delta q,\Delta T_L=0,\Delta T_R=0) \nonumber \\
    &\quad  |p+\Delta p, q+\Delta q, T_L, T_L^z, Y_L-2, T_R, T_R^z,Y_R-2\rangle\nonumber \\
    &\quad \bra{p,q,T_L,T_L^z,Y_L,T_R,T_R^z,Y_R},
\end{align}
where the tuple $(\Delta p, \Delta q)$ is summed over the elements in $\{ (1,0),(-1,1),(0,1) \}$. Hereafter, we will omit some of the arguments of $f_{\alpha \beta}$ for brevity, whenever the context is clear. As in SU($2$), we split the sums over all the quantum numbers into series of even and odd quantum numbers, and define the operators $\hat{U}_{\alpha \beta}$ using the diagonal operators,
\begin{align}
    &\quad \hat{D}_{\alpha \beta}(p,q,T_L,T_L^z,Y_L,T_R,T_R^z,Y_R,\Delta p,\Delta q,\Delta T_L,\Delta T_R) \nonumber \\
    &= \sum_{\substack{p_{\eta-1}...p_1 \\ q_{\eta-1}...q_1} = 0}^1 \sum_{\substack{T_{L;\eta}...T_{L;1} \\ T_{R;\eta}...T_{R;1}} = 0}^1
    \sum_{\substack{T^z_{L;\eta+1}...T^z_{L;1} \\ T^z_{R;\eta+1}...T^z_{R;1}} = 0}^1
    \sum_{\substack{Y_{L;\eta+2}...Y_{L;2}Y_{L;k} \\ Y_{R;\eta+2}...Y_{R;2}Y_{R;k}} = 0}^1 \nonumber \\
    & \quad f_{\alpha \beta}(p,q,T_L,T_L^z,Y_L,T_R,T_R^z,Y_R,\Delta p,\Delta q,\Delta T_L,\Delta T_R)  \nonumber \\
    &\quad \ketbra{p_{\eta-1}...p_1}{p_{\eta-1}...p_1}
    \otimes \ketbra{q_{\eta-1}...q_1}{q_{\eta-1}...q_1} \nonumber \\
    &\quad \otimes \ketbra{T_{L;\eta}...T_{L;1}}{T_{L;\eta}...T_{L;1}}\otimes \ketbra{T^z_{L;\eta+1}...T^z_{L;1}}{T^z_{L;\eta+1}...T^z_{L;1}}\nonumber \\
    &\quad \otimes \ketbra{Y_{L;\eta+2}...Y_{L;2}Y_{L;k}}{Y_{L;\eta+2}...Y_{L;2}Y_{L;k}}\otimes \ketbra{T_{R;\eta}...T_{R;1}}{T_{R;\eta}...T_{R;1}} \nonumber \\
    &\quad \otimes \ketbra{T^z_{R;\eta+1}...T^z_{R;1}}{T^z_{R;\eta+1}...T^z_{R;1}}\otimes \ketbra{Y_{R;\eta+2}...Y_{R;2}Y_{R;k}}{Y_{R;\eta+2}...Y_{R;2}Y_{R;k}},
    \label{eq:SU3_Diag_op}
\end{align}
where $\alpha, \beta \in \{1,2,3 \}$, $\Delta p,\Delta q,\Delta T_L,\Delta T_R \in \{1,0,-1 \}$, and
\begin{equation}
    Y_{L/R;k} = \begin{cases}
        Y_{L/R;1} \mbox{ if }\alpha/\beta = 1,2,\\
        Y_{L/R;0} \mbox{ if }\alpha/\beta = 3.
    \end{cases}
    \label{eq:SU3_Ybits}
\end{equation}
Because we consider all possible bit-strings that consist the zeroth digits of $p,q,T_L,T_L^z,T_R,T_R^z$ and $k$th digits of $Y_L,Y_R$ separately one at a time, these bit-strings are classically known, and the remaining digits are obtained from the qubits, and thus $f_{\alpha \beta}$ can be evaluated on a quantum computer. In particular, $f_{\alpha \beta}$, whose formula can be straightforwardly obtained from (\ref{eq:SU3_U_enc}), are products of the Clebsch-Gordan coefficients in Table \ref{tb:CG_SU3}, and normalization factors in (\ref{eq:SU3_N}). For brevity, we hereafter write the operator and diagonal elements in (\ref{eq:SU3_Diag_op}) as $\hat{D}_{\alpha \beta}(\Delta p,\Delta q,\Delta T_L,\Delta T_R)$, whenever the context is clear. Next, similar to the SU($2$) kinetic term, we express the operators $\hat{U}_{\alpha \beta}$ in terms of the diagonal operators in (\ref{eq:SU3_Diag_op}), Pauli ladder operators, which act on the zeroth qubits of $p,q,T_L,T_L^z,T_R,T_R^z$ and $k$th qubits of $Y_L,Y_R$, and binary incrementers and decrementers acting on the eight subregisters as follows:
\begin{align}
    \hat{U}_{\alpha \beta}+h.c. &= \sum_{n_1,n_2,...,n_8=0}^{1}(\hat{P}^{+})^{n_1} (\hat{Q}^{+})^{n_2}(\hat{T}_L^{+})^{n_3}(\hat{T}_L^{z+})^{n_4}
    (\hat{Y}_L^{+})^{\delta_{\alpha,3}+n_5}(\hat{T}_R^{+})^{n_6}(\hat{T}_R^{z+})^{n_7}(\hat{Y}_R^{+})^{\delta_{\beta,3}+n_8}\nonumber\\
    &\quad [\sum_{(\Delta p,\Delta q)}\sum_{\Delta T_L,\Delta T_R} \hat{D}_{\alpha \beta}{(\Delta p,\Delta q,\Delta T_L,\Delta T_R)}
    \hat{P}_{\alpha \beta}{(\Delta p,\Delta q,\Delta T_L,\Delta T_R)}+h.c.]\nonumber\\
    &\quad (\hat{P}^{-})^{n_1} (\hat{Q}^{-})^{n_2}(\hat{T}_L^{-})^{n_3}(\hat{T}_L^{z-})^{n_4}(\hat{Y}_L^{-})^{\delta_{\alpha,3}+n_5}
    (\hat{T}_R^{-})^{n_6}(\hat{T}_R^{z-})^{n_7}(\hat{Y}_R^{-})^{\delta_{\beta,3}+n_8},
    \label{eq:SU3_U_exp}
\end{align}
where $(\Delta p,\Delta q)\in \{(1,0),(-1,1),(0,1) \}$, $\Delta T_L,\Delta T_R \in \{-1,1\}$ for $\alpha, \beta = 1,2$, respectively, and $\Delta T_L, \Delta T_R=0$ if $\alpha, \beta = 3$. Further,
\begin{equation}
     \hat{P}_{\alpha \beta}{(\Delta p,\Delta q,\Delta T_L,\Delta T_R)} = \otimes_{i=0}^{7}  \hat{\sigma}_i,
     \label{eq:SU3_kin_pauli}
\end{equation}
where the operator $\hat{\sigma}_i \in \{ \hat{\sigma}^+,\hat{\sigma}^-,\hat{I} \}$ depends on $\alpha$, $\beta$, $\Delta p$, $\Delta q$, $\Delta T_L$, and $\Delta T_R$. In particular, $\hat{\sigma}_0,\hat{\sigma}_1,\hat{\sigma}_2,\hat{\sigma}_5=\hat{\sigma}^+, \hat{\sigma}^-, \hat{I}$ if $\Delta p,\Delta q,\Delta T_L,\Delta T_R = 1,-1,0$, respectively. Moreover, $\hat{\sigma}_3\hat{\sigma}_4$ $(\hat{\sigma}_6\hat{\sigma}_7)= \hat{\sigma}^+\hat{\sigma}^+$, $\hat{\sigma}^-\hat{\sigma}^+$, $\hat{I}\hat{\sigma}^+$ if $\alpha$ $(\beta) = 1,2,3$, respectively. Note that when $\hat{\sigma}_i = \hat{I}$, the corresponding gauge-field subregister's incrementer and decrementer cancel out. When $\alpha$ or $\beta = 3$, $\hat{U}_{\alpha \beta}$ lowers $Y_L$ or $Y_R$ by two, respectively. Hence, when $\alpha / \beta = 3$, $\hat{U}_{\alpha \beta}$ has an extra pair of incrementer and decrementer, and the Pauli ladder operators act on the first bit of $Y_{L/R}$.

Using $\alpha,r$ and $\beta,r+1$ to denote two colors $\alpha$, $\beta$ at two sites $r, r+1$, without loss of generality, we write
\begin{align}
    \hat{U}_{\alpha \beta}\hat{\sigma}_{\alpha,r}^{-}\hat{\sigma}_{\beta,r+1}^{+} + h.c. &= \sum_{n_1,n_2,...,n_8=0}^{1}(\hat{P}^{+})^{n_1} (\hat{Q}^{+})^{n_2}(\hat{T}_L^{+})^{n_3}(\hat{T}_L^{z+})^{n_4}
    (\hat{Y}_L^{+})^{n_5}(\hat{T}_R^{+})^{n_6}(\hat{T}_R^{z+})^{n_7}(\hat{Y}_R^{+})^{n_8}\nonumber\\
    &\quad [\sum_{(\Delta p,\Delta q)}\sum_{\Delta T_L,\Delta T_R} \hat{D}_{\alpha \beta}{(\Delta p,\Delta q,\Delta T_L,\Delta T_R)}
    \hat{F}_{\alpha \beta}{(\Delta p,\Delta q,\Delta T_L,\Delta T_R)}]\nonumber\\
    &\quad (\hat{P}^{-})^{n_1} (\hat{Q}^{-})^{n_2}(\hat{T}_L^{-})^{n_3}(\hat{T}_L^{z-})^{n_4}(\hat{Y}_L^{-})^{\delta_{\alpha,3}+n_5}
    (\hat{T}_R^{-})^{n_6}(\hat{T}_R^{z-})^{n_7}(\hat{Y}_R^{-})^{\delta_{\beta,3}+n_8},
    \label{eq:SU3_U_w_f}
\end{align}
where $\hat{F}_{\alpha \beta}= \hat{P}_{\alpha\beta}\hat{\sigma}_{\alpha,r}^{-}\hat{\sigma}_{\beta,r+1}^{+} + h.c.$ act on both the gauge field and fermionic registers, of which the gauge field part is defined via $\hat{P}_{\alpha \beta}$. Further, for notational convenience, we have dropped, and will hereafter drop the arguments of $\hat{D}_{\alpha \beta}$, whenever the context is clear, since they are implied by those of $\hat{P}_{\alpha \beta}$ or $\hat{F}_{\alpha \beta}$. We approximate the kinetic evolution $ e^{it \hat{T}_r^{(K)} }$ as
\begin{align}
    &\quad e^{it \hat{T}_{r}^{(K)}}
    \nonumber \\
    & \approx \prod_{(n_1,n_2,...,n_8)=GC(0)}^{GC(2^8 -1)} (\hat{P}^{+})^{n_1} (\hat{Q}^{+})^{n_2}(\hat{T}_L^{+})^{n_3}(\hat{T}_L^{z+})^{n_4}(\hat{Y}_L^{+})^{n_5} 
    (\hat{T}_R^{+})^{n_6}(\hat{T}_R^{z+})^{n_7}(\hat{Y}_R^{+})^{n_8}\prod_{\alpha=1}^{2+(1\oplus OR(n_3,n_4))}\nonumber \\
    &\quad \prod_{\beta=1}^{2+(1\oplus OR(n_6,n_7))} \{(\hat{Y}_L^{+})^{\delta_{\alpha,3}}
    (\hat{Y}_R^{+})^{\delta_{\beta,3}} [\prod_{\substack{(\Delta p, \Delta q),\\ \Delta T_L, \Delta T_R}} e^{i\frac{t}{2a}\hat{D}_{\alpha \beta}{(\Delta p,\Delta q,\Delta T_L,\Delta T_R)}
    \hat{F}_{\alpha \beta}{(\Delta p,\Delta q,\Delta T_L,\Delta T_R)}}]\nonumber \\
    &\quad(\hat{Y}_L^{-})^{\delta_{\alpha,3}}
    (\hat{Y}_R^{-})^{\delta_{\beta,3}} \}
    (\hat{P}^{-})^{n_1} (\hat{Q}^{-})^{n_2}(\hat{T}_L^{-})^{n_3}(\hat{T}_L^{z-})^{n_4}(\hat{Y}_L^{-})^{n_5}
    (\hat{T}_R^{-})^{n_6}(\hat{T}_R^{z-})^{n_7}(\hat{Y}_R^{-})^{n_8},
    \label{eq:SU3_gray}
\end{align}
where we have minimized the number of binary incrementers and decrementers using the Gray code encoding. Furthermore, the evolution operators with $\alpha=3$, $\beta=3$ are included only when $1\oplus OR(n_3, n_4)=1$, $1\oplus OR(n_6, n_7)=1$, respectively, where $OR(\cdot)$ is the bitwise OR function, and $\oplus$ is the binary addition. This is because if $\alpha/ \beta = 3$, then $\Delta T_{L/R}, \Delta T_{L/R}^z = 0$, meaning that $\ket{T_{L/R,0}},\ket{T_{L/R,0}^{z}}$ are not acted on by Pauli ladder operators. Here we adopt the same implementation strategy for the SU(2) kinetic term. Briefly, the strategy diagonalizes the Pauli ladder operators with CNOT and Hadamard gates, and then implements the diagonalized operators as a phase oracle using either qRAM or quantum arithmetic circuits. The number of oracle queries is determined by the number of terms in each $\hat{U}_{\alpha \beta}$, whereas the T-gate count is decided by the number of binary incrementers and decrementers. Conjugated by the incrementers and decrementers is a product of operators of the form
\begin{equation}
    e^{i\frac{t}{2a}\hat{D}_{\alpha \beta}
    \hat{F}_{\alpha \beta}{(\Delta p,\Delta q,\Delta T_L,\Delta T_R)}}.
    \label{eq:SU3_kin_DF}
\end{equation}
Once again, as in the SU(2) case, we diagonalize the Pauli part $\hat{F}_{\alpha \beta}$ using a CNOT network and two Hadamard gates, mapping (\ref{eq:SU3_kin_DF}) to
\begin{equation}
    e^{i\frac{t}{2a} \hat{D}_{\alpha \beta}\hat{\mathcal{D}}_{\alpha \beta}(\Delta p, \Delta q, \Delta T_L, \Delta T_R)},
    \label{eq:SU3_kin_diag}
\end{equation}
where $\hat{\mathcal{D}}_{\alpha \beta}(\Delta p, \Delta q, \Delta T_L, \Delta T_R)$ refers to the diagonalized Pauli part. As in the SU(2) kinetic term implementation, we implement the diagonalized operator as a phase oracle. Before defining the oracle, we define the following useful notations and function. We denote a set of qubits of the gauge-field registers by $S_0 \equiv \{\ket{p_0}$, $\ket{q_0}$, $\ket{T_{L,0}}$, $\ket{T_{L,0}^z}$, $\ket{Y_{L,k}}$, $\ket{T_{R,0}}$, $\ket{T_{R,0}^z}$, $\ket{Y_{R,k}} \}$, where $Y_{L/R,k}$ are defined in (\ref{eq:SU3_Ybits}), and the fermionic states by $\ket{f_{\alpha,r}}$, $\ket{f_{\beta,r+1}}$. Further, we define
\begin{align}
    B(\hat{\mathcal{D}}_{\alpha \beta}(\Delta p, \Delta q, \Delta T_L, \Delta T_R))=(-1)^{f_{\alpha,r}}f_{\beta,r+1} \prod_{j}b_j,
\end{align}
where $\{ \ket{b_j} \} \subseteq S_0$ is the set of qubits in the gauge-field registers that are acted on nontrivially by $\hat{\mathcal{D}}_{\alpha \beta}(\Delta p, \Delta q, \Delta T_L, \Delta T_R)$. Now we can define the phase oracle used to implement (\ref{eq:SU3_kin_diag}) as
\begin{multline}
    \hat{O}_{\alpha \beta}^{(\Delta p, \Delta q, \Delta T_L, \Delta T_R)} \ket{p_0^\prime}\ket{q_0^\prime}\ket{T_{L,0}^\prime}\ket{T_{L,0}^{z\prime}}
    \ket{Y_{L,k}^\prime}\ket{T_{R,0}^\prime}\ket{T_{R,0}^{z\prime}}\ket{Y_{R,k}^\prime}\ket{f_{\alpha,r}^\prime}\ket{f_{\beta,r+1}^\prime}\\
    \mapsto e^{if_{\alpha \beta}[\frac{t}{2a}B(\hat{\mathcal{D}}_{\alpha \beta}(\Delta p, \Delta q, \Delta T_L, \Delta T_R))]}
    \ket{p_0^\prime}\ket{q_0^\prime}
    \ket{T_{L,0}^\prime}\ket{T_{L,0}^{z\prime}}\ket{Y_{L,k}^\prime}
    \ket{T_{R,0}^\prime}\ket{T_{R,0}^{z\prime}}\ket{Y_{R,k}^\prime}
    \ket{f_{\alpha,r}^\prime}\ket{f_{\beta,r+1}^\prime},
    \label{eq:SU3_kin_oracle}
\end{multline}
where the function $f_{\alpha \beta}$ is defined in (\ref{eq:SU3_Diag_op}), and the primes are used to indicate that the states have been acted on by the diagonalization circuit consisting of CNOT and Hadamard gates. Note that we have chosen $f_{\alpha,r}$ to be the target of the diagonalization circuit for concreteness. Further, we note that the bit-value product, i.e. $f_{\beta,r+1} \prod_{j}b_j$, from the function $B(\cdot)$ in the phase implies that the qubits $\ket{f_{\beta,r+1}}$ and $\ket{b_j} \in S_0$ are the controls of the oracle, since the phase is zero if any of $\ket{f_{\beta,r+1}}$ and $\ket{b_j} \in S_0$ is in the zero state. This phase oracle can be implemented either using qRAM or quantum arithmetic circuits. As in the SU(2) case, we choose to synthesize this oracle directly as a diagonal gate, which imparts the phase $(-1)^{f_{\alpha,r}}f_{\alpha \beta}\frac{t}{2a}$, controlled by $f_{\beta,r+1}, \{ b_j \}$. In particular, for each link, $f_{\alpha \beta}$ can be computed efficiently using fixed point arithmetic circuits \cite{bhaskar2016quantum}, and the phases can be induced using $R_z$ gates. We refer readers to sec.~\ref{subsubsec:oracle_SU3} for the detailed implementation.

We proceed with the cost analysis of the kinetic term. We assume that the oracle incurs the same costs $\mathcal{C}^{(K)}$. We first consider a pair of the nearest-neighbor fermionic sites and the link joining them. The number of binary incrementers and decrementers required is $2^8 + 4\cdot 2 \cdot 2^6 + 4 \cdot 2^4 = 832$, out of which $2^8$ come from the outermost incrementers and decrementers in (\ref{eq:SU3_gray}), $4\cdot 2 \cdot 2^6$ are due to the fact that there are $4$ tuples $(\alpha, \beta)$ with either $\alpha$ or $\beta=3$, each of which leads to $2$ additional $\hat{Y}_{L/R}^{\pm}$ for each of the $2^6$ tuples $(n_1,n_2,...,n_8)$ that satisfy either $OR(n_3,n_4)$ or $OR(n_6,n_7)$, respectively, and $4\cdot 2^4$ is due to the fact that there is a tuple $(\alpha, \beta)$ with $\alpha, \beta=3$, which requires $2$ extra $\hat{Y}_{L}^{\pm}$ and $\hat{Y}_{R}^{\pm}$, for each of the $2^4$ tuples $(n_1,n_2,...,n_8)$ that satisfy $OR(n_3,n_4)$ and $OR(n_6,n_7)$. Without affecting complexity arguments, we assume that all the incrementers or decrementers act on the largest registers of size $\eta+3$ qubits, and hence, each of them costs $4(\eta+1)$ T gates \cite{shaw2020quantum}. In total, the incrementers and decrementers require at most $3328(\eta+1)$ T gates and $\eta+3$ reusable ancilla qubits. Now, we calculate the number of oracle queries needed. $12\cdot 2^8$ queries are required for when both $\alpha$ and $\beta \neq 3$, since in these cases, there are twelve combinations of $(\Delta p, \Delta q), \Delta T_L, \Delta T_R$ for each $(n_1, n_2, ..., n_8)$. When either $\alpha$ or $\beta=3$, $6 \cdot 2^6$ queries are needed because there are six combinations of $(\Delta p, \Delta q), \Delta T_L, \Delta T_R$ for each $(n_1, n_2, ..., n_8)$ satisfying one of the two conditions, $1\oplus OR(n_3, n_4)=1$ and $1\oplus OR(n_6, n_7)=1$. When both $\alpha$ and $\beta=3$, $3 \cdot 2^4$ queries are needed because there are three combinations of $(\Delta p, \Delta q), \Delta T_L, \Delta T_R$ for each $(n_1, n_2, ..., n_8)$ satisfying both $1\oplus OR(n_3, n_4)=1$ and $1\oplus OR(n_6, n_7)=1$. Therefore, the number of oracle queries is $4\cdot 12\cdot 2^8 + 4\cdot6 \cdot 2^6 + 3 \cdot 2^4=13872$. Finally, we multiply both the T-gate count and the number of oracle queries by the number of links $dL^d$ in order to obtain the costs of the entire kinetic term.

\subsubsection{Magnetic term \texorpdfstring{$e^{i\hat{L}_{\vec{n}}^{(B)}t}$}{}}
\label{sec:SU3_mag}
Here, we provide the implementation of the magnetic term, which is similar to that of the SU(2) magnetic term. Once again, we drop the location indices for brevity. The magnetic term for a single plaquette is given by
\begin{align}
    \hat{L}_{r}^{(B)}=-\frac{1}{2a^{4-d}g^2}\sum_{\alpha \beta \delta \gamma =1}^{3} (\hat{U}_{\alpha \beta}\hat{U}_{ \beta \gamma}\hat{U}_{\gamma \delta}^{\dag}\hat{U}_{\delta \alpha}^{\dag} + h.c.),
    \label{eq:SU3_plaq}
\end{align}
where $\hat{U}_{\alpha \beta}$ is defined in (\ref{eq:SU3_U_exp}). We now write express a single term in (\ref{eq:SU3_plaq}) as
\begin{align}
    \hat{U}_{\alpha \beta}\hat{U}_{\beta\gamma}\hat{U}_{\gamma \delta}^\dag \hat{U}_{\delta \alpha}^\dag &= \sum_{n_1,n_2,...,n_{32}=0}^{1}(\hat{P}_{1}^{+})^{n_1} (\hat{Q}_{1}^{+})^{n_2}(\hat{T}_{L,1}^{+})^{n_3}(\hat{T}_{L,1}^{z+})^{n_4}
    (\hat{Y}_{L,1}^{+})^{\delta_{\alpha,3}+n_5}...(\hat{T}_{R,4}^{+})^{n_{30}}(\hat{T}_{R,4}^{z+})^{n_{31}}\nonumber\\
    &\quad (\hat{Y}_{R,4}^{+})^{\delta_{\alpha,3}+n_{32}}\left[\sum_{\vec{\theta}_1,\vec{\theta}_2,\vec{\theta}_3,\vec{\theta}_4} \hat{D}_{\alpha \beta \gamma \delta}{(\vec{\theta}_1,\vec{\theta}_2,\vec{\theta}_3,\vec{\theta}_4)}
    \hat{P}_{\alpha \beta \gamma \delta}{(\vec{\theta}_1,\vec{\theta}_2,\vec{\theta}_3,\vec{\theta}_4)} \right]\nonumber\\
    &\quad (\hat{P}^{-}_{1})^{n_1} (\hat{Q}_{1}^{-})^{n_2}(\hat{T}_{L,1}^{-})^{n_3}(\hat{T}_{L,1}^{z-})^{n_4}(\hat{Y}_{L,1}^{-})^{\delta_{\alpha,3}+n_5}...
    (\hat{T}_{R,4}^{-})^{n_{30}}(\hat{T}_{R,4}^{z-})^{n_{31}}(\hat{Y}_{R,4}^{-})^{\delta_{\alpha,3}+n_{32}},
    \label{eq:SU3_mag_plaq}
\end{align}
where the subindices $i=1,2,3,4$ denote the four links around a plaquette, $\vec{\theta}_{i}$ are the parameters $(\Delta p, \Delta q, \Delta T_L, \Delta T_R)$ for link $i$, and the definitions of $\hat{D}_{\alpha \beta \gamma \delta}$ and $\hat{P}_{\alpha \beta \gamma \delta}$ are given by
\begin{equation}
    \hat{D}_{\alpha \beta \gamma \delta}(\vec{\theta}_1,\vec{\theta}_2,\vec{\theta}_3,\vec{\theta}_4) = \hat{D}_{\alpha \beta}(\vec{\theta}_{1})\hat{D}_{\beta \gamma}(\vec{\theta}_{2})\hat{D}_{\gamma \delta}(\vec{\theta}_{3})\hat{D}_{\delta \alpha}(\vec{\theta}_{4}),
    \label{eq:SU3_Dabcd}
\end{equation}
where $\hat{D}_{\alpha \beta}$ is defined in (\ref{eq:SU3_Diag_op}), and
\begin{equation}
    \hat{P}_{\alpha \beta \gamma \delta}(\vec{\theta}_1,\vec{\theta}_2,\vec{\theta}_3,\vec{\theta}_4) = \hat{P}_{\alpha \beta}(\vec{\theta}_{1})\hat{P}_{\beta \gamma}(\vec{\theta}_{2})\hat{P}_{\gamma \delta}(\vec{\theta}_{3})\hat{P}_{\delta \alpha}(\vec{\theta}_{4})+h.c.,
\end{equation}
where $\hat{P}_{\alpha \beta}$ is a Pauli operator of the form $\otimes_i \hat{\sigma}_i^\pm $, which acts on at most eight qubits, as defined in (\ref{eq:SU3_kin_pauli}). Similar to the kinetic term, we can minimize the number of binary incrementers and decrementers in the implementation of $e^{i\hat{L}_r^{(B)}t}$ using the Gray code ordering, i.e.,
\begin{align}
    &\quad e^{i\hat{L}_r^{(B)}t} \nonumber \\
    &\approx \prod_{(n_1,n_2,...,n_{32})=GC(0)}^{GC(2^{32} - 1)}(\hat{P}_{1}^{+})^{n_1} (\hat{Q}_{1}^{+})^{n_2}(\hat{T}_{L,1}^{+})^{n_3}(\hat{T}_{L,1}^{z+})^{n_4}
    (\hat{Y}_{L,1}^{+})^{n_5}...(\hat{T}_{R,4}^{+})^{n_{30}}(\hat{T}_{R,4}^{z+})^{n_{31}}(\hat{Y}_{R,4}^{+})^{n_{32}} \nonumber \\
    &\quad  \prod_{\alpha=1}^{N(n_{30},n_{31},n_3,n_4)}\prod_{\beta=1}^{N(n_{6},n_{7},n_{11},n_{12})}\prod_{\gamma=1}^{N(n_{14},n_{15},n_{19},n_{20})} \prod_{\delta=1}^{N(n_{22},n_{23},n_{27},n_{28})} \nonumber \\
    &\quad \{ (\hat{Y}_{L,1}^{+})^{\delta_{\alpha,3}}(\hat{Y}_{R,1}^{+})^{\delta_{\beta,3}}(\hat{Y}_{L,2}^{+})^{\delta_{\beta,3}}(\hat{Y}_{R,2}^{+})^{\delta_{\gamma,3}}(\hat{Y}_{L,3}^{+})^{\delta_{\gamma,3}}(\hat{Y}_{R,3}^{+})^{\delta_{\delta,3}}(\hat{Y}_{L,4}^{+})^{\delta_{\delta,3}}(\hat{Y}_{R,4}^{+})^{\delta_{\alpha,3}} \nonumber \\
    &\quad \bigg[\prod_{\vec{\theta}_{1},\vec{\theta}_{2},\vec{\theta}_{3},\vec{\theta}_{4}}  e^{i\frac{-t}{2a^{4-d}g^2} \hat{D}_{\alpha \beta \gamma \delta} \hat{P}_{\alpha \beta \gamma \delta}{(\vec{\theta}_1,\vec{\theta}_2,\vec{\theta}_3,\vec{\theta}_4)}}\bigg] \nonumber \\
    &\quad (\hat{Y}_{L,1}^{-})^{\delta_{\alpha,3}}(\hat{Y}_{R,1}^{-})^{\delta_{\beta,3}}(\hat{Y}_{L,2}^{-})^{\delta_{\beta,3}}(\hat{Y}_{R,2}^{-})^{\delta_{\gamma,3}}(\hat{Y}_{L,3}^{-})^{\delta_{\gamma,3}}(\hat{Y}_{R,3}^{-})^{\delta_{\delta,3}}(\hat{Y}_{L,4}^{-})^{\delta_{\delta,3}}(\hat{Y}_{R,4}^{-})^{\delta_{\alpha,3}} \} \nonumber \\
    &\quad (\hat{P}^{-}_{1})^{n_1} (\hat{Q}_{1}^{-})^{n_2}(\hat{T}_{L,1}^{-})^{n_3}(\hat{T}_{L,1}^{z-})^{n_4}(\hat{Y}_{L,1}^{-})^{n_5}...
    (\hat{T}_{R,4}^{-})^{n_{30}}(\hat{T}_{R,4}^{z-})^{n_{31}}(\hat{Y}_{R,4}^{-})^{n_{32}},
    \label{eq:SU3_mag_trot}
\end{align}
where the function $N(n_i,n_j,n_k,n_l)=2+(1\oplus OR(OR(n_i,n_j),OR(n_k,n_l)))$ evaluates to $3$ if and only if all of $n_i, n_j, n_k, n_l=0$. Since the operator $\hat{P}_{\alpha \beta \gamma \delta}$ is a Pauli operator of the form $\otimes_i \hat{\sigma}_i^\pm + h.c.$, $e^{it\frac{-1}{2a^{4-d}g^2}\hat{D}_{\alpha \beta \gamma \delta}\hat{P}_{\alpha \beta \gamma \delta}}$ can be diagonalized efficiently using CNOT and Hadamard gates into an operator of the form
\begin{equation}
    e^{it\frac{-1}{2a^{4-d}g^2}\hat{D}_{\alpha \beta \gamma \delta}\hat{\mathcal{D}}_{\alpha \beta \gamma \delta}(\vec{\theta}_1,\vec{\theta}_2,\vec{\theta}_3,\vec{\theta}_4)},
\end{equation}
where $\hat{\mathcal{D}}_{\alpha \beta \gamma \delta}$ is the diagonalized $\hat{P}_{\alpha \beta \gamma \delta}$. As in the case of SU($2$), we implement this diagonal operator, using either qRAM or quantum arithmetic circuits. Before defining the oracle, we define some useful notations. We denote the state of a plaquette as
\begin{equation}
    \otimes_{i=1}^4 \ket{p_i,q_i,{T}_{L,i},{T}_{L,i}^z,{Y}_{L,i},{T}_{R,i},{T}_{R,i}^z, {Y}_{R,i}},
\end{equation}
where the values of $i$ represent the four links around a plaquette. We denote a set of qubits of the subregisters by $S_0^\Box \equiv \{\ket{p_{i,0}}$, $\ket{q_{i,0}}$, $\ket{T_{L,i,0}}$, $\ket{T_{L,i,0}^z}$, $\ket{Y_{L,i,k}}$, $\ket{T_{R,i,0}}$, $\ket{T_{R,i,0}^z}$, $\ket{Y_{R,i,k}} \}_{i=1}^4$, where $\ket{Y_{L/R,i,k}}$ is the $k$th qubit of $\ket{Y_{L/R,i}}$ as defined in (\ref{eq:SU3_Ybits}). Further, we define
\begin{align}
    B(\hat{\mathcal{D}}_{\alpha \beta \gamma \delta}(\vec{\theta}_1,\vec{\theta}_2,\vec{\theta}_3,\vec{\theta}_4))=(-1)^{b_i} \prod_{j}b_j,
\end{align}
where $\{ \ket{b_i}, \ket{b_j} \} \subseteq S_0^\Box$ are acted on nontrivially by $\hat{\mathcal{D}}_{\alpha \beta \gamma \delta}(\vec{\theta}_1,\vec{\theta}_2,\vec{\theta}_3,\vec{\theta}_4)$, and
\begin{equation}
    f_{\alpha \beta \gamma \delta} = f_{\alpha \beta}(\vec{\theta}_1) f_{\beta \gamma}(\vec{\theta}_2)
    f_{\gamma \delta}(\vec{\theta}_3) f_{\delta \alpha}(\vec{\theta}_4).
    \label{eq:SU3_fabcd}
\end{equation}
As such, the phase oracle is defined as
\begin{align}
    &\quad \hat{O}_{\alpha \beta \gamma \delta}^{(\vec{\theta}_1,\vec{\theta}_2,\vec{\theta}_3,\vec{\theta}_4)} \otimes_{i=1}^4 \ket{p_i^\prime,q_i^\prime,{T}_{L,i}^\prime,{T}_{L,i}^{z\prime},{Y}_{L,i}^\prime,{T}_{R,i}^\prime,{T}_{R,i}^{z\prime}, {Y}_{R,i}^\prime} \nonumber \\
    &\mapsto e^{if_{\alpha \beta \gamma \delta}[(\frac{-1}{2a^{4-d}g^2} B(\hat{\mathcal{D}}_{\alpha \beta \gamma \delta}(\vec{\theta}_1,\vec{\theta}_2,\vec{\theta}_3,\vec{\theta}_4)) )]} \otimes_{i=1}^4 \ket{p_i^\prime,q_i^\prime,{T}_{L,i}^\prime,{T}_{L,i}^{z\prime},{Y}_{L,i}^\prime,{T}_{R,i}^\prime,{T}_{R,i}^{z\prime}, {Y}_{R,i}^\prime},
    \label{eq:SU3_mag_oracle}
\end{align}
where the primed states are the states that have been acted on by the diagonalization circuit consisting of CNOT and Hadamard gates. As in the kinetic term implementation, we can implement this diagonal operator as a phase oracle using qRAM, or directly synthesize it using quantum arithmetic circuits. Once again, we compute $\pm f_{\alpha \beta \gamma \delta}$ into the ancilla register, conditioned upon the values of the control bits, and induce the approximate phases using $R_z$ gates. See \ref{subsubsec:oracle_SU3} for the implementation details.

We proceed with the resource analysis of magnetic term. We first focus on a single plaquette. The number of binary incrementers and decrementers outside the braces in (\ref{eq:SU3_mag_trot}) is $2^{32}$, while the number of incrementers and decrementers inside the brackets depends on the values of $\alpha,\beta,\gamma,\delta$. Let $c(n)$ be the number of tuples $(\alpha,\beta,\gamma,\delta)$ with $n$ entries equal to $3$. Then, the number of incrementers and decrementers inside the bracket for a given $n$ is given by $c(n)\cdot 2^{32-4n} \cdot 4 n$. This is so because for every $(\alpha,\beta,\gamma,\delta)$ with $n$ entries equal to $3$, there are $2n$ incrementers and decrementers with exponents equal to one, whereas the rest have exponents equal to zero, and they are iterated over $2^{32-4n}$ tuples $(n_1,n_2,...,n_{32})$ that satisfy the $N(\cdot)$ functions in the upperbound of the products over $\alpha, \beta, \gamma, \delta$. As such, the number of binary incrementers and decrementers is given by $2^{32} + \sum_{n} c(n)\cdot 2^{32-4n} \cdot 4 n = 2^{32} + 32\cdot 2^{28} \cdot 4 + 24\cdot 2^{24} \cdot 8 + 8 \cdot 2^{20} \cdot 12 + 2^{16} \cdot 16 = 41977643008$. We assume without affecting complexity arguments that all the incrementers and decrementers act on the largest register with $\eta+3$ qubits. Then, each incrementer or decrementer costs $4(\eta+1)$ T gates. Thus, the binary incrementers and decrementers required for a single plaquette incurs at most $167910572032(\eta+1)$ T gates, and $\eta+3$ reusable ancilla qubits. Assuming the costs of the oracle, $\mathcal{C}^{(B)}$, are the same for all parameters, we compute the number of oracle queries. Once again, we consider separate cases based on the number of entries, $n$, in $(\alpha, \beta, \gamma, \delta)$ that are equal to $3$ because the number of parameters $\vec{\theta}_i$ and tuples $(n_1, n_2,...,n_{32})$ that are iterated over depend on $n$. For $n=0$, there are $16$ tuples $(\alpha,\beta,\gamma,\delta)$. For each tuple, the pairs $(\alpha,\beta)$, $(\beta,\gamma)$, $(\gamma,\delta)$ and $(\delta,\alpha)$ lead to $12$ $\vec{\theta}_i$ with $i\in \{1,2,3,4\}$, as explained in (\ref{eq:SU3_Diag_op}) and (\ref{eq:SU3_U_exp}), and hence, $12^4$ tuples $(\vec{\theta}_1,\vec{\theta}_2,\vec{\theta}_3,\vec{\theta}_4)$. As such, there are $12^4$ distinct $e^{i\frac{-t}{2a^{4-d}g^2} \hat{D}_{\alpha \beta \gamma \delta} \hat{P}_{\alpha \beta \gamma \delta}{(\vec{\theta}_1,\vec{\theta}_2,\vec{\theta}_3,\vec{\theta}_4)}}$ that need to be applied, which are then iterated over $2^{32}$ tuples $(n_1,n_2,...,n_{32})$. Thus, for $n=0$, $16\cdot (12\cdot 2^8)^4$ oracle queries are needed. For $n=1$, there are $32$ tuples $(\alpha,\beta,\gamma,\delta)$, each of which gives rise to $6^2\cdot 12^2$ tuples $(\vec{\theta}_1,\vec{\theta}_2,\vec{\theta}_3,\vec{\theta}_4)$ and hence, $e^{i\frac{-t}{2a^{4-d}g^2} \hat{D}_{\alpha \beta \gamma \delta} \hat{P}_{\alpha \beta \gamma \delta}{(\vec{\theta}_1,\vec{\theta}_2,\vec{\theta}_3,\vec{\theta}_4)}}$ that are iterated over $2^{28}$ tuples $(\alpha,\beta,\gamma,\delta)$. Thus, for $n=1$, $32\cdot (6\cdot 2^6)^2 \cdot (12 \cdot 2^8)^2$ oracle queries are needed. For $n=2$, there are $16$ tuples $(\alpha,\beta,\gamma,\delta)$ such that one of the pairs $(\alpha,\beta)$, $(\beta,\gamma)$, $(\gamma,\delta)$ and $(\delta,\alpha)$ has both entries equal to $3$, in which case there are $3\cdot 6^2 \cdot 12$ tuples $(\vec{\theta}_1,\vec{\theta}_2,\vec{\theta}_3,\vec{\theta}_4)$. Further, there are $8$ tuples $(\alpha,\beta,\gamma,\delta)$ such that all $(\alpha,\beta)$, $(\beta,\gamma)$, $(\gamma,\delta)$ and $(\delta,\alpha)$ have one entry equal to $3$, in which case there are $6^4$ tuples $(\vec{\theta}_1,\vec{\theta}_2,\vec{\theta}_3,\vec{\theta}_4)$. All $e^{i\frac{-t}{2a^{4-d}g^2} \hat{D}_{\alpha \beta \gamma \delta} \hat{P}_{\alpha \beta \gamma \delta}{(\vec{\theta}_1,\vec{\theta}_2,\vec{\theta}_3,\vec{\theta}_4)}}$ with $n=2$ are iterated over $2^{24}$ tuples $(n_1,n_2,...,n_{32})$. Thus, for $n=2$, $(16\cdot 3  \cdot 6^2 \cdot 12 + 8\cdot 6^4)\cdot 2^{24}$ oracle queries are required. Similarly, we obtain the number of oracle queries for $n=3,4$ as $8\cdot 3^2\cdot 6^2\cdot 2^{20}$, $3^4\cdot 2^{16}$, respectively. We sum up the number of oracle queries for $n=1-4$ to obtain that for a single plaquette: $1470021852266496$. In order to obtain the T-gate count and oracle queries required for the entire magentic term, we multiply the costs of a single plaquette by the number of plaquettes, $L^d\frac{d(d-1)}{2}$.

\subsection{Resource requirement estimates}
\label{sec:SU3_res}

In this section, we analyze the algorithmic and synthesis errors for our simulations. In Sec.~\ref{subsubsec:Trotter_SU3} we compute the algorithmic error for the Suzuki-Trotter formula for our SU(3) Hamiltonian. Therein we show our result first, then show a full derivation of it for completeness. In Sec.~\ref{subsubsec:Synth_SU3} we compute the $R_z$ synthesis error. In Sec.~\ref{subsubsec:Analysis_SU3} we combine the two errors discussed in Secs.~\ref{subsubsec:Trotter_SU3} and~\ref{subsubsec:Synth_SU3} to report the gate and query complexity, and ancilla requirements.

\subsubsection{Trotter errors}
\label{subsubsec:Trotter_SU3}

As in the simulation algorithms for U(1) and SU(2), we choose to use the second-order PF as our simulation algorithm, and evaluate the commutator bound for the error given in (\ref{eq:U1_trotter_err}). The result is
\begin{align}
    || e^{-i\hat{H}T}-\hat{U}_2^r(t)||
    \leq r\left(\frac{T}{r}\right)^3 \rho
    \equiv \epsilon_{Trotter},
\end{align}
where
\begin{align}
    \rho &= \frac{1}{12} \Big[ \frac{36 dL^d m^2}{a}+\frac{dL^d g^4(4 + 3\Lambda)^2}{4a^{2d-3}}+\frac{18 L^d (d^2-d)g^2(4 + 3 \Lambda)^2}{a^d}+\frac{L^d}{a^{6-d}g^2} 8989056(d^2-d) \nonumber\\
    &\quad +\frac{(332928 d^2 + 832320 d)L^d}{a^3} +\frac{L^d (476287080134344704 d^3 - 1190717700335861760 d^2 + 714430620201517056 d)}{a^{12-3d}g^6}\Big]\nonumber\\
    &\quad +\frac{1}{24} \Big[\frac{3 mdL^d g^2(4 +3\Lambda)}{a^{d-1}}+\frac{(432 d^2 + 108 d)mL^d}{a^2}+ \frac{5832 mL^d (d^2-d)}{a^{5-d}g^2}+\frac{(36 d^2 + 9 d) L^d g^2(4\Lambda + 3)}{a^d}\nonumber \\
    &\quad+\frac{486 d(d-1)L^d (4\Lambda + 3)}{a^3}+\frac{1944 d(d-1)L^d (3\Lambda +4)}{a^3}+ \frac{(34992 d^3 - 83106 d^2 + 48114 d)L^d(3\Lambda + 4)}{a^{6-d}g^2} \nonumber \\
    &\quad + (143824896 d^3 -170792064 d^2 + 85396032 d)\frac{L^d}{a^3}+(125846784 d^3 - 197759232 d^2 + 71912448 d)\frac{L^d}{a^{6-d}g^2}  \nonumber \\
    &\quad +(53934336 d^3 + 44945280 d^2 - 98879616 d)\frac{L^d}{a^{6-d}g^2} +(10193589504 d^3- 24755860224 d^2 + 14562270720 d)\frac{L^d}{a^{9-2d}g^4}\nonumber \\
    &\quad +( 308634027927055368192\frac{d^5}{5} - 102878009309018456064 d^4 +295774276763428061184 d^3 \nonumber \\
    &\quad - 1932177612335002877952 d^2 + 7230395091749453365248 \frac{d}{5} -154317013963527684096) \frac{L^d}{a^{12-3d}g^6}\Big].
\end{align}
For completeness, we show below a full derivation of the results shown above. Readers interested in how the results compare with the size of the synthesis error and how, together, they affect our simulation gate complexity should proceed to Secs.~\ref{subsubsec:Synth_SU3} and~\ref{subsubsec:Analysis_SU3}.

We start our derivation by first ordering the terms in the Hamiltonian $\hat{H}$. As in the U(1) and SU(2) simulations, we implement the diagonal mass and electric terms first, then the off-diagonal kinetic and magnetic terms. Due to the similarity between the SU(2) and SU(3) simulation algorithms, we adopt the same notation used in Sec.~\ref{subsubsec:Trotter_SU2}. We first discuss the kinetic terms. The ordering of the kinetic term is given by the ordered list, i.e.
\begin{equation}
    \mathbbm{T} = \{ (p, l) \} \times \{ n_1,n_2,...,n_8 \} \times \{ (\Delta p, \Delta q), \Delta T_L, \Delta T_R \} \times \{\alpha, \beta \},
\end{equation}
where $\times$ is an order-preserving Cartesian product. Further, the exponents of the incrementers and decrementers $n_1,n_2,...,n_8$ are ordered by the Gray code, and the color indices $\alpha, \beta \in \{1,2,3\}$ are ordered simply in the ascending order in the ternary representation of a three-digit number. Here, the values that the parameters $(\Delta p, \Delta q), \Delta T_L, \Delta T_R$ and $n_1,n_2,...,n_8$ can assume depend on the values of $\alpha, \beta$. Suppose $\alpha, \beta \neq 3$. There are $4$ of such tuples $(\alpha, \beta)$. The elements $\hat{h}_{T}$ that satisfy $\alpha, \beta \neq 3$ can assume $12$ combinations of $(\Delta p, \Delta q), \Delta T_L, \Delta T_R$, and $2^8$ combinations of $n_1,n_2,...,n_8$. Thus, the number of elements $\hat{h}_{T}$ in $\mathbbm{T}$ with $\alpha, \beta \neq 3$ is
\begin{equation}
    |\mathbbm{T}|_{\alpha, \beta \neq 3} = 2d\cdot 2^8 \cdot 12 \cdot 4 = 24576d. 
\end{equation}
Suppose either $\alpha$ or $\beta$ is $3$. There are $4$ such tuples. The elements $\hat{h}_{T}$ that satisfy $\alpha$ or $\beta = 3$ can assume $6$ combinations of $(\Delta p, \Delta q), \Delta T_L, \Delta T_R$, and $2^6$ combinations of $n_1,n_2,...,n_8$. Thus, the number of elements in $\mathbbm{T}$ with $\alpha$ or $\beta = 3$ is
\begin{equation}
    |\mathbbm{T}|_{\alpha / \beta = 3} = 2d\cdot 2^6 \cdot 6 \cdot 4 = 3072 d. 
\end{equation}
Suppose $(\alpha, \beta) = (3,3)$. The elements $\hat{h}_{T}$ that satisfy $(\alpha, \beta) = (3,3)$ can assume $3$ combinations of $(\Delta p, \Delta q), \Delta T_L, \Delta T_R$, and $2^4$ combinations of $n_1,n_2,...,n_8$. Thus, the number of elements in $\mathbbm{T}$ with $(\alpha, \beta) = (3,3)$ is
\begin{equation}
    |\mathbbm{T}|_{\alpha, \beta = 3} = 2d\cdot 2^4 \cdot 3 = 96 d. 
\end{equation}
Therefore, the total number of elements in $\mathbbm{T}$ is given by
\begin{equation}
    |\mathbbm{T}| = |\mathbbm{T}|_{\alpha, \beta \neq 3}+ |\mathbbm{T}|_{\alpha / \beta = 3}+ |\mathbbm{T}|_{\alpha, \beta = 3} = 27744d.
\end{equation}

The ordering of the magnetic terms is given by the following ordered list,
\begin{equation}
    \mathbbm{L} = \{ (p, j, k) \} \times \{n_1, n_2, ..., n_{32} \} \times \{ \Delta \vec{\theta}_1,\Delta \vec{\theta}_2,\Delta \vec{\theta}_3,\Delta \vec{\theta}_4 \} \times \{ \alpha, \beta, \gamma, \delta \},
\end{equation}
where the exponents of the incrementers and decrementers $q_1, q_2, ..., q_{32}$ are ordered using the Gray code, and the color indices $\alpha, \beta, \gamma, \delta$ are ordered in the ascending order in the ternary representation of a three-digit number. We remind the readers that $\Delta \vec{\theta}_i$ denotes the parameters $(\Delta p, \Delta q), \Delta T_L, \Delta T_R$ for the $i$th link on each plaquette. Similar to the kinetic terms, the values that the parameters $\Delta \vec{\theta}_i$, with $i=1,2,3,4$, and $n_1,n_2,...,n_{32}$ can assume depend on the values of the color indices $\alpha,\beta,\gamma,\delta$. Let $n$ be the number of color indices that are equal to $3$. First, we consider the tuples $(\alpha,\beta,\gamma,\delta)$ with $n=0$. The number of such tuples is $16$. Each element $\hat{h}_{L}$ with $n=0$ can assume $12^4$ combinations of $\Delta \vec{\theta}_i$ with $i=1,2,3,4$, and $2^{32}$ combinations of $n_1,n_2,...,n_{32}$. Thus, the number of elements $\hat{h}_{L}$ in $\mathbbm{L}$ with $n=0$ is
\begin{equation}
    |\mathbbm{L}|_{n=0} = 2\cdot \frac{d(d-1)}{2} \cdot 16\cdot 12^4 \cdot 2^{32}.
\end{equation}
The number of tuples $(\alpha,\beta,\gamma,\delta)$ with $n=1$ is $32$. Each element $\hat{h}_{L}$ with $n=1$ can assume $6^2\cdot 12^2$ combinations of $\Delta \vec{\theta}_i$ with $i=1,2,3,4$, and $2^{28}$ combinations of $n_1,n_2,...,n_{32}$. Thus, the number of elements $\hat{h}_{L}$ in $\mathbbm{L}$ with $n=1$ is
\begin{equation}
    |\mathbbm{L}|_{n=1} = 2\cdot \frac{d(d-1)}{2}\cdot 32 \cdot 6^2\cdot 12^2 \cdot 2^{28}.
\end{equation}
We now consider the tuples $(\alpha,\beta,\gamma,\delta)$ with $n=2$, which can be separated into two cases. The first case consists of the tuples, where one of the pairs $(\alpha,\beta)$, $(\beta,\gamma)$, $(\gamma,\delta)$, and $(\delta,\alpha)$ has both entries equal to $3$. There are $16$ of such tuples. Moreover, each element $\hat{h}_{L}$ labelled by these tuples can assume $3\cdot 6^2 \cdot 12$ combinations of $\Delta \vec{\theta}_i$ with $i=1,2,3,4$, and $2^{24}$ combinations of $n_1,n_2,...,n_{32}$. For each tuples in the second case, each pair $(\alpha,\beta)$, $(\beta,\gamma)$, $(\gamma,\delta)$, and $(\delta,\alpha)$ has one entry equal to $3$. There are $8$ of such tuples. Moreover, each element $\hat{h}_{L}$ labelled by these tuples can assume $6^4$ combinations of $\Delta \vec{\theta}_i$ with $i=1,2,3,4$, and $2^{24}$ combinations of $n_1,n_2,...,n_{32}$. Thus, the number of elements $\hat{h}_{L}$ in $\mathbbm{L}$ with $n=2$ is
\begin{equation}
    |\mathbbm{L}|_{n=2} = 2\cdot \frac{d(d-1)}{2}\cdot (16\cdot 3 \cdot 6^2 \cdot 12 + 8\cdot 6^4 ) \cdot 2^{24}.
\end{equation}
There are $8$ tuples with $n=3$. Each element labelled by these tuples can assume $3^2\cdot 6^2$ combinations $\Delta \vec{\theta}_i$ with $i=1,2,3,4$, and $2^{20}$ combinations of $n_1,n_2,...,n_{32}$. Then, the number of elements $\hat{h}_{L}$ in $\mathbbm{L}$ with $n=3$ is
\begin{equation}
    |\mathbbm{L}|_{n=3} = 2\cdot \frac{d(d-1)}{2}\cdot 8 \cdot 3^2\cdot 6^2 \cdot 2^{20}.
\end{equation}
Lastly, there is $1$ tuple with $n=4$. Each element labelled by this tuple can assume $3^4$ combinations $\Delta \vec{\theta}_i$ with $i=1,2,3,4$, and $2^{16}$ combinations of $n_1,n_2,...,n_{32}$. Then, the number of elements $\hat{h}_{L}$ in $\mathbbm{L}$ with $n=4$ is
\begin{equation}
    |\mathbbm{L}|_{n=4} = 2\cdot \frac{d(d-1)}{2} \cdot 3^4 \cdot 2^{16}.
\end{equation}
Thus, the number of elements in $\mathbbm{L}$ is given by
\begin{align}
    |\mathbbm{L}| &= \sum_{k = 0}^{4} |\mathbbm{L}|_{n=k} = 2\cdot \frac{d(d-1)}{2} \cdot 735010926133248.
\end{align}
Finally, the ordering of the terms in the Hamiltonian $\hat{H}$ is given by the following ordered list:
\begin{equation}
    \{ \hat{H}_x \} = \{ \sum_{\vec{n}} \hat{D}_{\vec{n}}^{(M)}, \sum_{\vec{n}} \hat{D}_{\vec{n}}^{(E)} \} \cup \mathbbm{T} \cup \mathbbm{L},
\end{equation}
where $\cup$ is denotes the order-preserving union.

We now evaluate the Trotter error incurred by the second-order PF, which is given in (\ref{eq:U1_trotter_err}). Due to the similarities between the SU(2) and SU(3) simulation algorithms, throughout the following analysis, we will make use of the notations and relevant results from the SU(2) analysis, and highlight the differences whenever necessary.

In the following, we derive useful expressions, similar to (\ref{eq:SU2_useful_k1}-\ref{eq:SU2_useful_m4}) in the SU(2) case. As in SU(2) LGTs, we have
\begin{gather}
    ||\hat{U}_{\alpha \beta}|| \leq 1,
    \label{eq:SU3_useful_k1} \\
    ||\hat{U}_{\alpha \beta}+\hat{U}_{\alpha \beta}^\dag|| \leq 2, \label{eq:SU3_useful_k2} \\
    ||\hat{U}_{\alpha \beta}\hat{\sigma}_{\alpha}^- \hat{\sigma}_{\beta}^+ +\hat{U}_{\alpha \beta}^\dag \hat{\sigma}_{\alpha}^+ \hat{\sigma}_{\beta}^-|| \leq 2. \label{eq:SU3_useful_k3}
\end{gather}
We note that the norm of each term in the block-diagonal decomposition of $\hat{U}_{\alpha \beta}\hat{\sigma}_{\alpha}^- \hat{\sigma}_{\beta}^+ +\hat{U}_{\alpha \beta}^\dag \hat{\sigma}_{\alpha}^+ \hat{\sigma}_{\beta}^-$, as shown in (\ref{eq:SU3_U_w_f}), is upper-bounded by $2$. The reason is that, for each term in the decomposition, the norms of the incrementers, decrementers and Pauli ladder operators are bounded from above by one, and that of the diagonal part $\hat{D}_{\alpha \beta}$, of which the elements are defined via $f_{\alpha \beta}$, is also upper-bounded by one. This implies that 
\begin{equation}
    ||\hat{h}_{T}(\vec{n}_p,l)|| \leq \frac{1}{a} \label{eq:SU3_useful_k4},
\end{equation}
because $\hat{h}_{T}(\vec{n}_p,l)$ is, up to a multiplicative factor of $\frac{1}{2a}$, a term in the block-diagonal decomposition. 

Furthermore, we consider the terms $\hat{h}_T(\vec{n},l)$, which act on a pair of nearest-neighbor sites and the link that connects them, i.e., $(\vec{n},l)$. We denote this set of terms as $\mathbbm{T}|_{(\vec{n},l)}$. Then,
\begin{equation}
    \sum_{\hat{h}_T(\vec{n},l)\in \mathbbm{T}|_{(\vec{n},l)}} \hat{h}_T(\vec{n},l) = \frac{1}{2a}\sum_{\alpha,\beta=1}^{3}\hat{U}_{\alpha \beta}(\vec{n},l)\hat{\sigma}_{\alpha}^-(\vec{n}) \hat{\sigma}_{\beta}^+(\vec{n}+\hat{l}) +h.c.,
    \label{eq:SU3_useful_k5}
\end{equation}
which we combine with (\ref{eq:SU3_useful_k3}) to get
\begin{equation}
    ||\sum_{\substack{\hat{h}_T(\vec{n},l)\in S;\\ S \subseteq \mathbbm{T}|_{(\vec{n},l)}}} \hat{h}_T(\vec{n},l)|| \leq || \frac{1}{2a}\sum_{\alpha,\beta=1}^{3}\hat{U}_{\alpha \beta}(\vec{n},l)\hat{\sigma}_{\alpha}^-(\vec{n}) \hat{\sigma}_{\beta}^+(\vec{n}+\hat{l}) +h.c.|| \leq \frac{9}{a}. \label{eq:SU3_useful_k6}
\end{equation}

Similarly, for the magnetic term, we obtain the following bound:
\begin{equation}
    ||\hat{U}_{\alpha\beta}\hat{U}_{\beta\gamma}\hat{U}_{\gamma\delta}^\dag\hat{U}_{\delta\alpha}^\dag + h.c.|| \leq 2||\hat{U}_{\alpha\beta}\hat{U}_{\beta\gamma}\hat{U}_{\gamma\delta}^\dag\hat{U}_{\delta\alpha}^\dag|| \leq 2 ||\hat{U}_{\alpha\beta}||^4 \leq 2, \label{eq:SU3_useful_m1}
\end{equation}
As in the kinetic term, the norm of each term in the block-diagonal decomposition of $\hat{U}_{\alpha\beta}\hat{U}_{\beta\gamma}\hat{U}_{\gamma\delta}^\dag\hat{U}_{\delta\alpha}^\dag+h.c.$, as shown in (\ref{eq:SU3_U_w_f}), is bounded by $2$. The reason is that the norms of the incrementers, decrementers and Pauli ladder operators are bounded from above by one, and that of the diagonal part $\hat{D}_{\alpha \beta \gamma \delta}$, of which the elements are defined via $f_{\alpha \beta \gamma \delta}$, as defined in (\ref{eq:SU3_fabcd}), is also upper-bounded by one. This implies that
\begin{equation}
    ||\hat{h}_L(\vec{n}_p,i,j)|| \leq \frac{1}{a^{4-d}g^2} \label{eq:SU3_useful_m2}
\end{equation}
because each $\hat{h}_L(\vec{n}_p,i,j)$ is a product between $\frac{1}{2a^{4-d}g^2}$ and a term in (\ref{eq:SU3_U_w_f}).

Lastly, we consider the terms $\hat{h}_L(\vec{n},i,j)$, which act on a single plaquette, denoted by $(\vec{n}, i,j)$. Let these terms form a set $\mathbbm{L}|_{(\vec{n},i, j)}$. Then,
\begin{equation}
    \sum_{\hat{h}_L(\vec{n},i,j) \in \mathbbm{L}|_{(\vec{n}, i,j)}}\hat{h}_L{(\vec{n},i,j)}=\frac{1}{2a^{4-d}g^2}\sum_{\alpha,\beta,\gamma,\delta=1}^{3}\hat{U}_{\alpha \beta}(\vec{n},i)\hat{U}_{ \beta \gamma}(\vec{n}+\hat{i},j)\hat{U}_{\gamma \delta}^{\dag}(\vec{n}+\hat{j},i)\hat{U}_{\delta \alpha}^{\dag}(\vec{n},j) + h.c..
    \label{eq:SU3_useful_m3}
\end{equation}
Using the above relation and (\ref{eq:SU3_useful_m1}), we obtain
\begin{align}
    || \sum_{\substack{\hat{h}_L(\vec{n},i,j)\in S;\\ S\subseteq \mathbbm{L}|_{(\vec{n}, i,j)}}}\hat{h}_L{(\vec{n},i,j)}||&\leq ||\frac{1}{2a^{4-d}g^2}\sum_{\alpha,\beta,\gamma,\delta=1}^{3}\hat{U}_{\alpha \beta}(\vec{n},i)\hat{U}_{ \beta \gamma}(\vec{n}+\hat{i},j)\hat{U}_{\gamma \delta}^{\dag}(\vec{n}+\hat{j},i)\hat{U}_{\delta \alpha}^{\dag}(\vec{n},j) + h.c.|| \nonumber \\
    &\leq \frac{1}{2a^{4-d}g^2}\sum_{\alpha,\beta,\gamma,\delta=1}^{3}||\hat{U}_{\alpha \beta}(\vec{n},i)\hat{U}_{ \beta \gamma}(\vec{n}+\hat{i},j)\hat{U}_{\gamma \delta}^{\dag}(\vec{n}+\hat{j},i)\hat{U}_{\delta \alpha}^{\dag}(\vec{n},j) + h.c.|| \nonumber \\
    &\leq \frac{1}{2a^{4-d}g^2} 81\cdot 2 = \frac{81}{a^{4-d}g^2}.
    \label{eq:SU3_useful_m4}
\end{align}
Whenever these expressions are used, we use them without explicit references for brevity. 

Next, we analyze the first sum in (\ref{eq:U1_trotter_err}), which is a sum of eight terms, i.e., $||C_{1,n}||$ with $n=1,...,8$, according to (\ref{eq:SU2_trot1},\ref{eq:SU2_trotter_err_1}). First, $C_{1,1}$ and $C_{1,3}$ both evaluate to zero because the mass term commutes with both the electric and mass terms. $C_{1,2}$ is bounded by
\begin{align}
    &\quad ||[[ \sum_{\vec{n}} \hat{D}_{\vec{n}}^{(M)} , \sum_{\vec{n}'} \hat{T}_{\vec{n}'}^{(K)} ]  , \sum_{\vec{n}} \hat{D}_{\vec{n}}^{(M)} ]|| \nonumber \\
    &\leq \sum_{\vec{n}} \sum_{l=1}^{d} || [[\hat{D}_{\vec{n}}^{(M)}+\hat{D}_{\vec{n}+\hat{l}}^{(M)},\frac{1}{2a}\sum_{\alpha, \beta=1}^{3}(\hat{U}_{\alpha \beta}(\vec{n},l)\hat{\sigma}_{\alpha}^{-}(\vec{n})\hat{\sigma}_{\beta}^{+}(\vec{n}+\hat{l}) + h.c.)], \hat{D}_{\vec{n}}^{(M)}+\hat{D}_{\vec{n}+\hat{l}}^{(M)}]|| \nonumber \\
    &\leq dL^d \sum_{\alpha, \beta=1}^{3}||[[ \frac{m}{2}((-1)^{\vec{n}} \hat{Z}_{\alpha}(\vec{n})+(-1)^{\vec{n}+\hat{l}} \hat{Z}_{\beta}(\vec{n}+\hat{l})) , \frac{1}{2a}(\hat{U}_{\alpha \beta}(\vec{n},l)\hat{\sigma}_{\alpha}^{-}(\vec{n})\hat{\sigma}_{\beta}^{+}(\vec{n}+\hat{l}) + h.c.) ]  \nonumber \\
    &\quad , \frac{m}{2}((-1)^{\vec{n}} \hat{Z}_{\alpha}(\vec{n})+(-1)^{\vec{n}+\hat{l}} \hat{Z}_{\beta}(\vec{n}+\hat{l})) ] \leq dL^d \cdot 4 ||m||^2 \cdot \frac{9}{a}=\frac{36 dL^d m^2}{a},
\end{align}
where in the first two inequalities, we have used the fact that only the fermionic sites of colors $\alpha, \beta$ at $\vec{n},\vec{n}+\hat{l}$, respectively, acted on by the kinetic term with color indices $\alpha \beta$.

Before evaluating $C_{1,4}$, we provide some useful properties about the kinetic operators $\frac{1}{2a}$ $(\hat{U}_{\alpha \beta}(\vec{n},l)\hat{\sigma}_{\alpha}^{-}(\vec{n})\hat{\sigma}_{\beta}^{+}(\vec{n}+\hat{l}) + h.c.)$. At the fermionic sites $\vec{n}$ and $\vec{n}+\hat{l}$ of color $\alpha$ and $\beta$, respectively, the kinetic operator takes a computational basis state to another basis state. Acting on the gauge field on the link $(\vec{n},l)$, it maps the subregisters $\ket{p}\ket{q} \mapsto \{ \ket{p\pm 1}\ket{q} , \ket{p\mp 1}\ket{q \pm 1}, \ket{p}\ket{q\pm 1} \}$, where $j\in [0, \Lambda]$, up to a multiplicative constant. Therefore, if we evaluate the commutator between an electric and kinetic operator acting on the same link, we obtain
\begin{align}
    &\quad || [\frac{g^2}{2a^{d-2}}\hat{E}^2(\vec{n},l),\frac{1}{2a}\sum_{\alpha,\beta=1}^{3}(\hat{U}_{\alpha \beta}(\vec{n},l)\hat{\sigma}_{\alpha}^{-}(\vec{n})\hat{\sigma}_{\beta}^{+}(\vec{n}+\hat{l}) + h.c.)] ||  \nonumber \\
    &\mapsto 
    \begin{cases}
        \frac{g^2}{6 a^{d-2}} [(p\pm 1 + q)(p\pm 1 +q +3) - (p\pm 1)q - (p+q)(p+q+3)+pq] \\
        \cdot\sum_{\alpha,\beta=1}^{3} ||\frac{1}{2a}(\hat{U}_{\alpha \beta}(\vec{n},l)\hat{\sigma}_{\alpha}^{-}(\vec{n})\hat{\sigma}_{\beta}^{+}(\vec{n}+\hat{l}) + h.c.)||, \mbox{ if } \ket{p}\ket{q} \mapsto  \ket{p\pm 1}\ket{q} \\
        \frac{g^2}{6 a^{d-2}} [(p\pm 1 + q\mp 1)(p\pm 1 +q\mp 1 +3) - (p\pm 1)(q\mp 1) - (p+q)(p+q+3)+pq] \\
        \cdot \sum_{\alpha,\beta=1}^{3} ||\frac{1}{2a}(\hat{U}_{\alpha \beta}(\vec{n},l)\hat{\sigma}_{\alpha}^{-}(\vec{n})\hat{\sigma}_{\beta}^{+}(\vec{n}+\hat{l}) + h.c.)||, \mbox{ if }\ket{p}\ket{q} \mapsto  \ket{p\pm 1}\ket{q\mp 1}\\
        \frac{g^2}{6 a^{d-2}} [(p + q\pm 1)(p +q\pm 1 +3) - p(q\pm 1) - (p+q)(p+q+3)+pq]\\
        \cdot \sum_{\alpha,\beta=1}^{3} ||\frac{1}{2a}(\hat{U}_{\alpha \beta}(\vec{n},l)\hat{\sigma}_{\alpha}^{-}(\vec{n})\hat{\sigma}_{\beta}^{+}(\vec{n}+\hat{l}) + h.c.)||, \mbox{ if }
        \ket{p}\ket{q} \mapsto  \ket{p}\ket{q \pm 1}
    \end{cases}
    \nonumber \\
    &\leq
    \begin{cases}
        \frac{g^2}{6 a^{d-1}} ||4+2p+q|| \cdot \frac{9}{a}, \mbox{ if } \ket{p}\ket{q} \mapsto  \ket{p + 1}\ket{q}\\
        \frac{g^2}{6 a^{d-1}} ||4+p+2q|| \cdot \frac{9}{a}, \mbox{ if } \ket{p}\ket{q} \mapsto  \ket{p}\ket{q+1}
    \end{cases}
    \nonumber \\
    &= \frac{3g^2}{2 a^{d-1}}(4+3\Lambda),
    \label{eq:SU3_EK}
\end{align}
where for the first inequality, we have used (\ref{eq:SU3_elec_eig}) and listed the cases where the norm is maximized. Using this equation, we evaluate the bound for $C_{1,4}$, and obtain
\begin{align}
    &\quad || [[\sum_{\vec{n}}\hat{D}_{\vec{n}}^{(E)}, \sum_{ \vec{n} } \hat{T}_{\vec{n}'}^{(K)}] ,\sum_{\vec{n}}\hat{D}_{\vec{n}}^{(E)}] || \nonumber \\
    &\leq \sum_{\vec{n}}\sum_{l=1}^{d}\sum_{\alpha, \beta=1}^{3} ||[ [\frac{g^2}{2a^{d-2}}\hat{E}^2(\vec{n},l),\frac{1}{2a}(\hat{U}_{\alpha \beta}(\vec{n},l)\hat{\sigma}_{\alpha}^{-}(\vec{n})\hat{\sigma}_{\beta}^{+}(\vec{n}+\hat{l}) + h.c.)],\frac{g^2}{2a^{d-2}}\hat{E}^2(\vec{n},l)] ||\nonumber \\
    &\mapsto \sum_{\vec{n}}\sum_{l=1}^{d}\sum_{\alpha, \beta=1}^{3}  \frac{g^2}{6a^{d-2}} (4+3\Lambda) ||[\frac{g^2}{2a^{d-2}}\hat{E}^2(\vec{n},l),\frac{1}{2a}(\hat{U}_{\alpha \beta}(\vec{n},l)\hat{\sigma}_{\alpha}^{-}(\vec{n})\hat{\sigma}_{\beta}^{+}(\vec{n}+\hat{l}) + h.c.)]|| \nonumber \\
    &\leq dL^d \frac{g^2}{6a^{d-2}} (4+3\Lambda)\frac{3g^2}{2 a^{d-1}}(4+3\Lambda) \nonumber \\
    &= \frac{dL^d g^4(4 + 3\Lambda)^2}{4a^{2d-3}},
\end{align}
where we have used (\ref{eq:SU3_EK}) for the last inequality.

In order to compute the bound for $C_{1,5}$, it is useful to evaluate the bound for the commutator between the electric and magnetic operators acting on a single plaquette, which is given by
\begin{align}
    &\quad || [ \frac{g^2}{2a^{d-2}} (\hat{E}^2(\vec{n},i)+\hat{E}^2(\vec{n}+\hat{i},j)+\hat{E}^2(\vec{n}+\hat{j},i)+\hat{E}^2(\vec{n},j) )  \nonumber \\
    &\quad, -\frac{1}{2a^{4-d}g^2} \sum_{\alpha, \beta, \delta, \gamma =1}^{3} (\hat{U}_{\alpha \beta}(\vec{n},i)\hat{U}_{ \beta \gamma}(\vec{n}+\hat{i},j)\hat{U}_{\gamma \delta}^{\dag}(\vec{n}+\hat{j},i)\hat{U}_{\delta \alpha}^{\dag}(\vec{n},j) + h.c.) ] || \nonumber \\
    &\leq 4 \sum_{\alpha, \beta, \delta, \gamma =1}^{3} || [ \frac{g^2}{2a^{d-2}} \hat{E}^2 ,-\frac{1}{2a^{4-d}g^2}( \hat{U}_{\alpha \beta}\hat{U}_{ \beta \gamma}\hat{U}_{\gamma \delta}^{\dag}\hat{U}_{\delta \alpha}^{\dag}+h.c.) ] || \nonumber \\
    &\leq 4 \sum_{\alpha, \beta, \delta, \gamma =1}^{3} ||\frac{4 + 3\Lambda}{12a^2}  (\hat{U}_{\alpha \beta}\hat{U}_{ \beta \gamma}\hat{U}_{\gamma \delta}^{\dag}\hat{U}_{\delta \alpha}^{\dag}+h.c.) || \nonumber \\
    &\leq \frac{54}{a^2}(4 + 3\Lambda),
    \label{eq:SU3_elec_mag}
\end{align}
where in the first inequality, we have dropped the location indices for brevity, since the commutator between any of the four electric terms and the magnetic terms shares the same bound, in the second inequality, we have used (\ref{eq:SU3_EK}), and in the last inequality, we have used (\ref{eq:SU3_useful_m4}). Now we use the above relation to compute the bound for $C_{1,5}$, and obtain
\begin{align}
    &\quad ||[[ \sum_{\vec{n}} \hat{D}_{\vec{n}}^{(E)} , \sum_{\vec{n}'} \hat{L}_{\vec{n}'}^{(B)} ]  , \sum_{\vec{n}} \hat{D}_{\vec{n}}^{(E)} ]|| \nonumber \\
    &\leq || \sum_{\vec{n}}\sum_{i=1}^{d}\sum_{j\neq i;j=1}^{d} [[\frac{g^2}{2a^{d-2}}(\hat{E}^2(\vec{n},i)+\hat{E}^2(\vec{n}+\hat{i},j)+\hat{E}^2(\vec{n}+\hat{j},i)+\hat{E}^2(\vec{n},j) ) \nonumber \\
    &\quad , -\frac{1}{2a^{4-d}g^2} \sum_{\alpha, \beta, \delta, \gamma =1}^{3} (\hat{U}_{\alpha \beta}(\vec{n},i)\hat{U}_{ \beta \gamma}(\vec{n}+\hat{i},j)\hat{U}_{\gamma \delta}^{\dag}(\vec{n}+\hat{j},i)\hat{U}_{\delta \alpha}^{\dag}(\vec{n},j) + h.c.) ] \nonumber \\
    &\quad , \frac{g^2}{2a^{d-2}}(\hat{E}^2(\vec{n},i)+\hat{E}^2(\vec{n}+\hat{i},j)+\hat{E}^2(\vec{n}+\hat{j},i)+\hat{E}^2(\vec{n},j) )] || \nonumber \\
    &\leq L^d\frac{d(d-1)}{2} 4|| \frac{g^2}{6a^{d-2}}(4+ 3 \Lambda) [\frac{g^2}{2a^{d-2}}(\hat{E}^2(\vec{n},i)+\hat{E}^2(\vec{n}+\hat{i},j)+\hat{E}^2(\vec{n}+\hat{j},i)+\hat{E}^2(\vec{n},j) ) \nonumber \\
    &\quad , -\frac{1}{2a^{4-d}g^2} \sum_{\alpha, \beta, \delta, \gamma =1}^{3} (\hat{U}_{\alpha \beta}(\vec{n},i)\hat{U}_{ \beta \gamma}(\vec{n}+\hat{i},j)\hat{U}_{\gamma \delta}^{\dag}(\vec{n}+\hat{j},i)\hat{U}_{\delta \alpha}^{\dag}(\vec{n},j) + h.c.) ] || \nonumber \\
    &\leq L^d\frac{d(d-1)}{2}4 \frac{g^2}{6a^{d-2}}(4+ 3 \Lambda) \frac{54}{a^2}(4 + 3\Lambda) = \frac{18 L^d d(d-1)g^2(4 + 3 \Lambda)^2}{a^d},
\end{align}
where in the second inequality, we have used the fact that there are $L^d\frac{d(d-1)}{2}$ plaquettes on the lattice, and in the last inequality, we have used (\ref{eq:SU3_elec_mag}).

Next, we evaluate the bound for $C_{1,6}$, and obtain
\begin{align}
    &\quad \sum_{\hat{h}_T \in \mathbbm{T}} ||  [[ \hat{h}_T , \sum_{\vec{n}}\hat{L}_{\vec{n}}^{(B)} ], \hat{h}_T ]  || \nonumber \\
    &\leq \sum_{\hat{h}_T \in \mathbbm{T}} ||  [[ \sum_{\vec{n}_p}\hat{h}_T(\vec{n}_p,l) , \sum_{\vec{n}}\hat{L}_{\vec{n}}^{(B)} ], \sum_{\vec{n}_p} \hat{h}_T(\vec{n}_p,l) ]  ||\nonumber \\
    &\leq 27744 d\frac{L^d}{2} 4|| \hat{h}_{T}(\vec{n}_p,l)|| \cdot || \frac{2(d-1)}{2a^{4-d}g^2}\sum_{\alpha, \beta, \gamma, \delta=1}^{3} (\hat{U}_{\alpha\beta}\hat{U}_{\beta\gamma}\hat{U}_{\gamma\delta}^\dag\hat{U}_{\delta\alpha}^\dag + h.c.) || \cdot || \hat{h}_{T}(\vec{n}_p,l) || \nonumber \\
    &\leq 27744 d \cdot 2 L^d ||\frac{1}{a}||^2 \cdot || 162\frac{d-1}{a^{4-d}g^2} || = \frac{8989056 d(d-1)L^d}{a^{6-d}g^2},
\end{align}
where in the second inequality, the factor of $27744 d$ outside the norm expression is the cardinality of $\mathbbm{T}$, $\frac{L^d}{2}$ is the number of even or odd sites $\vec{n}_p$, $2(d-1)$ is due to the fact that each link $(\vec{n}_p,l)$ is acted on concurrently by a kinetic term and $2(d-1)$ plaquette operators, and we have used (\ref{eq:comm_bound}).

Next, we analyze $C_{1,7}$. As in the SU(2) case, we divide the commutators up into two types; those between (i) terms acting on the same link, and (ii) terms acting on neighboring links that are connected via the sites. We denote the subset of $\mathbbm{T}$ that consists of elements with a fixed parity and direction by $\mathbbm{T}|_{(p,l)}$. The number of elements in $\mathbbm{T}|_{(p,l)}$ is the number of free parameters: $13872$. Then, for type (i), the bound is given by
\begin{align}
    &\quad 13872 || \sum_{p=e}^{o} \sum_{l=1}^{d} [[ \sum_{\vec{n}_p} \hat{h}_T(\vec{n}_p,l),\sum_{\hat{h}_{T'}\in \mathbbm{T}|_{(p,l)}; T'>T}\hat{h}_{T'}],\hat{h}_T(\vec{n}_p,l)]  || \nonumber \\
    &\leq 13872\cdot 2d \cdot 4\cdot ||\hat{h}_T(\vec{n}_p,l)||^2 \cdot ||\sum_{\hat{h}_{T'}\in \mathbbm{T}|_{(p,l)}; T'>T}\hat{h}_{T'}|| \nonumber \\
    &\leq 13872\cdot 2d \cdot 4\cdot ||\hat{h}_T(\vec{n}_p,l)||^2 \cdot || \frac{1}{2a}\sum_{\alpha, \beta=1}^{3}(\hat{U}_{\alpha \beta}(\vec{n}_p,l)\hat{\sigma}_{\alpha}^{-}(\vec{n}_p)\hat{\sigma}_{\beta}^{+}(\vec{n}_p+\hat{l}) + h.c.) || \nonumber \\
    &\leq 13872\cdot 2d \cdot 4\cdot ||\frac{1}{a}||^2\cdot ||\frac{9}{a}|| = \frac{998784 d}{a^3},
\end{align}
where in the second inequality, we used (\ref{eq:SU3_useful_k6}), and the fact that each $\hat{h}_T(\vec{n}_p,l)$ collides on a link with at most nine kinetic operators of different color indices to obtain the second norm expression. Next, we analyze type (ii), which we further divide into two cases. Case (i) consists of $d$ commutators, where $\hat{h}_{T}, \hat{h}_{T'}$ act on links in the same direction, but sites of different parities, whereas $\hat{h}_{T}, \hat{h}_{T'}$, in case (ii), act on links in different directions. There are $\sum_{p',p=e}^{o}\sum_{l=1}^{d}\sum_{l'>l,l'=1}^{d}=2d^2-2d$ commutators in case (ii). In both types, $\hat{h}_{T}$ and $\hat{h}_{T'}$ could collide on one or two sites, depending on their respective color indices. If $\hat{h}_{T}$ and $\hat{h}_{T'}$ are labelled by $\alpha \beta$ and $\beta \alpha$, respectively, then each $\hat{h}_{T}$ collides with two $\hat{h}_{T'}$ on sites of both colors $\alpha, \beta$. If $\hat{h}_{T}$ is lablled by $\alpha \beta$, and $\hat{h}_{T'}$ is labelled by $\beta \gamma$ with $\gamma \neq \beta$, or $\gamma \alpha$ with $\gamma \neq \alpha$, then each $\hat{h}_{T}$ collides with one $\hat{h}_{T'}$, on the site labelled by $\beta$ or $\alpha$, respectively. For each $\hat{h}_{T}$ labelled by $\alpha \beta$, there is one $\beta \alpha$, two $\beta \gamma$ with $\gamma \neq \alpha$, and two $\gamma \alpha$ with $\gamma \neq \beta$ that label $\hat{h}_{T'}$. Thus, the bound for each commutator, where there is a collision on two sites, is
\begin{equation}
    4\cdot ||\hat{h}_T(\vec{n}_p,l)||^2 \cdot ||2 \cdot \frac{1}{2a}(\hat{U}_{\beta\alpha }\hat{\sigma}_{\beta}^{-}\hat{\sigma}_{\alpha}^{+} + h.c.) || \leq \frac{8}{a^3},
\end{equation}
The bound for each commutator, where there is a collision on one site, is
\begin{equation}
    4\cdot ||\hat{h}_T(\vec{n}_p,l)||^2 \cdot || \frac{1}{2a}\sum_{\alpha' \beta'=\beta \gamma}^{\gamma \alpha}(\hat{U}_{\alpha' \beta' }\hat{\sigma}_{\alpha'}^{-}\hat{\sigma}_{\beta'}^{+} + h.c.) || \leq \frac{16}{a^3}.
\end{equation}
The bound for the commutators in type (ii) is thus
\begin{equation}
    \frac{8+16}{a^3} |\mathbbm{T}|_{(p,l)} \cdot(2d^2-2d + d) \frac{L^d}{2}=\frac{24}{a^3}13872\cdot(2d^2-d) \frac{L^d}{2} = \frac{(332928 d^2 - 166464 d)L^d}{a^3}.
\end{equation}
As such, the bound for $C_{1,7}$ is
\begin{equation}
    \frac{(332928 d^2 + 832320 d)L^d}{a^3}.
\end{equation}

Finally, we consider $C_{1,8}$, which consists of the commutators between magnetic terms. As in the SU(2) case, we separate the terms into two cases: case (i) and (ii) consists of intra- and inter-plaquette commutators. We first examine case (i). We denote the subset of $\mathbbm{L}$ that consists of elements with a fixed parity and two-dimensional plane by $\mathbbm{L}|_{(p,j,k)}$. As such, the number of elements in $\mathbbm{L}|_{(p,j,k)}$, $735010926133248$, is the number of remaining free parameters. We compute the bound of the commutators in case (i), and obtain
\begin{align}
    &\quad 735010926133248 L^d\frac{d(d-1)}{2} ||[[\hat{h}_{L}(\vec{n}_p,j,k), \sum_{\hat{h}_{L'}\in \mathbbm{L}|_{(p,j,k)};L'>L}\hat{h}_{L'}],\hat{h}_{L}(\vec{n}_p,j,k)]|| \nonumber \\
    &\leq 735010926133248 L^d\frac{d(d-1)}{2} \cdot 4 \cdot ||\hat{h}_{L}(\vec{n}_p,j,k)||^2 ||\frac{-1}{2a^{4-d}g^2}\sum_{\alpha,\beta,\gamma,\delta=1}^{3}\hat{U}_{\alpha \beta}\hat{U}_{\beta \gamma}\hat{U}_{\gamma\delta}^\dag \hat{U}_{\delta \alpha}^\dag  + h.c.|| \nonumber \\
    &= 735010926133248 L^d\frac{d(d-1)}{2} \cdot \frac{4\cdot 81}{a^{12-3d}g^6}= \frac{119071770033586176 (d^2-d)L^d }{a^{12-3d}g^6},
\end{align}
where in the second norm expression in the first inequality, we have used the fact that each $\hat{h}_{L}(\vec{n}_p,j,k)$ can collide with at most $81$ magnetic operators with different color indices. 

We divide case (ii) into two types; those where (i) $\hat{h}_L$ and $\hat{h}_{L'}$ act on neighboring plaquettes with different parities on the same two-dimensional plane, and those where (ii) $\hat{h}_L$ and $\hat{h}_{L'}$ act on plaquettes that share one common dimension. In type (i), there are $\frac{d(d-1)}{2}$ pairs of $\hat{h}_L$ and $\hat{h}_{L'}$, which is the same as the number of two-dimensional planes. Since we have chosen to implement even terms before odd terms, each commutator is bounded by
\begin{equation}
    4\cdot ||\hat{h}_L (\vec{n}_e,j,k)||^2 \cdot ||4\cdot \sum_{\hat{h}_{L'}\in \mathbbm{L}|_{(o,j,k)};L'>L} \hat{h}_{L'} (\vec{n}_o,j,k)|| = \frac{16\cdot 81}{a^{12-3d}g^6}=\frac{1296}{a^{12-3d}g^6},
\end{equation}
where in the second norm expression, the factor of $4$ is due to the fact that there are four $\hat{h}_{L'} (\vec{n}_o)$ terms acting on the four links, which form the plaquette that $\hat{h}_L (\vec{n}_e)$ acts on, and the factor of $81$ is the number of different color indices of the magnetic operators. There are $2d^3-6d^2+4d$ type-(ii) pairs of $\hat{h}_L$ and $\hat{h}_{L'}$. See Table \ref{tb:SU2_plaq1}. Each commutator is bounded by
\begin{equation}
    4\cdot ||\hat{h}_L (\vec{n}_p,j,k)||^2 \cdot ||2\cdot \sum_{\hat{h}_{L'}\in \mathbbm{L}|_{(p',j',k')};L'>L} \hat{h}_{L'} (\vec{n}_p',j',k')|| = \frac{8\cdot 81}{a^{12-3d}g^6}= \frac{648}{a^{12-3d}g^6},
\end{equation}
where in the second norm expression, the factor of $2$ is due to the fact that there are two $\hat{h}_{L'}$ that collide with $\hat{h}_L$ on a link. Therefore, the bound for case-(ii) terms is given by
\begin{align}
    &\quad 735010926133248\frac{L^d}{2}[\frac{1296}{a^{12-3d}g^6}\frac{d(d-1)}{2} + \frac{648}{a^{12-3d}g^6}(2d^3-6d^2+4d)] \nonumber \\
    &= \frac{L^d}{a^{12-3d}g^6}238143540067172352(2 d^3 - 5 d^2 + 3 d).
\end{align}
In total, $C_{1,8}$ is bounded by
\begin{equation}
    \frac{L^d}{a^{12-3d}g^6}(476287080134344704 d^3 - 1190717700335861760 d^2 + 714430620201517056 d).
\end{equation}

Next, we analyze the second sum in (\ref{eq:U1_trotter_err}), which is a sum of twelve terms, i.e. $||C_{2,n}||$ with $n=1,...,12$, according to (\ref{eq:SU2_trot2},\ref{eq:SU2_trotter_err_2}). For $C_{2,1}$, we obtain the bound
\begin{align}
    &\quad ||[[ \sum_{\vec{n}} \hat{D}_{\vec{n}}^{(M)} , \sum_{\vec{n}'} \hat{T}_{\vec{n}'}^{(K)} ]  , \sum_{\vec{n}''} \hat{D}_{\vec{n}''}^{(E)} ]|| \nonumber \\
    &\leq || \sum_{\vec{n}}\sum_{l=1}^{d}\sum_{\alpha, \beta=1}^{3}[[\frac{m}{2}((-1)^{\vec{n}}\hat{Z}_{\alpha}(\vec{n})+(-1)^{\vec{n}+\hat{l}}\hat{Z}_{\beta}(\vec{n}+\hat{l})),\frac{1}{2a}(\hat{U}_{\alpha \beta}(\vec{n},l)\hat{\sigma}_{\alpha}^-(\vec{n})\hat{\sigma}_{\beta}^+(\vec{n}) +h.c.)],\frac{g^2}{2a^{d-2}}\hat{E}^2(\vec{n},l)]  ||\nonumber \\
    &\leq dL^d \frac{g^2}{6a^{d-2}}(4+3\Lambda)\sum_{\alpha, \beta=1}^{3}||[\frac{m}{2}((-1)^{\vec{n}}\hat{Z}_{\alpha}(\vec{n})+(-1)^{\vec{n}+\hat{l}}\hat{Z}_{\beta}(\vec{n}+\hat{l})),\frac{1}{2a}(\hat{U}_{\alpha \beta}(\vec{n},l)\hat{\sigma}_{\alpha}^-(\vec{n})\hat{\sigma}_{\beta}^+(\vec{n}) +h.c.)]||\nonumber \\
    &\leq \frac{dL^d g^2(4 +3\Lambda)}{6a^{d-2}} 2||m|| \cdot \frac{9}{a} = \frac{3 mdL^d g^2(4 +3\Lambda)}{a^{d-1}},
\end{align}
where in the second inequality, we used the fact that the mass terms at $\vec{n},\vec{n}+\hat{l}$, of which the respective colors are not $\alpha,\beta$, commute with the kinetic term with color indices $\alpha\beta$.

For $C_{2,2}$, we divide the commutators up into two types as in the SU(2) analysis. We evaluate the bound for type (i), where the kinetic terms act on the same links, and obtain 
\begin{align}
    &\quad || \sum_{\vec{n}}\sum_{l=1}^{d}\sum_{\alpha,\beta=1}^{3} [[\frac{m}{2}((-1)^{\vec{n}}\hat{Z}_{\alpha}(\vec{n})+(-1)^{\vec{n}+\hat{l}}\hat{Z}_{\beta}(\vec{n}+\hat{l})),\frac{1}{2a}(\hat{U}_{\alpha \beta}(\vec{n},l)\hat{\sigma}_{\alpha}^-(\vec{n})\hat{\sigma}_{\beta}^+(\vec{n}+\hat{l}) +h.c.)] \nonumber \\
    &,\frac{1}{2a}\sum_{\alpha',\beta'=1}^{3}(\hat{U}_{\alpha', \beta'}(\vec{n},l)\hat{\sigma}_{\alpha'}^-(\vec{n})\hat{\sigma}_{\beta'}^+(\vec{n}+\hat{l}) +h.c.)] || \nonumber \\
    &\leq dL^d \cdot 4 ||m|| \cdot ||\frac{9}{a}|| \cdot ||\frac{9}{a}|| = \frac{324 mdL^d}{a^2}.
\end{align}

We split type (ii) into two cases. Suppose the color indices of $\hat{T}_{\vec{n}'}^{(K)}$ is $\alpha\beta$. Then, in case (i), $\hat{T}_{\vec{n}''}^{(K)}$ has color indices $\beta \alpha$, and in case (ii), it has color indices $\beta\gamma$ with $\gamma\neq \alpha$, or $\gamma \alpha$ with $\gamma\neq \beta$. The bound for case (i) is as follows:
\begin{align}
    &\quad || \sum_{\vec{n}}\sum_{l=1}^{d}\sum_{\alpha,\beta=1}^{3} [[\frac{m}{2}((-1)^{\vec{n}}\hat{Z}_{\alpha}(\vec{n})+(-1)^{\vec{n}+\hat{l}}\hat{Z}_{\beta}(\vec{n}+\hat{l})),\frac{1}{2a}(\hat{U}_{\alpha \beta}(\vec{n},l)\hat{\sigma}_{\alpha}^-(\vec{n})\hat{\sigma}_{\beta}^+(\vec{n}+\hat{l}) +h.c.)] \nonumber \\
    &,\frac{4d-2}{2a}(\hat{U}_{\beta\alpha}\hat{\sigma}_{\beta}^-\hat{\sigma}_{\alpha}^+ +h.c.)] || \nonumber \\
    &\leq 4dL^d \sum_{\alpha,\beta=1}^{3} ||m|| \cdot ||\frac{1}{2a}(\hat{U}_{\alpha \beta}(\vec{n},l)\hat{\sigma}_{\alpha}^-(\vec{n})\hat{\sigma}_{\beta}^+(\vec{n}+\hat{l}) +h.c.)|| \cdot ||\frac{4d-2}{2a}(\hat{U}_{ \beta\alpha}(\vec{n},l)\hat{\sigma}_{\beta}^-(\vec{n})\hat{\sigma}_{\alpha}^+(\vec{n}+\hat{l}) +h.c.)|| \nonumber \\
    &= (16d-8)dL^dm \sum_{\alpha,\beta=1}^{3} ||\frac{1}{2a}(\hat{U}_{\alpha \beta}(\vec{n},l)\hat{\sigma}_{\alpha}^-(\vec{n})\hat{\sigma}_{\beta}^+(\vec{n}+\hat{l}) +h.c.)||^2 \nonumber \\
    &= (16d-8)dL^dm \frac{9}{a^2} = \frac{(144 d^2 - 72 d)mL^d}{a^2},
\end{align}
where the factor of $4d-2$, in front of the third term of the commutator, is the number of $\hat{U}_{\beta \alpha}\hat{\sigma}_{\beta}^-\hat{\sigma}_{\alpha}^+ +h.c.$ that collide with each $\hat{U}_{ \alpha\beta}\hat{\sigma}_{\alpha}^-\hat{\sigma}_{\beta}^+ +h.c.$. The bound for case (ii) is
\begin{align}
    &\quad || \sum_{\vec{n}}\sum_{l=1}^{d}\sum_{\alpha,\beta=1}^{3} [[\frac{m}{2}((-1)^{\vec{n}}\hat{Z}_{\alpha}(\vec{n})+(-1)^{\vec{n}+\hat{l}}\hat{Z}_{\beta}(\vec{n}+\hat{l})),\frac{1}{2a}(\hat{U}_{\alpha \beta}(\vec{n},l)\hat{\sigma}_{\alpha}^-(\vec{n})\hat{\sigma}_{\beta}^+(\vec{n}+\hat{l}) +h.c.)] \nonumber \\
    &,\frac{2d-1}{2a}\sum_{\alpha'\beta'=\beta\gamma}^{\gamma\alpha}(\hat{U}_{\alpha'\beta'}\hat{\sigma}_{\alpha'}^-\hat{\sigma}_{\beta'}^+ +h.c.)] || \nonumber \\
    &= dL^d \cdot 4(2d-1) ||m|| \sum_{\alpha,\beta=1}^{3} ||\frac{1}{2a}(\hat{U}_{\alpha \beta}(\vec{n},l)\hat{\sigma}_{\alpha}^-(\vec{n})\hat{\sigma}_{\beta}^+(\vec{n}+\hat{l}) +h.c.)||\cdot ||\frac{1}{2a} \sum_{\alpha'\beta'=\beta\gamma}^{\gamma\alpha}(\hat{U}_{\alpha'\beta'}\hat{\sigma}_{\alpha'}^-\hat{\sigma}_{\beta'}^+ +h.c.)|| \nonumber \\
    &=mL^d(8d^2-4d)\cdot 9 \cdot \frac{1}{a}\frac{4\cdot 2}{2a} = \frac{mL^d(288 d^2 - 144 d)}{a^2}
\end{align}
where the factor of $2d-1$, in front of the third term of the nested commutator, is the number of $\hat{U}_{\alpha'\beta'}\hat{\sigma}_{\alpha'}^-\hat{\sigma}_{\beta'}^+ +h.c.$, with $\alpha'\beta'=\beta\gamma$ or $\gamma\alpha$, that collide with each $\hat{U}_{ \alpha\beta}\hat{\sigma}_{\alpha}^-\hat{\sigma}_{\beta}^+ +h.c.$.  In the last equality, we have accounted for the different bounds of $\hat{U}_{ \alpha\beta}\hat{\sigma}_{\alpha}^-\hat{\sigma}_{\beta}^+ +h.c.$ for each $\alpha \beta$, and the number of $\alpha'\beta'=\beta\gamma$ or $\gamma\alpha$, where $\gamma \neq \alpha, \beta$. Adding up the bounds for type (i) and (ii), we obtain the bound for the $C_{2,2}$ as follows:
\begin{equation}
    \frac{(432 d^2 + 108 d)mL^d}{a^2}.
\end{equation}

The bound for $C_{2,3}$ is given by
\begin{align}
    &\quad ||[[ \sum_{\vec{n}} \hat{D}_{\vec{n}}^{(M)} , \sum_{\vec{n}'} \hat{T}_{\vec{n}'}^{(K)} ]  , \sum_{\vec{n}''} \hat{L}_{\vec{n}''}^{(B)} ]|| \nonumber \\
    &\leq ||\sum_{\vec{n}}\sum_{l=1}^{d}\sum_{\alpha, \beta = 1}^{3} [[\frac{m}{2}((-1)^{\vec{n}}\hat{Z}_{\alpha}(\vec{n})+(-1)^{\vec{n}+\hat{l}}\hat{Z}_{\beta}(\vec{n}+\hat{l})),\frac{1}{2a}(\hat{U}_{\alpha \beta}(\vec{n},l)\hat{\sigma}_{\alpha}^-(\vec{n})\hat{\sigma}_{\beta}^+(\vec{n}+\hat{l}) +h.c.)] , \sum_{\vec{n}''} \hat{L}_{\vec{n}''}^{(B)} ]|| \nonumber \\
    &\leq dL^d\cdot 4||m||\cdot \frac{9}{a} \cdot ||\frac{2(d-1)}{2a^{4-d}g^2} \sum_{\alpha,\beta,\gamma,\delta=1}^{3} (\hat{U}_{\alpha \beta}\hat{U}_{\beta \gamma}\hat{U}_{\gamma \delta}^\dag \hat{U}_{\delta \alpha}^\dag + h.c.) || \leq \frac{5832 mL^d (d^2-d)}{a^{5-d}g^2}.
\end{align}

Now we consider $C_{2,4}$. We divide the commutators up into two cases the same way we did $C_{2,2}$. The bound for case (i), where the kinetic terms act on the same links, is given by
\begin{align}
    &\quad || \sum_{\vec{n}}\sum_{l=1}^{d}\sum_{\alpha,\beta=1}^{3} [[\frac{g^2}{2a^{d-2}}\hat{E}^2(\vec{n},l),\frac{1}{2a}(\hat{U}_{\alpha \beta}(\vec{n},l)\hat{\sigma}_{\alpha}^-(\vec{n})\hat{\sigma}_{\beta}^+(\vec{n}+\hat{l}) +h.c.)] ,\frac{1}{2a}\sum_{\alpha',\beta'=1}^{3}(\hat{U}_{\alpha' \beta'}(\vec{n},l)\hat{\sigma}_{\alpha'}^-(\vec{n})\hat{\sigma}_{\beta'}^+(\vec{n}+\hat{l}) +h.c.)] || \nonumber \\
    &\leq dL^d \cdot 2\sum_{\alpha,\beta=1}^{3}|| [\frac{g^2}{2a^{d-2}}\hat{E}^2(\vec{n},l),\frac{1}{2a}(\hat{U}_{\alpha \beta}(\vec{n},l)\hat{\sigma}_{\alpha}^-(\vec{n})\hat{\sigma}_{\beta}^+(\vec{n}+\hat{l}) +h.c.)]|| \nonumber \\
    &\quad \cdot || \frac{1}{2a}\sum_{\alpha',\beta'=1}^{3}(\hat{U}_{\alpha' \beta'}(\vec{n},l)\hat{\sigma}_{\alpha'}^-(\vec{n})\hat{\sigma}_{\beta'}^+(\vec{n}+\hat{l}) +h.c.) || \leq 2dL^d \frac{3g^2}{2a^{d-1}}(3\Lambda + 4) \cdot \frac{9}{a} = \frac{27 g^2 dL^d (3\Lambda + 4)}{a^d},
\end{align}
where we have used (\ref{eq:SU3_EK}) to evaluate the first norm expression in the first inequality. As in $C_{2,2}$, we separate case (ii), where the kinetic terms act on different links, into two types. The bound for type (i), where the color indices for the outer kinetic term is $\beta \alpha$, is given by
\begin{align}
    &\quad || \sum_{\vec{n}}\sum_{l=1}^{d}\sum_{\alpha,\beta=1}^{3} [[\frac{g^2}{2a^{d-2}}\hat{E}^2(\vec{n},l),\frac{1}{2a}(\hat{U}_{\alpha \beta}(\vec{n},l)\hat{\sigma}_{\alpha}^-(\vec{n})\hat{\sigma}_{\beta}^+(\vec{n}+\hat{l}) +h.c.)]  ,\frac{4d-2}{2a} (\hat{U}_{\beta \alpha}\hat{\sigma}_{\beta}^-\hat{\sigma}_{\alpha}^+ +h.c.)] || \nonumber \\
    &\leq dL^d \cdot 2 \frac{g^2}{6 a^{d-2}}(4+3\Lambda) \cdot \sum_{\alpha,\beta = 1}^{3}||\frac{1}{2a}(\hat{U}_{\alpha \beta}\hat{\sigma}_{\alpha}^- \hat{\sigma}_{\beta}^+ +h.c.)|| \cdot ||\frac{4d-2}{2a} (\hat{U}_{\beta \alpha}\hat{\sigma}_{\beta}^-\hat{\sigma}_{\alpha}^+ +h.c.)|| \nonumber \\
    &\leq \frac{(4d^2-2d)L^d g^2(3\Lambda + 4)}{3a^{d-2}}\sum_{\alpha,\beta = 1}^{3}||\frac{1}{2a}(\hat{U}_{\alpha \beta}\hat{\sigma}_{\alpha}^- \hat{\sigma}_{\beta}^+ +h.c.)||^2\nonumber \\
    &\leq \frac{(4d^2-2d)L^d g^2(3\Lambda + 4)}{3a^{d-2}}\frac{9}{a^2} =\frac{(12 d^2 - 6 d)L^d g^2(3\Lambda + 4)}{3a^d}.
\end{align}
For type (ii), where the color indices for the outer kinetic term is $\beta \gamma$ with $\gamma \neq \alpha$, or $\gamma \alpha$ with $\gamma \neq \beta$, we obtain the following bound:
\begin{align}
    &\quad || \sum_{\vec{n}}\sum_{l=1}^{d}\sum_{\alpha,\beta=1}^{3} [[\frac{g^2}{2a^{d-2}}\hat{E}^2(\vec{n},l),\frac{1}{2a}(\hat{U}_{\alpha \beta}(\vec{n},l)\hat{\sigma}_{\alpha}^-(\vec{n})\hat{\sigma}_{\beta}^+(\vec{n}+\hat{l}) +h.c.)]  ,\frac{2d-1}{2a}\sum_{\alpha'\beta'=\beta\gamma}^{\gamma\alpha}(\hat{U}_{\alpha'\beta'}\hat{\sigma}_{\alpha'}^-\hat{\sigma}_{\beta'}^+ +h.c.)] || \nonumber \\
    &\leq dL^d \cdot 2 \frac{g^2}{6 a^{d-2}}(4+3\Lambda) \cdot \sum_{\alpha,\beta = 1}^{3}||\frac{1}{2a}(\hat{U}_{\alpha \beta}\hat{\sigma}_{\alpha}^- \hat{\sigma}_{\beta}^+ +h.c.)|| \cdot ||\frac{2d-1}{2a}\sum_{\alpha'\beta'=\beta\gamma}^{\gamma\alpha}(\hat{U}_{\alpha'\beta'}\hat{\sigma}_{\alpha'}^-\hat{\sigma}_{\beta'}^+ +h.c.)|| \nonumber \\
    &\leq \frac{(2d^2-d)L^d g^2}{3 a^{d-2}}(4+3\Lambda) \cdot \sum_{\alpha,\beta = 1}^{3}||\frac{1}{2a}(\hat{U}_{\alpha \beta}\hat{\sigma}_{\alpha}^- \hat{\sigma}_{\beta}^+ +h.c.)|| \cdot ||\frac{1}{2a}\sum_{\alpha'\beta'=\beta\gamma}^{\gamma\alpha}(\hat{U}_{\alpha'\beta'}\hat{\sigma}_{\alpha'}^-\hat{\sigma}_{\beta'}^+ +h.c.)|| \nonumber \\
    &\leq \frac{(2d^2-d)L^d g^2}{3 a^{d-2}}(4+3\Lambda) \cdot 9 \cdot \frac{1}{a}\cdot \frac{4}{a} = \frac{(24 d^2 - 12 d)L^d g^2}{a^d}(4+3\Lambda).
\end{align}
Therefore, the bound for $C_{2,4}$ is
\begin{equation}
    \frac{(36 d^2 + 9 d) L^d g^2(4\Lambda + 3)}{a^d}.
\end{equation}

We proceed to evaluate the bound for $C_{2,5}$ as follows:
\begin{align}
    &\quad [[ \sum_{\vec{n}} \hat{D}_{\vec{n}}^{(E)} , \sum_{\vec{n}'} \hat{T}_{\vec{n}'}^{(K)} ]  , \sum_{\vec{n}''} \hat{L}_{\vec{n}''}^{(B)} ] \nonumber \\
    &\leq ||\sum_{\vec{n}}\sum_{l=1}^{d}\sum_{\alpha, \beta = 1}^{3} [[\frac{g^2}{2a^{d-2}}\hat{E}^2(\vec{n},l),\frac{1}{2a}(\hat{U}_{\alpha \beta}(\vec{n},l)\hat{\sigma}_{\alpha}^-(\vec{n})\hat{\sigma}_{\beta}^+(\vec{n}+\hat{l}) +h.c.)], \sum_{\vec{n}''} \hat{L}_{\vec{n}''}^{(B)}] || \nonumber \\
    &\leq dL^d\cdot 2\sum_{\alpha, \beta = 1}^{3} ||[\frac{g^2}{2a^{d-2}}\hat{E}^2(\vec{n},l),\frac{1}{2a}(\hat{U}_{\alpha \beta}(\vec{n},l)\hat{\sigma}_{\alpha}^-(\vec{n})\hat{\sigma}_{\beta}^+(\vec{n}+\hat{l}) +h.c.)]|| \nonumber \\
    &\quad \cdot ||\frac{2(d-1)}{2a^{4-d}g^2}\sum_{\alpha,\beta,\gamma,\delta=1}^{3}(\hat{U}_{\alpha \beta}\hat{U}_{\beta \gamma}\hat{U}_{\gamma \delta}^\dag \hat{U}_{\delta \alpha}^\dag + h.c. )||\nonumber \\
    &\leq dL^d\cdot 2 \frac{3g^2}{2a^{d-1}}(3\Lambda + 4) \cdot \frac{162(d-1)}{a^{4-d}g^2} = \frac{486 d(d-1)L^d (4\Lambda + 3)}{a^3},
\end{align}
where in the second inequality, we have used (\ref{eq:SU3_EK}) and the fact that there are $2(d-1)$ plaquettes consisting of the link $(\vec{n},l)$ to evaluate the first norm expression. The bound for the $C_{2,6}$ is given by
\begin{align}
    &\quad [[ \sum_{\vec{n}} \hat{D}_{\vec{n}}^{(E)} , \sum_{\vec{n}'} \hat{L}_{\vec{n}'}^{(B)} ]  , \sum_{\vec{n}''} \hat{T}_{\vec{n}''}^{(K)} ] \nonumber \\
    &= ||  [\sum_{\vec{n}} \sum_{i=1}^{d}\sum_{j\neq i;j=1}^{d} [ \frac{g^2}{2a^{d-2}} (\hat{E}^2(\vec{n},i) + \hat{E}^2(\vec{n}+\hat{i},j)+\hat{E}^2(\vec{n}+\hat{j},i) +\hat{E}^2(\vec{n},j) ) \nonumber \\
    &\quad , \frac{-1}{2a^{4-d}g^2} \sum_{\alpha,\beta,\gamma,\delta=1}^{3} (\hat{U}_{\alpha \beta}(\vec{n},i)\hat{U}_{ \beta \gamma}(\vec{n}+\hat{i},j)\hat{U}_{\gamma \delta}^{\dag}(\vec{n}+\hat{j},i)\hat{U}_{\delta \alpha}^{\dag}(\vec{n},j) + h.c.) ]\nonumber \\
    &\quad , \frac{1}{2a} \sum_{\alpha',\beta'=1}^{3}[\hat{U}_{\alpha'\beta'}(\vec{n},i) \hat{\sigma}^{-}_{\alpha'}(\vec{n})\hat{\sigma}^{+}_{\beta'}(\vec{n}+\hat{i}) +\hat{U}_{\alpha'\beta'}(\vec{n}+\hat{i},j) \hat{\sigma}^{-}_{\alpha'}(\vec{n}+\hat{i})\hat{\sigma}^{+}_{\beta'}(\vec{n}+\hat{i}+\hat{j}) \nonumber \\
    &\quad  + \hat{U}_{\alpha'\beta'}(\vec{n}+\hat{j},i) \hat{\sigma}^{-}_{\alpha'}(\vec{n}+\hat{j})\hat{\sigma}^{+}_{\beta'}(\vec{n}+\hat{j}+\hat{i}) +\hat{U}_{\alpha'\beta'}(\vec{n},j) \hat{\sigma}^{-}_{\alpha'}(\vec{n})\hat{\sigma}^{+}_{\beta'}(\vec{n}+\hat{j})
    +h.c.] ]  || \nonumber \\
    &\leq L^d \frac{d(d-1)}{2} \cdot 2 \cdot || \frac{54}{a^2}(4 + 3\Lambda) ||  \cdot 4|| \frac{1}{2a}\sum_{\alpha',\beta'=1}^{3} (\hat{U}_{\alpha'\beta'}\hat{\sigma}^{-}_{\alpha'}\hat{\sigma}^{+}_{\beta'} +h.c.) || \nonumber \\
    &\leq L^d \frac{d(d-1)}{2} \cdot 2 \cdot || \frac{54}{a^2}(4 + 3\Lambda) || \cdot 4|| \frac{9}{a}  || = \frac{1944 d(d-1)L^d (3\Lambda +4)}{a^3},
\end{align}
where we have used (\ref{eq:SU3_elec_mag}) to evaluate the first norm expression of the first inequality. The bound for $C_{2,7}$ is given by
\begin{align}
    &\quad ||[[ \sum_{\vec{n}} \hat{D}_{\vec{n}}^{(E)} , \sum_{\vec{n}'} \hat{L}_{\vec{n}'}^{(B)} ]  , \sum_{\vec{n}''} \hat{L}_{\vec{n}''}^{(B)} ]|| \nonumber \\
    &\leq ||  [\sum_{\vec{n}} \sum_{i=1}^{d}\sum_{j\neq i;j=1}^{d} [ \frac{g^2}{2a^{d-2}} (\hat{E}^2(\vec{n},i) + \hat{E}^2(\vec{n}+\hat{i},j)+\hat{E}^2(\vec{n}+\hat{j},i) +\hat{E}^2(\vec{n},j) ) \nonumber \\
    &\quad , \frac{-1}{2a^{4-d}g^2}\sum_{\alpha,\beta,\gamma,\delta=1}^{3} (\hat{U}_{\alpha \beta}(\vec{n},i)\hat{U}_{ \beta \gamma}(\vec{n}+\hat{i},j)\hat{U}_{\gamma \delta}^{\dag}(\vec{n}+\hat{j},i)\hat{U}_{\delta \alpha}^{\dag}(\vec{n},j) + h.c.) ], \sum_{\vec{n}''} \hat{L}_{\vec{n}''}^{(B)} ]||\nonumber \\
    &\leq L^d\frac{d(d-1)}{2} \cdot 2 ||\frac{54}{a^2}(4 + 3\Lambda)||\cdot || \frac{8d-11}{2a^{4-d}g^2} \sum_{\alpha,\beta,\gamma,\delta=1}^{3} (\hat{U}_{\alpha \beta}\hat{U}_{\beta \gamma}\hat{U}_{\gamma \delta}^\dag \hat{U}_{\delta \alpha}^\dag +h.c.) ||\nonumber\\
    &\leq L^d\frac{d(d-1)54(3\Lambda + 4)}{a^2} \cdot \frac{81(8d-11)}{a^{4-d}g^2} = \frac{L^d(3\Lambda + 4)}{a^{6-d}g^2} (34992 d^3 - 83106 d^2 + 48114 d),
\end{align}
where in the second inequality, we have used (\ref{eq:SU3_elec_mag}) to evaluate the first norm expression, and the factor of $8d-11$ in the second norm term is the number of plaquettes that collide on links with the plaquette acted on by the magnetic operators in the inner commutator, as explained in the paragraph below (\ref{eq:U1_EBB}).

Next we consider $C_{2,8}$. As in the case of SU(2), we divide the commutators up into five cases: (i) $\hat{h}_{T}$, $\hat{h}_{T'}$ and $\hat{h}_{T''}$ all act on the same links, (ii) $\hat{h}_{T}$ and $\hat{h}_{T'}$ act on the same links, while $\hat{h}_{T''}$ acts on neighboring links that are connect via the fermionic sites, (iii) $\hat{h}_{T}$ and $\hat{h}_{T''}$ act on the same links, while $\hat{h}_{T'}$ acts on neighboring links that are connect via the fermionic sites, (iv) $\hat{h}_{T'}$ and $\hat{h}_{T''}$ act on the same links, while $\hat{h}_{T}$ acts on neighboring links that are connect via the fermionic sites, (v) $\hat{h}_{T}$, $\hat{h}_{T'}$ and $\hat{h}_{T''}$ all act on different but connected links. We begin with case (i) by considering $\hat{h}_{T}$, $\hat{h}_{T'}$ and $\hat{h}_{T''}$ that act on one link $(\vec{n}_p,l)$ only. There are $13872$ such $\hat{h}_{T}$ terms, each with different parameters $a,b,c,\Delta j, \alpha, \beta$ , in $\mathbbm{T}$. Thus, the bound for case (i) is given by
\begin{align}
    &\quad 13872 \sum_{p=e}^{o}\sum_{l=1}^{d}||[[ \sum_{\vec{n}_p}\hat{h}_T(\vec{n}_p,l), \sum_{\hat{h}_{T'}\in \mathbbm{T};T'>T} \sum_{\vec{n}_p} \hat{h}_{T'}(\vec{n}_p,l) ]  , \sum_{\hat{h}_{T''}\in \mathbbm{T};T''>T} \sum_{\vec{n}_p} \hat{h}_{T''}(\vec{n}_p,l) ]|| \nonumber \\
    &\leq  13872 \cdot 2d \sum_{\vec{n}_p} 4|| \hat{h}_T(\vec{n}_p,l) || \cdot ||\sum_{\hat{h}_{T'}\in \mathbbm{T};T'>T} \hat{h}_{T'}(\vec{n}_p,l)|| \cdot || \sum_{\hat{h}_{T''}\in \mathbbm{T};T''>T} \hat{h}_{T''}(\vec{n}_p,l) ]|| \nonumber \\
    &\leq 55488 dL^d ||\frac{1}{2a}|| \cdot ||\frac{1}{a}\sum_{\alpha,\beta = 1}^{3}(\hat{U}_{\alpha \beta}(\vec{n}_p,l)\hat{\sigma}_{\alpha}^-(\vec{n}_p)\hat{\sigma}_{\beta}^+(\vec{n}_p+\hat{l}) +h.c.)||^2 \nonumber \\
    &\leq \frac{55488 dL^d}{a}(\frac{9}{a})^2 = \frac{4494528 dL^d}{a^3}.
\end{align}
 The bound for case (ii) is
\begin{align}
    &\quad 13872 \sum_{p=e}^{o}\sum_{l=1}^{d}||[[ \sum_{\vec{n}_p}\hat{h}_T(\vec{n}_p,l), \sum_{\hat{h}_{T'}\in \mathbbm{T};T'>T} \sum_{\vec{n}_p} \hat{h}_{T'}(\vec{n}_p,l) ]  , \sum_{\hat{h}_{T''}\in \mathbbm{T};T''>T} \hat{h}_{T''} ]|| \nonumber \\
    &\leq  13872\cdot 2d \sum_{\vec{n}_p} 4|| \hat{h}_T(\vec{n}_p,l) || \cdot ||\sum_{\hat{h}_{T'}\in \mathbbm{T};T'>T} \hat{h}_{T'}(\vec{n}_p,l)|| \cdot || \sum_{\hat{h}_{T''}\in \mathbbm{T};T''>T} \hat{h}_{T''} || \nonumber \\
    &\leq  \frac{55488 dL^d}{a} \cdot ||\frac{1}{2a}\sum_{\alpha,\beta = 1}^{3}(\hat{U}_{\alpha \beta}(\vec{n}_p,l)\hat{\sigma}_{\alpha}^-(\vec{n}_p)\hat{\sigma}_{\beta}^+(\vec{n}_p+\hat{l}) +h.c.)|| \cdot ||\frac{4d-1}{2a}\sum_{\alpha,\beta = 1}^{3}(\hat{U}_{\alpha \beta}\hat{\sigma}_{\alpha}^-\hat{\sigma}_{\beta}^+ +h.c.)|| \nonumber \\
    &\leq \frac{55488 dL^d}{a}\frac{9}{a}\frac{9(4d-1)}{a} = \frac{4494528(4d^2-d)L^d}{a^3},
\end{align}
where the factor of $4d-1$ in the third norm expression of the second inequality is the number of links connected to $(\vec{n}_p,l)$, via the fermionic sites on both its ends.

We separate case (iii) into two types: type (i) consists of commutators where $\hat{h}_{T'}$ act on links in the same direction, but of different parity, as $\hat{h}_{T},\hat{h}_{T''}$; type (ii) consists of commutators where $\hat{h}_{T'}$ act on links in different directions from $\hat{h}_{T},\hat{h}_{T''}$. Consider type (i), since we implement the even terms before the odd terms, the commutator bound is
\begin{align}
    &\quad 13872\sum_{l=1}^{d}||[[ \sum_{\vec{n}_e}\hat{h}_T(\vec{n}_e,l), \sum_{\hat{h}_{T'}\in \mathbbm{T};T'>T} \sum_{\vec{n}_o} \hat{h}_{T'}(\vec{n}_o,l) ]  , \sum_{\hat{h}_{T''}\in \mathbbm{T};T''>T} \sum_{\vec{n}_e} \hat{h}_{T''}(\vec{n}_e,l) ]|| \nonumber \\
    &\leq  13872 d \sum_{\vec{n}_e} 4|| \hat{h}_T(\vec{n}_e,l) || \cdot ||2 \frac{1}{2a}\sum_{\alpha,\beta = 1}^{3}(\hat{U}_{\alpha \beta}(\vec{n}_o,l)\hat{\sigma}_{\alpha}^-(\vec{n}_o)\hat{\sigma}_{\beta}^+(\vec{n}_o+\hat{l}) +h.c.) || \nonumber \\
    &\quad \cdot || 3 \frac{1}{2a}\sum_{\alpha,\beta = 1}^{3}(\hat{U}_{\alpha \beta}(\vec{n}_e,l)\hat{\sigma}_{\alpha}^-(\vec{n}_e)\hat{\sigma}_{\beta}^+(\vec{n}_e+\hat{l}) +h.c.) ]|| \nonumber \\
    &\leq 55488 d\frac{L^d}{2} ||\frac{1}{a}|| \cdot ||\frac{18}{a}||\cdot || \frac{27}{a}|| = \frac{13483584 dL^d}{a^3},
\end{align}
where in the first inequality, the factor of $2$ in the second norm expression is the number of odd links that are connected to each $(\vec{n}_e,l)$, and the factor of $3$ in the third norm expression is the number of links that collide with $(\vec{n}_e,l)$ or are connected with the two odd links connected to $(\vec{n}_e,l)$. Similarly, we obtain the bound for type (ii)
\begin{align}
    &\quad 13872 \sum_{p,p'=e}^{o} \sum_{l'>l} \sum_{l=1}^{d}||[[ \sum_{\vec{n}_p}\hat{h}_T(\vec{n}_p,l), \sum_{\hat{h}_{T'}\in \mathbbm{T};T'>T} \sum_{\vec{n}_{p'}} \hat{h}_{T'}(\vec{n}_{p'},l') ]  , \sum_{\hat{h}_{T''}\in \mathbbm{T};T''>T} \sum_{\vec{n}_p} \hat{h}_{T''}(\vec{n}_p,l) ]|| \nonumber \\
    &\leq  13872\cdot 4\frac{d(d-1)}{2} \sum_{\vec{n}_p} 4|| \hat{h}_T(\vec{n}_p,l) || \cdot ||2 \frac{1}{2a}\sum_{\alpha,\beta = 1}^{3}(\hat{U}_{\alpha \beta}(\vec{n}_p',l')\hat{\sigma}_{\alpha}^-(\vec{n}_p')\hat{\sigma}_{\beta}^+(\vec{n}_p'+\hat{l}') +h.c.) || \nonumber \\
    &\quad \cdot || \frac{1}{2a}\sum_{\alpha,\beta = 1}^{3}(\hat{U}_{\alpha \beta}(\vec{n}_p,l)\hat{\sigma}_{\alpha}^-(\vec{n}_p)\hat{\sigma}_{\beta}^+(\vec{n}_p+\hat{l}) +h.c.) ]|| \nonumber \\
    &\leq 55488 d(d-1)L^d ||\frac{1}{a}|| \cdot ||\frac{18}{a}||\cdot || \frac{9}{a}||= \frac{8989056 (d^2-d)L^d}{a^3}.
\end{align}
Therefore, the bound for case (iii) is
\begin{equation}
    \frac{(8989056 d^2 + 22472640 d)L^d}{a^3}. 
\end{equation}

We divide case (iv) into two types: type-(i) commutators are those where $\hat{h}_{T}$, $\hat{h}_{T'}$ and $\hat{h}_{T''}$ act on links in the same direction; and type-(ii) commutators are those where $\hat{h}_{T}$, $\hat{h}_{T'}$ and $\hat{h}_{T''}$ act on links in different directions. We consider type (i) first. Since even terms are implemented before odd terms, we obtain its bound as follows
\begin{align}
    &\quad 13872 \sum_{l=1}^{d}||[[ \sum_{\vec{n}_e}\hat{h}_T(\vec{n}_e,l), \sum_{\hat{h}_{T'}\in \mathbbm{T};T'>T} \sum_{(\vec{n}_o,l)} \hat{h}_{T'}(\vec{n}_o,l) ]  , \sum_{\hat{h}_{T''}\in \mathbbm{T};T''>T} \sum_{(\vec{n}_o,l)} \hat{h}_{T''}(\vec{n}_o,l) ]|| \nonumber \\
    &\leq 13872 d \cdot \sum_{\vec{n}_e}4 || \hat{h}_T(\vec{n}_e,l) || \cdot || \frac{2}{2a}\sum_{\alpha, \beta=1}^{3} (\hat{U}_{\alpha \beta}(\vec{n}_o,l)\hat{\sigma}_{\alpha}^-(\vec{n}_o)\hat{\sigma}_{\beta}^+(\vec{n}_o+\hat{l}) +h.c.)||^2 \nonumber \\
    &\leq \frac{13872 dL^d}{2} 4||\frac{1}{a}||\cdot ||\frac{18}{a}||^2 = \frac{8989056 dL^d}{a^2},
\end{align}
where the factor of $2$ in the numerator of the second norm expression of the second line is due to the fact that $(\vec{n}_e,l)$ is connected to two $(\vec{n}_o,l)$. Next, we evaluate the bound for type (ii) as follows:
\begin{align}
    &\quad 13872 \sum_{p,p'=e}^{o}\sum_{l'>l}\sum_{l=1}^{d}||[[ \sum_{\vec{n}_p}\hat{h}_T(\vec{n}_p,l), \sum_{\hat{h}_{T'}\in \mathbbm{T};T'>T} \sum_{(\vec{n}_{p'},l')} \hat{h}_{T'}(\vec{n}_{p'},l') ]  , \sum_{\hat{h}_{T''}\in \mathbbm{T};T''>T} \sum_{(\vec{n}_{p'},l')} \hat{h}_{T''}(\vec{n}_{p'},l') ]|| \nonumber \\
    &\leq 13872 \cdot 4\frac{d(d-1)}{2} \cdot 4 \sum_{\vec{n}_p}|| \hat{h}_{T}(\vec{n}_p,l) ||\cdot || \frac{2}{2a}\sum_{\alpha, \beta=1}^{3} (\hat{U}_{\alpha \beta}(\vec{n}_{p'},l')\hat{\sigma}_{\alpha}^-(\vec{n}_{p'})\hat{\sigma}_{\beta}^+(\vec{n}_{p'}+\hat{l}') +h.c.)||^2 \nonumber \\
    &\leq 27744 d(d-1)\frac{L^d}{2} 4 ||\frac{1}{a}|| \cdot ||\frac{18}{a}||^2 = \frac{17978112 (d^2-d)L^d}{a^3},
\end{align}
where the factor of $2$ in the numerator of the second norm expression of the second line is due to the fact that the link $(\vec{n}_p,l)$, acted on by $\hat{h}_T$, is connected to two links $(\vec{n}_{p'},l')$, acted on by $\hat{h}_{T'}$ and $\hat{h}_{T''}$. Therefore, case (iv) is bounded by
\begin{equation}
    \frac{8989056 (2d^2-d)L^d}{a^3}.
\end{equation}
Lastly, we obtain the bound for case (v)
\begin{align}
    &\quad 13872 \sum_{\substack{(p'',l'');\\(p'',l'')>(p',l')}}\sum_{\substack{(p',l');\\(p',l')>(p,l)}}\sum_{(p,l)}||[[ \sum_{\vec{n}_p}\hat{h}_T(\vec{n}_p,l), \sum_{\substack{\hat{h}_{T'}\in \mathbbm{T};\\T'>T}} \sum_{(\vec{n}_{p'},l')} \hat{h}_{T'}(\vec{n}_{p'},l') ]  , \sum_{\substack{\hat{h}_{T''}\in \mathbbm{T};\\T''>T}} \sum_{(\vec{n}_{p''},l'')} \hat{h}_{T''}(\vec{n}_{p''},l'') ]|| \nonumber \\
    &\leq 13872 \frac{4}{3}(2d^3-3d^2+d) \cdot 4 \sum_{\vec{n}_p} || \hat{h}_T(\vec{n}_p,l)|| \cdot || \frac{2}{2a}\sum_{\alpha, \beta=1}^{3} (\hat{U}_{\alpha \beta}(\vec{n}_{p'},l')\hat{\sigma}_{\alpha}^-(\vec{n}_{p'})\hat{\sigma}_{\beta}^+(\vec{n}_{p'}+\hat{l}') +h.c.)|| \nonumber \\
    &\quad \cdot || \frac{4}{2a}\sum_{\alpha, \beta=1}^{3} (\hat{U}_{\alpha \beta}(\vec{n}_{p''},l'')\hat{\sigma}_{\alpha}^-(\vec{n}_{p''})\hat{\sigma}_{\beta}^+(\vec{n}_{p''}+\hat{l}'') +h.c.)||\nonumber \\
    &= \frac{110976 L^d}{a}(2d^3-3d^2+d) ||\frac{18}{a}||\cdot ||\frac{36}{a}|| = \frac{L^d}{a^3}71912448(2 d^3 - 3 d^2 +  d),
\end{align}
where $(p',l')>(p,l)$ means that $(p',l')$ appears after $(p,l)$ in $\mathbbm{T}$, and since there are $2d$ $(p,l)$, the triple sum outside the norm expression in the first line evaluates to
\begin{equation}
    \sum_{q=1}^{2d} (2d-q)(2d-q-1) = \frac{4}{3}(2d^3-3d^2+d).
\end{equation}
In the second inequality, the factor of $2$ in the numerator of the second norm expression is because of the fact that $(\vec{n}_p,l)$ is connected to at most $2$ $(\vec{n}_{p'},l')$, and thus, each of the inner commutators acts on at most three links and four sites. Further, each of these four sites are connected to at most one $(\vec{n}_{p''},l'')$, hence the factor of $4$ in the numerator of the third norm expression. Note that constants can be further tightened by considering the color indices of the fermionic operators. Finally, adding the bounds for all cases, we arrive at the bound for $C_{2,8}$
\begin{equation}
    (143824896 d^3 - 170792064 d^2 + 85396032 d)\frac{L^d}{a^3}.
\end{equation}

Now for $C_{2,9}$, we separate the commutators into three cases. Case-(i) commutators consists of $\hat{h}_{T}$ and $\hat{h}_{T'}$ that act on the same links. Case-(ii) commutators consists of $\hat{h}_{T}$ and $\hat{h}_{T'}$ that act on links in the same directions, but of different parities. Case-(iii) commutators consists of $\hat{h}_{T}$ and $\hat{h}_{T'}$ that act on links in different directions, but connected via fermionic sites. The bound for case (i) is given by
\begin{align}
    &\quad 13872 \sum_{p=e}^{o} \sum_{l=1}^{d} [[ \sum_{\vec{n}_p}\hat{h}_T(\vec{n}_p,l) , \sum_{\hat{h}_{T'}\in \mathbbm{T};T'>T} \sum_{\vec{n}_p} \hat{h}_{T'}(\vec{n}_p,l) ]  , \sum_{\vec{n}} \hat{L}_{\vec{n}}^{(B)} ]\nonumber \\
    &\leq 13872\cdot 2d \sum_{\vec{n}_p} 4||\hat{h}_T(\vec{n}_p,l)|| \cdot || \frac{1}{2a}\sum_{\alpha, \beta=1}^{3} (\hat{U}_{\alpha, \beta}(\vec{n}_{p},l)\hat{\sigma}_{\alpha}^-(\vec{n}_{p})\hat{\sigma}_{\beta}^+(\vec{n}_{p}+\hat{l}) +h.c.) || \nonumber \\
    &\quad \cdot || \frac{2(d-1)}{2a^{4-d}g^2}\sum_{\alpha,\beta,\gamma,\delta=1}^{3} (\hat{U}_{\alpha \beta}\hat{U}_{\beta \gamma}\hat{U}_{\gamma \delta}^\dag \hat{U}_{\delta \alpha}^\dag +h.c.) ||\nonumber \\
    &\leq 13872 dL^d \cdot 4||\frac{1}{a}||\cdot ||\frac{9}{a}||\cdot ||\frac{162(d-1)}{a^{4-d}g^2} || = \frac{80901504(d^2 - d)L^d}{a^{6-d}g^2},
\end{align}
where in the first inequality, the numerator $2(d-1)$ in the third norm term is the number of magnetic operators that act on link $(\vec{n}_p, l)$.

For case (ii), there are two types of commutators; those where (i) $\hat{h}_T$ and $\hat{h}_{T'}$ have color indices $\alpha \beta$ and $\beta \alpha$, respectively, and thus, collide on two fermionic sites, and where (ii) $\hat{h}_T$ and $\hat{h}_{T'}$ have color indices $\alpha \beta$ and $\beta \gamma$ with $\gamma \neq \alpha$ or $\gamma \alpha$ with $\gamma \neq \beta$, respectively, and thus, collide on one fermionic sites. Thus, considering type (i), the inner commutators act on three links, which collide with $3\cdot 2(d-1)$ magnetic operators. Since we implement the even terms before the odd terms, we obtain the bound for the type-(i) commutators as follows:
\begin{align}
    &\quad 13872 \sum_{l=1}^{d} 4 \sum_{\vec{n}_e} ||\hat{h}_T(\vec{n}_e,l)|| \cdot || \frac{2}{2a} (\hat{U}_{ \beta\alpha}(\vec{n}_{o},l)\hat{\sigma}_{\beta}^-(\vec{n}_{o})\hat{\sigma}_{\alpha}^+(\vec{n}_{o}+\hat{l}) +h.c.) || \nonumber \\
    &\quad \cdot || \frac{6(d-1)}{2a^{4-d}g^2} \sum_{\alpha,\beta,\gamma,\delta=1}^{3} (\hat{U}_{\alpha \beta}\hat{U}_{\beta \gamma}\hat{U}_{\gamma \delta}^\dag \hat{U}_{\delta \alpha}^\dag +h.c.) || \nonumber \\
    &\leq 27744 dL^d ||\frac{1}{a}|| \cdot ||\frac{2}{a}||\cdot ||\frac{486 (d-1)}{a^{4-d}g^2}|| = \frac{26967168 (d^2-d)L^d}{a^{6-d}g^2}.
\end{align}
For type (ii), the inner commutators act on two links, and thus collide with $2\cdot 2(d-1)$ magnetic operators. We obtain the bound for the type-(ii) commutators as follows:
\begin{align}
    &\quad 13872 \sum_{l=1}^{d} 4 \sum_{\vec{n}_e} ||\hat{h}_T(\vec{n}_e,l)|| \cdot || \frac{1}{2a} \sum_{\alpha' \beta' = \beta \gamma}^{\gamma \alpha}(\hat{U}_{ \alpha' \beta'}(\vec{n}_{o},l)\hat{\sigma}_{\alpha'}^-(\vec{n}_{o})\hat{\sigma}_{\beta'}^+(\vec{n}_{o}+\hat{l}) +h.c.) || \nonumber \\
    &\quad \cdot || \frac{4(d-1)}{2a^{4-d}g^2} \sum_{\alpha,\beta,\gamma,\delta=1}^{3} (\hat{U}_{\alpha \beta}\hat{U}_{\beta \gamma}\hat{U}_{\gamma \delta}^\dag \hat{U}_{\delta \alpha}^\dag +h.c.) || \nonumber \\
    &\leq 27744 dL^d ||\frac{1}{a}|| \cdot ||\frac{4}{a}||\cdot ||\frac{324(d-1)}{a^{4-d}g^2}|| = \frac{35956224(d^2-d)L^d}{a^{6-d}g^2},
\end{align}
where in the last inequality, the numerator $4$ in the second norm expression in the number of $\alpha' \beta'=\beta \gamma$, or $\gamma \alpha$ with $\gamma \neq \alpha, \beta$. Thus, for case (ii), we obtain the bound
\begin{equation}
    \frac{62923392 (d^2-d)L^d}{a^{6-d}g^2}.
\end{equation}

In the third case, $\hat{h}_{T}$ and $\hat{h}_{T'}$ act on links in different directions. As in the second case, we divide up case (iii) based on the color indices of $\hat{h}_{T}$ and $\hat{h}_{T'}$. Focusing on the first type, the inner commutators act on three links, which collide with $2+3\cdot2(d-2)=6d-10$ magnetic operators, where two of them act on all three links, and there are $2(d-2)$ magnetic operators acting on each one link, but not the other two links. Thus, we obtain the bound for type (i)
commutators as follows:
\begin{align}
    &\quad 13872 \sum_{p,p'=e}^{o}\sum_{l'>l}\sum_{l=1}^{d} 4 \sum_{\vec{n}_p} ||\hat{h}_T(\vec{n}_p,l)|| \cdot || \frac{2}{2a} (\hat{U}_{ \beta\alpha}(\vec{n}_{p'},l')\hat{\sigma}_{\beta}^-(\vec{n}_{p'})\hat{\sigma}_{\alpha}^+(\vec{n}_{p'}+\hat{l}') +h.c.) || \nonumber \\
    &\quad \cdot || \frac{6d-10}{2a^{4-d}g^2} \sum_{\alpha,\beta,\gamma,\delta=1}^{3} (\hat{U}_{\alpha \beta}\hat{U}_{\beta \gamma}\hat{U}_{\gamma \delta}^\dag \hat{U}_{\delta \alpha}^\dag +h.c.) || \nonumber \\
    &\leq 13872 \cdot 4 \frac{d(d-1)}{2} \cdot 4\frac{L^d}{2} ||\frac{1}{a}|| \cdot ||\frac{2}{a}||\cdot ||\frac{81(6d-10)}{a^{4-d}g^2}||\nonumber \\
    &=(53934336 d^3 - 143824896 d^2 + 89890560 d) \frac{L^d}{a^{6-d}g^2}.
\end{align}
Moving onto the second type, the inner commutators act on two links, which collide with $1+2\cdot2(d-2)=4d-7$ magnetic operators, where one of them acts on both links, and there are $2(d-2)$ magnetic operators acting on each one link, but not the other. Hence, we evaluate the bound for type (ii), and obtain,
\begin{align}
    &\quad 13872 \sum_{p,p'=e}^{o}\sum_{l'>l}\sum_{l=1}^{d} 4 \sum_{\vec{n}_p} ||\hat{h}_T(\vec{n}_p,l)|| \cdot || \frac{1}{2a} \sum_{\alpha' \beta' = \beta \gamma}^{\gamma \alpha}(\hat{U}_{ \alpha' \beta'}(\vec{n}_{p'},l)\hat{\sigma}_{\alpha'}^-(\vec{n}_{p'})\hat{\sigma}_{\beta'}^+(\vec{n}_{p'}+\hat{l}) +h.c.) || \nonumber \\
    &\quad \cdot || \frac{4d-7}{2a^{4-d}g^2} \sum_{\alpha,\beta,\gamma,\delta=1}^{3} (\hat{U}_{\alpha \beta}\hat{U}_{\beta \gamma}\hat{U}_{\gamma \delta}^\dag \hat{U}_{\delta \alpha}^\dag +h.c.) || \nonumber \\
    &\leq 13872 \cdot 4 \frac{d(d-1)}{2} \cdot 4\frac{L^d}{2} ||\frac{1}{a}|| \cdot ||\frac{4}{a}||\cdot ||\frac{81(4d-7)}{a^{4-d}g^2}|| \nonumber \\
    &=(71912448 d^3 - 197759232 d^2 + 125846784 d) \frac{L^d}{a^{6-d}g^2}.
\end{align}
Therefore, the bound for case (iii) is
\begin{equation}
    (125846784 d^3 - 341584128 d^2 + 215737344 d) \frac{L^d}{a^{6-d}g^2}.
\end{equation}
Summing up the bounds for all three cases, we obtain the bound for $C_{2,9}$
\begin{equation}
    (125846784 d^3 - 197759232 d^2 + 71912448 d)\frac{L^d}{a^{6-d}g^2}.
\end{equation}

Now for $C_{2,10}$, we divide the commutators up into cases and types, as we have done for $C_{2,9}$. The bound for case (i) of both $C_{2,9}$ and $C_{2,10}$ is the same, and is given by
\begin{equation}
    \frac{80901504 (d^2-d)L^d}{a^{6-d}g^2}.
\end{equation}
The bound for case (ii) can be obtained from the case-(ii) bounds for $C_{2,9}$ after some slight modifications. First, since the kinetic operator $\hat{h}_T$ only act on one link, there are only $2(d-1)$ plaquette operators in the inner commutator that do not commute with each $\hat{h}_T$ because the plaquettes may lie on $d-1$ two-dimensional planes and can be of two different parities. Second, the kinetic operators $\hat{h}_{T'}$ not only collide with $\hat{h}_{T}$ via fermionic sites, but also with the magnetic operators on links. Thus, we obtain the bound for type (i) of case (ii) 
\begin{align}
    &\quad 13872  \sum_{l=1}^{d} 4 \sum_{\vec{n}_e} ||\hat{h}_T(\vec{n}_e,l)||  \cdot || \frac{2(d-1)}{2a^{4-d}g^2} \sum_{\alpha,\beta,\gamma,\delta=1}^{3} (\hat{U}_{\alpha \beta}\hat{U}_{\beta \gamma}\hat{U}_{\gamma \delta}^\dag \hat{U}_{\delta \alpha}^\dag +h.c.) ||\nonumber \\
    &\quad \cdot || \frac{4}{2a} (\hat{U}_{ \beta\alpha}(\vec{n}_{o},l)\hat{\sigma}_{\beta}^-(\vec{n}_{o})\hat{\sigma}_{\alpha}^+(\vec{n}_{o}+\hat{l}) +h.c.) || \nonumber \\
    &\leq 27744 d(d-1) L^d ||\frac{1}{a}|| \cdot ||\frac{162}{a^{4-d}g^2}||\cdot ||\frac{4}{a}|| = \frac{17978112 (d^2-d)L^d}{a^{6-d}g^2},
\end{align}
where the numerator $4$ in the third norm expression is due to the fact that two $\hat{h}_{T'}$ collides with each of $\hat{h}_{T}$, and the pair of magnetic operators that lie on the same plane. The bound for type (ii) of case (ii) is
\begin{align}
    &\quad 13872  \sum_{l=1}^{d} 4 \sum_{\vec{n}_e} ||\hat{h}_T(\vec{n}_e,l)||   \cdot || \frac{2(d-1)}{2a^{4-d}g^2} \sum_{\alpha,\beta,\gamma,\delta=1}^{3} (\hat{U}_{\alpha \beta}\hat{U}_{\beta \gamma}\hat{U}_{\gamma \delta}^\dag \hat{U}_{\delta \alpha}^\dag +h.c.) ||\nonumber \\
    &\quad \cdot || \frac{3}{2a} \sum_{\alpha' \beta' = \beta \gamma}^{\gamma \alpha}(\hat{U}_{ \alpha' \beta'}(\vec{n}_{o},l)\hat{\sigma}_{\alpha'}^-(\vec{n}_{o})\hat{\sigma}_{\beta'}^+(\vec{n}_{o}+\hat{l}) +h.c.) ||\nonumber \\
    &\leq 27744 d(d-1)L^d ||\frac{1}{a}|| \cdot ||\frac{162}{a^{4-d}g^2}||\cdot ||\frac{3\cdot 4}{a}|| = \frac{53934336 (d^2-d)L^d}{a^{6-d}g^2},
\end{align}
where the numerator $3$ in the third norm expression is due to the fact that for a fixed pair of color indices $\alpha' \beta' $, one $\hat{h}_{T'}$ collides with $\hat{h}_{T}$, and two with the pair of magnetic operators that lie on the same plane. Thus, for case (ii), the bound is given by
\begin{equation}
    \frac{71912448 (d^2-d)L^d}{a^{6-d}g^2}.
\end{equation}
Now we consider the third case. Once again, we modify the case-(iii) bounds of $C_{2,9}$. Thus, we obtain the respective bounds for type (i) and (ii)
commutators as follows:
\begin{align}
    &\quad 13872\sum_{p,p'=e}^{o}\sum_{l'>l}\sum_{l=1}^{d} 4 \sum_{\vec{n}_p} ||\hat{h}_T(\vec{n}_p,l)|| \cdot || \frac{2(d-1)}{2a^{4-d}g^2} \sum_{\alpha,\beta,\gamma,\delta=1}^{3} (\hat{U}_{\alpha \beta}\hat{U}_{\beta \gamma}\hat{U}_{\gamma \delta}^\dag \hat{U}_{\delta \alpha}^\dag +h.c.) || \nonumber \\
    &\quad \cdot || \frac{2}{2a} (\hat{U}_{ \beta\alpha}(\vec{n}_{p'},l')\hat{\sigma}_{\beta}^-(\vec{n}_{p'})\hat{\sigma}_{\alpha}^+(\vec{n}_{p'}+\hat{l}') +h.c.) || \nonumber \\
    &\leq 13872 \cdot 4 \frac{d(d-1)}{2} \cdot 4\frac{L^d}{2} ||\frac{1}{a}||\cdot ||\frac{162(d-1)}{a^{4-d}g^2}||  \cdot ||\frac{2}{a}|| =17978112( d^3 - 2 d^2 +  d)\frac{L^d}{a^{6-d}g^2},
\end{align}
and
\begin{align}
    &\quad 13872 \sum_{p,p'=e}^{o}\sum_{l'>l}\sum_{l=1}^{d} 4 \sum_{\vec{n}_p} ||\hat{h}_T(\vec{n}_p,l)|| \cdot || \frac{2(d-1)}{2a^{4-d}g^2} \sum_{\alpha,\beta,\gamma,\delta=1}^{3} (\hat{U}_{\alpha \beta}\hat{U}_{\beta \gamma}\hat{U}_{\gamma \delta}^\dag \hat{U}_{\delta \alpha}^\dag +h.c.) || \nonumber \\
    &\quad \cdot || \frac{1}{2a} \sum_{\alpha' \beta' = \beta \gamma}^{\gamma \alpha}(\hat{U}_{ \alpha' \beta'}(\vec{n}_{p'},l)\hat{\sigma}_{\alpha'}^-(\vec{n}_{p'})\hat{\sigma}_{\beta'}^+(\vec{n}_{p'}+\hat{l}) +h.c.) ||  \nonumber \\
    &\leq 13872 \cdot 4 \frac{d(d-1)}{2} \cdot 4\frac{L^d}{2} ||\frac{1}{a}||\cdot ||\frac{162(d-1)}{a^{4-d}g^2}||\cdot ||\frac{4}{a}|| =35956224( d^3 - 2 d^2 +  d) \frac{L^d}{a^{6-d}g^2}.
\end{align}
Therefore, the bound for case (iii) is
\begin{equation}
    53934336( d^3 - 2 d^2 +  d) \frac{L^d}{a^{6-d}g^2}.
\end{equation}
Summing up the bounds for all cases, we obtain the bound for $C_{2,10}$
\begin{equation}
    (53934336 d^3 + 44945280 d^2 - 98879616 d)\frac{L^d}{a^{6-d}g^2}.
\end{equation}

We compute the bound for $C_{2,11}$, and obtain
\begin{align}
    &\quad \sum_{\hat{h}_{T}\in \mathbbm{T}} ||[[ \hat{h}_T , \sum_{\vec{n}} \hat{L}_{\vec{n}}^{(B)} ]  , \sum_{\vec{n}'} \hat{L}_{\vec{n}'}^{(B)} ]|| \nonumber \\  
    &\leq 13872 dL^d\cdot 4||\hat{h}_T(\vec{n}_p,l)||\cdot ||\frac{2(d-1)}{2a^{4-d}g^2} \sum_{\alpha,\beta,\gamma,\delta=1}^{3} (\hat{U}_{\alpha \beta}\hat{U}_{\beta \gamma}\hat{U}_{\gamma \delta}^\dag \hat{U}_{\delta \alpha}^\dag +h.c.) ||  \cdot ||\frac{14d-20}{2a^{4-d}g^2}\sum_{\alpha,\beta,\gamma,\delta=1}^{3} (\hat{U}_{\alpha \beta}\hat{U}_{\beta \gamma}\hat{U}_{\gamma \delta}^\dag \hat{U}_{\delta \alpha}^\dag +h.c.)|| \nonumber \\
    &\leq \frac{55488 d{L^d}}{a}\frac{162(d-1)}{a^{4-d}g^2}\frac{162(7d-10)}{a^{4-d}g^2} = \frac{(10193589504 d^3 - 24755860224 d^2 + 14562270720 d)L^d}{a^{9-2d}g^4},
\end{align}
where in the first inequality, the factors of $2(d-1)$ and $14d-20$ are explained in the paragraph below (\ref{eq:U1_KBB}).

Lastly, we consider $C_{2,12}$, which consists of commutators between only magnetic operators. The commutators are either \textit{intra-plaquette} or \textit{inter-plaquette}, where $\hat{h}_{L}$, $\hat{h}_{L'}$ and $\hat{h}_{L''}$ act on the same or different plaquettes, respectively. We consider intra-plaquette terms first. We remind the readers that there are $N_L \equiv 735010926133248$ different $\hat{h}_{L}(\vec{n}_p,j,k)$ terms acting on each plaquette $(\vec{n}_p,j,k)$. The bound for the intra-plaquette commutators is given by
\begin{align}
    &\quad N_L \sum_{k\neq j; k=1}^{d} \sum_{j=1}^{d}\sum_{\vec{n}_p}\sum_{p=e}^o||[[ \hat{h}_L(\vec{n}_p,j,k) , \sum_{\hat{h}_{L'}\in \mathbbm{L};L'>L} \hat{h}_{L'}(\vec{n}_p,j,k) ]  , \sum_{\hat{h}_{L''}\in \mathbbm{L};L''>L} \hat{h}_{L''}(\vec{n}_p,j,k) ]|| \nonumber \\
    &\leq \frac{N_L d(d-1)L^d}{2} 4||\hat{h}_L(\vec{n}_p,j,k)|| \cdot ||\frac{1}{2a^{4-d}g^2}\sum_{\alpha,\beta,\gamma,\delta=1}^{3} (\hat{U}_{\alpha \beta}\hat{U}_{\beta \gamma}\hat{U}_{\gamma \delta}^\dag \hat{U}_{\delta \alpha}^\dag +h.c.)||^2 \nonumber \\
    &\leq 2N_L d(d-1)L^d ||\frac{1}{a^{4-d}g^2}||\cdot ||\frac{81}{a^{4-d}g^2}||^2 \nonumber \\
    &= \frac{L^d}{a^{12-3d}g^6}9644813372720480256(d^2-d).
\end{align}

We now proceed to analyze the inter-plaquette commutators. Since each $\hat{h}_{L''}$ operator is non-zero, the inner commutator $[ \hat{h}_L , \sum_{\hat{h}_{L'}\in \mathbbm{L};L'>L} \hat{h}_{L'} ]$ must be non-zero to guarantee a non-trivial triple commutator
\begin{equation}
    [[ \hat{h}_L , \sum_{\hat{h}_{L'}\in \mathbbm{L};L'>L} \hat{h}_{L'} ]  , \sum_{\hat{h}_{L''}\in \mathbbm{L};L''>L} \hat{h}_{L''} ]. \nonumber
\end{equation}
Given a non-zero inner commutator, we further divide the inter-plaquette commutators into three types. Type (i), (ii) and (iii) commutators satisfy (\ref{eq:SU_BBB1}),(\ref{eq:SU_BBB2}) and (\ref{eq:SU_BBB3}), respectively.

For type (i), the plaquettes acted on by $\hat{h}_L$, collide with those acted on by $\hat{h}_{L'}$ and $\hat{h}_{L''}$, and those acted on by $\hat{h}_{L'}$ also collide with those acted on by $\hat{h}_{L''}$. Suppose $\hat{h}_L$ is labelled by $(p,k,l)$. The possible parity-location tuples that label $\hat{h}_{L'}$ and $\hat{h}_{L''}$ are given in Table \ref{tb:SU2_bbb_1} and \ref{tb:SU2_bbb_2} for $p=$ $even$ and $odd$, respectively. Consider first the case where $p=even$, and $\hat{h}_{L}$ and $\hat{h}_{L'}$ are labelled by items $1-4$, $6$ and $7$ in Table \ref{tb:SU2_bbb_1}. Then, each plaquette acted on by $\hat{h}_{L}$ is acted on by two $\hat{h}_{L'}$ as they share one dimension. Further, the plaquettes acted on by $\hat{h}_{L}$ and $\hat{h}_{L'}$ are acted on by either $2,4,6$ or $8$ $\hat{h}_{L''}$. In particular, if the plaquettes acted on by $\hat{h}_{L''}$ (i) share one common dimension with $\hat{h}_{L}$, and two common dimensions and parity with $\hat{h}_{L'}$, or (ii) share one dimension with $\hat{h}_{L}$ and a different dimension with $\hat{h}_{L'}$, then $\hat{h}_{L}$ and $\hat{h}_{L'}$ collide with two $\hat{h}_{L''}$. We compute the number of combinations of parity-location labels that satisfy these conditions using Table \ref{tb:SU2_bbb_1}, and obtain 
\begin{equation}
    \sum_{l>k}\sum_{k=1}^{d-1} 12(d-l)+2(l-k-1) = \frac{7}{3}(d^3-3d^2+2d).
\end{equation}
The bound in this case is given by
\begin{align}
    &\quad N_L ||[[ \sum_{\vec{n}_e}\hat{h}_L(e,k,l) ,\sum_{\hat{h}_{L'}\in \mathbbm{L};L'>L} \hat{h}_{L'} ]  , \sum_{\hat{h}_{L''}\in \mathbbm{L};L''>L} \hat{h}_{L''} ]|| \nonumber \\
    &\leq \frac{N_L \cdot 7}{3}(d^3-3d^2+2d) \sum_{\vec{n}_e} 4\cdot || \hat{h}_L(e,k,l) || \cdot ||\frac{2}{2a^{4-d}g^2}\sum_{\alpha,\beta,\gamma,\delta=1}^{3} (\hat{U}_{\alpha \beta}\hat{U}_{\beta \gamma}\hat{U}_{\gamma \delta}^\dag \hat{U}_{\delta \alpha}^\dag +h.c.)|| \nonumber \\
    &\quad  \cdot || \frac{2}{2a^{4-d}g^2}\sum_{\alpha,\beta,\gamma,\delta=1}^{3} (\hat{U}_{\alpha \beta}\hat{U}_{\beta \gamma}\hat{U}_{\gamma \delta}^\dag \hat{U}_{\delta \alpha}^\dag +h.c.)||\nonumber \\
    &\leq \frac{L^d}{a^{12-3d}g^6}90018258145391149056(d^3-3d^2+2d),
\end{align}
where $N_L$ is the number of $\hat{h}_L$ terms per plaquette.

If the plaquettes acted on by $\hat{h}_{L''}$ share only one dimension with both $\hat{h}_{L}$ and $\hat{h}_{L'}$, then $\hat{h}_{L}$ and $\hat{h}_{L'}$ collide with four $\hat{h}_{L''}$. Using Table \ref{tb:SU2_bbb_1}, we find the number of combinations of parity-location labels that satisfy this condition, i.e.
\begin{align}
    &\quad \sum_{l>k}\sum_{k=1}^{d-1} (d-l)[8(d-l-1)+4(l-k-1)] + (l-k-1)[4(d-l)+4(l-k-2)] \nonumber \\
    &= \frac{4}{3}(d^4-6d^3+11d^2-6d).
\end{align}
The bound in this case is given by
\begin{align}
    &\quad N_L ||[[ \sum_{\vec{n}_e}\hat{h}_L(e,k,l) ,\sum_{\hat{h}_{L'}\in \mathbbm{L};L'>L} \hat{h}_{L'} ]  , \sum_{\hat{h}_{L''}\in \mathbbm{L};L''>L} \hat{h}_{L''} ]|| \nonumber \\
    &\leq \frac{N_L \cdot 4}{3}(d^4-6d^3+11d^2-6d) \sum_{\vec{n}_e} 4\cdot || \hat{h}_L(e,k,l) || \cdot ||\frac{2}{2a^{4-d}g^2}\sum_{\alpha,\beta,\gamma,\delta=1}^{3} (\hat{U}_{\alpha \beta}\hat{U}_{\beta \gamma}\hat{U}_{\gamma \delta}^\dag \hat{U}_{\delta \alpha}^\dag +h.c.)|| \nonumber \\
    &\quad  \cdot || \frac{4}{2a^{4-d}g^2}\sum_{\alpha,\beta,\gamma,\delta=1}^{3} (\hat{U}_{\alpha \beta}\hat{U}_{\beta \gamma}\hat{U}_{\gamma \delta}^\dag \hat{U}_{\delta \alpha}^\dag +h.c.)||\nonumber \\
    &\leq \frac{L^d}{a^{12-3d}g^6}102878009309018456064(d^4-6d^3+11d^2-6d).
\end{align}

If the plaquettes acted on by $\hat{h}_{L''}$ share only one dimension with $\hat{h}_{L'}$, and share both dimensions, but not the parity, with $\hat{h}_{L}$, then $\hat{h}_{L}$ and $\hat{h}_{L'}$ collide with six $\hat{h}_{L''}$. Once again, we use Table \ref{tb:SU2_bbb_1} to obtain the number of combinations of parity-location labels that satisfy this condition, i.e.
\begin{equation}
    \sum_{l>k}\sum_{k=1}^{d-1} 4(d-l)+2(l-k-1) = (d^3-3d^2+2d).
\end{equation}
The bound in this case is given by
\begin{align}
    &\quad N_L ||[[ \sum_{\vec{n}_e}\hat{h}_L(e,k,l) ,\sum_{\hat{h}_{L'}\in \mathbbm{L};L'>L} \hat{h}_{L'} ]  , \sum_{\hat{h}_{L''}\in \mathbbm{L};L''>L} \hat{h}_{L''} ]|| \nonumber \\
    &\leq N_L (d^3-3d^2+2d) \sum_{\vec{n}_e} 4\cdot || \hat{h}_L(e,k,l) || \cdot ||\frac{2}{2a^{4-d}g^2}\sum_{\alpha,\beta,\gamma,\delta=1}^{3} (\hat{U}_{\alpha \beta}\hat{U}_{\beta \gamma}\hat{U}_{\gamma \delta}^\dag \hat{U}_{\delta \alpha}^\dag +h.c.)|| \nonumber \\
    &\quad  \cdot || \frac{6}{2a^{4-d}g^2}\sum_{\alpha,\beta,\gamma,\delta=1}^{3} (\hat{U}_{\alpha \beta}\hat{U}_{\beta \gamma}\hat{U}_{\gamma \delta}^\dag \hat{U}_{\delta \alpha}^\dag +h.c.)||\nonumber \\
    &\leq \frac{L^d}{a^{12-3d}g^6}115737760472645763072(d^3-3d^2+2d).
\end{align}

If the plaquettes acted on by $\hat{h}_{L''}$ share only one dimension with $\hat{h}_{L}$, and share both dimensions, but not the parity, with $\hat{h}_{L'}$, then $\hat{h}_{L}$ and $\hat{h}_{L'}$ collide with eight $\hat{h}_{L''}$. Once again, we use Table \ref{tb:SU2_bbb_1} to obtain the number of combinations of parity-location labels that satisfy this condition, i.e.
\begin{equation}
    \sum_{l>k}\sum_{k=1}^{d-1} 4(d-l)+2(l-k-1) = (d^3-3d^2+2d).
\end{equation}
The bound in this case is given by
\begin{align}
    &\quad N_L ||[[ \sum_{\vec{n}_e}\hat{h}_L(e,k,l) ,\sum_{\hat{h}_{L'}\in \mathbbm{L};L'>L} \hat{h}_{L'} ]  , \sum_{\hat{h}_{L''}\in \mathbbm{L};L''>L} \hat{h}_{L''} ]|| \nonumber \\
    &\leq N_L (d^3-3d^2+2d) \sum_{\vec{n}_e} 4\cdot || \hat{h}_L(e,k,l) || \cdot ||\frac{2}{2a^{4-d}g^2}\sum_{\alpha,\beta,\gamma,\delta=1}^{3} (\hat{U}_{\alpha \beta}\hat{U}_{\beta \gamma}\hat{U}_{\gamma \delta}^\dag \hat{U}_{\delta \alpha}^\dag +h.c.)|| \nonumber \\
    &\quad  \cdot || \frac{8}{2a^{4-d}g^2}\sum_{\alpha,\beta,\gamma,\delta=1}^{3} (\hat{U}_{\alpha \beta}\hat{U}_{\beta \gamma}\hat{U}_{\gamma \delta}^\dag \hat{U}_{\delta \alpha}^\dag +h.c.)||\nonumber \\
    &\leq \frac{L^d}{a^{12-3d}g^6}154317013963527684096(d^3-3d^2+2d).
\end{align}

If $\hat{h}_{L''}$ acts on plaquettes that share only one dimension with $\hat{h}_{L}$ and $\hat{h}_{L'}$, then $\hat{h}_{L}$ and $\hat{h}_{L'}$ collide with eight $\hat{h}_{L''}$. There are
\begin{equation}
    \sum_{l>k}\sum_{k=1}^{d-1} 4(d-l)+2(l-k-1) = d^3-3d^2+2d
\end{equation}
combinations of parity-location labels that satisfy this condition. The bound in this case is given by
\begin{align}
    &\quad N_L ||[[ \sum_{\vec{n}_e}\hat{h}_L(e,k,l) ,\sum_{\hat{h}_{L'}\in \mathbbm{L};L'>L} \hat{h}_{L'} ]  , \sum_{\hat{h}_{L''}\in \mathbbm{L};L''>L} \hat{h}_{L''} ]|| \nonumber \\
    &\leq N_L (d-1) \sum_{\vec{n}_e} 4\cdot || \hat{h}_L(e,k,l) || \cdot ||\frac{4}{2a^{4-d}g^2}\sum_{\alpha,\beta,\gamma,\delta=1}^{2} (\hat{U}_{\alpha \beta}\hat{U}_{\beta \gamma}\hat{U}_{\gamma \delta}^\dag \hat{U}_{\delta \alpha}^\dag +h.c.)|| \nonumber \\
    &\quad  \cdot || \frac{8}{2a^{4-d}g^2}\sum_{\alpha,\beta,\gamma,\delta=1}^{2} (\hat{U}_{\alpha \beta}\hat{U}_{\beta \gamma}\hat{U}_{\gamma \delta}^\dag \hat{U}_{\delta \alpha}^\dag +h.c.)||\nonumber \\
    &\leq \frac{L^d}{a^{12-3d}g^6}308634027927055368192(d^3-3d^2+2d).
\end{align}

Consider now the case where $\hat{h}_{L}$ and $\hat{h}_{L'}$ collide on two dimensions, but have different parities, i.e., item 5 in Table \ref{tb:SU2_bbb_1}. Since we implement even terms before odd ones, the parities of $\hat{h}_{L}$ and $\hat{h}_{L'}$ are even and odd, respectively. Moreover, $\hat{h}_{L}$ acting on a plaquette collides with four $\hat{h}_{L'}$ on the four links. If $\hat{h}_{L''}$ act on plaquettes that share one dimension with $\hat{h}_{L}$, and both dimensions and the parity with those acted on by $\hat{h}_{L'}$, then $\hat{h}_{L}$ and $\hat{h}_{L'}$ collide with four $\hat{h}_{L''}$. There are
\begin{equation}
    \sum_{k=1}^{d-1} 1 = d-1
\end{equation}
combinations of parity-location labels that satisfy this condition. The bound in this case is given by
\begin{align}
    &\quad N_L ||[[ \sum_{\vec{n}_e}\hat{h}_L(e,k,l) ,\sum_{\hat{h}_{L'}\in \mathbbm{L};L'>L} \hat{h}_{L'} ]  , \sum_{\hat{h}_{L''}\in \mathbbm{L};L''>L} \hat{h}_{L''} ]|| \nonumber \\
    &\leq N_L (d-1) \sum_{\vec{n}_e} 4\cdot || \hat{h}_L(e,k,l) || \cdot ||\frac{4}{2a^{4-d}g^2}\sum_{\alpha,\beta,\gamma,\delta=1}^{3} (\hat{U}_{\alpha \beta}\hat{U}_{\beta \gamma}\hat{U}_{\gamma \delta}^\dag \hat{U}_{\delta \alpha}^\dag +h.c.)|| \nonumber \\
    &\quad  \cdot || \frac{4}{2a^{4-d}g^2}\sum_{\alpha,\beta,\gamma,\delta=1}^{3} (\hat{U}_{\alpha \beta}\hat{U}_{\beta \gamma}\hat{U}_{\gamma \delta}^\dag \hat{U}_{\delta \alpha}^\dag +h.c.)||\nonumber \\
    &\leq \frac{L^d}{a^{12-3d}g^6}154317013963527684096(d-1).
\end{align}

Therefore, type (i) commutators, where $\hat{h}_{L}$ acts on even plaquettes are bounded by
\begin{align}
    &\quad(102878009309018456064 d^4 + 51439004654509228032 d^3-874463079126656876544 d^2 + 874463079126656876544 d\nonumber\\
    &-154317013963527684096)\frac{L^d}{a^{12-3d}g^6}.
\end{align}
Similarly, we obtain the bound for the commutators where $\hat{h}_{L}$ acts on odd plaquettes, i.e.,
\begin{align}
    &\quad (102878009309018456064 d^4 - 64298755818136535040 d^3  \nonumber \\
    &-527249797708719587328 d^2+488670544217837666304 d)\frac{L^d}{a^{12-3d}g^6}.
\end{align}
by considering separately the cases, in which $\hat{h}_L$ and $\hat{h}_{L'}$ collide with 2, 4, or 8 $\hat{h}_{L''}$, listed in Table \ref{tb:SU2_bbb_2}. Thus, the bound for all type-(i) commutators is
\begin{align}
    &\quad(205756018618036912128 d^4 - 12859751163627307008 d^3 -1401712876835376463872 d^2
    + 1363133623344494542848 d\nonumber \\ &-154317013963527684096)\frac{L^d}{a^{12-3d}g^6}.
\end{align}

We proceed to analyze type-(ii) commutators. By definition, $\hat{h}_{L}$ does not commute with both $\hat{h}_{L'}$ and $\hat{h}_{L''}$, but $\hat{h}_{L'}$ and $\hat{h}_{L''}$ commute with each other. On the lattice, this implies that $\hat{h}_{L}$ share one common dimension each with $\hat{h}_{L'}$ and $\hat{h}_{L''}$, but $\hat{h}_{L'}$ and $\hat{h}_{L''}$ share no common dimension. Thus, each plaquette acted on by $\hat{h}_{L}$ is also acted on by two $\hat{h}_{L'}$ and $\hat{h}_{L''}$. Using table \ref{tb:SU2_bbb_3}, we obtain the number of combinations of parity-location labels that satisfy this condition as follows
\begin{equation}
    \sum_{l>k}\sum_{k=1}^{d-1} 2(d-l) [4(l-k-1)+8(d-l-1)] + 8(l-k-1)(d-l) = 2d^4 -12d^3+11d^2-6d.
\end{equation}
Hence, the bound for all type-(ii) commutators is
\begin{align}
    &\quad N_L \sum_{p=e}^{o} ||[[ \sum_{\vec{n}_p}\hat{h}_L(p,k,l) ,\sum_{\hat{h}_{L'}\in \mathbbm{L};L'>L} \hat{h}_{L'} ]  , \sum_{\hat{h}_{L''}\in \mathbbm{L};L''>L} \hat{h}_{L''} ]|| \nonumber \\
    &\leq 2N_L (2d^4 -12d^3+11d^2-6d)  \sum_{\vec{n}_p} 4\cdot || \hat{h}_L(p,k,l) || \cdot ||\frac{2}{2a^{4-d}g^2}\sum_{\alpha,\beta,\gamma,\delta=1}^{3} (\hat{U}_{\alpha \beta}\hat{U}_{\beta \gamma}\hat{U}_{\gamma \delta}^\dag \hat{U}_{\delta \alpha}^\dag +h.c.)|| \nonumber \\
    &\quad  \cdot || \frac{2}{2a^{4-d}g^2}\sum_{\alpha,\beta,\gamma,\delta=1}^{3} (\hat{U}_{\alpha \beta}\hat{U}_{\beta \gamma}\hat{U}_{\gamma \delta}^\dag \hat{U}_{\delta \alpha}^\dag +h.c.)||\nonumber \\
    &\leq \frac{L^d}{a^{12-3d}g^6}( 154317013963527684096 d^4 - 925902083781166104576 d^3 \nonumber \\
    &\quad+ 848743576799402262528 d^2  - 462951041890583052288 d).
\end{align}

Last but not least, for type (iii) commutators, $\hat{h}_{L'}$ does not commute with both $\hat{h}_{L}$ and $\hat{h}_{L''}$, but $\hat{h}_{L}$ and $\hat{h}_{L''}$ commute with each other. On the lattice, this implies that $\hat{h}_{L'}$ share one common dimension each with $\hat{h}_{L}$ and $\hat{h}_{L''}$, but $\hat{h}_{L}$ and $\hat{h}_{L''}$ share no common dimension. Thus, each plaquette acted on by $\hat{h}_{L}$ is also acted on by two $\hat{h}_{L'}$, and each plaquette acted on by $\hat{h}_{L'}$ is in turn acted on by two $\hat{h}_{L''}$. Using table \ref{tb:SU2_bbb_4}, we evaluate the number of combinations of parity-location labels that satisfy this condition, and obtain
\begin{equation}
    \sum_{j>l}\sum_{l>k}\sum_{k=1}^{d-1} 16(d-l)(d-j+l-k-1) + 8(l-k-1)[(l-j-1)+(j-k-1)] = \frac{2}{5}(2d^5 - 15d^4 + 40d^3 - 45 d^2 + 18d).
\end{equation}
Hence, the bound for all type-(iii) commutators is
\begin{align}
    &\quad N_L \sum_{p=e}^{o} ||[[ \sum_{\vec{n}_p}\hat{h}_L(p,k,l) ,\sum_{\hat{h}_{L'}\in \mathbbm{L};L'>L} \hat{h}_{L'} ]  , \sum_{\hat{h}_{L''}\in \mathbbm{L};L''>L} \hat{h}_{L''} ]|| \nonumber \\
    &\leq \frac{4N_L}{5}(2d^5 - 15d^4 + 40d^3 - 45 d^2 + 18d) \sum_{\vec{n}_p} 4\cdot || \hat{h}_L(p,k,l) || \cdot ||\frac{2}{2a^{4-d}g^2}\sum_{\alpha,\beta,\gamma,\delta=1}^{2} (\hat{U}_{\alpha \beta}\hat{U}_{\beta \gamma}\hat{U}_{\gamma \delta}^\dag \hat{U}_{\delta \alpha}^\dag +h.c.)|| \nonumber \\
    &\quad  \cdot || \frac{2}{2a^{4-d}g^2}\sum_{\alpha,\beta,\gamma,\delta=1}^{2} (\hat{U}_{\alpha \beta}\hat{U}_{\beta \gamma}\hat{U}_{\gamma \delta}^\dag \hat{U}_{\delta \alpha}^\dag +h.c.)||\nonumber \\
    &= ( 308634027927055368192\frac{d^5}{5}- 462951041890583052288 d^4 +1234536111708221472768 d^3 \nonumber \\ &-1388853125671749156864 d^2 +2777706251343498313728 \frac{d}{5} )\frac{L^d}{a^{12-3d}g^6}.
\end{align}

Finally, summing up the bounds for the intra-plaquette and all three types of inter-plaquette commutators, we obtain the bound for $C_{2,12}$,
\begin{align}
 &\quad ( 308634027927055368192\frac{d^5}{5} -
 102878009309018456064 d^4+ 295774276763428061184 d^3 
  \nonumber\\
 &-1932177612335002877952 d^2 + 7230395091749453365248 \frac{d}{5} -154317013963527684096) \frac{L^d}{a^{12-3d}g^6}.
\end{align}

\subsubsection{Oracle errors}
\label{subsubsec:oracle_SU3}

Here, we describe the direct syntheses of the kinetic and magnetic oracles, and compute the errors incurred by the fixed point arithmetic circuits.

{\bf Syntheses of the kinetic oracles:} The kinetic oracle, defined in (\ref{eq:SU3_kin_oracle}), can be directly synthesized as two controlled-diagonal gates, which impart the phases $f_{\alpha \beta}\frac{t}{2a}(-1)^{f_{\alpha,r}}f_{\beta,r+1} \prod_{j}b_j$, for $f_{\alpha,r}=0,1$, if the control bits $f_{\beta,r+1}$ and $b_j$'s are all ones. Recall that the function 
\begin{equation}
    f_{\alpha \beta}(p,q,T_L,T_L^z,Y_L,T_R,T_R^z,Y_R,\Delta p,\Delta q,\Delta T_L,\Delta T_R) \nonumber    
\end{equation}
are the matrix elements of the diagonal operator defined in (\ref{eq:SU3_Diag_op}). The number of control bits, and costs of implementation for each oracle depend on the parameters of the function $f_{\alpha \beta}$. In order to estimate the implementation costs, we first construct $f_{\alpha \beta}$ from the Clebsch-Gordan coefficients in Table \ref{tb:CG_SU3}, and normalization factors in (\ref{eq:SU3_N}), and then, obtain its arguments using (\ref{eq:SU3_inv_shift}) to invert the encoding of the quantum numbers onto the registers. Here, for the sake of complexity analysis, we only estimate the costs of the most expensive oracle, where $\alpha\beta=11$, and $(\Delta p,\Delta q,\Delta T_L,\Delta T_R) = (0,-1,1,1)$. In terms of SU(3) Clebsch-Gordan coefficients and normalization factor, we write
\begin{equation}
    f_{11}(p,q,T_L,T_L^z,Y_L,T_R,T_R^z,Y_R,0,-1,1,1) = C^L_{31}C^R_{31}N_3 = I_{31}^L c_{11}^L I_{31}^R c_{11}^R N_3,
\end{equation}
where $C_{31}$ is a SU(3) Clebsch-Gordan coefficient from Table \ref{tb:CG_SU3}, $I_{31}$ is an isoscalar factor defined in Table (\ref{tb:iso}), $c_{11}$ is a SU(2) Clebsch-Gordan coefficient from Table \ref{tb:CG_SU2}, and $N_3$ is a normalization factor in (\ref{eq:SU3_N}). Hereafter, we denote this function as $f_{11}(0,-1,1,1)$. Now we apply the mapping in (\ref{eq:SU3_inv_shift}) to obtain the relations
\begin{gather}
    I_{31}^i = \sqrt{\frac{(4p+2q-3T_i-Y_i+3\Lambda)(2p-2q+3T_i+Y_i-3\Lambda+6)(2p+4q-3T_i+Y_i-3\Lambda)}{(1+q)(2+p+q)(432+216T_i)}}\\
    c_{11}^i=\sqrt{\frac{\frac{(T_i + T_i^z)}{2}-\Lambda+1}{T_i+1}} \\
    N_3 = \sqrt{\frac{(1+q)(2+p+q)}{q(1+p+q)}}.
\end{gather}

Without loss of generality, we consider the case where $f_{\alpha,r}=0$. Then, we implement the diagonal phase gate by computing $f_{11}(0,-1,1,1)$ into an ancilla register, conditioned upon the values of the control bits. Then, by applying $R_z$ gates to the ancilla state $\ket{f_{11}(0,-1,1,1)}$, we induce the correct phase. Finally we uncompute $\ket{f_{11}(0,-1,1,1)}$. The computation of $f_{11}(0,-1,1,1)$, can be broken down into five steps. In the first and second steps, we compute the numerator and denominator, respectively. In the third step, we approximate the inverse of the denominator, using the fixed-point circuits in \cite{bhaskar2016quantum}. In the fourth step, we approximate the argument of the square-root by multiplying together the numerator and the inverse of the denominator. Lastly, we approximate the square-root, using the fixed-point circuits in \cite{bhaskar2016quantum}. Only the third and last steps incur approximation errors. Hereafter, we use the logarithmic depth out-of-place adder developed in \cite{draper2006logarithmic}, unless one of the inputs is classically known in which case we use the adder proposed in \cite{gidney2018halving}, and the multiplier proposed in \cite{shaw2020quantum}.

We consider the computation of the numerator. We compute the numerators of $I_{31}^L$, $c_{11}^L$, $I_{31}^R$, and $c_{11}^R$, separately. The numerator of $N_3$ cancels out with the part of the denominator of $I_{31}^L$. First, we compute the numerator of $I_{31}^L$. $\ket{4p}$ and $\ket{2q}$ can be computed by two ancilla qubits to $\ket{p}$ and one to $\ket{q}$, costing no T gates. $\ket{3T_L}$ can be computed by adding $\ket{2T_L}$ and $\ket{T_L}$, using an $(\eta+2)$-bit adder since $T_L$ is an $(\eta+1)$-bit number. Computing $4p+2q$ and $3T_L-Y_L$ requires an $(\eta+2)$ and $(\eta+3)$-bit adder. Computing $4p+2q-3T_L-Y_L$ costs one $(\eta+4)$-bit adder, and adding $3\Lambda$ requires $4(\eta+3)$ T gates. Therefore, computing $(4p+2q-3T_L-Y_L+3\Lambda)$ costs $84\eta-12(2\lfloor \log(\eta+2)\rfloor+\lfloor \log(\eta+3)\rfloor+\lfloor \log(\eta+4)\rfloor)+216$ T gates, $5\eta+21$ storage ancilla qubits, and $\eta-\lfloor \log(\eta+4)\rfloor+4$ workspace ancilla qubits. The computation of $(2p+4q-3T_L+Y_L-3\Lambda)$ has the same costs. Similarly, we compute $(2p+2q+3T_L+Y_L-3\Lambda+6)$, using $84\eta-12(\lfloor \log(\eta+1)\rfloor+\lfloor \log(\eta+2)\rfloor+\lfloor \log(\eta+3)\rfloor+\lfloor \log(\eta+4)\rfloor)+196$ T gates, $5\eta+20$ storage ancilla qubits, and $\eta-\lfloor \log(\eta+4)\rfloor+4$ workspace ancilla qubits. Next, we multiply together these three $(\eta+5)$-bit numbers, using two multipliers, costing $(12\eta+60)+(4\eta+16)(36\eta-3\lfloor \log(\eta+5)\rfloor-3\lfloor \log(2\eta+10)\rfloor+154$ T gates, $5\eta+25$ storage ancilla qubits, and $6\eta-\lfloor\log(2\eta+10)\rfloor + 29$ workspace ancilla qubits. The computation of the numerator for $I_{31}^R$ incurs the same costs. The computation of the numerator of $c^L_{11}$ requires an $(\eta+2)$-bit adder, since $T_L^z$ is an $(\eta+2)$-bit number. The division by 2 is done by shifting the decimal point, and the addition of $-\Lambda+1$ requires $\eta+3$ ancilla qubits, and $4(\eta+1)$ T gates. Therefore, the total costs are $24\eta-12\lfloor\log(\eta+2)\rfloor+40$ T gates, $2\eta+6$ storage ancilla qubits, and $\eta-\lfloor\log(\eta+2)\rfloor+2$ workspace ancilla qubits. The computation of the numerator for $c_{11}^R$ incurs the same costs. We now have to multiply together two $(3\eta+15)$-bit numbers and two $(\eta+3)$-bit numbers to obtain the numerator of $f_{11}(0,-1,1,1)$, conditioned upon the fact that they are all positive. The positivity condition guarantees that the Clebsch-Gordan coefficients and isoscalar factors are real \cite{de1963octet,kaeding1995tables}, since in their definitions, the denominators in the square roots are positive. This can be implemented by controlling one of the three multiplications to be done by the sign bits of the four inputs. The first multiplication is between two $(3\eta+15)$-bit numbers; the second one is between two $(\eta+3)$-bit numbers; and the last one is between the outputs from the previous two multiplications, which result in a $(6\eta+30)$ and $(2\eta+6)$-bit number. The total costs of the three multipliers are $40\eta+192+4(2\eta+5)(12(6\eta+30)-3\lfloor\log(6\eta+30)\rfloor-13)+4(3\eta+14)(12(3\eta+15)-3\lfloor\log(3\eta+15)\rfloor-13)+4(\eta+3)(12(\eta+2)-12\lfloor\log(\eta+3)\rfloor-13)$ T gates, $16\eta+72$ storage ancilla qubits, and $3(6\eta+30)-\lfloor\log(6\eta+30)\rfloor-1$ workspace ancilla qubits. We control the multiplication of the second multiplier by the four sign bits. As discussed in the SU(2) oracle implementation, the quadruply-controlled multiplication additionally costs a quadruply-controlled Toffoli gate and $\eta+3$ Toffoli gates, which amount to $4\eta+35$ T gates, $\eta+3$ storage ancilla qubits, and a reusable workspace ancilla qubit.

Next, we consider the computation of the denominator, i.e., $(1+q)(2+p+q)(432+216T_L)(432+216T_R)(T_L+1)(T_R+1)q(1+p+q)$. Computing $p+q$ requires an $\eta$-bit adder, and then, adding 1 requires $\eta+2$ storage ancilla qubits and $4(\eta-1)$ T gates. We then obtain $q(1+p+q)$ by multiplying $(1+p+q)$ by $q$. The computation of $q(1+p+q)$ costs $4(\eta+2)+4(\eta-1)(12\eta-3\lfloor\log(\eta+2)\rfloor+11)$ T gates, $2\eta+2$ storage ancilla qubits, and $3\eta-\lfloor\log(\eta+2)\rfloor+5$ workspace ancilla qubits. We then compute $(1+q)(2+p+q)$ by adding $2+p+2q$ to $q(1+p+q)$. Computing $p+2q$ costs an $(\eta+1)$-bit adder, and the addition of 2 costs $(\eta+3)$ ancilla qubits and $4(\eta+1)$ T gates. Thus, adding $2+p+2q$ to $q(1+p+q)$ costs a $(2\eta+2)$-bit adder. Then, we multiply together $q(1+p+q)$ and $(1+q)(2+p+q)$. As such, calculating $q(1+p+q)(1+q)(2+p+q)$ incurs $(8\eta+4)(24\eta-3\lfloor\log(2\eta+3)\rfloor +23)+(4\eta-4)(12\eta-3\lfloor\log(\eta+2)\rfloor+11)+100\eta-12(\lfloor\log(\eta)\rfloor+\lfloor\log(\eta+1)\rfloor+\lfloor\log(2\eta+2)\rfloor)+68$ T gates, $12\eta+18$ storage ancilla qubits, and $6\eta-\lfloor\log(2\eta+3)\rfloor+8$ workspace ancilla qubits. Since $(432+216 T_L) = 2^3(2^5-2^2-2^0)(2+T_L)$, we obtain $(432+216 T_L)$ in three steps. First, we compute $2+T_L$, incurring $\eta+2$ storage ancilla qubits and $4(\eta-1)$ T gates. Then, we compute $2^5(2+T_L)$ and $2^2(2+T_L)$ by copying $2+T_L$ to two ancilla registers, and then appending extra ancilla qubits for the multiplications. Finally, we use an $(\eta+4)$-bit adder and an $(\eta+7)$-bit adder to compute $(2^5-2^2-2^0)(2+ T_L)$, and then, append three ancilla qubits to obtain the desired output. Thus, calculating $(432+216 T_L)$ costs $48\eta-12(\lfloor\log(\eta+4)\rfloor+\lfloor\log(\eta+7)\rfloor)+204$ T gates, $5\eta+26$ storage ancilla qubits, and $\eta-\lfloor\log(\eta+7)\rfloor+7$ workspace ancilla qubits. Computing $(432+216 T_R)$ incurs the same costs. $(432+216 T_L)$ and $(432+216 T_R)$ are both $(\eta+11)$-bit numbers. Thus, evaluating their product costs $4\eta+44+(4\eta+40)(12\eta-3\lfloor\log(\eta+11)\rfloor+119)$ T gates, $2\eta+22$ storage ancilla qubits, and $3\eta-\lfloor\log(\eta+11)\rfloor+30$ workspace ancilla qubits. Calculating $(T_L+1)$ and $(T_R+1)$ cost $2\eta+4$ storage ancilla qubits, and $8(\eta-1)$ T gates. We compute their product using $4\eta+8+(4\eta+4)(12\eta-3\lfloor\log(\eta+2)\rfloor+11)$ T gates, $2\eta+4$ storage ancilla qubits, and $3\eta-\lfloor\log(\eta+2)\rfloor+5$ workspace ancilla qubits. We then multiply together $(432+216 T_L)(432+216 T_R)$, a $(2\eta+22)$-bit number, and $(T_L+1)(T_R+1)$, a $(2\eta+4)$-bit number, which costs $(8\eta+12)(24\eta-3\lfloor\log(2\eta+22)\rfloor+251)+8\eta+88$ T gates, $4\eta+26$ storage ancilla qubits, and $6\eta-\lfloor\log(2\eta+22)\rfloor+65$ workspace ancilla qubits. Finally, we obtain the denominator by multiplying the $(4\eta+26)$-bit output to $q(1+p+q)(1+q)(2+p+q)$, a $(4\eta+5)$-bit number, which requires $(4\eta+16)(48\eta -3\lfloor\log(4\eta+26)\rfloor+299)+16\eta+104$ T gates, $8\eta+31$ storage ancilla qubits, and $12\eta-\lfloor\log(4\eta+26)\rfloor+77$ workspace ancilla qubits.

We proceed to compute the inverse of the denominator using the algorithm in \cite{bhaskar2016quantum}. We refer the readers to the SU(2) kinetic oracle implementation in Sec. \ref{subsubsec:oracle_SU2} for a detailed overview of the algorithm. Here, we simply state the algorithmic parameters and costs. The input $w$ is at most an $(8\eta+31)$-bit integer. Truncating at $b\geq 8\eta+31$ bits after the decimal point, the approximation error is bounded from above by $\frac{2+\log(b)}{2^b}$. The T-gate count is $\lceil\log(b)\rceil\cdot (48b^2+768b\eta-(12b+360\eta-12)\lfloor\log(b)\rfloor+2824b-256\eta -12\lfloor\log(2b+8\eta+31)\rfloor-96\lfloor\log(2b)\rfloor-888) + 32\eta + 135$. The required number of storage ancilla qubits is $\lceil\log(b)\rceil\cdot(7b+16\eta+64)$, and that of workspace ancilla qubits is $6b-\lfloor\log(2b)\rfloor-1$. Next, we multiply the numerator, an $8\eta+36$-bit number, and the inverse of the denominator, a $b$-bit number to obtain the fraction. Assuming that $b \geq 8\eta+36$, this costs $384b\eta-96\eta\lfloor\log(b)\rfloor+1684b-416\eta-420\lfloor\log(b)\rfloor-1820$ T gates, $b+8\eta+36$ storage ancilla qubits, and $3b-\lfloor\log(b)\rfloor-1$ workspace ancilla qubits. Lastly, we compute the square-root function. The output of the previous step is a $(b+8\eta+36)$-bit fraction, bounded above and below by 1 and 0, respectively. Since the algorithm require an input that is larger than 1, we shift it by $b+8\eta+36$ bits to obtain an integer input $w$. Once we obtain the root, we shift it back by $(b+8\eta+36)/2$ bits, assuming without loss of generality that $b$ is even. Let $c\geq 2b+16\eta+72$. Then, we approximate $\sqrt{w}$, up to $\left(\frac{3}{4}\right)^{c-(2b+16\eta+72)}(2+c+\log(c))$ error. This costs $\lceil\log(c)\rceil\cdot[528c^2+96bc+768c\eta-c(12\lfloor \log(c)\rfloor+12\lfloor \log(\lceil \frac{3c}{2}\rceil)\rfloor+12\lfloor \log(\lceil \frac{5c}{2}\rceil)\rfloor+18\lfloor \log(1+4c)\rfloor)-12b\lfloor \log(2c)\rfloor-96n\lfloor \log(2c)\rfloor+12(\lfloor \log(c)\rfloor+\lfloor \log(\lceil \frac{3c}{2}\rceil)\rfloor-35\lfloor \log(2c)\rfloor+\lfloor \log(\lceil \frac{5c}{2}\rceil)\rfloor+\lfloor \log(1+4c)\rfloor-\lfloor \log(2c+b+8n+36)\rfloor)+2930c-32b-256\eta-988]
+\lceil \frac{19c}{2}\rceil+4b+32\eta+163$ T gates, $\lceil\log(c)\rceil\cdot(\lceil \frac{49c}{2}\rceil+2b+16\eta+77)$ storage ancilla qubits, and $12c-\lfloor \log(4c+1)\rfloor+5$ workspace ancilla qubits.

We impart the phase by applying $R_z(2^{k-c} \theta)$, where $c$ is the number of digits after the decimal point in the output of the square-root functions, and $\theta = \frac{t}{2a}$, to the $k$th qubit of the ancilla state $\ket{f_{\alpha \beta}}$. In order to implement the controlled version of this phase gate, we control each $R_z$ gate by nine control bits, as required by the costliest kinetic oracle, and the ancilla bit that checks the realness of the Clebsch-Gordan coefficients and isoscalar factors. Each $R_z$ gate with ten controls requires two Toffoli gates with ten controls, which cost 71 T gates each \cite{maslov2016advantages}, two $R_z$ gates and one ancilla qubit~\cite{wang2020resource}. Since the state has at most $\lceil \frac{3c}{2}\rceil$ qubits and there are $dL^d$ links on the lattice, there are $\lceil \frac{3c}{2}\rceil dL^d$ multi-controlled $R_z$ gates to be applied.

We sum up the T-gate requirements for all the steps, and multiply the outcome by two to account for the uncomputation costs. Conjugating each controlled $R_z$ gate with a pair of CNOT gates, where $f_{\alpha,r}$ is the control bit, addresses both $f_{\alpha,r}=0,1$. As such, the T-gate count for one oracle call, i.e.,
\begin{align}
    \mathcal{T}^{(K)}&=2\lceil\log(c)\rceil\cdot[528c^2+96bc+768c\eta-c(12\lfloor \log(c)\rfloor+12\lfloor \log(\lceil \frac{3c}{2}\rceil)\rfloor+12\lfloor \log(\lceil \frac{5c}{2}\rceil)\rfloor\nonumber \\
    &\quad +18\lfloor \log(1+4c)\rfloor)-12b\lfloor \log(2c)\rfloor-96\eta\lfloor \log(2c)\rfloor+12(\lfloor \log(c)\rfloor+\lfloor \log(\lceil \frac{3c}{2}\rceil)\rfloor\nonumber\\
    &\quad -35\lfloor \log(2c)\rfloor+\lfloor \log(\lceil \frac{5c}{2}\rceil)\rfloor+\lfloor \log(1+4c)\rfloor-\lfloor \log(2c+b+8\eta+36)\rfloor)+2930c\nonumber \\
    &\quad -32b-256\eta-980]
    +19c+8b+64\eta+326\nonumber \\
    &\quad + 768b\eta-192\eta\lfloor\log(b)\rfloor+3368b-832\eta-840\lfloor\log(b)\rfloor-3640\nonumber \\
    &\quad +2\lceil\log(b)\rceil\cdot (48b^2+768b\eta-(12b+360\eta-12)\lfloor\log(b)\rfloor+2824b-256\eta \nonumber \\
    &\quad -12\lfloor\log(2b+8\eta+31)\rfloor-96\lfloor\log(2b)\rfloor-888) + 64\eta + 270 \nonumber\\
    &\quad + 4128\eta^2-\eta(48\lfloor\log(2+\eta)\rfloor+96\lfloor\log(3+\eta)\rfloor+48\lfloor\log(5+\eta)\rfloor+24\lfloor\log(11+\eta)\rfloor\nonumber\\
    &\quad+48\lfloor\log(3+2\eta)\rfloor+48\lfloor\log(10+2\eta)\rfloor+48\lfloor\log(22+2\eta)\rfloor+72\lfloor\log(15+3\eta)\rfloor\nonumber\\
    &\quad+24\lfloor\log(26+4\eta)\rfloor+48\lfloor\log(30+6\eta)\rfloor) -(24\lfloor\log(\eta)\rfloor+72\lfloor\log(1+\eta)\rfloor\nonumber\\
    &\quad+288\lfloor\log(2+\eta)\rfloor+432\lfloor\log(3+\eta)\rfloor+192\lfloor\log(4+\eta)\rfloor+192\lfloor\log(5+\eta)\rfloor\nonumber\\
    &\quad+48\lfloor\log(7+\eta)\rfloor+240\lfloor\log(11+\eta)\rfloor+24\lfloor\log(2+2\eta)\rfloor+24\lfloor\log(3+2\eta)\rfloor\nonumber\\
    &\quad +192\lfloor\log(10+2\eta)\rfloor+72\lfloor\log(22+2\eta)\rfloor+336\lfloor\log(15+3\eta)\rfloor+96\lfloor\log(26+4\eta)\rfloor\nonumber\\
    &\quad +120\lfloor\log(30+6\eta)\rfloor)+30952 \eta+54156+426c+96\eta+80.
    \label{eq:SU3_kin_oracle_T}
\end{align}
The number of storage ancilla qubits required is
\begin{align}
    \lceil \log(c)\rceil \cdot (\lceil \frac{49c}{2}\rceil+2b+16\eta+77)+\lceil\log(b)\rceil\cdot(16\eta+7b+64)+b+89\eta+667,
\end{align}
and the number of workspace ancilla qubits required is
\begin{equation}
    12c-\lfloor\log(4c+1)\rfloor+5.
\end{equation}

There are two types of syntheses errors, i.e., arithmetic approximation errors and $R_z$ syntheses errors. The latter will be analyzed in Sec. \ref{subsubsec:Synth_SU3}. The arithmetic approximation errors per step is given by
\begin{equation}
    \frac{2+\log(b)}{2^b}+\left(\frac{3}{4}\right)^{c-(2b+16\eta+72)}(2+c+\log(c)).
\end{equation}
In total, the approximation errors are 
\begin{equation}
    \epsilon^{(K)}=r\cdot 13872dL^d[\frac{2+\log(b)}{2^b}+\left(\frac{3}{4}\right)^{c-(2b+16\eta+72)}(2+c+\log(c))],
\end{equation}
where $r$ is the number of Trotter steps, and $13872dL^d$ is the number of oracle calls. We divide the approximation errors evenly such that
\begin{equation}
    \epsilon^{(K)}_b=\frac{\epsilon^{(K)}}{2r\cdot 13872dL^d},\:\epsilon^{(K)}_c=\frac{\epsilon^{(K)}}{2r\cdot 13872dL^d},
\end{equation}
and
\begin{gather}
    \epsilon^{(K)}_b\geq \frac{2+\log(b)}{2^b}, \label{eq:SU3_eps_kinb}\\
    \epsilon^{(K)}_c\geq \left(\frac{3}{4}\right)^{c-( 2b+16\eta+72)}(2+c+\log(c)).\label{eq:SU3_eps_kinc}
\end{gather}
Note that this is not the optimal division of approximation errors. We let $b = \log(\frac{8}{\epsilon^{(K)}_b})$. Then, (\ref{eq:SU3_eps_kinb}) is always satisfied for $0<\epsilon^{(K)}_b<1$. We proceed to compute the upper bound for $c$. Since $c\geq 2b+16\eta+72\geq 32\eta+144$, and $\eta\geq 1$, $\frac{186}{176}c \geq 2+c+\log(c)$. Inserting this relation and our choice of $b$ into (\ref{eq:SU3_eps_kinc}), we obtain
\begin{align}
    \epsilon^{(K)}_c\geq \left(\frac{3}{4}\right)^{c-( \log(8/\epsilon^{(K)}_b)+16\eta+72)}\frac{186}{176}c.
\end{align}
Let
\begin{equation}
    \tilde{\epsilon}^{(K)}_c = \epsilon^{(K)}_c\cdot \left(\frac{3}{4}\right)^{ \log({8}/{\epsilon^{(K)}_b})+16\eta+72}\frac{176}{186}.
\end{equation}
Then, we want to find a $c$ such that 
\begin{align}
    c\left(\frac{3}{4}\right)^c \leq \tilde{\epsilon}^{(K)}_c.
    \label{eq:SU3_kin_c}
\end{align}
Assuming that $0<\tilde{\epsilon}^{(K)}_c<1$, the choice
\begin{equation}
    c=\log_{\frac{3}{4}}\left(\frac{\frac{\tilde{\epsilon}^{(K)}_c}{2.28}}{\log_{\frac{3}{4}}(\frac{\tilde{\epsilon}^{(K)}_c}{2.28})}\right)
    \label{eq:SU3_chosen_kin_c}
\end{equation}
satisfies (\ref{eq:SU3_kin_c}). We have verified numerically with Mathematica that for $0<\tilde{\epsilon}_{\rm oracle}<1$, $\tilde{\epsilon}_{\rm oracle}-c\left(\frac{3}{4}\right)^c$ never exceeds $0.0431937$, given our choice of $c$.

We mention in passing that the phase-inducing step can be parallelized. As in U(1), we first divide the kinetic terms up into bulk and edge terms. Then, for each direction of the bulk or edge terms, we implement $O(L^{d}-L^{d-1})$ or $O(L^{d-1})$ $R_z(2^{k-c} \theta)$ in parallel, using the weight-sum trick, and thus, exponentially reducing the number of $R_z$ gates required to $O(c d \log(L^d))$. In this work, we focus on the resource analysis of the serial implementation, and will leave that of the parallel implementation for future work.

{\bf Syntheses of the magnetic oracles:}
Similar to the implementation of the kinetic oracle, we directly synthesize the magnetic oracle, defined in (\ref{eq:SU3_mag_oracle}), as two controlled-diagonal gates, which impart the phases $\frac{-f_{\alpha \beta \gamma \delta}}{2a^{4-d}g^2}(-1)^{b_k}$ for $b_k=0,1$, if the control bits $\in \{\ket{b_j}\}$ are all ones. Recall that  $\{\ket{b_k},\ket{b_j}\}\subseteq S_0^\Box \equiv \{\ket{p_{i,0}'}$, $\ket{q_{i,0}'}$, $\ket{T_{L,i,0}'}$, $\ket{T_{L,i,0}^{z'}}$, $\ket{Y_{L,i,k}'}$, $\ket{T_{R,i,0}'}$, $\ket{T_{R,i,0}^{z'}}$, $\ket{Y_{R,i,k}'} \}_{i=1}^4$, where $i$ denotes the $i$th link on a plaquette, and $\ket{Y_{L/R,i,k}'}$ is the $k$th qubit of $\ket{Y_{L/R,i}'}$ as defined in (\ref{eq:SU3_Ybits}). The function $f_{\alpha \beta \gamma \delta}$ is defined in (\ref{eq:SU3_fabcd}) as a product of four functions, $f_{\alpha \beta}$, $f_{\beta \gamma}$, $f_{\gamma \delta}$, and $f_{\delta \alpha}$. Here, we consider the costliest $f_{\alpha \beta \gamma \delta}$, which is the product of four $f_{11}(0,-1,1,1)$ considered in the kinetic oracle. In this case, all registers are acted on, and thus, $\{\ket{b_k},\ket{b_j}\}= S_0^\Box$. We choose $b_k = p_{1,0}'$. Without loss of generality, we consider the case where $p_{1,0}'=0$. We begin by computing $f_{\alpha \beta \gamma \delta}$ into an ancilla register, conditioned upon the values of the control bits. Then, we induce the correct phase by applying $R_z$ gates to the ancilla state $\ket{f_{\alpha \beta \gamma \delta}}$. Finally, we uncompute $\ket{f_{\alpha \beta \gamma \delta}}$. The computation of $f_{\alpha \beta \gamma \delta}$ can be broken down into five steps. In the first and second steps, we compute the numerator and denominator, respectively. In the third step, we approximate the inverse of the denominator, using the circuits in \cite{bhaskar2016quantum}. In the fourth step, we approximate the argument of the square-root by multiplying together the numerator and the inverse of the denominator. Lastly, we approximate the square-root, using the circuits in \cite{bhaskar2016quantum}. Only the third and last steps incur approximation errors. Hereafter, we use the logarithmic depth out-of-place adder developed in \cite{draper2006logarithmic}, unless one of the inputs is classically known in which case we use the adder proposed in \cite{gidney2018halving}, and the multiplier proposed in \cite{shaw2020quantum}.

First, we consider the computation of the numerator. We begin by computing the numerators in the four functions that constitute $f_{\alpha \beta \gamma \delta}$. This costs four times the T gates and storage ancilla qubits required for computing the numerator in a kinetic oracle, i.e.,
\begin{align}
&5376\eta^2-\eta(192\lfloor\log(\eta+3)\rfloor+96\lfloor\log(\eta+5)\rfloor+96\lfloor\log(2\eta+10)\rfloor+144\lfloor\log(3\eta+15)\rfloor\nonumber\\
&+96\lfloor\log(6\eta+30)\rfloor)+40656\eta-96\lfloor\log(\eta+1)\rfloor-576\lfloor\log(\eta+2)\rfloor-864\lfloor\log(\eta+3)\rfloor \nonumber\\
&-288\lfloor\log(\eta+4)\rfloor-384\lfloor\log(\eta+5)\rfloor-384\lfloor\log(2\eta+10)\rfloor-672\lfloor\log(3\eta+15)\rfloor\nonumber\\
&-240\lfloor\log(6\eta+30)\rfloor+73340+96\eta + 224 \nonumber
\end{align}
and $41\eta+174$, respectively. We then multiply the four outputs with two $(8\eta+36)$-bit multipliers, and one $(16\eta+72)-$bit multiplier. The multipliers cost $18432\eta^2-192\eta(\lfloor\log(8\eta+36)\rfloor+\lfloor\log(16\eta+72)\rfloor)-840\lfloor\log(8\eta+36)\rfloor-852\lfloor\log(16\eta+72)\rfloor+162816\eta+359580$ T gates, $64\eta+288$ storage ancilla qubits, and $48\eta-\lfloor \log(16\eta+72)\rfloor+215$ workspace ancilla qubits. We proceed to compute the denominator. First, we compute the denominators in the four functions that constitute $f_{\alpha \beta \gamma \delta}$. This costs four times the T gates and storage ancilla qubits required for computing the denominator in a kinetic oracle, i.e., 
\begin{align}
    &720\eta^2-\eta(24\lfloor\log(\eta+2)\rfloor+12\lfloor\log(\eta+11)\rfloor+\lfloor\log(2\eta+3)\rfloor+24\lfloor\log(2\eta+22)\rfloor\nonumber\\
    &+12\lfloor\log(4\eta+26)\rfloor)+5312\eta-12\lfloor\log(\eta)\rfloor-12\lfloor\log(1+\eta)\rfloor-24\lfloor\log(\eta+4)\rfloor\nonumber\\
    &-24\lfloor\log(\eta+7)\rfloor-120\lfloor\log(\eta+11)\rfloor-12\lfloor\log(2\eta+2)\rfloor-12\lfloor\log(2\eta+3)\rfloor\nonumber\\
    &-36\lfloor\log(2\eta+22)\rfloor-48\lfloor\log(4\eta+26)\rfloor+8743, \nonumber
\end{align}
and $160\eta+628$, respectively. We then multiply the four outputs with two $(8\eta+31)$-bit multipliers, and one $(16\eta+62)-$bit multiplier. The multipliers cost $18432\eta^2-192\eta(\lfloor\log(8\eta+31)\rfloor+\lfloor\log(16\eta+62)\rfloor)-720\lfloor\log(8\eta+31)\rfloor-732\lfloor\log(16\eta+62)\rfloor+139776\eta+265020+96\eta-96$ T gates, $64\eta+248$ storage ancilla qubits, and $48\eta-\lfloor \log(16\eta+62)\rfloor+185$ workspace ancilla qubits.

We now proceed to compute the inverse of the denominator using the algorithm in \cite{bhaskar2016quantum}. We refer the readers to the SU(2) kinetic oracle implementation in Sec. \ref{subsubsec:oracle_SU2} for a detailed overview of the algorithm. Here, we simply state the algorithmic parameters and costs. The input $w$ is at most an $(32\eta+124)$-bit integer. Truncating at $b\geq 32\eta+124$ bits after the decimal point, the approximation error is bounded from above by $\frac{2+\log(b)}{2^{b}}$. The T-gate count is $\lceil\log(b)\rceil\cdot (48b^2+3072b\eta-(12b+384\eta-12)\lfloor\log(b)\rfloor+11752b-1024\eta -12\lfloor\log(2b+32\eta+124)\rfloor-1476\lfloor\log(2b)\rfloor-3864) + 128\eta + 507$. The required number of storage ancilla qubits is $\lceil\log(b)\rceil\cdot(7b+64\eta+250)$, and that of workspace ancilla qubits is $6b-\lfloor\log(2b)\rfloor-1$. Next, we multiply together the numerator, an $32\eta+144$-bit number, and the inverse of the denominator, a $b$-bit number to obtain the fraction. Assuming that $b \geq 32\eta+144$, this costs 
$1536 b\eta- 384\eta\lfloor\log(b)\rfloor+6868b-1664\eta-1716\lfloor\log(b)\rfloor-7436$ T gates, $b+32\eta+144$ storage ancilla qubits, and $3b-\lfloor\log(b)\rfloor-1$ workspace ancilla qubits. Lastly, we compute the square-root function. The output of the previous step is a $(b+32\eta+144)$-bit fraction, bounded above and below by 1 and 0, respectively. As in the kinetic oracle implementation, we shift it by $b+32\eta+144$ bits to obtain an integer input $w$, and shift it back by $(b+32\eta+144)/2$ bits once we obtain the root, assuming without loss of generality that $b$ is even. Let $c\geq 2b+64\eta+288$. Then, we approximate $\sqrt{w}$, up to $\left(\frac{3}{4}\right)^{c-2m}(2+c+\log(c))$ error. This costs $\lceil\log(c)\rceil\cdot[528c^2+96bc+3072c\eta-c(12\lfloor \log(c)\rfloor+12\lfloor \log(\lceil \frac{3c}{2}\rceil)\rfloor+12\lfloor \log(\lceil \frac{5c}{2}\rceil)\rfloor+18\lfloor \log(1+4c)\rfloor)-12b\lfloor \log(2c)\rfloor-384\eta\lfloor \log(2c)\rfloor+12\lfloor \log(c)\rfloor+12\lfloor \log(\lceil \frac{3c}{2}\rceil)\rfloor-1716\lfloor \log(2c)\rfloor+12\lfloor \log(\lceil \frac{5c}{2}\rceil)\rfloor+12\lfloor \log(1+4c)\rfloor-12\lfloor \log(2c+b+32n+144)\rfloor+13298c-32b-1024\eta-4444]+\lceil \frac{19c}{2}\rceil+4b+128\eta+595$ T gates, $\lceil\log(c)\rceil\cdot(23c+2b+64\eta+293)$ storage ancilla qubits, and $12c-\lfloor \log(4c+2)\rfloor+5$ workspace ancilla qubits.

We impart the phase by applying $R_z(2^{k-c}\theta)$, where $c$ is the number of digits after the decimal in the output of the square-root function, and $\theta = \frac{-1}{2a^{4-d}g^2}$, to the $k$th qubit of the ancilla state $\ket{f_{\alpha\beta\gamma\delta}}$. In order to implement the controlled version of this phase gate, we control each $R_z$ gate by 31 control bits, as required by the costliest magnetic oracle, and the four ancilla bits that check the realness of the Clebsch-Gordan coefficients and isoscalar factors. Each multi-controlled $R_z$ gate requires two Toffoli gates with 31 controls, which cost 239 T gates each \cite{maslov2016advantages}, two $R_z$ gates and one ancilla qubit~\cite{wang2020resource}. Since the state has at most $\lceil \frac{3c}{2}\rceil$ qubits and there are $L^d\frac{d(d-1)}{2}$ plaquettes on the lattice, there are $3c L^d\frac{d(d-1)}{2}$ multi-controlled $R_z$ gates to be applied.

We sum up the T-gate requirements for all the steps, and multiply the outcome by two to account for the uncomputation costs. Conjugating each controlled $R_z$ gate with a pair of CNOT gates, where $p_{1,0}$ is the control bit, addresses both $p_{1,0}=0,1$. As such, the T-gate count for one oracle call, i.e.,
\begin{align}
    \mathcal{T}^{(B)}&=2\cdot[42960\eta^2-\eta(192\lfloor\log(\eta+3)\rfloor+96\lfloor\log(\eta+5)\rfloor+96\lfloor\log(2\eta+10)\rfloor+144\lfloor\log(3\eta+15)\rfloor\nonumber\\
    &\quad +96\lfloor\log(6\eta+30)\rfloor+192\lfloor\log(8\eta+36)\rfloor+192\lfloor\log(16\eta+72)\rfloor+24\lfloor\log(\eta+2)\rfloor\nonumber\\
    &\quad +12\lfloor\log(\eta+11)\rfloor+\lfloor\log(2\eta+3)\rfloor+24\lfloor\log(2\eta+22)\rfloor+12\lfloor\log(4\eta+26)\rfloor\nonumber\\
    &\quad +192\lfloor\log(8\eta+31)\rfloor+192\lfloor\log(16\eta+62)\rfloor)+ 348560\eta-96\lfloor\log(\eta+1)\rfloor-576\lfloor\log(\eta+2)\rfloor\nonumber\\
    &\quad -864\lfloor\log(\eta+3)\rfloor-288\lfloor\log(\eta+4)\rfloor-384\lfloor\log(\eta+5)\rfloor-384\lfloor\log(2\eta+10)\rfloor\nonumber\\
    &\quad -672\lfloor\log(3\eta+15)\rfloor-240\lfloor\log(6\eta+30)\rfloor-840\lfloor\log(8\eta+36)\rfloor-852\lfloor\log(16\eta+72)\rfloor\nonumber\\
    &\quad -12\lfloor\log(\eta)\rfloor-12\lfloor\log(1+\eta)\rfloor-24\lfloor\log(\eta+4)\rfloor-24\lfloor\log(\eta+7)\rfloor-120\lfloor\log(\eta+11)\rfloor\nonumber\\
    &\quad -12\lfloor\log(2\eta+2)\rfloor-12\lfloor\log(2\eta+3)\rfloor-36\lfloor\log(2\eta+22)\rfloor-48\lfloor\log(4\eta+26)\rfloor\nonumber\\
    &\quad -720\lfloor\log(8\eta+31)\rfloor-732\lfloor\log(16\eta+62)\rfloor+706683]\nonumber\\
    &\quad+2\cdot\lceil\log(b)\rceil\cdot (48b^2+3072b\eta-(12b+384\eta-12)\lfloor\log(b)\rfloor+11752b-1024\eta \nonumber\\
    &\quad-12\lfloor\log(2b+32\eta+124)\rfloor-1476\lfloor\log(2b)\rfloor-3864) + 256\eta + 1014\nonumber\\
    &\quad+3072 b\eta- 768\eta\lfloor\log(b)\rfloor+13736 b-3328\eta-3432\lfloor\log(b)\rfloor-14872\nonumber\\
    &\quad +2\cdot\lceil\log(c)\rceil\cdot[528c^2+96bc+3072c\eta-c(12\lfloor \log(c)\rfloor+12\lfloor \log(\lceil \frac{3c}{2}\rceil)\rfloor+12\lfloor \log(\lceil \frac{5c}{2}\rceil)\rfloor\nonumber\\
    &\quad+18\lfloor \log(1+4c)\rfloor)-12b\lfloor \log(2c)\rfloor-384\eta\lfloor \log(2c)\rfloor+12\lfloor \log(c)\rfloor+12\lfloor \log(\lceil \frac{3c}{2}\rceil)\rfloor\nonumber\\
    &\quad -1716\lfloor \log(2c)\rfloor+12\lfloor \log(\lceil \frac{5c}{2}\rceil)\rfloor+12\lfloor \log(1+4c)\rfloor-12\lfloor \log(2c+b+32n+144)\rfloor\nonumber\\
    &\quad +13298c-32b-1024\eta-4444]+19c+8b+256\eta+694+372+384\eta+320.
    \label{eq:SU3_mag_oracle_T}
\end{align}
The number of storage ancilla qubits required is
\begin{equation}
    \lceil\log(c)\rceil\cdot(\lceil \frac{49c}{2}\rceil+2b+64\eta+293)+\lceil\log(b)\rceil\cdot(7b+64\eta+250)+b+361 \eta + 1482,
\end{equation}
and the number of workspace ancilla qubits required is
\begin{equation}
    12c-\lfloor \log(4c+1)\rfloor+5.
\end{equation}
There are two types of syntheses errors, i.e., arithmetic approximation errors and $R_z$ syntheses errors. The latter will be analyzed in Sec. \ref{subsubsec:Synth_SU2}. The arithmetic approximation errors per step is given by
\begin{equation}
    \frac{2+\log(b)}{2^b}+\left(\frac{3}{4}\right)^{c-(2b+64\eta+288)}(2+c+\log(c)).
\end{equation}
In total, the approximation errors are 
\begin{equation}
    \epsilon^{(B)}= r\cdot 1470021852266496\cdot \frac{d(d-1)}{2}L^d(\frac{2+\log(b)}{2^{b}}+\left(\frac{3}{4}\right)^{c-(2b+64\eta+288)}(2+c+\log(c))),
\end{equation}
where $r$ is the number of Trotter steps, and $1470021852266496\cdot \frac{d(d-1)}{2}L^d$ is the number of oracle calls. We divide the approximation errors evenly such that
\begin{equation}
    \epsilon^{(B)}_b=\frac{\epsilon^{(B)}}{r\cdot 1470021852266496 d(d-1)L^d},\:\epsilon^{(B)}_c=\frac{\epsilon^{(B)}}{r\cdot 1470021852266496 d(d-1)L^d},
\end{equation}
and
\begin{gather}
    \epsilon^{(B)}_b\geq \frac{2+\log(b)}{2^b}, \label{eq:SU3_eps_magb}\\
    \epsilon^{(B)}_c\geq \left(\frac{3}{4}\right)^{c-( 2b+64\eta+288)}(2+c+\log(c)).\label{eq:SU3_eps_magc}
\end{gather}
Note that this is not the optimal division of approximation errors. We let $b = \log(\frac{8}{\epsilon^{(B)}_b})$. Then, (\ref{eq:SU3_eps_magb}) is always satisfied for $0<\epsilon^{(K)}_b<1$. We proceed to compute the upper bound for $c$. Since $c\geq 2b+64\eta+288\geq 128\eta+576$, and $\eta\geq 1$, $\frac{716}{704}c \geq 2+c+\log(c)$. Inserting this relation and our choice of $b$ into (\ref{eq:SU3_eps_magc}), we obtain
\begin{align}
    \epsilon^{(B)}_c\geq \left(\frac{3}{4}\right)^{c-( \log({8}/{\epsilon^{(B)}_b})+64\eta+288)}\frac{716}{704}c.
\end{align}
Let
\begin{equation}
    \tilde{\epsilon}^{(B)}_c = \epsilon^{(B)}_c\cdot \left(\frac{3}{4}\right)^{ \log({8}/{\epsilon^{(B)}_b})+64\eta+288}\frac{704}{716}.
\end{equation}
Then, we want to find a $c$ such that 
\begin{align}
    c\left(\frac{3}{4}\right)^c \leq \tilde{\epsilon}^{(B)}_c.
    \label{eq:SU3_mag_c}
\end{align}
Assuming that $0<\tilde{\epsilon}^{(B)}_c<1$, the choice
\begin{equation}
    c=\log_{\frac{3}{4}}\left(\frac{\frac{\tilde{\epsilon}^{(B)}_c}{2.28}}{\log_{\frac{3}{4}}(\frac{\tilde{\epsilon}^{(B)}_c}{2.28})}\right)
    \label{eq:SU3_chosen_mag_c}
\end{equation}
satisfies (\ref{eq:SU3_mag_c}).

We mention in passing that the phase-inducing step can be parallelized. As in U(1), we can implement the magnetic terms acting on the odd and even plaquettes on a given two dimensional plane in parallel. In particular, we effect $O(L^d)$ same-angle $R_z$ gates, using the weight-sum trick, for each bit of the ancilla state $\ket{\pm f_{\alpha \beta \gamma \delta} }$. Thus, the parallel implementation exponentially reduces the $R_z$-gate count to $O(c d^2 \log(L^d) )$. In this work, we focus on the resource analysis of the serial implementation, and will leave that of the parallel implementation for future work.

We divide the oracle error $\epsilon_{oracle}$ evenly between the 
kinetic and magnetic oracles, i.e.,
\begin{equation}
    \epsilon^{(K)} = \frac{ \epsilon_{\rm oracle}}{2}, \:\epsilon^{(B)} = \frac{ \epsilon_{\rm oracle}}{2} \: \implies \: \epsilon_{\rm oracle}=\epsilon^{(K)}+\epsilon^{(B)}.
    \label{eq:SU3_oracle_err}
\end{equation}

\subsubsection{Synthesis errors}
\label{subsubsec:Synth_SU3}

Here, we compute the synthesis errors for $R_z$ gates required for the mass and electric term, separately from those for the oracles required for the kinetic and magnetic terms. To start, we consider the mass term. In this term, we have $\lfloor\log(3L^d)+1 \rfloor$ $R_z$ gates to implement. Therefore, we incur for each Trotter step $\lfloor\log(3L^d)+1 \rfloor \cdot \epsilon(R_z)$ amount of error, where $\epsilon(R_z)$ denotes the error per $R_z$ gate.

Next, we consider the electric term, which has $(2\eta + 4) \lfloor\log(dL^d)+1 \rfloor$ $R_z$ gates. Therefore, each Trotter step incurs $(2\eta + 4)\lfloor\log(dL^d)+1 \rfloor \cdot \epsilon(R_z)$ amount of error. If we instead use the phase gradient operation, once the gadget state $\ket{\psi_M}$ in (\ref{eq:phgradstate}) is prepared, each quantum adder call to implement the operation does not incur any synthesis error. We come back to the error incurred in preparing the gadget state itself in the next section.

The errors per Trotter step due to the $R_z$ gates for the kinetic and magnetic terms are $13872 dL^d\cdot \lceil \frac{3c^{(K)}}{2}\rceil\cdot \epsilon(R_z)$ and $1470021852266496\cdot \frac{d(d-1)}{2}L^d \cdot \lceil \frac{3c^{(B)}}{2}\rceil \cdot \epsilon(R_z)$, where we have and will continue to denote the approximation parameter $c$ for the kinetic and magnetic oracles as  $c^{(K)}$ and $c^{(B)}$, respectively.

We add the error incurred for the four terms to obtain the synthesis error $\epsilon_{\rm synthesis}$. Note that, as in the U(1) case, the implementation of the diagonal mass and electric terms can be optimized. As in the U(1) case, there are $r+1$ diagonal mass and electric terms, and $2r$ off-diagonal kinetic and magnetic terms to implement in total. Thus, $\epsilon_{\rm synthesis}$ is given by
\begin{align}
    \epsilon_{\rm synthesis} &= \{(r+1) \cdot [\lfloor\log(3L^d)+1 \rfloor + (2\eta + 4)\lfloor\log(dL^d)+1 \rfloor] + 2r\cdot[ 20808 c^{(K)}dL^d \nonumber\\
    &+ 1102516389199872 c^{(B)}d(d-1)L^d ]\}\cdot \epsilon(R_z),
    \label{eq:SU3_synth_err}
\end{align}
where $r$, to reiterate for the convenience of the readers, is the total number of Trotter steps.

\subsubsection{Complexity analysis}
\label{subsubsec:Analysis_SU3}

Having computed the Trotter, oracle and synthesis errors, we proceed to perform the complexity analysis for the SU(3) LGT.

The total error is given by
\begin{equation}
    \epsilon_{\rm total}=\epsilon_{\rm Trotter}+\epsilon_{\rm oracle}+\epsilon_{\rm synthesis}.
\end{equation}
We choose to evenly distribute the total error between the Trotter, oracle, and synthesis errors. Focusing on the Trotter error, we obtain the number of Trotter steps by
\begin{equation}
  \epsilon_{\rm Trotter} = \frac{\epsilon_{\rm total}}{3} \:\implies\: r = \lceil \frac{T^{3/2}3^{1/2}\rho^{1/2}}{\epsilon_{\rm total}^{1/2}} \rceil.
\end{equation}

Next, we use the above relation and $\epsilon_{\rm oracle} = \frac{\epsilon_{\rm total}}{3}$ to obtain the expressions for $c^{(K)}$ and $c^{(B)}$. $c^{(K)}$ is given by (\ref{eq:SU3_chosen_kin_c}), where
\begin{align}
    \tilde{\epsilon}_c^{(K)} = \frac{\epsilon_{\rm total}}{12r\cdot 13872 dL^d}\left(\frac{3}{4}\right)^{ 2 \log(8\frac{12r\cdot 13872 dL^d}{\epsilon_{\rm total}}) +16\eta+72}\frac{176}{186}.
\end{align}
$c^{(B)}$ is given by (\ref{eq:SU3_chosen_mag_c}), where
\begin{align}
    \tilde{\epsilon}_c^{(B)} = \frac{\epsilon_{\rm total}}{6r\cdot 1470021852266496 d(d-1)L^d}\left(\frac{3}{4}\right)^{ 2 \log(8\frac{6r\cdot 1470021852266496 d(d-1)L^d}{\epsilon_{\rm total}}) +64\eta+288}\frac{704}{716}.
\end{align}

Finally, we compute the error each $R_z$ gate can incur by
\begin{align}
    &\epsilon_{\rm synthesis} = \frac{\epsilon_{\rm total}}{3} \implies \nonumber \\
    &\epsilon(R_z) = \frac{\epsilon_{\rm total}}{3} \{(\lceil \frac{T^{3/2}3^{1/2}\rho^{1/2}}{\epsilon_{\rm total}^{1/2}} \rceil+1) \cdot [\lfloor\log(3L^d)+1 \rfloor + (2\eta + 4)\lfloor\log(dL^d)+1 \rfloor] + 2\lceil \frac{T^{3/2}3^{1/2}\rho^{1/2}}{\epsilon_{\rm total}^{1/2}} \rceil\cdot[ 20808 c^{(K)}dL^d \nonumber\\
    &+ 1102516389199872 c^{(B)}d(d-1)L^d ]\}^{-1}.
\end{align}
With this, we obtain the number of T gates required to synthesize each $R_z$ gate using RUS circuit \cite{bocharov2015efficient},
\begin{equation}
    \text{Cost}(R_z)=1.15\log(\frac{1}{\epsilon(R_z)}).
\end{equation}
Combining the T gates required for implementation of $R_z$ gates and the T gates used elsewhere in the circuit, we obtain the total number of T gates for the entire circuit as
\begin{align}
    &\quad (\lceil \frac{T^{3/2}3^{1/2}\rho^{1/2}}{\epsilon_{\rm total}^{1/2}} \rceil+1)\cdot [\lfloor\log(3L^d)+1 \rfloor + (2\eta + 4)\lfloor\log(dL^d)+1 \rfloor] + 2\lceil \frac{T^{3/2}3^{1/2}\rho^{1/2}}{\epsilon_{\rm total}^{1/2}} \rceil\cdot [20808 c^{(K)}dL^d \nonumber\\
    &+ 1102516389199872 c^{(B)}d(d-1)L^d ]\cdot \text{Cost}(R_z)+(\lceil \frac{T^{3/2}3^{1/2}\rho^{1/2}}{\epsilon_{\rm total}^{1/2}} \rceil+1)\cdot [4(3L^d-\text{Weight}(3L^d))+8dL^d[(8\eta - 8)\nonumber\\
    &+(5\eta - 3\lfloor\log(\eta)\rfloor-4)+(\eta - 1)(12\eta - 3\lfloor\log(\eta)\rfloor - 12)+1+\eta(12\eta-3\lfloor\log(\eta+2)\rfloor+10)\nonumber\\
    &+1+(10\eta-3\lfloor\log(2\eta+3)\rfloor+11)]+(8\eta+16)(dL^d-\text{Weight}(dL^d))+3328(\eta+1)dL^d]\nonumber\\
    &+2\lceil \frac{T^{3/2}3^{1/2}\rho^{1/2}}{\epsilon_{\rm total}^{1/2}} \rceil\cdot [13872dL^d \mathcal{T}^{(K)}+83955286016(\eta+1)L^d d(d-1)+735010926133248d(d-1)L^d\mathcal{T}^{(B)} ],
\end{align}
where $\mathcal{T}^{(K)}, \mathcal{T}^{(B)}$ are given in (\ref{eq:SU3_kin_oracle_T},\ref{eq:SU3_mag_oracle_T}), respectively. 

The size of the ancilla register is given by the maximum between the ancilla qubits required by the electric and magnetic Hamiltonian, i.e.,
\begin{multline}
    Q_{\max}=\max \{(8\eta+10)dL^d+3(\eta+2)-\lfloor\log(\eta+2)\rfloor+dL^d-\text{Weight}(dL^d),\\ \lceil\log(c)\rceil\cdot(\lceil \frac{49c}{2}\rceil+2b+64\eta+293)+\lceil\log(b)\rceil\cdot(7b+64\eta+250)+b+361\eta+12c-\lfloor \log(4c+1)\rfloor+1487 \}.
\end{multline}
Taking this into account, we obtain the total number of qubits required for the simulation by summing up those in the ancilla, fermionic and gauge-field registers, which is given by
\begin{equation}
    3L^d+(8\eta+12)dL^d+Q_{\max}.
\end{equation}

Note that in the case where the electric term is implemented using phase gradient operation, the T-gate count changes by 
\begin{align}
    &\quad (\lceil \frac{T^{3/2}3^{1/2}\rho^{1/2}}{\epsilon_{total}^{1/2}} \rceil+1)\cdot [4dL^d\log(\frac{6\pi a^{d-2}}{g^2 t})+O(dL^d) -(2\eta+4)\lfloor\log(dL^d)+1 \rfloor\cdot \text{Cost}(R_z)\nonumber \\
    &-4(2\eta+4)(dL^d - \text{Weight}(dL^d))]+\text{Cost}(\ket{\psi_M}),
\end{align} 
where the Cost$(R_z)$ needs to be modified, since $\epsilon(R_z)$ has changed to
\begin{align}
    &\epsilon(R_z) = \frac{\epsilon_{\rm total}}{3} \{(\lceil \frac{T^{3/2}3^{1/2}\rho^{1/2}}{\epsilon_{\rm total}^{1/2}} \rceil+1) \cdot \lfloor\log(3L^d)+1 \rfloor + 2\lceil \frac{T^{3/2}3^{1/2}\rho^{1/2}}{\epsilon_{\rm total}^{1/2}} \rceil\cdot[ 20808 c^{(K)}dL^d \nonumber\\
    &+ 1102516389199872 c^{(B)}d(d-1)L^d ]\}^{-1}.
\end{align}
Further, $\text{Cost}(\ket{\psi_M})$, which denotes the one-time synthesis costs of the phase gradient gadget state. Here, we choose to use the synthesis method delineated in \cite{nam2020approximate}. Briefly, we apply Hadamard gates to the register $\ket{00...0}$, and then apply gates $Z, Z^{-1/2},...,Z^{-1/2^{M-1}}$. Each $Z^\alpha$ gates are synthesized using RUS circuits \cite{bocharov2015efficient}. Let $\delta$ be the error of preparing the gadget state $\ket{\psi_M}$. Then, each gate can incur at most $M/\delta$ error, and thus, costs $1.15\log(M/\delta)$, using RUS circuits \cite{bocharov2015efficient}. Thus, the gadget state preparation costs $1.15M\log(M/\delta)$.

Finally, in this case, the ancilla-qubit count is given by that of the maximum between the mass term and the phase gradient state, and the magnetic term, i.e.,
\begin{multline}
    Q_{\max}=\max \{ (3L^d - \text{Weight}(3L^d)) + \log(\frac{6\pi a^{d-2}}{g^2 t}), \lceil\log(c)\rceil\cdot(\lceil \frac{49c}{2}\rceil+2b+64\eta+293)\\
    +\lceil\log(b)\rceil\cdot(7b+64\eta+250)+b+361\eta+12c-\lfloor \log(4c+1)\rfloor+1487 \}.
\end{multline}
As such, the total qubit count is given by
\begin{equation}
    3L^d+dL^d(8\eta+12)+Q_{\max}.
\end{equation}

\section{Improvements over previous work}
\label{sec:yamamoto}

In this section, we demonstrate the algorithmic improvements achieved in this work over previous works, \cite{byrnes2006simulating} and \cite{shaw2020quantum}. In particular, we use the asymptotic scaling, with respect to $L$, $\Lambda$, and $\epsilon$, assuming $d$, $a$, $g$, and $m$ are fixed, of T gates required for a single Trotter step as the comparison metric.

\subsection{U(1) case:}
\label{sec:U1compare}

Here we compare our algorithm with those proposed in \cite{byrnes2006simulating} and \cite{shaw2020quantum}. We first provide a brief overview of the simulation methods in \cite{byrnes2006simulating}. Similar to our method, as described in \ref{sec:SimCircSynth}, the authors applied a truncation $\Lambda$ to the electric eigenbasis $\ket{E}$ on each link such that $E\in \{-\Lambda, -\Lambda+1,...,\Lambda \}$, and the gauge-field operators $\hat{E}$, $\hat{U}$, and $\hat{U}^\dag$ become finite-dimensional. Then, they mapped $\ket{E}$ to qubits using an unary encoding. In particular, an integer $-\Lambda \leq j \leq \Lambda$ is represented on an unary $(2\Lambda + 1)-$qubit register as the state where the $j$th qubit is $\ket{0}$ and the remaining qubits are all $\ket{1}$. As such, the gauge-field operators are represented as follows:
\begin{gather}
    \hat{E} = \sum_{l = -\Lambda}^{\Lambda} l \frac{(\hat{Z}_{l}+\hat{I}_{l})}{2},\\
    \hat{U} = \sum_{l = -\Lambda}^{\Lambda-1} \hat{\sigma}_l^+ \hat{\sigma}_{l+1}^-,
\end{gather}
where the subscript $l$ denotes the qubit index for each link register $\ket{E}$. 

Using the above relations, the evolution of each link due to the electric Hamiltonian $e^{it\frac{g^2}{2a^{d-2}}\hat{E}^2}$ can be implemented, up to a global phase, as
\begin{equation}
    \prod_{l = -\Lambda}^{\Lambda} e^{it\frac{g^2}{2a^{d-2}} \frac{l^2}{2}\hat{Z}_{l} },
\end{equation}
which requires $O(\Lambda)$ $R_z$ gates. Therefore, for a $d$-dimensional cubic lattice with $L^d$ sites, the evolution due to the electric Hamiltonian costs $O(dL^d \Lambda)$ $R_z$ gates in total. Hereafter, we suppose an error budget $\epsilon$ is allocated for synthesizing all the required $R_z$ gates, and we use RUS circuits \cite{bocharov2015efficient} to synthesize them. Then, $O(dL^d \Lambda \log(dL^d \Lambda/\epsilon))$ T gates are needed. In comparison, our algorithm (see Sec. \ref{sec:U1_electric} for details) requires $O(\log(dL^d)\log(\Lambda))$ $R_z$ gates and $O(dL^d (\log(\Lambda))^2)$ T gates. This amounts to $O(\log(dL^d)\log(\Lambda) \log(\log(dL^d)\log(\Lambda)/\epsilon)+dL^d (\log(\Lambda))^2)$ total T gates. As a result, our algorithm reduces the $\Lambda$-dependence from linear to quadratic logarithmic.

Next, we discuss the evolution of each plaquette due to the magnetic Hamiltonian $e^{-i\frac{t}{2a^{4-d}g^2}(\hat{U} \hat{U} \hat{U}^{\dag} \hat{U}^{\dag} + h.c.)}$, which can be Trotterized to first order as,
\begin{equation}
    \prod_{j,k,l,m=-\Lambda}^{\Lambda-1} e^{-i\frac{t}{2a^{4-d}g^2}(\hat{\sigma}_j^+ \hat{\sigma}_{j+1}^- \hat{\sigma}_k^+ \hat{\sigma}_{k+1}^- \hat{\sigma}_l^- \hat{\sigma}_{l+1}^+ \hat{\sigma}_m^- \hat{\sigma}_{m+1}^+ + h.c.)}.
\end{equation}
This can be implemented as $O(\Lambda^4)$ unitary operations of the form $e^{-i\theta(\otimes_k \hat{\sigma}_k + h.c.)}$, each of which, as shown in Sec. \ref{sec:U1_mag}, requires $O(1)$ $R_z$ gate. The magnetic term for a $d$-dimensional lattice, where there are $O(d^2 L^d)$ plaquettes, costs $O(d^2 L^d \Lambda^4 \log(d^2 L^d \Lambda^4/\epsilon))$ T gates in total. In comparison, our algorithm (see Sec. \ref{sec:U1_mag} for details) requires $O(d^2\log(L^d))$ $R_z$ gates and $O(d^2\log(\Lambda)L^d)$ T gates, which amount to $O(d^2[\log(L^d)\log(d^2\log(L^d)/\epsilon)+\log(\Lambda)L^d])$ total T gates. As such, our algorithm scales exponentially better in $\Lambda$.

In addition to the improvements in T gate count, our algorithm also scales more favorably in terms of qubit count. In particular, their method requires $O(dL^d \Lambda)$ qubits, whereas our algorithm requires $O(dL^d \log(\Lambda))$ qubits.

Next, we apply our algorithm to a one-dimensional lattice, and compare it to that in \cite{shaw2020quantum}. Both algorithms encode the gauge-field operators to qubits using binary encoding, and use Jordan-Wigner transformation to map fermions to qubits. The main difference is the parallelization techniques, i.e. , the weight-sum trick, used in our algorithm, which reduce the number of $R_z$ gates incurred. Given the similarity between our algorithms, we simply report the T gate counts for a Trotter step due to each term of the Hamiltonian. Their algorithm requires $O(L\log(L/\epsilon))$,  $O(L [\log(\Lambda) \log(L \log(\Lambda)/\epsilon)+(\log(\Lambda))^2])$, and $O(L[\log(L/\epsilon)+\log(\Lambda)])$ T gates for the mass, electric, and kinetic terms, respectively. Our algorithm requires $O(L+\log(L)\log(\log(L)/\epsilon))$, $O(\log(L)\log(\Lambda) \log(\log(L)\log(\Lambda)/\epsilon)+ L (\log(\Lambda))^2)$, and $O(\log(L)\log(\log(L)/\epsilon)+L \log(\Lambda))$ T gates for the mass, electric, and kinetic terms, respectively.
\subsection{SU(2) case:}

Here we provide a comparison between our algorithm and that in \cite{byrnes2006simulating}. Briefly, we describe the method in \cite{byrnes2006simulating}. Similar to our algorithm, as described in Sec. \ref{sec:SimCircSynth_SU2}, a truncation is applied to the basis $\ket{j,m^L, m^R}$, such that, for a given link, $j\in \{ 0,\frac{1}{2}, ..., \Lambda \}$ and $m^L, m^R \in \{ -\Lambda, -\Lambda+\frac{1}{2}, ..., \Lambda \}$. As in the U(1) case, they map $\ket{j}$, $\ket{m^L}$, and $\ket{m^R}$ to qubit using an unary encoding. In particular, $k \in \{ -\Lambda, -\Lambda+\frac{1}{2}, ..., \Lambda \}$ is represented on an unary $(4\Lambda + 1)$-qubit register as the state where the $2k$th qubit is $\ket{0}$ and the remaining qubits are all $\ket{1}$. $k \in \{ 0,\frac{1}{2}, ..., \Lambda \}$ is represented similarly on a $(2\Lambda + 1)$-qubit register. 

Using this unary encoding, the operator $\hat{E}^2$, which satisfies the relation in (\ref{eq:SU2_eig1}), i.e., $\hat{E}^2 \ket{j,m^L, m^R} = j(j+1) \ket{j,m^L, m^R}$, can be represented as
\begin{equation}
    \hat{E}^2 = \sum_{l=0}^{2\Lambda} l(l+1) \frac{(\hat{Z}_l + \hat{I}_l)}{2},
\end{equation}
where the subscript $l$ denotes the qubit index in the register $\ket{j}$ for a given link. Then, the evolution of each link due to the electric Hamiltonian $e^{it\frac{g^2}{2a^{d-2}}\hat{E}^2}$ can be implemented, up to a global phase, with $O(\Lambda)$ $R_z$ gates. Therefore, for a $d$-dimensional cubic lattice with $L^d$ sites, the evolution due to the electric Hamiltonian costs $O(dL^d \Lambda)$ $R_z$ gates in total, which, using RUS circuits, translate to $O(dL^d \Lambda \log(dL^d \Lambda/\epsilon))$ total T gates. In comparison, our algorithm (see Sec.\ref{sec:SimCircSynth_SU2} for details) requires $O(\log(dL^d)\log(\Lambda))$ $R_z$ gates and $O(dL^d (\log(\Lambda))^2)$ T gates, which amounts to $O(\log(dL^d)\log(\Lambda) \log(\log(dL^d)\log(\Lambda)/\epsilon)+dL^d (\log(\Lambda))^2)$ T gates in total. As in the U(1) case, our algorithm reduces the $\Lambda$-dependence from linear to quadratic logarithmic.

We now proceed to discuss the magnetic term for each plaquette, i.e., $e^{-i\frac{t}{2a^{4-d}g^2}(\hat{U}_{\alpha \beta} \hat{U}_{\beta \gamma} \hat{U}^{\dag}_{\gamma \delta} \hat{U}^{\dag}_{\delta \alpha} + h.c.)}$. Briefly, the $\hat{U}_{\alpha \beta}$ operators in (\ref{eq:SU2_U_enc}) are defined in terms of raising and lowering operators on $\ket{j}$, and $\ket{m^{L/R}}$, which, in the unary encoding, are given by
\begin{gather}
    \hat{J}^+ = \sum_{l=0}^{2\Lambda-1} \hat{\sigma}_{l}^{+} \hat{\sigma}_{l+1}^{-}, \: \hat{J}^- = (\hat{J}^+)^\dag \\
    \hat{M}^{L/R}_{1} = \sum_{l=-2\Lambda}^{2\Lambda-1} \hat{\sigma}_{l}^{+} \hat{\sigma}_{l+1}^{-}, \: \hat{M}^{L/R}_{2} = (\hat{M}^{L/R}_{1})^{\dag}
\end{gather}
respectively, where, $l$ denotes the qubit index, and diagonal operators, $\hat{N}_\alpha$ in (\ref{eq:SU2_norm_op}) and $\hat{c}_{\alpha \beta}^{L/R}$ in (\ref{eq:SU2_CG_op}) that encode the normalization factors and SU(2) Clebsch-Gordan coefficients, respectively. Then, the $\hat{U}_{\alpha \beta}$ operators can be expressed as
\begin{align}
    \hat{U}_{\alpha \beta} &= \hat{M}^{L}_{\alpha}\hat{M}^{R}_{\beta}[\hat{J}^+  \hat{c}^{L}_{1 \alpha}  \hat{c}^{R}_{1 \beta} \hat{N}_{1} + \hat{J}^-  \hat{c}^{L}_{2 \alpha}  \hat{c}^{R}_{2 \beta} \hat{N}_{2}] \nonumber \\
    &= \sum_{i, k=-2\Lambda}^{2\Lambda-1} \sum_{l=0}^{2\Lambda-1} \hat{\sigma}_{i}^{s} \hat{\sigma}_{i+1}^{s} \hat{\sigma}_{k}^{q} \hat{\sigma}_{k+1}^{q}[ \hat{\sigma}_{l}^{+} \hat{\sigma}_{l+1}^{-}\hat{c}^{L}_{1 \alpha}  \hat{c}^{R}_{1 \beta} \hat{N}_{1} + \hat{\sigma}_{l}^{-} \hat{\sigma}_{l+1}^{+}   \hat{c}^{L}_{2 \alpha}  \hat{c}^{R}_{2 \beta} \hat{N}_{2} ] \nonumber \\
    &= \sum_{i, k=-2\Lambda}^{2\Lambda-1} \sum_{l=0}^{2\Lambda-1} \hat{\sigma}_{i}^{s} \hat{\sigma}_{i+1}^{s} \hat{\sigma}_{k}^{q} \hat{\sigma}_{k+1}^{q}[{c}^{L}_{1 \alpha}(\frac{l}{2},\frac{i}{2}) {c}^{R}_{1 \beta}(\frac{l}{2},\frac{k}{2})\sqrt{\frac{l+1}{l+2}}  \hat{\sigma}_{l}^{+} \hat{\sigma}_{l+1}^{-}  + {c}^{L}_{2 \alpha}(\frac{l}{2},\frac{i}{2})  {c}^{R}_{2 \beta}(\frac{l}{2},\frac{k}{2}) \sqrt{\frac{l+1}{l}} \hat{\sigma}_{l}^{-} \hat{\sigma}_{l+1}^{+}],
\end{align} 
where $i,k,l$ are qubit indices for registers $\ket{m^L}$, $\ket{m^R}$, $\ket{j}$, respectively, and $s,q \in \{ +,- \}$ depend on $\alpha, \beta$, respectively. Moreover, in the third equality, we are able deduce the unique state each string of Pauli operators, i.e., $\bigotimes_{r}\hat{\sigma}^r$ with $r\in \{ +,- \}$, acts on, and thus, the Clebsch-Gordan coefficient and normalization factor induced by the diagonal operators. Using this expression, there are $O(\Lambda^{12})$ $\bigotimes_{r}\hat{\sigma}^r + h.c.$ operators in each $\hat{U}_{\alpha \beta} \hat{U}_{\beta \gamma} \hat{U}^{\dag}_{\gamma \delta} \hat{U}^{\dag}_{\delta \alpha} + h.c.$ operator. Since each $e^{it(\otimes_{r}\hat{\sigma}^r + h.c.)}$ costs $O(1)$ $R_z$ gate, each evolution operator $e^{-i\frac{t}{2a^{4-d}g^2}(\hat{U}_{\alpha \beta} \hat{U}_{\beta \gamma} \hat{U}^{\dag}_{\gamma \delta} \hat{U}^{\dag}_{\delta \alpha} + h.c.)}$ costs $O(\Lambda^{12})$ $R_z$ gates. Therefore, the magnetic term for the entire $d$-dimensional lattice costs $O(d^2 L^d \Lambda^{12} \log(d^2 L^d \Lambda^{12}/\epsilon))$ T gates in total. In comparison, our algorithm requires $O(d^2 L^d \log(\Lambda))$ T gates, $O(cd^2 L^d)$ $R_z$ gates, and $O(d^2 L^d)$ queries to arithmetic oracles, each of which costs $O(c^2\log(c)+(\log(\Lambda))^2)$ T gates. From Sec. \ref{subsubsec:oracle_SU2}, we obtain that $c = O(\log(\frac{d^2 L^d \Lambda}{\epsilon}) + \log(\log(\frac{d^2 L^d \Lambda}{\epsilon})))$. Hence, omitting $O(\log\log(\epsilon^{-1}))$ and $O(\log\log(\Lambda))$ factors, our algorithm has a quadratically worse $\epsilon^{-1}$-dependence, i.e., $(\log(\epsilon^{-1}))^2$ versus $\log(\epsilon^{-1})$, but a superpolynomial improvement in the $\Lambda$-dependence, i.e, $(\log(\Lambda))^2$ versus $\Lambda^{12}$.

\subsection{SU(3) case:}
Once again, we provide a comparison between our algorithm and that in \cite{byrnes2006simulating}. Briefly, we describe the method in \cite{byrnes2006simulating}. Similar to our algorithm, as described in Sec. \ref{sec:SimCircSynth_SU3}, a truncation is applied to the basis $\ket{p,q,T_L,T_L^z,Y_L,T_R,T_R^z,Y_R}$, such that, for a given link, $p, q \in \{0,1,...,\Lambda \}$, $T_i \in \{0,\frac{1}{2},...,\Lambda \}$, $T_i^z \in \{-\Lambda, -\Lambda+\frac{1}{2},...,\Lambda \}$, $Y_i = \{-\Lambda, -\Lambda + \frac{1}{3},...,\Lambda  \}$, where $i=L,R$. As in the U(1) and SU(2) cases, they map $\ket{p}$, $\ket{q}$, $\ket{T_i}$, $\ket{T_i^z}$, and $\ket{Y_i}$ to qubit using an unary encoding. In particular, $k \in \{ 0,1, ..., \Lambda \}$ is represented on an unary $(\Lambda + 1)$-qubit register as the state where the $k$th qubit is $\ket{0}$ and the remaining qubits are all $\ket{1}$. Similarly, $k \in \{ -\Lambda, -\Lambda+\frac{1}{2}, ..., \Lambda \}$ is represented on an unary $(4\Lambda + 1)$-qubit register as the state where the $2k$th qubit is $\ket{0}$ and the remaining qubits are all $\ket{1}$, while $k \in \{ 0,\frac{1}{2}, ..., \Lambda \}$ is represented similarly on a $(2\Lambda + 1)$-qubit register. Finally, $k \in \{ -\Lambda, -\Lambda+\frac{1}{3}, ..., \Lambda \}$ is represented on an unary $(6\Lambda + 1)$-qubit register as the state where the $3k$th qubit is $\ket{0}$ and the remaining qubits are all $\ket{1}$.

Using this unary encoding, the operator $\hat{E}^2$, which satisfies the relation in \ref{eq:SU3_elec}, i.e., $\hat{E}^2 \ket{p,q} = \frac{1}{3}[p^2+q^2+pq+3(p+q)]\ket{p,q}$, can be represented as
\begin{equation}
    \hat{E}^2 = \sum_{l,m=0}^{\Lambda} \frac{1}{3}[l^2+m^2+lm+3(l+m)]\frac{(\hat{Z}_l + \hat{I}_l)}{2} \frac{(\hat{Z}_m + \hat{I}_m)}{2},
\end{equation}
where the subscripts $l,m$ denote the qubit indices in the registers $\ket{p}$, $\ket{q}$, respectively, for a given link. Then, the evolution of each link due to the electric Hamiltonian $e^{it \frac{g^2}{2a^{d-2}}\hat{E}^2}$ can be implemented, up to a global phase, with $O(\Lambda^2)$ $R_z$ gates. Therefore, for a $d$-dimensional cubic lattice with $L^d$ sites, the electric term costs $O(dL^d \Lambda^2)$ $R_z$ gates in total, which using RUS circuits, translate to $O(dL^d \Lambda^2 \log(dL^d \Lambda^2/\epsilon))$ total T gates. In comparison, our algorithm (see Sec.\ref{sec:SimCircSynth_SU3} for details) requires $O(\log(dL^d)\log(\Lambda))$ $R_z$ gates and $O(dL^d (\log(\Lambda))^2)$ T gates, which amounts to $O(\log(dL^d)\log(\Lambda) \log(\log(dL^d)\log(\Lambda)/\epsilon)+dL^d (\log(\Lambda))^2)$ T gates in total. Our algorithm reduces the $\Lambda$-dependence from quadratic to quadratic logarithmic.

Finally, we discuss the magnetic term for each plaquette, i.e., $e^{-i\frac{t}{2a^{4-d}g^2}(\hat{U}_{\alpha \beta} \hat{U}_{\beta \gamma} \hat{U}^{\dag}_{\gamma \delta} \hat{U}^{\dag}_{\delta \alpha} + h.c.)}$. Briefly, the $\hat{U}_{\alpha \beta}$ operators in (\ref{eq:SU3_U_enc}) are defined in terms of raising and lowering operators on $\ket{p}$, $\ket{q}$, $\ket{T_i}$, $\ket{T_i^z}$, and $\ket{Y_i}$, where $i=L,R$, and diagonal operators that encode the normalization factors and SU(3) Clebsch-Gordan coefficients. As in the U(1) and SU(2) case, in the unary encoding, each raising and lowering operator is a sum $O(\Lambda)$ $\hat{\sigma}^+ \hat{\sigma}^-$ operators. Similar to the SU(2) case, the $\hat{U}_{\alpha \beta}$ operators consist of $O(\Lambda^8)$ strings of Pauli operators, i.e., $\bigotimes_{r}\hat{\sigma}^r$ with $r\in \{ +,- \}$. Hence, there are $O(\Lambda^{32})$ $\bigotimes_{r}\hat{\sigma}^r + h.c.$ in each $\hat{U}_{\alpha \beta} \hat{U}_{\beta \gamma} \hat{U}^{\dag}_{\gamma \delta} \hat{U}^{\dag}_{\delta \alpha} + h.c.$ operator. Since each $e^{it(\otimes_{r}\hat{\sigma}^r + h.c.)}$ costs $O(1)$ $R_z$ gate, each evolution operator $e^{-i\frac{t}{2a^{4-d}g^2}(\hat{U}_{\alpha \beta} \hat{U}_{\beta \gamma} \hat{U}^{\dag}_{\gamma \delta} \hat{U}^{\dag}_{\delta \alpha} + h.c.)}$ costs $O(\Lambda^{32})$ $R_z$ gates. Therefore, the magnetic term for the entire $d$-dimensional lattice costs $O(d^2 L^d \Lambda^{32} \log(d^2 L^d \Lambda^{32}/\epsilon))$ T gates in total. Our algorithm has the same scaling as in the SU(2) case. As such, it requires $O(d^2 L^d \log(\Lambda))$ T gates, $O(cd^2 L^d)$ $R_z$ gates, and $O(d^2 L^d)$ queries to arithmetic oracles, each of which costs $O(c^2\log(c)+(\log(\Lambda))^2)$ T gates, where $c = O(\log(\frac{d^2 L^d \Lambda}{\epsilon}) + \log(\log(\frac{d^2 L^d \Lambda}{\epsilon})))$. Hence, omitting $O(\log\log(\epsilon^{-1}))$ and $O(\log\log(\Lambda))$ factors, our algorithm has a quadratically worse $\epsilon^{-1}$-dependence, i.e., $(\log(\epsilon^{-1}))^2$ versus $\log(\epsilon^{-1})$, but a superpolynomial improvement in the $\Lambda$-dependence, i.e, $(\log(\Lambda))^2$ versus $\Lambda^{32}$.
\end{document}